%% file: review.tex
\documentclass{ws-ijmpa}

\usepackage{url,eucal,dsfont,boxedminipage,setspace}
\usepackage{graphicx,wrapfig,subfigure}
\usepackage{amsxtra,amstext,amssymb,amsmath,mathrsfs,latexsym,esint}

\usepackage{color,bm,hhline}
\usepackage{array,tabularx,multirow,afterpage}
\usepackage{footmisc} % provides \footref and \mpfootnotemark commands
\usepackage{xspace}
\usepackage{epsfig}

\def\gsim{\ \rlap{\raise 3pt \hbox{$>$}}{\lower 3pt \hbox{$\sim$}}\ }
 \def\lsim{\ \rlap{\raise 3pt \hbox{$<$}}{\lower 3pt \hbox{$\sim$}}\ }
\newcommand{\be}{\begin{equation}}
\newcommand{\ee}{\end{equation}}
\newcommand{\bea}{\begin{eqnarray}}
\newcommand{\eea}{\end{eqnarray}}
\newcommand{\eqn}[1]{eq.~(\ref{#1})}
\newcommand{\Eqn}[1]{Eq.~(\ref{#1})}

\newcommand{\Eqns}[2]{Eqs.~(\ref{#1})-(\ref{#2})}
 \newcommand{\fh}{f_{H_u}^{eq}}
 \newcommand{\fht}{f_{\widetilde{H}_u}^{eq}}
\newcommand{\fLeq}{f_{\ell}^{eq}} 
\newcommand{\fLteq}{f_{\widetilde{\ell}}^{eq}}

\newcommand{\muh}{\mu}
\newcommand{\submuh}{\scriptstyle \mu_{\scriptscriptstyle\tilde H}}

\graphicspath{{figures/}}

\begin{document}

\markboth{Chee Sheng Fong, M.C. Gonzalez-Garcia and Enrico Nardi} {Soft Leptogenesis}

\title{
\vspace{-3.0cm} 
\begin{flushright}
\textnormal{IFT-UAM/CSIC-11-56\\
            FTUAM-11-53\\ 
            YITP-SB-11-27\\ 
            ICCUB-11-161
} 
\end{flushright} 
\vspace{2.5cm}
LEPTOGENESIS FROM SOFT SUPERSYMMETRY BREAKING  \\ [5pt]
(Soft Leptogenesis)}

\author{CHEE SHENG FONG} 
\address{C.N. Yang Institute for Theoretical Physics\\
  State University of New York at Stony Brook\\
  Stony Brook, NY 11794-3840, USA.\\
  E-mail:fong@insti.physics.sunysb.edu}
\author{M.~C.~GONZALEZ-GARCIA}
\address {
C.N. Yang Institute for Theoretical Physics\\
  State University of New York at Stony Brook\\
  Stony Brook, NY 11794-3840, USA.\\
{\rm and:}  \\ 
Instituci\'o Catalana de Recerca i Estudis Avan\c{c}ats (ICREA), \\
  Departament d'Estructura i Constituents de la Mat\`eria and ICC-UB, \\
  Universitat de Barcelona, 
  Diagonal 647, E-08028 Barcelona, Spain.\\
  E-mail:concha@insti.physics.sunysb.edu
}
\author{ENRICO NARDI}
\address{
INFN, Laboratori Nazionali di Frascati,\\
  C.P. 13, 100044 Frascati, Italy.\\
{\rm and:}  \\ 
Departamento de F\'{\i}sica Te\'orica,  
   C-XI, Facultad de Ciencias,  \\    
                Universidad Aut\'onoma de Madrid,       
                C.U. Cantoblanco, 28049 Madrid, Spain \\ 
{\rm and:}  \\ 
 Instituto de F\'{\i}sica Te\'orica UAM/CSIC,  \\ 
Nicolas Cabrera 15, C.U. Cantoblanco, 28049 Madrid, Spain  \\
  E-mail:enrico.nardi@lnf.infn.it}

\maketitle
\begin{abstract}
  Soft leptogenesis is a scenario in which the cosmic baryon asymmetry
  is produced from a lepton asymmetry generated in the decays of heavy
  sneutrinos (the partners of the singlet neutrinos of the seesaw) and
  where the relevant sources of CP violation are the complex phases of
  soft supersymmetry-breaking terms.  We explain the motivations for
  soft leptogenesis, and review its basic ingredients: the different
  CP-violating contributions, the crucial role played by thermal
  corrections, and the enhancement of the efficiency from lepton
  flavour effects.  We also discuss the high temperature regime $T >
  10^7\,$GeV in which the cosmic baryon asymmetry originates from
  an initial asymmetry of an anomalous $R$-charge, and soft
  leptogenesis reembodies in $R$-genesis.
\end{abstract}
\ccode{PACS numbers: 13.30.Fs, 14.60.St, 12.60.Jv, 14.80.Ly}

\include{introduction}

\setcounter{footnote}{0}
 \include{slq}

\setcounter{footnote}{0}
 \include{flavour}

\setcounter{footnote}{0}
 \include{nonseq}

\setcounter{footnote}{0}
\include{conclusions}

\section*{Acknowledgements}
This work is supported by USA-NSF grant PHY-0653342, 
by consolider-ingenio 2010 program CSD-2008-0037, by CUR Generalitat de
 Catalunya grant 2009SGR502 and by MICINN grant FPA2010-20807. 
CSF would like to thank Juan Racker for delightful collaboration
and discussion in leptogenesis.
He is also grateful to CNYITP, Stony Brook for the generous support.
EN would like to thank Diego Aristizabal, Enrico Bertuzzo, Pasquale Di
 Bari, Sacha Davidson, Guy Engelhard, Ferruccio Feruglio, Yuval
 Grossman, Marta Losada, Luis Alfredo Mu\~noz, Yosef Nir, Jorge
 Nore\~na, Marco Peloso, Juan Racker and Esteban Roulet for fruitful
 collaborations in leptogenesis, and for the pleasure of working with
 them.

\appendix
\include{boltzmanneqs}

\include{se}

%%%%%%%%%%%%%%%%%%%%%%%%%%%%%%%%%%%%%%%%%%%%%%%%%%%
% \bibliographystyle{JHEP} %plain,unsrt,ieeetr,abbrv,JHEP
 \bibliographystyle{elsarticle-num} % ws citation style
 \bibliography{references}
%%%%%%%%%%%%%%%%%%%%%%%%%%%%%%%%%%%%%%%%%%%%%%%%%%%

\end{document}

%% file: introduction.tex
\section{The Baryon Asymmetry of the Universe}
\label{sec:intro}

%%%%%%%%%%%%%%%%%%%%%%%%%%%%
\subsection{Observations}
\label{sec:observations}

Up to date no traces of cosmological antimatter have been observed.
The presence of a small amount of antiprotons and positrons in cosmic
rays can be consistently explained by their secondary origin in
energetic cosmic particles collisions or in highly energetic
astrophysical processes, but no antinuclei, even as light as
anti-deuterium or as tightly bounded as anti-$\alpha$ particles, has ever
been detected.

The absence of annihilation radiation $p\bar p\to \dots \pi^0\to \dots
2\gamma$ excludes significant matter-antimatter admixtures in objects
up to the size of galactic clusters\cite{Steigman:1976ev} 
$\sim 20\,$Mpc, while observational limits on anomalous contributions to the
cosmic diffuse $\gamma$-ray background and the absence of distortions
in the cosmic microwave background allows to conclude that little
antimatter is to be found within $\sim 1\,$Gpc, and that within our
horizon an equal amount of matter and antimatter is empirically
excluded.\cite{Cohen:1997ac}  Of course, at larger super-horizon
scales the vanishing of the average asymmetry cannot be excluded, and
this would indeed be the case if the fundamental Lagrangian is $C$ and
$CP$ symmetric and charge invariance is broken
spontaneously.\cite{Dolgov:1991fr}

%%%%%%%%%%%%%%%%%

Quantitatively, the value of baryon asymmetry of the Universe is
inferred from observations in two independent ways.  The first way is
by confronting the abundances of the light elements, $D$, $^3{\rm
  He}$, $^4{\rm He}$, and $^7${\rm Li}, with the predictions of Big
Bang nucleosynthesis
(BBN).\cite{Iocco:2008va,Steigman:2007xt,Nakamura:2010zzi,%
  Steigman:2005uz,Cyburt:2004yc,Olive:1999ij} The crucial time for
primordial nucleosynthesis is when the thermal bath temperature falls
below $T\lsim 1\,$MeV.  With the assumption of only three light
neutrinos, these predictions depend on essentially a single parameter,
that is the difference between the number of baryons and anti-baryons
normalized to the number of photons:
\be
\eta\equiv \frac{n_B- n_{\bar{B}}}{n_{\gamma }} {\Big |}_{0}, 
% = (6.21 \pm 0.16)  \times 10^{-10}, 
\label{etaB}
\ee
where the subscript $0$ means ``at present time''.  By using only
the abundance of deuterium, that is particularly sensitive to $\eta$,
Ref.~\refcite{Iocco:2008va} quotes:
\begin{equation}
  \label{eq:2H}
10^{10}\,\eta = 5.7\pm 0.6\quad \qquad\qquad(95\%\> {\rm c.l.}) \,.
\end{equation}
In this same range there is also an acceptable agreement among the
various abundances, once theoretical uncertainties as well as
statistical and systematic errors are accounted for.\cite{Nakamura:2010zzi}

The second way is from measurements of the cosmic microwave background
(CMB) anisotropies (for pedagogical reviews, see
Refs.~\refcite{Hu:2001bc,Dodelson:2003ft}).  The crucial time for CMB
is that of recombination, when the temperature dropped low enough
that, in spite of the extremely large entropy, protons and electrons
could form neutral hydrogen, which happened at $T\lsim 1\,$eV.  CMB
observations measure the relative baryon contribution to the energy
density of the Universe multiplied by the square of the (reduced)
Hubble constant $h\equiv H_0/(100\,{\rm km}\>{\rm sec}^{-1}\,{\rm
  Mpc}^{-1})$:
\begin{equation}
\label{eq:Omegab}
\Omega_Bh^2\equiv h^2\frac{\rho_B}{\rho_{\rm crit}}\,,
\end{equation}
that is related to $\eta$ through $10^{10}\eta=274\; \Omega_B\ h^2$.
The physical effect of the baryons at the onset of matter domination,
which occurs quite close to the recombination epoch, is to provide
extra gravity which enhances the compression into potential wells. The
consequence is enhancement of the compressional phases which
translates into enhancement of the odd peaks in the spectrum. Thus, a
measurement of the odd/even peak disparity constrains the baryon
energy density. A fit to the most recent observations (WMAP7 data
only, assuming a $\Lambda$CDM model with a scale-free power spectrum
for the primordial density fluctuations) gives at 68\%
c.l.\cite{Larson:2010gs}
\begin{equation}
\label{eq:omecmb}
10^2\, \Omega_B h^2 = 2.258^{+0.057}_{-0.056}\,.
\end{equation}

There is a third way to express the baryon asymmetry of the Universe,
that is by normalizing the baryon asymmetry to the entropy density $s
=g_*(2\pi^2/45)T^3$, where $g_*$ is the number of degrees of freedom
in the plasma, and $T$ is the temperature:
\be
Y_{\Delta B} \equiv  \frac{n_B- n_{\bar{B}}}{s} {\Big |}_{0}\,. 
% = ( 8.75 \pm 0.23)  \times 10^{-11}
\label{eq:YB}
\ee
The relation with the previous definitions is given by the
conversion factor $s_0/n_{\gamma0}= 7.04$.  $Y_{\Delta B}$ is a
convenient quantity in theoretical studies of the generation of the
baryon asymmetry from very early times, because it is conserved
throughout the thermal evolution of the Universe.

In terms of $Y_{\Delta B}$ the BBN results \eqref{eq:2H} and the CMB
measurement \eqref{eq:omecmb} (at $95\% c.l.$)  read:
\begin{equation}
\label{eq:YB_CMB}
Y_{\Delta B}^{BBN} =  (8.10 \pm 0.85) \times 10^{-11},
\qquad 
Y_{\Delta B}^{CMB}=(8.79 \pm 0.44) \times 10^{-11}. 
\end{equation}
The impressive consistency between the determinations of the baryon
density of the Universe from BBN and CMB that, besides being
completely independent, also refer to epochs with a six orders of
magnitude difference in temperature, provides a striking confirmation
of the hot Big Bang cosmology.

%%%%%%%%%%%%%%%%%%%%%%%%%%%%
\subsection{Theory}
\label{sec:theory}

From the theoretical point of view, the question is where the Universe
baryon asymmetry comes from.  Could it simply be the result of a fine
tuned initial condition, one that would require just one quark in
excess over 6,000,000 antiquarks and an exactly conserved baryon (or
more appropriately $B-L$) number?  The inflationary cosmological model
excludes this possibility, and since we do not know any other way to
construct a consistent cosmology without inflation, this veto is a
very strong one. The argument goes as follows: the inflationary stage,
that is the epoch in which the volume of the Universe undergoes
exponential expansion, can only be successful if it lasts at least 65
Hubble times $H_I t_I\gsim 65$. During this epoch the energy density
of relativistic baryons would drop exponentially as $\exp(-4 H_It)$.
However, exponential expansion requires that the total energy density
is (approximately) constant.  From Eq.~\eqref{eq:YB_CMB} we see that
just about seven Hubble times backward from the end of inflation
$\rho_B$ would become ${\cal O}(1)$ and dominate the (non-constant)
Universe energy density, and this would destroy inflation.  This
simple argument implies that baryon number cannot be conserved, which
opens the way to the possibility of generating the Universe baryon
asymmetry dynamically, a scenario that is known as {\em baryogenesis}.
In fact, as Sakharov pointed out,\cite{Sakharov:1967dj} the
ingredients required for baryogenesis are three:

\begin{enumerate}
\item  Baryon number violation:
  This condition is required in order to evolve from an initial state
  with $Y_{\Delta B}=0$ to a state with $Y_{\Delta  B}\neq0$.
\item C and CP violation:
  If either C or CP were conserved, then processes involving baryons
  would proceed at precisely the same rate as the C- or CP-conjugate
  processes involving antibaryons, with the overall effects that no
  baryon asymmetry is generated.
\item Out of equilibrium dynamics:
Equilibrium distribution functions are determined 
solely by the particle energy $E$ and chemical potential $\mu$
\begin{equation}
  \label{eq:neq}
  n_{\rm eq}=\left(e^{(E-\mu)/T}\pm 1\right)^{-1}, 
\end{equation}
and when charges (such as $B$) are not conserved, the corresponding
chemical potentials vanish. On the other hand, because of the CPT
theorem masses of particles and antiparticles are the same, and thus
their equilibrium distributions must also be the same, which yields:
\begin{equation}
  \label{eq:nB}
n_B=n_{\bar B} = \int \frac{d^3p}{(2\pi^3)} n_{\rm eq}. 
\end{equation}
\end{enumerate}
Although these ingredients are all present in the Standard Model (SM),
so far all attempts to reproduce quantitatively the observed baryon
asymmetry have failed.
\begin{enumerate}
\item Baryon number is violated in the SM, and baryon number violating
  processes (sphalerons) are fast in the early
  Universe.\cite{Kuzmin:1985mm} $B$ violation is due to the triangle
  anomaly, and leads to processes that involve nine left-handed quarks
  (three from each generation) and three left-handed leptons (one from
  each generation). Sphaleron processes cannot mediate proton decay
because of the selection rule
\begin{equation}
\Delta B=\Delta L = \pm3.
\end{equation}
At zero temperature, the amplitude of the baryon number violating
processes is proportional to\cite{'tHooft:1976up} 
$e^{-8 \pi^2/g^2}$,  
which is too small to have any observable effect. At high
temperatures, however, these transitions become
unsuppressed,\cite{Kuzmin:1985mm} the first condition is then 
quantitatively realized, and would not impede successful baryogenesis.
\item The weak interactions of the SM violate C maximally while CP is
  violated by the Kobayashi-Maskawa complex phase of the Yukawa
  couplings.\cite{Kobayashi:1973fv} CP violation in the SM can be
  parametrized by the Jarlskog invariant\cite{Jarlskog:1985ht} which
  is of order $10^{-20}$. Since there are practically no kinematic
  enhancement factors in the thermal 
bath,\cite{Gavela:1994ds,Gavela:1994dt,Huet:1994jb} it is then
  impossible to generate $Y_{\Delta B}\sim10^{-10}$.
\item Departures from thermal equilibrium occur in the SM at the
  electroweak phase transition.\cite{Rubakov:1996vz,Trodden:1998ym}
  Here, the non-equilibrium condition is provided by the interactions
  of particles with the bubble wall, as it sweeps through the
  plasma. The experimental lower bound on the Higgs mass implies,
  however, that this transition is not strongly first order, as
  required for successful baryogenesis.\cite{Kajantie:1995kf}
\end{enumerate}
This shows that baryogenesis requires new physics that extends the SM
in at least two ways: It must introduce new sources of CP violation
and it must either provide a departure from thermal equilibrium in
addition to the electroweak phase transition (EWPT) or modify the EWPT
itself.  Some possible new physics mechanisms for baryogenesis are the
following:

{\bf GUT  baryogenesis}\cite{Ignatiev:1978uf,Yoshimura:1978ex,%
Toussaint:1978br,Dimopoulos:1978kv,Ellis:1978xg,Weinberg:1979bt,%
Yoshimura:1979gy,Barr:1979ye,Nanopoulos:1979gx,Yildiz:1979gx}
generates the baryon asymmetry in the out-of-equilibrium decays of
heavy bosons in Grand Unified Theories (GUTs).  The GUT baryogenesis
scenario has difficulties with the non-observation of proton decay,
which puts a lower bound on the mass of the decaying boson, and
therefore on the reheat temperature after inflation.  Simple inflation
models do not give such a high reheat temperature.  Furthermore, in
the simplest GUTs, $B+L$ is violated but $B-L$ is not. Consequently,
the $B+L$ violating SM sphalerons, which are in equilibrium at $T
\lsim 10^{12}\,$GeV, would destroy this asymmetry.

{\bf Electroweak baryogenesis}\cite{Rubakov:1996vz,Riotto:1999yt,%
  Cline:2006ts} is a scenario in which the departure from thermal
equilibrium is provided by the EWPT.  Models for electroweak
baryogenesis need a modification of the SM scalar potential such that
the EWPT  becomes first order, as well as new sources of CP
violation. One example\cite{Losada:1996ju} is the 2HDM (two Higgs
doublet model), where the Higgs potential has more parameters and,
unlike the SM potential, violates CP. Another well known example is
the Minimal Supersymmetric SM (MSSM), where a light stop modifies the
Higgs potential in the required
way\cite{Carena:1996wj,Delepine:1996vn} and where there are new,
flavour-diagonal, CP-violating phases. Electroweak baryogenesis and,
in particular, MSSM baryogenesis, might soon be subject to
experimental tests at the CERN Large Hadron Collider (LHC).

{\bf Affleck-Dine mechanism}.\cite{Affleck:1984fy,Dine:1995kz}
The asymmetry arises in a classical scalar field, which later decays
to particles. The field starts with a large expectation value, and
rolls towards the origin. At the initial configuration displaced from
the origin there can be contributions to the potential from baryon or
lepton number violating interactions, that impart a net asymmetry to
the rolling field.  Generically, this mechanism could produce an
asymmetry in any combination of $B$ and $L$.

{\bf Spontaneous Baryogenesis}.\cite{Cohen:1987vi,Cohen:1988kt} In this
scenario, baryon number is an approximate symmetry spontaneously
broken at some large scale.  A baryon asymmetry can develop while
baryon violating interactions are still in thermal equilibrium by
using the effective breaking of CPT invariance caused by the Universe
expansion, which breaks time-invariance.  Furthermore, both the ground
state and fundamental interactions in these theories can be CP
conserving: the Universe as a whole is CP symmetric, but a period of
exponential expansion blew domains of antimatter well outside our
horizon. No sacred principles are violated, and although at first
sight the mechanism could seem quite exotic, it is in fact rather
natural.

{\bf Leptogenesis}. This scenario was first proposed by Fukugita and
Yanagida in Ref.~\refcite{Fukugita:1986hr}, and in its simplest and
theoretically best motivated realization is intrinsically related to
the seesaw mechanism for neutrino masses.\cite{Minkowski:1977sc,%
  Yanagida:1979as,Glashow,GellMann:1980vs,Mohapatra:1980yp} To implement 
the seesaw, new Majorana $SU(2)_L$ singlet neutrinos with a large mass scale
$M$ are added to the SM particle spectrum.  The complex Yukawa
couplings of these new particles provide new sources of CP
violation, departure from thermal equilibrium can occur if their
lifetime is not much shorter than the age of the Universe when $T\sim
M$, and their Majorana masses imply that lepton number is not
conserved. A lepton asymmetry can then be generated dynamically, and
SM sphalerons will partially convert it  into a
baryon asymmetry.\cite{Khlebnikov:1988sr}
A popular and well studied possibility is ``thermal leptogenesis''
where the heavy Majorana neutrinos are produced by scatterings in the
thermal bath starting from a vanishing initial abundance, so that
their number density can be calculated solely in terms of the seesaw
parameters and of the reheat temperature of the Universe.

\subsection{Prerequisites}
\label{sec:prereq}

This review focuses on a particular realization of thermal
leptogenesis, that was first proposed in
Refs.~\refcite{Grossman:2003,DAmbrosio:2003}, in which the lepton
asymmetry is generated in the decays of heavy sneutrinos (the
supersymmetric partners of the Majorana neutrinos of the seesaw) and
where the relevant sources of CP violation are the complex phases of
soft supersymmetry-breaking terms.\footnote{The idea of utilizing soft
  supersymmetry-breaking terms to realize low scale leptogenesis was
  first put forth in Ref.~\refcite{Boubekeur:2002jn}.}  It is then
clear that for reading this review some acquaintance with standard
leptogenesis as well as with its supersymmetric version is necessary.
Thermal leptogenesis has been studied in detail by many people, and
many general papers and pedagogical reviews are available.  Early
studies that mainly focused on hierarchical singlet neutrinos include
Refs.~\refcite{Luty:1992un,Gherghetta:1993kn,Plumacher:1996kc,Plumacher:1997ru}.
The importance of including the wave function renormalization of the
decaying singlet neutrinos in calculating the CP asymmetry was
recognized in Ref.~\refcite{Covi:1996wh}.  Various reviews were
written at this stage, and a pedagogical presentation that introduces
the Boltzmann equations for thermal leptogenesis can be found in
Ref.~\refcite{Buchmuller:2000as}.  A partial set of thermal
corrections to leptogenesis processes were first given in
Ref.~\refcite{Covi:1997dr}, while more complete and detailed
calculations can be found in Refs.~\refcite{Giudice:2003jh}.

All these studies did not include flavour effects that were first
discussed in Refs.~\refcite{Barbieri:2000,Endoh:2004}, but whose
importance was fully recognized only later in
Refs.~\refcite{Abada:2006a,Nardi:2006b,Abada:2006b}.  They can play an
even more important role in soft leptogenesis (see Section~\ref{sec:flavor}) than in
standard leptogenesis.  A pedagogical introduction to flavour effects
can be found in the review Ref.~\refcite{Davidson:2008} together with
all technical details.  Short but self-contained resumes are also
given in TASI lectures\cite{Chen:2007fv} as well as in conference
proceedings.\cite{Davidson:2007xu,Nardi:2007fs,Nardi:2007cf} Finally,
a comprehensive study of supersymmetric leptogenesis in which the
effects of non-superequilibration (see Section~\ref{sec:nse}) have
been included for the first time can be found in
Ref.~\refcite{Fong:2010qh}.

%%%%%%%%%%%%

\subsection{Reading this review}
\label{sec:reading}

This review is organized as follows: in Section~\ref{sec:SL} the basis
of soft leptogenesis (SL) are reviewed and the main results are
recapped.  The relevant Lagrangian for SL is introduced in
Section~\ref{sec:lag}.  The CP asymmetries are derived in
Section~\ref{sec:cp_asymmetries} by using two different approaches.
In Section~\ref{sec:QFT} a field theoretical approach is followed,
while in Section~\ref{sec:QM} the same quantities are evaluated with a
quantum mechanical approach.  Beyond this section only the results of
the field theoretical approach are used, and thus the reader can skip
the details of the quantum mechanical approach, without affecting the
understanding of the rest of the review.  In SL thermal effects are
needed to prevent a vanishing total CP asymmetry. This is a
fundamental issue and is reviewed in detail in
Section~\ref{sec:vertexCP}.

Section~\ref{sec:unflavored} begins with a general discussion
(Section~\ref{sec:effective}) of how the appropriate effective theory
to study dynamical processes in the early Universe can be formulated.
Its aim is to render clear the different steps taken in studying SL
with increasing degree of precision.  The first step is discussed in
Section~\ref{sec:unflavoredBE} where the dynamics of SL in the 
so-called `one flavour approximation' is addressed, and an initial set of
Boltzmann equations is derived, in which flavour as well as other
important effects are left out.  This Section is crucial to understand
the dynamics of SL and to follow the qualitative discussion presented
in Section~\ref{sec:qualitative}, although the quantitative results,
that are given in Section~\ref{sec:quantitative}, can give at best a
rough estimate of the baryon asymmetry yield of SL.

The resonant enhancement of the CP asymmetries from 
self-energy contributions is an important ingredient of SL, 
and for this type of contributions quantum corrections 
to the dynamical equations can be important.  
This issue is reviewed in Section~\ref{sec:quantum_role}
that, however, being a bit technical can be skipped at a first reading. 

The inclusion of lepton flavour effects in SL studies is mandatory,
because SL always occurs in the flavoured regime.  The role of lepton
flavours is reviewed in Section~\ref{sec:flavor}.  The flavoured CP
asymmetries are introduced in Section~\ref{sec:flavored_CP}, and two
flavour structures representative of different soft supersymmetry
breaking patterns are discussed in Section~\ref{sec:fla_scenarios}.
Lepton flavour violation from soft breaking slepton masses is part of
the phenomenology of supersymmetry, and if the related processes are
sufficiently rapid all flavour effects would be efficiently
damped. This issue is discussed in Section~\ref{sec:lfe}, and it is
addressed again in relation with low energy data in
Section~\ref{sec:num_lfe}.  The network of flavoured Boltzmann
equations, including also Higgs and other spectator effects, is
presented in Section~\ref{sec:flavored_BE}, and in
Section~\ref{sec:flavored_results} the numerical results obtained with
these equations are discussed.  Finally, the impact that flavour
enhancements of the final baryon asymmetry can have on the soft
supersymmetry-breaking parameter space is discussed in
Section~\ref{sec:natural_B}.

In the high temperature regime ($T\gsim 10^7\,$GeV) SL, as described
in the previous sections, is no more the appropriate theory. Important
modifications take place, that are related with the fact that reactions
that depend on the soft gaugino masses and on the higgsino mixing
parameter $\mu$ become irrelevant, and a new effective theory, that
has been named $R$-genesis,\cite{Fong:2010bv} should be considered
instead. This is the topic of Section~\ref{sec:nse}. Various details
of the construction of $R$-genesis are discussed in
Sections~\ref{sec:symmetries} and \ref{sec:chem}, and the
corresponding Boltzmann equations are given in
Section~\ref{sec:nse_BE}. A simplified scenario that illustrates
rather clearly what is new in $R$-genesis with respect to SL is
presented in Section~\ref{sec:simple}, and numerical
results are discussed in Section~\ref{sec:nse_results}.

%!!%
The prospect of (not) being able to experimentally verify the standard 
SL scenario is briefly discussed in Section \ref{sec:verifications}. 
The variations of SL with their possible experimental signatures
are reviewed in Section \ref{sec:variations}.

The main topics discussed in the review are resumed in the conclusions in 
Section~\ref{sec:conclusions}, while the more technical details are
collected in two appendices.
% \ref{app:boltzmanneqs_new} and \ref{app:se_new}

%%%%%%%%%%%%%%%%%%%%%%%%%%%%%%%%%%%%%%

%%% Local Variables: 
%%% mode: latex
%%% TeX-master: t
%%% End: 

%% file: slq.tex
\section{Soft Leptogenesis: the Basic Ingredients}
\label{sec:SL}

The basic ingredients for generating a lepton asymmetry in 
SL are the CP asymmetries induced in the decays of the
right-handed sneutrinos (RHSN) by the complex phases of the soft
supersymmetry(SUSY)-breaking terms.  Starting from the relevant soft
leptogenesis Lagrangian, in this section we compute the CP asymmetries
following first a field theoretical approach, and then a quantum
mechanical approach. In spite of minor differences between the results
obtained with the two approaches, it is found that in both cases to an
excellent approximation the total CP asymmetries for decays into
scalars and into fermions vanish in the zero temperature limit. In
fact, a general proof for the vanishing of the one-loop CP asymmetries
in decays can be given without resorting to explicit computations, and
will be presented in  Section~\ref{sec:vertexCP}.

\subsection{Lagrangian for soft leptogenesis}
\label{sec:lag}
The superpotential for the supersymmetric seesaw model is:
\begin{eqnarray}
W &=& W_{\rm MSSM}+Y_{i\alpha}
\epsilon_{ab} {\hat N}^c_i {\hat \ell}_\alpha^a {\hat H_u^b} 
+ \frac{1}{2}M_{i}{\hat N^c}_i{\hat N^c}_i,
\label{eq:superpotential}
\end{eqnarray}
where $a,b=0,1$ are the $SU(2)_L$ indices with
$\epsilon_{01}=-\epsilon_{10}=1$, $\alpha=e,\mu,\tau$ is the lepton
flavour index and $i=1,2,...$ labels the generations of right-handed
neutrinos (RHN) chiral superfields defined according to usual
convention in terms of their left-handed Weyl spinor components
(${\hat N_i^c}$ contains scalar component $\widetilde N_i \equiv
\widetilde \nu_{R_i}^*$ and fermion component
$\left(\nu_{R_i}\right)^c\,$).  $\,{\hat \ell}_\alpha =
\left(\hat{\nu}_{L_\alpha},\hat{\alpha}_L^-\right)^T$, $\,{\hat H_u} =
\left(\hat{H}_u^+,\hat{H}_u^0\right)^T$ are the left chiral
superfields of the lepton and up-type Higgs $SU(2)_L$ doublets
respectively.  Without loss of generality, one can work in the basis
where the Majorana mass matrix $M$ is diagonal.  Notice that due to
the Majorana mass term, one cannot consistently assign lepton number
to $\hat{N}_i\,$ such that the superpotential
\eqref{eq:superpotential} remains invariant under global
$U(1)_{L_\alpha}$.  In other words, both $L\,$ and $L_\alpha\,$ are
broken by the superpotential \eqref{eq:superpotential}.

Starting from Eq.~\eqref{eq:superpotential}, the interaction
Lagrangian density involving $N_i \equiv \nu_{R_i} +
\left(\nu_{R_i}\right)^c$ and $\widetilde{N}_i$ can be written as
follows: 
\begin{eqnarray}
\!\!\!-\mathcal{L}_{\rm int} = 
Y_{i\alpha}\epsilon_{ab}\left(M_{i}^{*}
\widetilde{N}_{i}^{*}\widetilde{\ell}_{\alpha}^{a}H_{u}^{b}
+\overline{\widetilde{H}_u^{c,b}}
P_{L}\ell_{\alpha}^{a}\widetilde{N}_{i} \right.
%\nonumber \\
%&  & 
\left.+\overline{\widetilde{H}_u^{c,b}}
P_{L}N_{i}\widetilde{\ell}_{\alpha}^{a}
+\overline{N}_{i}P_{L}\ell_{\alpha}^{a}H_{u}^{b}\right)\!+{\rm h.c.},
\; \; \; \;
\label{eq:int_terms}
\end{eqnarray}
where $P_{L,R}=\frac{1}{2}\left(1\mp\gamma_{5}\right)$ are
respectively the left and right chiral projectors.
In Eq.~\eqref{eq:int_terms} the $SU(2)_L$ doublets are 
$\widetilde{\ell}_{\alpha}=\left(\widetilde{\nu}_{L_\alpha},
\widetilde{\alpha}_L^{-}\right)^T$,
$H_{u}=\left(H_{u}^{+},H_{u}^{0}\right)^T$, 
and
$\widetilde{H}_u^{c}=\left(\widetilde{H}_u^{+,c},
\widetilde{H}_u^{0,c}\right)^T$. Notice that
since $\widetilde{H}_u^{+} = \widetilde{H}_{u,L}^+$ 
is the left-handed positively charged Weyl higgsino, 
$\widetilde{H}_u^{+,c} = \widetilde{H}_{u,R}^{-}$ 
is the right-handed negatively charged Weyl higgsino.

The relevant soft SUSY-breaking terms involving
$\widetilde{N}_{i}$, the $SU(2)_L$ gauginos
$\widetilde{\lambda}_{2}^{\pm,0}$, the $U(1)_Y$ gauginos
$\widetilde{\lambda}_{1}$ and the three sleptons
$\widetilde{\ell}_\alpha$ in the basis in which the charged lepton
Yukawa couplings are diagonal, are given by
\begin{eqnarray}
\! -\mathcal{L}_{\rm soft} 
& = & \widetilde{M}_{ij}^{2}\widetilde{N}_{i}^{*}\widetilde{N}_{j}
+\left(A Y_{i\alpha}\epsilon_{ab}\widetilde{N}_{i}
\widetilde{\ell}_{\alpha}^{a}H_{u}^{b}+\frac{1}{2}BM_{i}
\widetilde{N}_{i}\widetilde{N}_{i}+{\rm h.c.}\right)
\nonumber \\
&  & +\frac{1}{2}\left(m_{2}\overline{\widetilde{\lambda}_{2}^{\pm,0}}
P_{L}\widetilde{\lambda}_{2}^{\pm,0}
%+m_{2}^{*}\overline{\widetilde{\lambda}_{2}^{a}}P_{R}\widetilde{\lambda}_{2}^{a}
+m_{1}\overline{\widetilde{\lambda}_{1}}P_{L}\widetilde{\lambda}_{1}
%+m_{1}^{*}\overline{\widetilde{\lambda}_{1}}P_{R}\widetilde{\lambda}_{1}
+{\rm h.c.}\right), 
%+m_{\widetilde{\ell}_{\alpha\beta}}^2
%\widetilde{\ell}^{(m)*}_\alpha \widetilde{\ell}^{(m)}_\beta,
\label{eq:soft_terms}
\end{eqnarray}
where for simplicity proportionality of the bilinear and trilinear
soft breaking terms to the corresponding SUSY invariant
couplings has been assumed: $B_i=BM_i$ and $A_{i\alpha}=AY_{i\alpha}$.
In Section~\ref{sec:flavor} this assumption will be dropped in favour
of a more general flavour structure for the trilinear couplings
$A_{i\alpha}=AZ_{i\alpha}$ and, as we will see, this can result in
important qualitative and quantitative differences.

Even if the off-diagonal terms in the soft breaking mass matrix 
$\widetilde M_{ij}$ are assumed to be 
negligible  $\widetilde M_{i\neq j} \ll \widetilde M_{ii}$,  
% for $i \neq j$, 
the presence of the $B$ term implies that the RHSN and anti-RHSN 
states mix in the mass matrix with mass eigenstates 
\begin{eqnarray}
\widetilde{N}_{+i}  &=&  
\frac{1}{\sqrt{2}}\left(e^{i\Phi_i/2}\widetilde{N}_{i}
+e^{-i\Phi_i/2}\widetilde{N}_{i}^{*}\right), \\ 
\widetilde{N}_{-i}  &=& 
-\frac{i}{\sqrt{2}}(e^{i\Phi_i/2}\widetilde{N}_{i}
-e^{-i\Phi_i/2}\widetilde{N}_{i}^{*}),
\label{eq:mass_eigenstates}
\end{eqnarray}
where $\Phi_i\equiv\arg\left(B M_{i}\right)$. 
The corresponding mass eigenvalues are
\begin{eqnarray}
M_{i\pm}^{2}  =  M_{i}^{2}+\widetilde{M}_{ii}^{2}
\pm\left|B M_{i}\right|.
\label{eq:mass_eigenvalues}
\end{eqnarray}

In the following we will set, without loss of generality, $\Phi_i=0$,
which is equivalent to assigning the phases only to $A$ and
$Y_{i\alpha}$. 
Including the soft terms from
Eq.~\eqref{eq:soft_terms}, the Lagrangian involving the interactions
of the (s)leptons and Higgs(inos) with the RHSN mass eigenstates
$\widetilde{N}_{\pm i}$, the RHN $N_i$, and with the $SU(2)_L$ and
$U(1)_Y$ gauginos, is given by
\begin{eqnarray}
-\mathcal{L}_{SL} \!\! & = &  
\frac{Y_{i\alpha}}{\sqrt{2}}\epsilon_{ab}\left\{ \widetilde{N}_{+i}
\left[\overline{\widetilde{H}_u^{c,b}}P_{L}\ell_{\alpha}^{a}
%+\left( \frac{A Z_{i\alpha}}{Y_{i\alpha}}+M_{i}\right)
+\left( A+M_{i}\right)
\widetilde{\ell}_{\alpha}^{a}H_{u}^{b}\right]\right.
\nonumber \\
&  & \left.+i\widetilde{N}_{-i}\left[
\overline{\widetilde{H}_u^{c,b}}P_{L}\ell_{\alpha}^{a}
%+\left(\frac{A Z_{i\alpha}}{Y_{i\alpha}}-M_{i}\right)
+\left( A-M_{i}\right)
\widetilde{\ell}_{\alpha}^{a}H_{u}^{b}\right]\right\} 
\nonumber \\
&  & +Y_{i\alpha}\epsilon_{ab}
\left(\overline{\widetilde{H}_u^{c,b}}P_{L}N_{i}
\widetilde{\ell}_{\alpha}^{a}
+\overline{N}_{i}P_{L}\ell_{\alpha}^{a}H_{u}^{b}\right)
\nonumber \\
&  & +g_{2}\left(\sigma_{\pm}\right)_{ab}
\left(\overline{\widetilde{\lambda}_{2}^{\pm}}P_{L}\ell_{\alpha}^{a}
\widetilde{\ell}_{\alpha}^{b*}
+\overline{\widetilde{H}_u^{c,a}}P_{L}\widetilde{\lambda}_{2}^{\pm}H_u^{b*}\right)
\nonumber \\
&  & +\frac{g_{2}}{\sqrt{2}}\left(\sigma_{3}\right)_{ab}
\left(\overline{\widetilde{\lambda}_{2}^{0}}P_{L}\ell_{\alpha}^{a}
\widetilde{\ell}_{\alpha}^{b*}
+\overline{\widetilde{H}_u^{c,a}}P_{L}
\widetilde{\lambda}_{2}^{0}H_u^{b*}\right)
\nonumber \\
&  & +\frac{g_{Y}}{\sqrt{2}}\delta_{ab}
\left[\overline{\widetilde{\lambda}_{1}}
\left(y_{\ell L}P_{L} \!-\! y_{\ell R}P_{R}\right)\ell_{\alpha}^{a}
\widetilde{\ell}_{\alpha}^{b*}
+\overline{\widetilde{H}_u^{c,a}}P_{L}
\widetilde{\lambda}_{1}H_u^{b*}\right] \!+\! {\rm h.c.},
\label{eq:mass_basis}
\end{eqnarray}
where $g_2$ and $g_Y$ are respectively the $SU(2)_L$ and $U(1)_Y$ 
gauge couplings, $y_{\ell L}=-1$ and $y_{\ell R}=2$ denote  
respectively the hypercharges of the left- and right-handed (s)leptons, 
and $\sigma_{\pm}=\left(\sigma_1 \pm i\sigma_2\right)/2$ 
with $\sigma_i$ the Pauli matrices.
%\footnote{The Feynman rules we 
%used are collected in \ref{app:phases}.}

All the parameters appearing in the superpotential
\eqref{eq:superpotential} and in the Lagrangian \eqref{eq:soft_terms}
(or equivalently in the first three lines of
Eq.~\eqref{eq:mass_basis}) are in principle complex quantities.
However, superfield phase redefinition allows to remove several
phases. In the following, for simplicity, we concentrate on SL arising
from a single RHSN generation $i=1$ and to simplify notations we will
drop that index ($Y_\alpha\equiv Y_{1\alpha},\, Z_{\alpha}\equiv
Z_{1\alpha},\, B=B_{11},$ etc.).\footnote{This simplification does not
  imply any crucial loss of generality. As it is explained in detail in
  Ref.~\refcite{Engelhard:2006yg}, the dynamics of the heavier
  leptogenesis states can become important only in temperature regimes
  in which the flavours of the leptons are not completely resolved by
  their Yukawa mediated interactions with the Higgs. The relevant
  temperature range falls in any case above $T\sim 2\times 10^9\,$GeV
  (see Section~\ref{sec:flavor}), while SL can proceed successfully
  only at lower temperatures.}
After superfield phase rotations, the relevant Lagrangian terms 
restricted to $i=1$ are characterized by only  
three independent physical phases:
\begin{eqnarray}
%\phi_{A_{\alpha}} & \equiv & \arg\left(AZ_{\alpha}Y_{\alpha}^{*}B^{*}\right),
\phi_{A} & \equiv & \arg\left(AB^{*}\right),\label{eq:CP_phase1} \\
\phi_{g_{2}} & \equiv & \frac{1}{2}\arg\left(Bm_{2}^{*}\right), \label{eq:CP_phase2} \\
\phi_{g_{Y}} & \equiv & \frac{1}{2}\arg\left(Bm_{1}^{*}\right), \label{eq:CP_phase3} 
\label{eq:CP_phase4}
\end{eqnarray}
which can be assigned to  $A$,   
and to the gaugino coupling operators $g_2,\,g_Y$ 
respectively.
%\footnote{For details of the phase convention 
%we use, please refer to \ref{app:phases}.}
Thus, in the calculation of the CP asymmetry described below
$M,\,B,\,m_2,\,m_1$ and $Y_\alpha$ correspond to  real and
positive parameters, while $A,\,g_2$ and $g_Y$ are complex quantities
with respective phases $\phi_{A},\,\phi_{g_2}$, and $\phi_{g_Y}$.

The tree-level RHSN decay width is given by
\be
\Gamma_{\widetilde{N}_\pm} = \frac{M}{4\pi}\sum_\alpha Y_\alpha^2
\left[1\pm\frac{\mbox{Re}(A)}{M}\left(1-\frac{B}{2M}\right)
+\frac{|A|^2}{2M^2}+\frac{B^2}{8M^2}+\mathcal{O}(\delta_S^3)\right],
\label{eq:decay_width_pm}
\ee
where 
\begin{eqnarray}
\delta_{S} & \equiv & \frac{A}{M},\frac{B}{M},\frac{m_2}{M},\frac{\widetilde M}{M},
\label{eq:delta_S}
\end{eqnarray}
and $\delta_S \ll 1$ is assumed.  Neglecting SUSY-breaking effects in
the RHSN masses and in the vertex, we have 
\begin{equation}
\Gamma_{\widetilde{N}_+}
\simeq \Gamma_{\widetilde{N}_-} \simeq \Gamma
\equiv \frac {M}{4\pi} {\displaystyle \sum_\alpha Y_{\alpha}^2}.
\label{eq:gamma}
\end{equation}

\subsection{CP asymmetries} 
\label{sec:cp_asymmetries}

\begin{figure}
\centering
\includegraphics[scale=0.6]{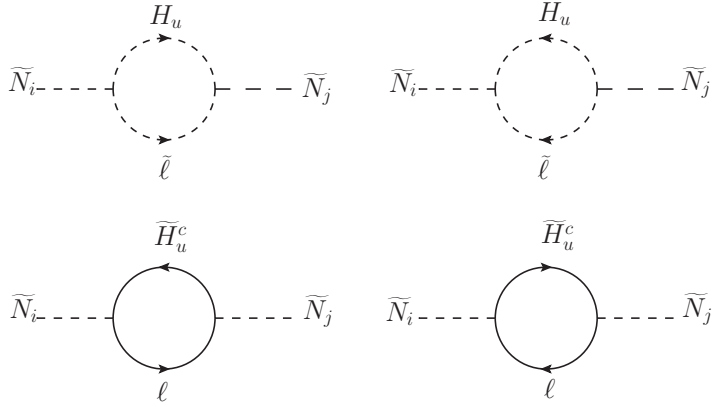}
\caption{Feynman diagrams contributing to the RHSN self-energies
at one-loop.}
\label{fig:self_energies}
\end{figure}

\begin{figure}
\centering
\includegraphics[scale=0.6]{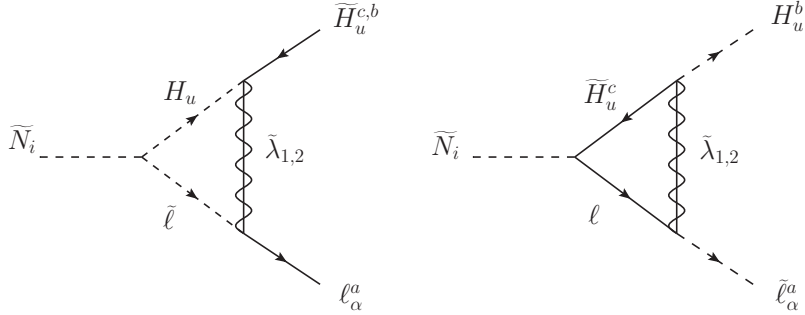}
\caption{Feynman diagrams contributing to the RHSN decay vertex 
at one-loop.}
\label{fig:vertex_1loop}
\end{figure}

The total CP asymmetry in the decays of 
$\widetilde N_{\pm}$ is defined as:
\be
\epsilon_\alpha = \frac{\displaystyle \sum_{i=\pm,a_\alpha} 
\left[\gamma(\widetilde{N}_i \rightarrow a_\alpha)
- \gamma(\widetilde{N}_i \rightarrow \bar{a}_\alpha)\right]}
{\displaystyle \sum_{i=\pm,a_\beta,\beta} 
\left[\gamma(\widetilde{N}_i \rightarrow a_\beta)
+ \gamma(\widetilde{N}_i \rightarrow \bar{a}_\beta)\right]} , 
\label{eq:cp_asym_total} 
\ee
where $\gamma(\widetilde{N}_i \to a_\alpha)$ 
is the thermally averaged decay rate\footnote{The thermally
averaged reaction density is defined in Eq.~\eqref{eq:therm_ave_rate}.}
for the decay of $\widetilde{N}_i$ into final state $a_\alpha$ 
($a_\alpha\equiv s_\alpha,f_\alpha$ with 
$s_\alpha=\tilde{\ell}^a_\alpha H_u^b$ and 
$f_\alpha=\ell_\alpha^a \widetilde H_u^{c,b}$).

Ignoring thermal effects and  taking into account  only 
the mass splitting in the decay width and amplitudes,
Eq.~\eqref{eq:cp_asym_total} becomes
\bea 
\epsilon_{\alpha} (T=0) &=& \frac{\displaystyle
\sum_{i=\pm,a_{\alpha}}
%\int \frac{d^3 p_{\widetilde N_i}}
%{2E_{\widetilde N_i}(2\pi)^3} f^{eq}(E_{\widetilde N_i})
\left(|\hat{\mathcal{A}}_{i}^{a_{\alpha}}|^{2}
-|\overline{\hat{\mathcal{A}}_{i}^{\bar{a}_{\alpha}}}|^{2}\right)/M_i}
{\displaystyle \sum_{i=\pm,a_{\beta},\beta}
%\int \frac{d^3 p_{\widetilde N_i}}
%{2E_{\widetilde N_i}(2\pi)^3} f^{eq}(E_{\widetilde N_i})
\left(|\hat{\mathcal{A}}_i^{a_{\beta}}|^{2}
+|\overline{\hat{\mathcal{A}}_{i}^{\bar{a}_{\beta}}}|^{2}\right)/M_i},
%\nonumber \\
%&=& \frac{\displaystyle
%\sum_{i=\pm,a_{\alpha}}
%M_i\mathcal{K}_1\left(\frac{M_i}{T}\right)
%\left(|\hat{\mathcal{A}}_{i}^{a_{\alpha}}|^{2}
%-|\overline{\hat{\mathcal{A}}_{i}^{\bar{a}_{\alpha}}}|^{2}\right)}
%{\displaystyle \sum_{i=\pm,a_{\beta},\beta}
%M_i\mathcal{K}_1\left(\frac{M_i}{T}\right)
%\left(|\hat{\mathcal{A}}_i^{a_{\beta}}|^{2}
%+|\overline{\hat{\mathcal{A}}_{i}^{\bar{a}_{\beta}}}|^{2}\right)},
\label{eq:cp_asym_0}
\eea
where $\hat{\mathcal{A}}_{i}^{{a}_{\alpha}}$ is 
the  amplitude for the decay of $\widetilde N_i$ into $a_\alpha$.

To fully account for finite temperature corrections 
several different effects must be considered:

\begin{enumerate}
\item[(A)] Thermal corrections to (s)leptons and Higgs(inos) propagators,
\item[(B)] Final state statistical factors, 
\item[(C)] Thermal masses of (s)leptons and Higgs(inos),
\item[(D)] Thermal corrections to gauge and Yukawa couplings,
\item[(E)] Particle motion in the thermal bath.
\end{enumerate}

In the two pioneering papers\cite{Grossman:2003,DAmbrosio:2003} it was
shown that the most relevant thermal effects in SL are those of type
(B) that arise from final state Bose-enhancement and Fermi-blocking
for RHSN decays respectively into scalars and fermions.  These effects
spoil the cancellation between the decay asymmetries into scalars and
fermions, and are large enough to render SL viable.  In
Ref.~\refcite{Grossman:2003} only effects of type (B) were taken into
account.  In Ref.~\refcite{DAmbrosio:2003} effects of types (C) and
(D) were also included, but it was found that they did not change
significantly the overall picture. However, in all these studies,
effect (E) was always ignored.  Later, the authors of
Ref.~\refcite{Giudice:2003jh} studied the full-fledged thermal effects
(A)-(D), and concluded that all the effects previously neglected did
not introduce significant changes.  As regards specifically the
effects of type (E), Refs.~\refcite{Covi:1997dr,Giudice:2003jh} showed that in the case of SM type I
leptogenesis the related corrections are at most $\sim 20\%$ with
respect to the $T=0$ case, which suggests that they can be neglected
also in SL.

Including only the main thermal effects (B), (C) and (D) 
the total CP asymmetry \eqref{eq:cp_asym_total} simplifies to:
\begin{equation}
\epsilon_\alpha = \epsilon_{+ \alpha}^{s}+\epsilon_{- \alpha}^{s}
+\epsilon_{+ \alpha}^{f}+\epsilon_{- \alpha}^{f},
\end{equation}
where 
\begin{eqnarray} 
\epsilon^{s}_{\pm \alpha} =\frac{\displaystyle
\left(|\hat{\mathcal{A}}_{\pm}^{s_{\alpha}}|^{2}
-|\overline{\hat{\mathcal{A}}_{\pm}^{\bar{s}_{\alpha}}}|^{2}\right)
c^{s_{\alpha}}_\pm /M_i}
{\displaystyle \sum_{i=\pm,a_{\beta},\beta}
\left(|\hat{\mathcal{A}}_i^{a_{\beta}}|^{2}
+|\overline{\hat{\mathcal{A}}_{i}^{\bar{a}_{\beta}}}|^{2}\right)
c^{a_{\beta}}_i /M_i}
\;,
\label{eq:asym_s} 
\\
\epsilon_{\pm \alpha}^{f} 
=\frac{\left(|\hat{\mathcal{A}}_{\pm}^{f_{\alpha}}|^{2}
-|\overline{\hat{\mathcal{A}}_{\pm}^{\bar{f}_{\alpha}}}|^{2}\right)
c^{f_{\alpha}}_\pm /M_i}
{\displaystyle \sum_{i=\pm,a_{\beta},\beta}
\left(|\hat{\mathcal{A}}_{i}^{a_{\beta}}|^{2}
+|\overline{\hat{\mathcal{A}}_{i}^{\bar{a}_{\beta}}}|^{2}\right)
c^{a_{\beta}}_i /M_i} \;. 
\label{eq:asym_f} 
\end{eqnarray}
In these equations the finite temperature corrections from thermal
phase-space, final state Bose-enhancement for decays into scalars and
Fermi-blocking for decays into fermions have been factored out in the
thermal coefficients $c^{s_\alpha}_\pm, c^{f_\alpha}_\pm$, so that
$\hat{\mathcal{A}}_{i}^{a_{\alpha}}$ and
$\overline{\hat{\mathcal{A}}_{i}^{\bar{a}_{\alpha}}}$ are the zero
temperature amplitudes.  Note that as long as the zero temperature
lepton and slepton masses and small neutrino Yukawa couplings are
neglected, the thermal coefficients are flavour independent, and if the
mass splitting between $\widetilde N_+$ and $\widetilde N_-$ is also
ignored, they are the same also for $i=\pm$.  In the approximation in
which $\widetilde N_\pm$ decay at rest the thermal coefficients are
given by:
\bea
c^f_\pm
% +(T)=c^f_-(T)
\equiv c^f(T)
&=&(1-x_{\ell} -x_{\widetilde{H}_u})
\lambda(1,x_{\ell},x_{\widetilde{H}_u})
\,\left[ 1-\fLeq\right] \left[ 1-\fht\right], 
\label{cfeq}\\
c^s_\pm % +(T)=c^s_-(T)
\equiv c^s(T)&=&\lambda(1,x_{H_u},x_{\tilde{\ell}})
\,\left[ 1+\fh\right] \left[ 1+\fLteq\right],
\label{cbeq}
\eea
where
\begin{equation}
\lambda(1,x,y)=\sqrt{(1+x-y)^2-4x}, \qquad x_a\equiv \frac{m_a(T)^2}{M^2}\,,
\end{equation}
and
\bea
f^{eq}_{H_u,\tilde{\ell}}=\frac{1}{\exp[E_{H_u,\widetilde{\ell}}/T]-1},
&\qquad\qquad & 
f^{eq}_{\widetilde H_u,\ell}= \frac{1}{\exp[E_{\widetilde H_u,\ell}/T]+1},  
\label{eq:fhHeq}
\eea
are the  Bose-Einstein and Fermi-Dirac equilibrium distributions,
respectively, with  
\be
E_{\ell,\widetilde H_u}=\frac{M}{2} (1+x_{\ell,\widetilde{H}_u}-
x_{\widetilde H_u,\ell}), \qquad
E_{H_u,\widetilde{\ell}}=\frac{M}{2} (1+x_{H_u ,\widetilde{\ell}}-
x_{\widetilde{\ell},H_u}). 
\ee

The CP asymmetry is generated at the loop level from the interference
between the tree-level and the one-loop diagrams shown in
Figs.~\ref{fig:self_energies} and \ref{fig:vertex_1loop}, that
correspond to different sources of CP violation: the first one arises
from the self-energy corrections (Fig.~\ref{fig:self_energies}) while
the second arises from vertex corrections
(Fig.~\ref{fig:vertex_1loop}).  In the following we describe how the
decay asymmetries are computed within two different approaches: the
first one relies on field theory, the second one on quantum mechanics.

\subsubsection{Field theoretical approach}
\label{sec:QFT}
When $\Gamma \gg \Delta M_\pm \equiv M_+ - M_-$, the two RHSN states
are not well-separated particles.\cite{DAmbrosio:2003}  In this case,
the result for the asymmetry depends on how the initial state is prepared.
%!!%
\footnote{The effects of initial conditions in SL have been studied 
in Ref.~\refcite{BahatTreidel:2008}.}
In what follows we assume that the RHSN are in a thermal
bath with a thermalization time $\Gamma^{-1}$ shorter than the typical
oscillation time $\Delta M_\pm^{-1}$. In this case coherence is lost,
and it is appropriate to compute the CP asymmetries in terms of the
mass eigenstates \eqref{eq:mass_eigenstates}.  The relevant decay
amplitudes can be obtained following the effective field-theoretical
approach described in Refs.~\refcite{Pilaftsis:1997,%
Pilaftsis:2004,Pilaftsis:2005a,Pilaftsis:2005b,Pilaftsis:2008},
which takes into account CP violation due to mixing and decay (as well
as their interference) of nearly degenerate states, by using resummed
propagators for unstable mass eigenstate particles.  The decay
amplitude $\hat{\mathcal A}_i^{a_\alpha}$ of the unstable external
state $\widetilde{N}_i$
%defined in Eq.~(\ref{eq:mass_eigenstates}) 
into final state $a_\alpha$ ($a_\alpha\equiv s_\alpha,f_\alpha$ with 
$s_\alpha=\tilde{\ell}^a_\alpha H_u^b$ and 
$f_\alpha=\ell_\alpha^a \widetilde H_u^{c,b}$)
is described by a superposition of amplitudes with stable
final states: 
\begin{eqnarray}
\hat{\mathcal{A}}_{\pm}^{a_{\alpha}} 
&\! =\! & \left(A_{\pm}^{a_\alpha}+i{\mathcal{V}_{\pm}^{a_\alpha}}^{{\rm abs}}(p^2)\right)
-\left(A_{\mp}^{a_\alpha}+i{\mathcal{V}_{\mp}^{a_\alpha}}^{{\rm abs}}
(p^2)\right)
\times\frac{i\Sigma_{\mp\pm}^{{\rm abs}}}{M_\pm^{2}-M_{\mp}^{2}
+i\Sigma_{\mp\mp}^{{\rm abs}}},\label{eq:amp}\\
\overline{\hat{\mathcal{A}}_{\pm}^{\bar{a}_{\alpha}}} 
&\! =\!& \left({A_{\pm}^{a_\alpha}}^{*}\!+i{\mathcal{V}_{\pm}^{a_\alpha}}^{{\rm abs}*}
(p^2)\right)
-\left({A_{\mp}^{a_\alpha}}^{*}\!\!
+i{\mathcal{V}_{\mp}^{a_\alpha}}^{{\rm abs}*}
(p^2)\right)\! 
\times\!\frac{i\overline{\Sigma}_{\mp\pm}^{{\rm abs}}}{M_\pm^{2}\!-M_{\mp}^{2}
\!+i\overline{\Sigma}_{\mp\mp}^{{\rm abs}}}\,.\;\;\;
\label{eq:amp_cp} 
\end{eqnarray}
In these equations $A_{\pm}^{a_\alpha}$ are the tree-level amplitudes:
\begin{eqnarray}
\!\!\! A_{+}^{s_\alpha} & = &  \frac{Y_{\alpha}}{\sqrt{2}}(A^{*}+M)\epsilon_{ab},
\;\;\;\;\;\;\;\;\;\;\ \ \quad
A_{-}^{s_\alpha} =  -i\frac{Y_{\alpha}}{\sqrt{2}}(A^{*}-M)\epsilon_{ab}, 
\label{eq:tree_amp_s} \\
\!\!\! A_{+}^{f_k} & = &  
\frac{Y_{\alpha}}{\sqrt{2}}[\bar{u}(p_{\ell})
P_{R}v(p_{\widetilde H_u^c})] \epsilon_{ab},
\;\;\;\quad
A_{-}^{f_\alpha} =  -i\frac{Y_{\alpha}}{\sqrt{2}}
[\bar{u}(p_{\ell})P_{R}v(p_{\widetilde H_u^c})]\epsilon_{ab}.
\label{eq:tree_amp_f}
\end{eqnarray}
$\Sigma_{ij}^{{\rm abs}}$ are the absorptive parts of the
$\widetilde{N}_{i}\to\widetilde{N}_{j}$ self-energies (see
Fig. \ref{fig:self_energies}) which can be obtain by directly
evaluating the imaginary part of the Feynman integral or by using
Cutkosky's cutting rules:\cite{Cutkosky:1960sp}
\begin{eqnarray}
\Sigma_{\mp\mp}^{(1){\rm abs}}& = & 
\Gamma\; M\left[\frac{1}{2}+\frac{M_\mp^{2}}{2M^{2}}+\frac{|A|^{2}}{2M^{2}}
\mp\frac{\mbox{Re}(A)}{M}\right], \label{eq:self_energies0} \\
\Sigma_{\mp\pm}^{(1){\rm abs}} & = & 
-\Gamma\;\mbox{Im}(A).
\label{eq:self_energies}
\end{eqnarray}
${\mathcal{V}_{\pm}^{a_\alpha}}^{{\rm abs}}$ are the absorptive parts
of the vertex corrections (see Fig. \ref{fig:vertex_1loop}):
\begin{eqnarray}
{\mathcal{V}_{+}^{s_\alpha}}^{{\rm abs}}\left(p^2\right)
& = & \epsilon_{ab}\frac{Y_{\alpha}}{\sqrt{2}}\frac{3m_2}{32\pi}(g_{2})^{2}
\ln\frac{m_{2}^{2}}{p^{2}+m_2^{2}}\; , \label{eq:vertexabs0} \\
{\mathcal{V}_{-}^{s_\alpha}}^{\rm abs}\left(p^2\right)
& = & -i\epsilon_{ab}\frac{Y_{\alpha}}{\sqrt{2}}
\frac{3 m_{2}}{32\pi}(g_{2})^{2}
\ln\frac{m_{2}^{2}}{p^{2}+m_2^{2}}\; , \\
{\mathcal{V}_{+}^{f_\alpha}}^{{\rm abs}}\left(p^2\right)
& = & \epsilon_{ab}\frac{Y_{\alpha}}{\sqrt{2}}
\frac{3 m_2}{32\pi p^{2}}(A^{*}+M)
(g_{2}^{*})^{2}\ln\frac{m_2^{2}}{p^{2}+m_2^{2}}
% \nonumber \\ &  & 
\times[\bar{u}(p_{\ell})P_{R}v(p_{\widetilde H_u^c})] ,\quad \\
{\mathcal{V}_{-}^{f_\alpha}}^{{\rm abs}}\left(p^2\right)
& = & -i\epsilon_{ab}\frac{Y_{\alpha}}{\sqrt{2}}
\frac{3 m_2}{32\pi p^{2}}(A^{*}-M)(g_{2}^{*})^{2}
\ln\frac{m_2^{2}}{p^{2}+m_2^{2}}
% \nonumber \\ &  & 
\times [\bar{u}(p_{\ell})P_{R}v(p_{\widetilde H_u^c})] ,\quad 
\label{eq:vertexabs}
\end{eqnarray}
where only the contribution from $SU(2)_L$ gauginos has been included. 
The contribution from $U(1)_Y$ gaugino can be obtained 
by simply substituting $\alpha_2 \to \alpha_Y \equiv \frac{|g_Y|^2}{4\pi}$ 
and $3 \to 1$ in Eqs.~\eqref{eq:vertexabs0}--\eqref{eq:vertexabs}.

Substituting Eqs.~\eqref{eq:amp} and \eqref{eq:amp_cp} into 
Eqs.~\eqref{eq:asym_s} and \eqref{eq:asym_f} and using that
$\Sigma_{\mp\mp}^{{\rm abs}}=\overline{\Sigma}_{\mp\mp}^{{\rm abs}}$
and $\Sigma_{\mp\pm}^{{\rm abs}*}=\overline{\Sigma}_{\mp\pm}^{{\rm abs}}$,
one gets: 
\begin{eqnarray}
|\hat{\mathcal{A}}_{\pm}^{a_{\alpha}}|^{2}-
|\overline{\hat{\mathcal{A}}_{\pm}^{\bar{a}_{\alpha}}}|^{2} 
&\simeq&
-4\left\{-\mbox{Im}\left[{A_{\pm}^{a_\alpha}}^{*}A_{\mp}^{a_\alpha}
\Sigma_{\mp\pm}^{{\rm abs}}\right]
\frac{M_\pm^{2}-M_{\mp}^{2}}
{(M_\pm^{2}-M_{\mp}^{2})^{2}+|\Sigma_{\mp\mp}^{{\rm abs}}|^{2}}
\right.\nonumber \\ 
&   &
+\mbox{Im}\left[{A_{\pm}^{a_\alpha}}^{*}{\mathcal{V}_{\pm}^{a_\alpha}}^{{\rm abs}}
(M^2_\pm)\right]
\nonumber \\ && 
 \left.+\mbox{Im}
\left[{\mathcal{V}_{\pm}^{a_\alpha}}^{{\rm abs}*}(M^2_\pm)
A_{\mp}^{a_\alpha}
\Sigma_{\mp\pm}^{{\rm abs}}
- 
{A_{\pm}^{a_\alpha}}^{*}
{\mathcal{V}_{\mp}^{a_\alpha}}^{{\rm abs}}(M^2_\pm)
\Sigma_{\mp\pm}^{{\rm abs}}\right]\right.
 \nonumber \\ &  & 
\left.\times\frac{\Sigma_{\mp\mp}^{{\rm abs}}}
{(M_\pm^{2}-M_{\mp}^{2})^{2}+|\Sigma_{\mp\mp}^{{\rm abs}}|^{2}}\right\},
\label{eq:cp_asym_num} 
\end{eqnarray}
where the $\simeq$ sign means that terms of order $\delta_{S}^{3}$ and
higher are ignored, with $\delta_S$ defined in Eq.~\eqref{eq:delta_S}.
The three terms inside curly brackets in Eq.~\eqref{eq:cp_asym_num}
correspond respectively to CP violation in $\widetilde N$ mixing from
the off-diagonal one-loop self-energies that will be denoted below
with $S$ (=`self-energy'),\footnote{This corresponds to the effects
originally considered in
Refs.~\refcite{Grossman:2003,DAmbrosio:2003}.}  CP violation due to
the gaugino-mediated one-loop vertex corrections to the $\widetilde N$
decay\cite{Grossman:2004} denoted by $V$ (=`vertex'), and CP violation
in the interference of vertex and self-energies denoted by $I$
(=`interference'). On the other hand, the amplitudes appearing in the denominators in 
Eqs.~\eqref{eq:asym_s} -- \eqref{eq:asym_f} verify the tree-level relations 
 $|\hat{\mathcal{A}}_{\pm}^{a_{\alpha}}|^{2}
 +|\overline{\hat{\mathcal{A}}_{\pm}^{\bar{a}_{\alpha}}}|^{2} =
 2|A_{\pm}^{a_\alpha}|^{2}$, with $|A_{\pm}^{s_\alpha}|^{2} =
 Y_{\alpha}^{2} \left[|A|^{2}+M^{2}\pm 2M\mbox{Re}(A)\right]$ and $
 |A_{\pm}^{f_\alpha}|^{2} = Y_{\alpha}^{2}M_{\pm}^{2}$.

Using the explicit forms in Eqs.~\eqref{eq:tree_amp_s} --
\eqref{eq:vertexabs} one can verify that the the three contributions
$S$, $V$ and $I$ to the CP asymmetry from scalar and fermion decays
satisfy :
\begin{eqnarray}
\epsilon_{\pm \alpha}^{sS} 
=\Delta^s(T) \epsilon^S_{\pm\alpha}, 
\;\;\;\; \;\;\;\; 
\epsilon_{\pm \alpha}^{fS} 
=-\Delta^f(T) \epsilon^S_{\pm \alpha},
\nonumber  \\
\epsilon_{\pm \alpha}^{sV} 
=\Delta^s(T) \epsilon^V_{\pm \alpha}, 
\;\;\;\; \;\;\;\; 
\epsilon_{\pm \alpha}^{fV} 
=-\Delta^f(T) \epsilon^V_{\pm \alpha}, 
\nonumber \\
\epsilon_{\pm \alpha}^{sI} 
=\Delta^s(T) \epsilon^I_{\pm \alpha}, 
\;\;\;\; \;\;\;\; 
\epsilon_{\pm \alpha}^{fI} 
=-\Delta^f(T) \epsilon^I_{\pm \alpha}, 
\label{eq:cancel}
\end{eqnarray}
with 
\begin{eqnarray}
\!\!\!
\epsilon_{\pm \alpha}^S 
& = & -P_{\alpha}\frac{A}{M}%\left(1\mp\frac{B}{2M}\right)
\sin\left(\phi_{A}\right)\frac{2B\Gamma}{4B^{2}+\Gamma^{2}},
\label{eq:cp_asym_S}\\
\!\!\!\!\!\!
\epsilon_{\pm \alpha}^V 
& = & -\frac{3P_{\alpha}\alpha_{2}}{8}\frac{m_2}{M}
\ln\frac{m_2^2}{m_2^2+M^2}
\left[\frac{A}{M}
\sin\left(\phi_{A}+2\phi_{g}\right)
%\right. \nonumber \\
%&& \left. 
-\frac{B}{M}
\sin\left(2\phi_{g}\right)\pm\sin\left(2\phi_{g}\right)\right]\!,
\quad \ \  
\label{eq:cp_asym_V}\\
\!\!\!\epsilon_{\pm \alpha}^{I} 
& = & \frac{3P_{\alpha}\alpha_{2}}{4}\frac{m_2}{M}\frac{A}{M}
\ln\frac{m_2^2}{m_2^2+M^2}\sin\left(\phi_{A}\right)
\cos\left(2\phi_{g}\right)\frac{\Gamma^{2}}{4B^{2}+\Gamma^{2}},
\label{eq:cp_asym_I}
\end{eqnarray}
and
\be
\Delta_{s,f}(T) \equiv \frac{c^{s,f}(T)}{c^s(T)+c^f(T)}. 
\ee
In the expressions Eqs.~\eqref{eq:cp_asym_S}-\eqref{eq:cp_asym_I} we
have introduced $\alpha_{2}=\frac{|g_2|^2}{4\pi}$, and the physical
complex phases $\phi_{A}$ and $\phi_g \equiv \phi_{g_2}$ have been
explicitly written, so that all the parameters $A$ and $Y_\alpha$
etc. are understood to be real and positive.  We will adopt this
convention also in the following, unless explicitly stated in the
text.  The flavour projectors are defined as
\begin{equation} 
P_\alpha= \frac{Y_{\alpha}^2}{{\displaystyle \sum_\beta}Y_{\beta}^2},
\label{eq:fla_proj}
\end{equation}
and satisfy the conditions
\begin{equation}
\sum_\alpha P_\alpha=1 \quad  {\rm and} \quad 0\leq P_\alpha\leq 1\,. 
\end{equation}

Summing up the contributions from the decays 
of $\widetilde{N}_{+}$ and $\widetilde{N}_{-}$ into scalars and fermions, 
one obtains the three contributions to the total CP 
asymmetry Eq.~\eqref{eq:cp_asym_total}:\cite{Fong:2009}
\begin{eqnarray}
\epsilon_{\alpha}^{S}(T) & = & 
P_\alpha \bar \epsilon^S \Delta_{BF}(T) ,
\label{eq:cp_asym_2}\\
\epsilon_{\alpha}^{V}(T) & = & 
P_\alpha \bar \epsilon^V \Delta_{BF}(T) ,
\label{eq:cp_asym_1} \\
\epsilon_{\alpha}^{I}(T) & = & 
P_\alpha \bar \epsilon^I \Delta_{BF}(T) ,
\label{eq:cp_asym_3}
\end{eqnarray}
where
\bea
\bar\epsilon^S &\equiv& -\frac{A}{M}\sin\left(\phi_{A}\right)
\frac{4B\Gamma}{4B^{2}+\Gamma^{2}}, 
\label{eq:cp0_S} \\
\bar\epsilon^V &\equiv& 
-\frac{3\alpha_{2}}{4}\frac{m_2}{M}
\ln\frac{m_2^2}{m_2^2+M^2}\left[\frac{A}{M}\sin\left
(\phi_{A}+2\phi_{g}\right)
-\frac{B}{M}\sin\left(2\phi_{g}\right)\right],
\label{eq:cp0_V} \\
\bar\epsilon^I &\equiv&
\frac{3\alpha_{2}}{2}\frac{m_2}{M}
\frac{A}{M}\ln\frac{m_2^2}{m_2^2+M^2}
\sin\left(\phi_{A}\right)
\mbox{cos}\left(2\phi_{g}\right)\frac{\Gamma^{2}}{4B^{2}+\Gamma^{2}},
\label{eq:cp0_I}
\eea
and the \emph{thermal factor} $\Delta_{BF}(T)$ is given by 
\begin{eqnarray}
\Delta_{BF}(T) 
& \equiv & \Delta^s(T)-\Delta^f(T) .
\label{eq:therm_factor}
\end{eqnarray}

Eq.~\eqref{eq:cp_asym_2} contains the contribution to the
asymmetry due to CP violation in RHSN mixing discussed in the original
works.\cite{Grossman:2003,DAmbrosio:2003} Eqs.~\eqref{eq:cp_asym_1}
and \eqref{eq:cp_asym_3} give respectively the contribution to the
asymmetry from CP violation in decay and in the interference between
mixing and decay. These last two contributions have parametric
dependence similar to the ones obtained in
Ref.~\refcite{Grossman:2004}.  However, as it is explicitly shown in
Eqs.~\eqref{eq:cancel}, the scalar and fermionic CP asymmetries cancel
each other at zero temperature,\cite{Fong:2009} because as $T \to 0$
both $c^s(T),\, c^f(T) \to 1$.  Consequently up to second order in the
soft parameters, all contributions to the SL CP lepton asymmetry
require thermal effects in order to be significant.  More precisely,
$\epsilon_\alpha^V(T)$ and $\epsilon_\alpha^I(T)$ vanish exactly in
the $T=0$ limit, in agreement with a general proof that will be
presented in Section~\ref{sec:vertexCP}. As regards
$\epsilon_\alpha^S(T)$, it does not vanish exactly; however, the
surviving terms are of order $\mathcal{O}(\delta_S^3)$ and thus
completely negligible.

\begin{figure}
\begin{center}
\includegraphics[width=0.6\textwidth]{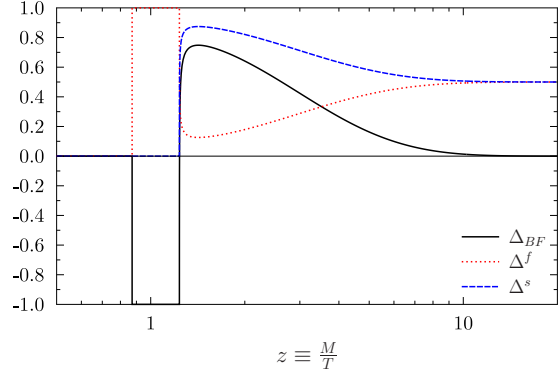}
\end{center}
\caption{The thermal factors $\Delta_{BF}$ (black solid curve),
  $\Delta^s$ (blue dashed curve) and $\Delta^f$ (red dotted curve) as
  a function of $z\equiv M/T$.}
\label{fig:ep_ratio}
\end{figure}

Fig. \ref{fig:ep_ratio} displays the thermal factors $\Delta_{BF}$
(black solid curve), $\Delta^s$ (blue dashed curve) and $\Delta^f$
(red dotted curve) as a function of $z \equiv M/T$. For $z \lesssim
0.8$, the decays of RHSN to scalars and fermions are kinematically
forbidden.  In the small interval $0.8 \lesssim z \lesssim 1.2$ the
fermionic channel becomes accessible although the scalar channel is
still closed; this is because the thermal masses for the fermions are
half than the ones for the scalars.  For $z \gtrsim 1.2$, the scalar
channel opens up as well, however because of thermal effects the
cancellation between $\Delta^s$ and $\Delta^f$ is not very effective,
and for relatively small values of $z$ a sizable total asymmetry
survives.  For $z \gsim 10$ thermal effects are strongly suppressed
and the cancellation becomes almost exact.

%%%%%%%%%%%%%%%%%%%%%%
%%%%%%%%%%%%%%%%%%%%%%

As a final remark, let us note that in this derivation thermal
corrections to the loop diagrams responsible for the CP asymmetries
have been neglected.  That is, the imaginary part of the one-loop
graphs has been obtained by directly evaluating the imaginary part of
the Feynman integrals or by Cutkosky's cutting rules at
$T=0$.\cite{Cutkosky:1960sp}

%%%%%%%%%%%%%%%%%%%%%%%%%%%%%%%%%%%%%%%%%%%%%%%%%%%%%%%%%%%%%%%%%%

\subsubsection{Quantum mechanical approach}
\label{sec:QM}
In this section we describe the computation of the CP asymmetry using
a quantum mechanical (QM) approach, based on an effective (non
Hermitian)
Hamiltonian.\cite{Grossman:2003,DAmbrosio:2003,Grossman:2004} In this
language an analogy can be drawn between the $\widetilde
N$--$\widetilde N^{\dagger}$ system and the system of neutral mesons
such as $K^0$--$\overline K^0$, for which the time evolution is
determined, in the non-relativistic limit, by the Hamiltonian:
\begin{eqnarray}
H & = & 
\left(\begin{array}{cc} M & \frac{B}{2}\\
\frac{B}{2} & M \end{array}
\right)
-\frac{i}{2}\left(\begin{array}{cc}
\Gamma & \frac{\Gamma A^{*}}{M}\\
\frac{\Gamma A}{M} &  \Gamma\end{array}\right),
\end{eqnarray}
with $\Gamma$ given in Eq.~\eqref{eq:gamma}.

In Refs. \refcite{Grossman:2003,DAmbrosio:2003,Grossman:2004} the QM
formalism was applied for weak initial states $\widetilde N$ and
$\widetilde N^{\dagger}$.  In practice, the formalism can be applied
to study the evolution of initial states that are either weak or mass
eigenstates. In order to illustrate the dependence of the results on
the choice of initial conditions, we compute the asymmetry
for both types of initial states. 
Let us define the basis: 
%
% \begin{eqnarray}
\be
\tilde{N}_{1}  =  \left(g\tilde{N}+h\tilde{N}^{\dagger}\right),
% \nonumber \\
\qquad 
\tilde{N}_{2}  =  e^{i\theta}\left(h\tilde{N}-g\tilde{N}^{\dagger}\right).
\label{eq:arbitrary_basis}
\ee
% \end{eqnarray}
%
The mass basis introduced in Eq.~\eqref{eq:mass_eigenstates},
corresponds to
$(g,h,\theta)=(\frac{1}{\sqrt{2}},\frac{1}{\sqrt{2}},-\frac{\pi}{2})$.
Pure $\widetilde N$ and $\widetilde N^{\dagger}$ initial states
correspond instead to $(g,h,\theta)=(1,0, \pi)$.

Including the one-loop contribution from gaugino exchange, the decay
amplitudes of $\widetilde N_1$ and $\widetilde N_2$ into fermions
$f_\alpha=\ell_\alpha^a \widetilde H_{u}^{c,b}$ are:
\begin{eqnarray}
A_{1}^{f_\alpha} 
& = & \left\{Y_{\alpha} h 
- \frac{3Y_{\alpha}}{2M^2}\left(g M + h A^{*}\right)
\left(g_{2}^{*}\right)^{2}\frac{m_2}{16\pi}I_{f} \right\}
[\overline{u}\left(p_{\ell}\right)P_{R}v (p_{\widetilde H_u^c} )] 
\epsilon_{ab},\nonumber \\
\overline{A_{1}^{\bar f_\alpha}} 
& = & \left\{Y_{\alpha} g
- \frac{3Y_{\alpha}}{2M^2} \left(h M + g A\right)
\left(g_{2}\right)^{2}\frac{m_2}{16\pi}I_{f} \right\}
[\overline{u} (p_{\widetilde H_u^c} )
P_{L}v\left(p_{\ell}\right)]
\epsilon_{ab}, \nonumber \\
A_{2}^{f_\alpha} 
& = & -e^{-i\theta} \left\{Y_{\alpha} g
- \frac{3Y_{\alpha}}{2 M^2}
\left(h M - g A^{*}\right)
\left(g_{2}^{*}\right)^{2}\frac{m_2}{16\pi}I_{f}\right\}
[\overline{u}\left(p_{\ell}\right)
P_{R}v(p_{\widetilde H_u^c})]\epsilon_{ab}
,\nonumber \\
\overline{A _{2}^{\bar f_\alpha}} 
& = & e^{-i\theta} \left\{ Y_{\alpha} h
- \frac{3Y_{\alpha}}{2 M^2}
\left(h A - g M\right)\left(g_{2}\right)^{2}\frac{m_2}{16\pi}I_{f}\right\}
% \nonumber \\ &  & 
 [\overline{u} (p_{\widetilde H_u^c})
P_{L}v\left(p_{\ell}\right)]\epsilon_{ab},
\label{eq:qmamplif}
\end{eqnarray}
where  $\overline{A}$ denotes the decay amplitudes into antifermions.
The corresponding decay amplitudes 
into scalar $s_\alpha=\widetilde{\ell}_\alpha^a H_u^b$ are:
\begin{eqnarray}
A_{1}^{s_\alpha} 
& = & \left\{Y_{\alpha}\left(g M + h A^{*}\right) 
-\frac{3Y_{\alpha}}{2} h\left(g_{2}\right)^{2}
\frac{m_2}{16\pi}I_{s} \right\}\epsilon_{ab}
,\nonumber \\
\overline{A_{1}^{\bar s_\alpha}} 
& = & \left\{Y_{\alpha}\left(h M + g A\right)
-\frac{3Y_{\alpha}}{2} g\left(g_{2}^{*}\right)^{2}\frac{m_2}{16\pi}I_{s}
\right\}\epsilon_{ab},\nonumber \\
A_{2}^{s_\alpha} 
& = & e^{-i\theta}\left\{Y_{\alpha}
\left(h M - g A^{*}\right)
+\frac{3Y_{\alpha}}{2} g\left(g_{2}\right)^{2}\frac{m_2}{16\pi}
I_{s} \right\}\epsilon_{ab},\nonumber \\
\overline{A_{2}^{\bar s_\alpha}} 
& = & e^{-i\theta}\left\{Y_{\alpha}
\left(h A - g M\right)
-\frac{3Y_{\alpha}}{2} h\left(g_{2}^{*}\right)^{2}
\frac{m_2}{16\pi}I_{s}\right\} \epsilon_{ab}.
\label{eq:qmamplis}
\end{eqnarray}
In Eqs.~\eqref{eq:qmamplif} and \eqref{eq:qmamplis}:
\begin{eqnarray}
\mbox{Re}(I_{f}) &\equiv& f_R  = -\frac{1}{\pi}
\left[\frac{1}{2}\left(\ln\frac{m_{2}^{2}}{m_{2}^{2}+M^{2}}
\right)^{2}+\mbox{Li}_{2}\left(\frac{m_{2}^{2}}{m_{2}^{2}+M^{2}}
\right)-\zeta(2)\right], 
\nonumber  \\
\mbox{Re}(I_s) &\equiv& s_R =  
\frac{1}{\pi}\left[\frac{1}{2} \left(\ln\frac{m_2^{2}}
{m_2^{2}+M^{2}}\right)^{2}+\mbox{Li}_{2}
\left(\frac{m_2^{2}}{m_2^{2}+M^{2}}\right)-\zeta(2) \right.
\nonumber \\
&  & +B_{0}\left(M^{2},m_2,0\right)
+B_{0}\left(M^{2},0,m_2\right)\Biggr],
\nonumber \\
\mbox{Im}(I_f) &\equiv& f_I=\mbox{Im}(I_s)\equiv s_I=
-\ln\frac{m_2^{2}}{m_2^{2}+M^{2}}.
\end{eqnarray}

In terms of $\widetilde N_1$ and $\widetilde N_2$ the eigenvectors of
the Hamiltonian are:
\begin{eqnarray}
\left|\tilde{N}_{L}\right\rangle  & = 
& \left(gp+hq\right)\left|\tilde{N}_{1}
\right\rangle +e^{-i\theta}\left(hp-gq\right)\left|\tilde{N}_{2}
\right\rangle ,\nonumber \\
\left|\tilde{N}_{H}\right\rangle  & = & 
\left(gp-hq\right)\left|\tilde{N}_{1}\right\rangle 
+e^{-i\theta}\left(hp+gq\right)\left|\tilde{N}_{2}\right\rangle ,
\label{eq:LH_to_arbitrary}
\end{eqnarray}
where
\begin{eqnarray}
\frac{q}{p}  & = &  -1-\frac{\Gamma A }{BM}\sin\left(\phi_{A}\right)
-\frac{\Gamma^{2} A^{2}}{M^{2}B^{2}}\cos^{2}\left(\phi_{A}\right)
-\frac{i}{2}\frac{\Gamma^{2} A^{2}}{M^{2}B^{2}}\sin\left(2\phi_{A}\right).
\label{eq:ratio_qp}
\end{eqnarray}

At time $t$ the states $\widetilde N_1$ and $\widetilde N_2$ 
evolve into 
\begin{eqnarray}
\left|\tilde{N}_{1,2}(t)\right\rangle  
 & = & \frac{1}{2}\left\{ \left[e_{L}(t)+e_{H}(t) \pm C_{0}
\left(e_{L}(t)-e_{H}(t)\right)\right]\left|\tilde{N}_{1,2}\right
\rangle \right.  \nonumber \\
&& \left.+e^{\mp i \theta}C_{1,2}\left(e_{L}(t)
-e_{H}(t)\right)\left|\tilde{N}_{2,1}\right\rangle \right\} \; , 
\label{eq:arbitrary_time_evolution}
\end{eqnarray}
where
\begin{eqnarray}
& & C_{0}  = g h \left(\frac{p}{q}+\frac{q}{p}\right),
\;\;\;\;
C_{1} = h^{2}\frac{p}{q}-g^{2}\frac{q}{p}, \;\;\;\;
C_{2}  =  h^{2}\frac{q}{p}-g^{2}\frac{p}{q}, \;\;\;\;
\label{eq:C_coeff}
\end{eqnarray}
and 
\begin{eqnarray}
e_{H,L}(t) & \equiv & e^{-i(M_{H,L}-\frac{i}{2}\Gamma_{H,L})t}.
\end{eqnarray}

The total time integrated CP asymmetry is
\begin{equation}
\epsilon_\alpha^{QM} = \frac{\displaystyle \sum_{i=1,2,a_\alpha} 
\Gamma(\widetilde{N}_i \rightarrow a_\alpha)
- \Gamma(\widetilde{N}_i \rightarrow \bar{a}_\alpha)}
{\displaystyle \sum_{i=1,2,a_\beta,\beta} 
\Gamma(\widetilde{N}_i \rightarrow a_\beta)
+ \Gamma(\widetilde{N}_i \rightarrow \bar{a}_\beta)} \ , 
\label{eq:qmcp_asym_total} 
\end{equation}
where the time integrated  rates 
$\Gamma(\widetilde{N}_i \rightarrow a_\alpha)$ 
can be obtained 
from Eq.~\eqref{eq:arbitrary_time_evolution}: 
\begin{eqnarray}
\Gamma(\widetilde{N}_i \rightarrow a_\alpha) 
& = & \frac{1}{4}\frac{c^{a_\alpha}}{16\pi M}
\biggl( \left|A^{a_\alpha}_{i}\right|^{2}G_{i+}+
\left|A_{j\neq i}^{a_\alpha}\right|^{2}G_{j-} 
\nonumber \\
&  & +2\left[\mbox{Re}
\left({A^{a_\alpha}_{i}}^{*}A^{a_\alpha}_{j\neq i}
\right)G_{ii}^{R} -\mbox{Im}
\left({A^{a_\alpha}_{i}}^{*}A^{a_\alpha}_{j\neq i}
\right)G_{ii}^{I}\right]\biggr), 
\label{eq:intdecrat}
\end{eqnarray}
and the rates 
$\Gamma(\widetilde{N}_i \rightarrow \bar a_\alpha)$ 
for  antiparticles are obtained 
% from Eq.~\eqref{eq:intdecrat} with the 
by replacing 
$A^{a_\alpha}_{i}\rightarrow  \overline{A^{\bar a_\alpha}_{i}}$.
In Eq.~\eqref{eq:intdecrat} we have introduced 
the time integrated projections 
\begin{eqnarray}
G_{1(2)+}  & = & 
2\left(\frac{1}{1-y^{2}}+\frac{1}{1+x^{2}}\right)+ 2
\left|C_{0}\right|^{2}\left(\frac{1}{1-y^{2}}
-\frac{1}{1+x^{2}}\right)\nonumber \\
 &  & \pm 8\left[\mbox{Re}
\left(C_{0}\right)\frac{y}{1-y^{2}}
-\mbox{Im}\left(C_{0}\right)\frac{x}{1+x^{2}}\right],\label{eq:G_plus}\\
G_{1(2)-}  & = & 2\left|C_{1,2}\right|^{2}
\left(\frac{1}{1-y^{2}}-\frac{1}{1+x^{2}}\right),
\label{eq:G_minus} \\
G_{11(22)}^{R} 
 & = & 2\left\{ \mbox{Re}
\left[e^{\mp i\theta}C_{1(2)}\right]
\frac{y}{1-y^{2}}-\mbox{Im}\left[e^{\mp i\theta}
C_{1(2)}\right]\frac{x}{1+x^{2}}\right\} \nonumber \\
 &  & \pm 2
\mbox{Re}\left[e^{\mp i\theta}
C_{0}^{*}C_{1(2)}\right]\left(\frac{1}{1-y^{2}}
-\frac{1}{1+x^{2}}\right),\label{eq:G_real}\\
G_{11(22)}^{I} 
 & = & 2\left\{ 
\mbox{Im}\left[e^{\mp i\theta}C_{1(2)}\right]
\frac{y}{1-y^{2}}+\mbox{Re}\left[e^{\mp i\theta}C_{1(2)}
\right]\frac{x}{1+x^{2}}\right\} \nonumber \\
 &  & \pm 2
\mbox{Im}\left[e^{\mp i\theta}C_{0}^{*}C_{1(2)}\right]
\left(\frac{1}{1-y^{2}}-\frac{1}{1+x^{2}}\right),
\label{eq:G_factors}
\end{eqnarray}
written in terms of the mass and width differences\footnote{ We use
  the expression of $\Gamma_H-\Gamma_L$ from
  Ref.~\refcite{Grossman:2004}.  Notice that with this definition
  $\Gamma_H-\Gamma_L\neq \Gamma_{\widetilde N_+}-\Gamma_{\widetilde
    N_-}$ where $\Gamma_{\widetilde N_\pm}$ is defined in
  Eq.~\eqref{eq:decay_width_pm}.}:
\begin{eqnarray}
\label{eq:xy}
x & = & \frac{M_H-M_L}{\Gamma}
=\frac{B}{\Gamma}
-\frac{1}{2}\frac{\Gamma A ^{2}}{BM^{2}}\sin^{2}\left(\phi_{A}\right), 
\nonumber \\
y & = & \frac{\Gamma_H-\Gamma_L}{2\Gamma}=\frac{ A}{M}
\cos\left(\phi_{A}\right)-\frac{B}{2M}.
\end{eqnarray}
Using Eqs.~\eqref{eq:intdecrat}--\eqref{eq:G_factors} one can 
write the numerator in Eq.~\eqref{eq:qmcp_asym_total} as
\begin{equation}
\sum_i \Gamma(\widetilde{N}_i \rightarrow a_\alpha)
- \Gamma(\widetilde{N}_i \rightarrow \bar{a}_\alpha)\equiv
\Delta\Gamma^{a_\alpha,R}
+\Delta\Gamma^{a_\alpha,NR}+\Delta\Gamma^{a_\alpha,I},
\end{equation}
with 
\begin{eqnarray}
\Delta\Gamma^{a_\alpha,R} 
& = & \frac{1}{2}\frac{c^{a_\alpha}}{16 \pi M} 
\frac{x^{2}+y^{2}}{\left(1-y^{2}\right)\left(1+x^{2}\right)}
%\nonumber \\
%&  & 
\Biggl\{\left|C_{0}\right|^{2} {\cal F}_{1+}
%\left(\left|A^{a_\alpha}_{1}\right|^{2}
%-\left|\overline{A^{\bar a_\alpha}_{1}}\right|^{2}
%+\left|A^{a_\alpha}_{2}\right|^{2}
%-\left|\overline{A^{\bar a_\alpha}_{2}}\right|^{2}
%\right)
%\nonumber \\
%&  & 
-\frac{\left(\left|C_{1}\right|^{2}-\left|C_{2}\right|^{2}\right)}{2}
{\cal F}_{1-}
%\left(\left|A^{a_\alpha}_{1}\right|^{2}
%-\left|\overline{A^{\bar a_\alpha}_{1}}\right|^{2}
%-\left|A^{a_\alpha}_{2}\right|^{2}
%+\left|\overline{A^{\bar a_\alpha}_{2}}\right|^{2}
%\right)
\nonumber \\
&  & +2 \left[\mbox{Re} {\cal F}_{2-}
%\left({A^{a_\alpha}_{1}}^{*}A^{a_\alpha}_{2}
%-\overline{A^{\bar a_\alpha}_{1}}^{*}\overline{A^{\bar a_\alpha}_{2}}\right)
\mbox{Re}\left(e^{-i\theta}C_{0}^{*}C_{1}\right) \right.
%\nonumber \\
%&  & 
\left. 
-\mbox{Re} {\cal F}_{3+}
%\left({A^{a_\alpha}_{2}}^{*}A^{a_\alpha}_{1}
%+\overline{A^{\bar a_\alpha}_{2}}^{*}\overline{A^{\bar a_\alpha}_{1}}\right)
\mbox{Re}\left(e^{i\theta}C_{0}^{*}C_{2}\right)\right]
\nonumber \\
&  & 
-2 \left[\mbox{Im} {\cal F}_{2-}
%\left({A^{a_\alpha}_{1}}^{*}A^{a_\alpha}_{2}
%-\overline{A^{\bar a_\alpha}_{1}}^{*}\overline{A^{\bar a_\alpha}_{2}}\right)
\mbox{Im}\left(e^{-i\theta}C_{0}^{*}C_{1}\right) \right.
%\nonumber \\
%&  & 
\left. -\mbox{Im} {\cal F}_{3+}
%\left({A^{a_\alpha}_{2}}^{*}A^{a_\alpha}_{1}
%+\overline{A^{\bar a_\alpha}_{2}}^{*}\overline{A^{\bar a_\alpha}_{1}}\right)
\mbox{Im}\left(e^{i\theta}C_{0}^{*}C_{2}
\right)\right] \Biggr\},
\label{eq:delgm}  
\end{eqnarray}
\begin{eqnarray}
\Delta\Gamma^{a_\alpha,NR}
& = & \frac{c^{a_\alpha}}{16 \pi M} 
\frac{1}{\left(1-y^{2}\right)}\Biggl\{
2 y \mbox{Re}(C_0) {\cal F}_{1-}
%\left(\left|A^{a_\alpha}_{1}\right|^{2}
%-\left|\overline{A^{\bar a_\alpha}_{1}}\right|^{2}
%-\left|A^{a_\alpha}_{2}\right|^{2}
%+\left|\overline{A^{\bar a_\alpha}_{2}}\right|^{2}
%\right)
%\nonumber \\
%&  & 
+ {\cal F}_{1+}
%\left(\left|A^{a_\alpha}_{1}\right|^{2}
%-\left|\overline{A^{\bar a_\alpha}_{1}}\right|^{2}
%+\left|A^{a_\alpha}_{2}\right|^{2}
%-\left|\overline{A^{\bar a_\alpha}_{2}}\right|^{2}
%\right)
\nonumber  \\
&&+ y\left[
\mbox{Re} {\cal F}_{2-}
%\left({A^{a_\alpha}_{1}}^{*}A^{a_\alpha}_{2}
%-\overline{A^{\bar a_\alpha}_{1}}^{*}\overline{A^{\bar a_\alpha}_{2}}\right)
\mbox{Re}\left(e^{-i\theta}C_{1}\right) \right.
%\nonumber \\
%&  & 
\left.+\mbox{Re}{\cal F}_{3-}
%\left({A^{a_\alpha}_{2}}^{*}A^{a_\alpha}_{1}
%-\overline{A^{\bar a_\alpha}_{2}}^{*}\overline{A^{\bar a_\alpha}_{1}}\right)
\mbox{Re}\left(e^{i\theta}C_{2}\right)\right]
\nonumber \\
&  & 
-y \left[
\mbox{Im}{\cal F}_{2-}
%\left({A^{a_\alpha}_{1}}^{*}A^{a_\alpha}_{2}
%-\overline{A^{\bar a_\alpha}_{1}}^{*}\overline{A^{\bar a_\alpha}_{2}}\right)
\mbox{Im}\left(e^{-i\theta}C_{1}\right) \right.
%\nonumber \\
%&  & 
\left. +\mbox{Im} {\cal F}_{3-}
%\left({A^{a_\alpha}_{2}}^{*}A^{a_\alpha}_{1}
%-\overline{A^{\bar a_\alpha}_{2}}^{*}\overline{A^{\bar a_\alpha}_{1}}\right)
\mbox{Im}\left(e^{i\theta}C_{2}
\right)\right]\Biggr\}, 
\label{eq:delgd}
\end{eqnarray}
\begin{eqnarray}
\Delta\Gamma^{a_\alpha,I} & = &
\frac{c^{a_\alpha}}{16 \pi M} 
\frac{x}{\left(1+x^{2}\right)}
\Biggl\{-2\mbox{Im}(C_0) {\cal F}_{1-}
%\left(\left|A^{a_\alpha}_{1}\right|^{2}
%-\left|\overline{A^{\bar a_\alpha}_{1}}\right|^{2}
%-\left|A^{a_\alpha}_{2}\right|^{2}
%+\left|\overline{A^{\bar a_\alpha}_{2}}\right|^{2}
%\right) 
\nonumber  \\ 
&  & - 
\left[\mbox{Re}{\cal F}_{2-}
%\left({A^{a_\alpha}_{1}}^{*}A^{a_\alpha}_{2}
%-\overline{A^{\bar a_\alpha}_{1}}^{*}\overline{A^{\bar a_\alpha}_{2}}\right)
\mbox{Re}\left(e^{-i\theta}C_{1}\right) \right.
%\nonumber \\
%&  & 
\left.+\mbox{Re}{\cal F}_{3-}
%\left({A^{a_\alpha}_{2}}^{*}A^{a_\alpha}_{1}
%-\overline{A^{\bar a_\alpha}_{2}}^{*}\overline{A^{\bar a_\alpha}_{1}}\right)
\mbox{Re}\left(e^{i\theta}C_{2}\right)
\right]
\nonumber \\
&  & 
- \left[
\mbox{Im}{\cal F}_{2-}
%\left({A^{a_\alpha}_{1}}^{*}A^{a_\alpha}_{2}
%-\overline{A^{\bar a_\alpha}_{1}}^{*}\overline{A^{\bar a_\alpha}_{2}}\right)
\mbox{Im}\left(e^{-i\theta}C_{1}\right) \right.
%\nonumber \\
%&  & 
\left.+\mbox{Im} {\cal F}_{3-}
%\left({A^{a_\alpha}_{2}}^{*}A^{a_\alpha}_{1}
%-\overline{A^{\bar a_\alpha}_{2}}^{*}\overline{A^{\bar a_\alpha}_{1}}\right)
\mbox{Im}\left(e^{i\theta}C_{2}
\right)\right]\Biggr\} 
\label{eq:delgi} \;. 
\end{eqnarray}
where
\begin{eqnarray}
{\cal F}_{1\pm}&=&
\left|A^{a_\alpha}_{1}\right|^{2}
-\left|\overline{A^{\bar a_\alpha}_{1}}\right|^{2}
\pm\left|A^{a_\alpha}_{2}\right|^{2} 
\mp\left|\overline{A^{\bar a_\alpha}_{2}}\right|^{2} \,, \\
{\cal F}_{2\pm}&=&
{A^{a_\alpha}_{1}}^{*}A^{a_\alpha}_{2}
\pm\overline{A^{\bar a_\alpha}_{1}}^{*}\overline{A^{\bar a_\alpha}_{2}} \,, \\ 
{\cal F}_{3\pm}&=&
{A^{a_\alpha}_{2}}^{*}A^{a_\alpha}_{1}
\pm\overline{A^{\bar a_\alpha}_{2}}^{*}\overline{A^{\bar a_\alpha}_{1}} \,.
\end{eqnarray}
In writing the above equations we have classified the contributions as
{\sl resonant} ($R$) if they include an overall factor
$\frac{x^2+y^2}{1+x^2}$ and {\sl non-resonant} ($NR$) if no factor of
$\frac{1}{1+x^2}$ is present, while the remainder has been labeled as
{\sl interference} ($I$).
 After substituting the explicit values for the amplitudes and the
coefficients, and neglecting all the terms that cancel in both
bases,  the following relations are obtained:
\begin{eqnarray}
\Delta\Gamma^{f_\alpha,R}=-c^{f} \Delta\Gamma_\alpha^{R},
&&\;\;\;\;\;\;
\Delta\Gamma^{s_\alpha,R}=c^{s} \Delta\Gamma_\alpha^{R}, \nonumber \\
\Delta\Gamma^{f_\alpha,NR}=-c^{f} \Delta\Gamma_\alpha^{NR},
&&\;\;\;\;\;\;
\Delta\Gamma^{s_\alpha,NR}=c^{s} \Delta\Gamma_\alpha^{NR}, \nonumber \\
\Delta\Gamma^{f_\alpha,I}=-c^{f} \Delta\Gamma_\alpha^{I},
&&\;\;\;\;\;\;
\Delta\Gamma^{s_\alpha,I}=c^{s} \Delta\Gamma_\alpha^{I},  
\label{eq:qmcancel}
\end{eqnarray}
with
\begin{eqnarray}  
\Delta\Gamma_\alpha^{R}
& = & -\frac{1}{4\pi}Y_{\alpha}^2 
\left[ (g^2 - h^2)^2+ (2gh)^2 \cos (2\theta)\right]
 A \sin(\phi_A) 
 \\ &  & 
 \times \frac{1}{x} 
\frac{x^2+y^2}{(1-y^2)(1+x^2)}, \\
\Delta\Gamma_\alpha^{NR} 
& = &\frac{3}{16\pi} Y_{\alpha}^2 \alpha_2  \ln\frac{m_2^2}{m_2^2+M^2} 
\frac{m_2}{M}  \frac{1}{1-y^2}
\left[- A \sin(\phi_A+2\phi_g) \right.\nonumber \\
&  & \;\; \left. + y M 
\left(2 (2gh)^2+ (g^2-h^2)^2\cos(2\theta)\right)
\sin(2\phi_g)\right], \\
\Delta\Gamma_\alpha^{I}
& = & \frac{3}{16\pi} Y_{\alpha}^2 
\alpha_2 \ln\frac{m_2^2}{m_2^2+M^2} 
\frac{m_2}{M} \frac{1}{1+x^2} A
\nonumber \\
&  & \times \sin(\phi_A)\cos(2\theta)\cos(2\phi_g). 
\end{eqnarray}      
Eqs.~\eqref{eq:qmcancel} explicitly show that the $T=0$ cancellation
of the CP asymmetries occurs also in the QM formalism in both cases of
RHSN as initial mass or weak eigenstates.  Given that the dependence
on the thermal factor $\Delta_{BF}(T)$ Eq.~\eqref{eq:therm_factor} is
the same as in the field-theoretical approach and, after normalizing
to the total decay width, the same projectors $P_\alpha$
Eq.~\eqref{eq:fla_proj} multiply the CP asymmetries, the results can
again be recast in terms of flavour and temperature independent
quantities $\bar \epsilon$ defined as:
\begin{equation}
  \label{eq:bar_epsilon}
\epsilon_\alpha^{(C)QM(is)}(T)=
P_\alpha \; \bar \epsilon^{(C)QM(is)}\;\Delta_{BF}(T),
\end{equation}
where the superscript $(C)=R,\,NR,\,I$ refers to the resonant,
non-resonant, and interference contributions, while $(is)=w,m$ refers
to the case of weak or mass RHSN initial states.  Substituting the
values for the coefficients for initial weak RHSN, together with the
expressions for $x$ and $y$ in Eqs.~\eqref{eq:xy} and expanding at order
$\delta_S^2$, one gets
\begin{eqnarray} 
\bar\epsilon^{R,QM,w} 
& = & -\frac{A}{M}\sin\left(\phi_{A}\right)
\frac{B\Gamma}{B^{2}+\Gamma^{2}},
\label{eq:qmw_asym_2}\\
\bar\epsilon^{NR,QM,w} 
& = & -\frac{3\alpha_{2}}{4}\frac{m_2}{M}
\ln\frac{m_2^2}{m_2^2+M^2}\Big[\frac{A}{M}\sin(\phi_{A})
\cos(2\phi_{g})
+\frac{B}{2M}\sin\left(2\phi_{g}\right)\Big],\quad 
\label{eq:qmw_asym_1} \\
\bar\epsilon^{I,QM,w} 
& = & \frac{3\alpha_{2}}{4}\frac{m_2}{M}
\frac{A}{M}\ln\frac{m_2^2}{m_2^2+M^2}
\sin\left(\phi_{A}\right)
\mbox{cos}\left(2\phi_{g}\right) 
\frac{\Gamma^{2}}{B^{2}+\Gamma^{2}} \;.\quad 
\label{eq:qmw_asym_3}
\end{eqnarray}
Correspondingly, for initial $\widetilde N_\pm$ states one gets: 
\begin{eqnarray} 
\bar\epsilon^{R,QM,m}&=&\frac{A}{M}\sin\left(\phi_{A}\right)
\frac{B\Gamma}{B^{2}+\Gamma^{2}},
\label{eq:qmm_asym_2}\\
\bar\epsilon^{NR,QM,m}&=&-\frac{3\alpha_{2}}{4}\frac{m_2}{M}
\ln\frac{m_2^2}{m_2^2+M^2}\Big[\frac{A}{M}\sin(\phi_{A})
\cos(2\phi_{g})+\frac{B}{2M}\sin\left(2\phi_{g}\right)\Big],\quad  
\label{eq:qmm_asym_1} \\
\bar\epsilon^{I,QM,m}&=&-\frac{3\alpha_{2}}{4}\frac{m_2}{M}
\frac{A}{M}\ln\frac{m_2^2}{m_2^2+M^2}\sin\left(\phi_{A}\right)
\mbox{cos}\left(2\phi_{g}\right)\frac{\Gamma^{2}}{B^{2}+\Gamma^{2}}\;.
\label{eq:qmm_asym_3}
\end{eqnarray}
Comparing Eqs.~\eqref{eq:qmm_asym_2}--\eqref{eq:qmm_asym_3} with
Eqs.~\eqref{eq:qmw_asym_2}--\eqref{eq:qmw_asym_3} and
Eqs.~\eqref{eq:cp0_S}--\eqref{eq:cp0_I} one sees that the parametric
dependence is very similar, although there are some differences in the
numerical coefficients.  In particular in either the weak or mass
basis $\bar\epsilon^{R,QM}$, $\bar\epsilon^{I,QM}$ and the
$B$-dependent (second term) in $\bar\epsilon^{NR,QM}$ coincide with
$\bar\epsilon^{S}$, $\bar\epsilon^{I}$ and the B-dependent term in
$\bar\epsilon^{V}$ derived in the previous section, modulo the
redefinition $A\rightarrow 2 A$, $B\rightarrow 2 B$ and
$\sin(\phi_A)\rightarrow\pm \sin(\phi_A)$.  There are however, some
differences in the phase combination which appears in the $B$
independent term in the asymmetries $\bar\epsilon_\alpha^{NR,QM}$ and
$\bar\epsilon_\alpha^{V}$ as seen in Eqs.~\eqref{eq:cp_asym_1},
\eqref{eq:qmw_asym_1} and \eqref{eq:qmm_asym_1}.  In other words, the
choice of initial state only leads to minor differences. But the
crucial role of thermal effects to avoid exact cancellations and to
allow for a non-vanishing CP asymmetry is the same in both the QM and
field-theoretical approaches, and is independent of the particular
basis chosen for the initial RHSN states.

%%%%%%%%%%%%%%%%%%%%%%%%%%%%%%%%%%%%%%%%%%%%%%%%%%%%%%%%%%%%%%%%%%

\subsection{The vanishing of the CP asymmetry in decays at $T=0$}
\label{sec:vertexCP}

As we have seen in the previous two sections, the original claim that
the sources of direct CP violation from vertex corrections involving
the gauginos do not require thermal effects to produce a sizable
lepton asymmetry in the plasma\cite{Grossman:2004} is incorrect, and
after including vertex corrections the CP asymmetries for
decays into scalars and into fermions still cancel in the $T=0$
limit.  This issue is of some interest, because if thermal corrections
are necessary for SL to work, then non-thermal scenarios, like the
ones in which RHSN are produced by inflaton decays and the
thermal bath remains at a temperature $T \ll M$ during the following
leptogenesis epoch, would be completely excluded. In the following, we
present a simple but general argument proving that at $T=0$ the direct
leptonic CP violation in RHSN decays vanishes at one loop, due
to an exact cancellation between the scalar and fermion contributions.

%%%%%%%%%%%%%%%%%%%%%%%%%%%%%%%%%%%%%%%%%%%%%%%%%%%%%%%%%%%%%%%%%%
% PANEL
\begin{figure}[t!]
\centering
\includegraphics[width=\textwidth]{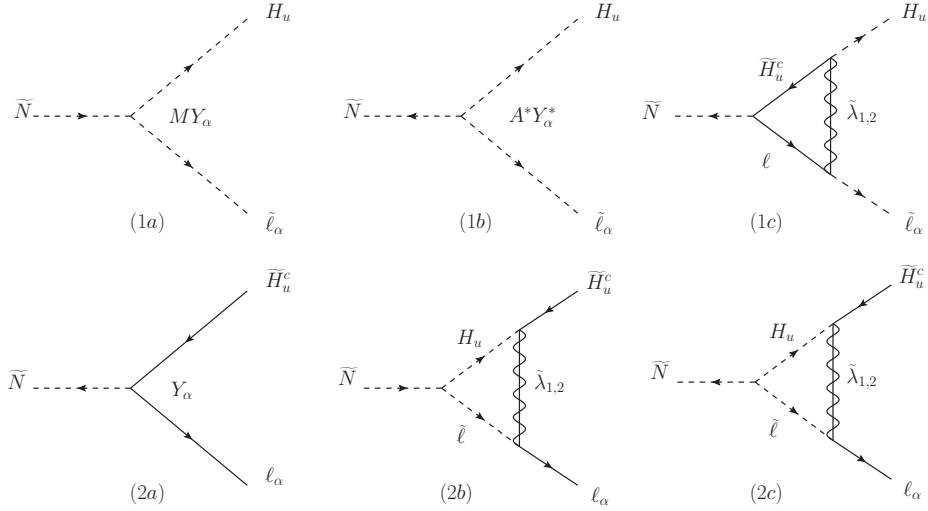}
\caption{ 
Soft leptogenesis diagrams for RHSN decays into
scalars $(1a)$, $(1b)$, $(1c)$ and into fermions 
$(2a)$, $(2b)$, $(2c)$.}
\label{fig:1}
\end{figure} 

%%%%%%%%%%%%%%%%%%%%%%%%%%%%%%%%%%%%%%%%%%%%%%%%%%%%%%%%%%%%%%%%%%

Let us take for simplicity $\Phi=0$ in Eq.~\eqref{eq:mass_eigenstates} 
(this amounts to assign the phases $\phi_{A}$ 
and $\phi_g$ in Eqs.~\eqref{eq:CP_phase1} 
and \eqref{eq:CP_phase2} respectively to $A$ and
$m_2$)\footnote{Here we only consider the contributions from
$SU(2)_L$ gauginos since for $U(1)_Y$ gaugino 
the proof proceeds in exactly the same way.}. 
Since the lepton flavour $\alpha$ will not play a role in this proof,
we will suppress in this section the corresponding label.
Let us introduce for the various amplitudes the shorthand
notation $A^{\pm}_\ell \equiv A(\widetilde{N}_\pm\to \ell 
\widetilde H_u^c)$, $A^{\widetilde{N}\, (\widetilde{N}^*)}_\ell \equiv
A\left(\widetilde{N}\,(\widetilde{N}^*) \to \ell 
\widetilde H_u^c\right)$ with similar expressions for the other final
states. From \Eqn{eq:mass_eigenstates} we can write
\begin{eqnarray}
  \label{eq:Apmell}
 2\, \left|A^\pm_\ell\right|^2 &=& 
\left|A^{\widetilde{N}}_\ell\right|^2+
\left|A^{\widetilde{N}^*}_\ell\right|^2 
\pm  2\, \mathrm{Re}
\left(A^{\widetilde{N}}_\ell
\cdot 
A^{\widetilde{N}}_{\bar\ell}\right)\,,
\\
  \label{eq:Apmbell}
 2\, \left|A^\pm_{\bar \ell}\right|^2 &=& 
\left|A^{\widetilde{N}}_{\bar \ell}\right|^2+
\left|A^{\widetilde{N}^*}_{\bar \ell}\right|^2 
\pm  2\, \mathrm{Re}
\left(A^{\widetilde{N}}_{\bar \ell}
\cdot 
A^{\widetilde{N}}_{\ell}\right)\,, 
\end{eqnarray}
where the complex conjugate amplitudes in the last terms of both these
equations have been rewritten as follows:
$(A^{\widetilde{N}^*}_\ell)^*= A_{\widetilde{N}^*}^\ell=
A^{\widetilde{N}}_{\bar\ell}$ and $(A^{\widetilde{N}^*}_{\bar
  \ell})^*= A_{\widetilde{N}^*}^{\bar \ell}= A^{\widetilde{N}}_{\ell}$
by using CPT invariance in the second step.  The direct CP asymmetry for
$\widetilde{N}_\pm$ decays into fermions is given by the difference
between Eqs.~\eqref{eq:Apmell} and \eqref{eq:Apmbell}:
\begin{equation}
  \label{eq:Af}
  2\,\left(
\left|A^\pm_{\ell}\right|^2-
\left|A^\pm_{\bar \ell}\right|^2\right)=
\left(\left|A^{\widetilde{N}^*}_\ell\right|^2-
\left|A^{\widetilde{N}^*}_{\bar \ell}\right|^2\right)
+
\left(\left|A^{\widetilde{N}}_\ell\right|^2-
\left|A^{\widetilde{N}}_{\bar \ell}\right|^2\right)\,.
\end{equation}
With the replacements $\ell \to \tilde\ell$ and 
$\bar\ell\to\tilde\ell^*$, a completely equivalent expression
holds also for the decays into scalars. 

The tree-level and one-loop diagrams for the various decay amplitudes
into scalars and fermions are given in Fig.~\ref{fig:1}.  We note at
this point that $A^{\widetilde N}_{\widetilde \ell}$ has no one-loop
amplitude to interfere with (see diagram $(1\,a)$) and thus, up to
one-loop, the full amplitude coincides with the tree-level result, and
is CP conserving.  $A^{\widetilde N}_{\ell}$ is a pure one-loop
amplitude (see diagram $(2\,c)$) and therefore is also CP
conserving. It follows that:
\begin{equation}
  \label{eq:tree}
\left| A^{\widetilde N}_{\widetilde \ell}\right|^2 =  
 \left|  A^{\widetilde N^*}_{\widetilde \ell^*}\right|^2\,, 
% \\    \label{eq:loop}
\qquad {\rm and} \qquad
\left|A^{\widetilde N}_{\ell}\right|^2  =
  \left| A^{\widetilde N^*}_{\bar \ell}\right|^2\,. 
\end{equation}
We can thus change simultaneously the signs of $\left|A^{\widetilde
    N}_{\ell}\right|^2$ and $\left| A^{\widetilde N^*}_{\bar
    \ell}\right|^2$ in~\eqn{eq:Af} without affecting the equality, and
the same we can do in the analogous equation for the 
scalars. This gives:
\begin{eqnarray}
  \label{eq:Af1}
  2\,\left(
\left|A^\pm_{\ell}\right|^2-
\left|A^\pm_{\bar \ell}\right|^2\right) &=& 
\left(\left|A^{\widetilde{N}^*}_\ell\right|^2+
\left|A^{\widetilde{N}^*}_{\bar \ell}\right|^2\right)
-
\left(\left|A^{\widetilde{N}}_\ell\right|^2+
\left|A^{\widetilde{N}}_{\bar \ell}\right|^2\right)\,, \\
  \label{eq:As1}
  2\,\left(
\left|A^\pm_{\tilde \ell}\right|^2-
\left|A^\pm_{\tilde \ell^*}\right|^2\right) &=&
\left(\left|A^{\widetilde{N}^*}_{\tilde\ell}\right|^2+
\left|A^{\widetilde{N}^*}_{\tilde \ell^*}\right|^2\right)
-
\left(\left|A^{\widetilde{N}}_{\tilde\ell}\right|^2+
\left|A^{\widetilde{N}}_{\tilde \ell^*}\right|^2\right)\,.
\end{eqnarray}
Using CPT invariance
\bea
\left|A^{\widetilde{N}^*}_\ell\right|^2+
\left|A^{\widetilde{N}^*}_{\bar \ell}\right|^2
& = & 
\left|A_{\widetilde{N}}^{\bar\ell}\right|^2+
\left|A_{\widetilde{N}}^{\ell}\right|^2,
\\
\left|A^{\widetilde{N}^*}_{\tilde\ell}\right|^2+
\left|A^{\widetilde{N}^*}_{\tilde \ell^*}\right|^2
& = &
\left|A_{\widetilde{N}}^{\tilde\ell^*}\right|^2+
\left|A_{\widetilde{N}}^{\tilde \ell}\right|^2,
\eea
and unitarity
\bea
\hspace{-4mm}
\left|A_{\widetilde{N}}^{\bar\ell}\right|^2+
\left|A_{\widetilde{N}}^{\ell}\right|^2 +
\left|A_{\widetilde{N}}^{\tilde\ell^*}\right|^2+
\left|A_{\widetilde{N}}^{\tilde \ell}\right|^2
& = & 
\left|A^{\widetilde{N}}_{\bar\ell}\right|^2+
\left|A^{\widetilde{N}}_{\ell}\right|^2 +
\left|A^{\widetilde{N}}_{\tilde\ell^*}\right|^2+
\left|A^{\widetilde{N}}_{\tilde \ell}\right|^2 \!,
\eea
we can readily see that the sum of zero temperature fermionic CP asymmetry 
Eq.~\eqref{eq:Af1} and scalar CP asymmetry Eq.~\eqref{eq:As1} vanishes.  
We have thus proved that for
$\widetilde{N}_+$ and $\widetilde{N}_-$ independently, at one loop
there is an exact cancellation between the scalar and fermion final
state contributions, and thus at $T=0$ the direct decay CP
asymmetries vanish.

\section{One-flavour Approximation and Superequilibration Regime}\label{sec:unflavored}

\subsection{Effective theories in the early Universe}
\label{sec:effective}

In the expanding early Universe, at each temperature $T$ is associated
a characteristic time scale given by the Universe age
$t_U(T)\sim H^{-1}(T)$ ($H(T)$ being the Hubble parameter at $T$). 
Particle reactions must be treated in a different way depending if
their characteristic time scale $\tau$ (given by inverse of their
their thermally averaged rates) is:
\begin{itemize} \itemsep=0pt
\item[(i)] Much shorter than the age of the Universe:\quad $\tau \ll t_U(T)$; 
\item[(ii)] Much larger than the age of the Universe:\quad $\tau \gg t_U(T)$; 
\item[(iii)] Comparable with the  Universe age:\quad $\tau \sim t_U(T)$.
\end{itemize}
The first type of reactions (i) occur very frequently during one
expansion time and their effects can be simply `resummed' by imposing
on the thermodynamic system the chemical equilibrium condition
appropriate for each specific reaction, that is $\sum_I \mu_I = \sum_F
\mu_F$, where $\mu_I$ denotes the chemical potential of an initial
state particle, and $\mu_F$ that of a final state particle. The
numerical values of the parameters that are responsible for these
reactions only determine the precise temperature $T$ when chemical
equilibrium is attained and the resummation of all effects into
chemical equilibrium conditions holds but, apart from this, have no
other relevance, and do not appear explicitly in the effective
formulation of the problem.

Reactions of the second type (ii) cannot have any effect on the
system, since they basically do not occur.  Then all physical
processes are blind to the corresponding parameters, that can be set
to zero in the effective Lagrangian.  By doing this, it is then easy
to read out if new global symmetries appear and, if no anomalies are
involved, these symmetries correspond to exactly conserved quantities.
The corresponding conservation laws must be respected by the
equations describing the dynamics of the system.

Reactions of the third type (iii) in general violate some symmetries,
and thus spoil the corresponding conservation conditions, but are not
fast enough to enforce chemical equilibrium conditions.  Only
reactions of this type appear explicitly in the formulation of the
problem (they generally enter into a set of Boltzmann equations for
the evolution of the system) and only the corresponding parameters
represent fundamental quantities in the specific effective theory.

Several examples of the importance of using the appropriate early
Universe effective theory can be found in leptogenesis studies.
Leptogenesis was first formulated in the so-called `one flavour
approximation' in which a single $SU(2)_L$ lepton doublet of an
unspecified flavour is assumed to couple to the lightest singlet
seesaw neutrino, and it is thus responsible for the generation of the
lepton asymmetry. Indeed, until the works in
Refs.~\refcite{Abada:2006a,Nardi:2006b}, most leptogenesis studies
were carried out within this framework.  Nowadays, it is well
understood that the `one flavour approximation' gives a rather rough
and often unreliable description of leptogenesis dynamics in the
regime when the flavours of the leptons are identified by
in-equilibrium charged leptons Yukawa reactions.  This is because such
an `approximation' has no control over the effects that are neglected,
and thus the related uncertainty cannot be estimated.  On the other
hand, if leptogenesis occurs above $T\sim 10^{12}\,$GeV, when all the
charged leptons Yukawa reactions have characteristic time scales
much larger than $t_U$, the `one flavour approximation' is not at all
an approximation. Rather, it is the correct high temperature effective
theory that must be used to compute the baryon asymmetry.  The
corresponding effective Lagrangian is obtained by setting to zero, in
the first place, all the charged lepton Yukawa couplings, so that the only
remaining flavour structure is determined by the Yukawa couplings of
the heavy Majorana neutrinos.

In supersymmetric leptogenesis instead, the effective theory that was
generally used was in fact only appropriate for temperatures much
lower than the typical temperatures $T\gg 10^8\,$GeV in which
leptogenesis can be successful, and only quite recently it was
clarified that in the relevant temperature range a completely
different effective theory holds instead.\cite{Fong:2010qh} More
specifically, it was always assumed that lepton-slepton reactions like
e.g. $\ell\ell \leftrightarrow \tilde\ell\tilde\ell$ that are induced
by soft SUSY-breaking gaugino masses are in thermal
equilibrium, and this implies equilibration between the leptons and
sleptons density asymmetries ({\sl superequilibration}).
Superequilibration (SE) instead, only occurs below $T\sim 10^7\,$GeV,
and thus supersymmetric leptogenesis always proceeds in the
non-superequilibration (NSE) regime.

As regards SL, it always occurs in a temperature regime in which the
charged lepton Yukawa couplings cannot be set to zero, and thus
flavour effects must be taken into account, while, since SL can be
successful from $T\sim 10^8\,$GeV downwards, the two possibilities
that it will occur in the SE or in the NSE regimes remain open.

Here, as it was done in the original
formulation,\cite{Grossman:2003,DAmbrosio:2003} we first describe SL
taking into account only reactions of type (iii). That is, we will
neglect all considerations about flavour effects, that are related to
reactions of type (i), as well as NSE effects, that are related to
reactions of type (ii). These two issues are addressed respectively in 
Section~\ref{sec:flavor} and in Section~\ref{sec:nse}.

\subsection{Boltzmann equations in the unflavoured approximation}
\label{sec:unflavoredBE} 

In order to quantify the parameter ranges in which SL is successful
one needs to solve the relevant set of Boltzmann equations (BE).  All
technical details about the BE for SL are given
in~\ref{app:boltzmanneqs_new}.

To eliminate the dependence on the expansion of the Universe it is
customary to recast the BE in terms of the variables $Y_X=n_X/s$, that
is in terms of particle number densities $n_X$ normalized to the
entropy density $s=\frac{2\pi^2}{45}g_{*}T^3$ where $g_*$
is the total number of relativistic degrees of freedom.
To account for  the sources of violation
of lepton number  one then needs to follow the evolution of
$Y_{\widetilde{N}_i}$ and, since RHN decays are also $\Delta L=1$ 
processes, the evolution of $Y_N$ must also be considered.

To simplify the understanding of how a sizable density asymmetry is
dynamically generated it is convenient to adopt a certain number of
approximations.

The first approximation is to neglect lepton flavours, and work in the
so-called `one flavour approximation'.  The relevant quantities one
wants to estimate in this case are the fermionic $Y_{\Delta\ell}$ and
scalar $Y_{\Delta\widetilde\ell}$ lepton asymmetries generated in the
leptonic states coupled to the RHSN (that in general correspond to a
superposition of the different lepton flavours).  They are defined
respectively as
$Y_{\Delta\ell}=\left(Y_\ell-Y_{\bar\ell}\right)/g_\ell$ and
$Y_{\Delta\widetilde\ell}=\left(Y_{\widetilde\ell}-Y_{\widetilde\ell^*}
\right)/g_{\widetilde\ell}$, that is, we define the density
asymmetries
%!!%
for single $SU(2)_L$ degree of freedom, with 
$g_\ell=g_{\widetilde\ell}=2$.
%where a sum over the $SU(2)_L$ degrees of freedom is understood.

The second approximation is to neglect all ``spectator
effects''.\cite{Buchmuller:2001,Nardi:2006a}  Of course, besides the
lepton density asymmetries, many other asymmetries related to the
finite chemical potentials of the Higgs, higgsinos, quarks and squarks,
$SU(2)_L$ singlet leptons and sleptons, are also present in the
plasma, and affect indirectly the outcome of SL through the so-called
spectator effects.\cite{Buchmuller:2001,Nardi:2006a}  In this
section all effects of this type will be neglected, which amounts to
assume that all particles except the heavy (s)neutrinos and the
$SU(2)_L$ doublet (s)leptons follow either Bose-Einstein or Fermi-Dirac
distribution with vanishing chemical potential $f = (e^{E/T}\mp
1)^{-1}$.

A third simplification arises from the fact that at relatively low
temperatures ($T \lsim 10^7\,{\rm GeV}$) reactions that transform
leptons into sleptons and vice versa are much faster than the Universe
expansion rate.  Consequently, the chemical potentials of lepton and
slepton equilibrate $\mu_\ell = \mu_{\widetilde\ell}$ or equivalently
$ \frac{Y_{\Delta\ell}}{Y_{\ell}^{eq}} =
\frac{{Y}_{\Delta\widetilde\ell}}{Y_{\widetilde\ell}^{eq}}$, 
a condition known as SE.  In the NSE regime $T
\gsim 10^7\,{\rm GeV}$ interesting new effects arise that, however,
introduce highly non-trivial modifications in the description of SL.
For this reason in this section SE is assumed even when the relevant
temperature regimes fall above $T \sim 10^7\,{\rm GeV}$.
% These three approximations will be dropped in the following sections:
% in Section~\ref{sec:flavor} a complete treatment of lepton flavours
% and spectator effects will be presented, while the NSE regime
% will be discussed in Section~\ref{sec:nse}.

Neglecting SUSY-breaking effects in the RHSN masses and in the
vertices, all the amplitudes for $N_+$ and $N_-$ decays are equal, as
well as their corresponding equilibrium number densities,
$n_{\widetilde{N}_{+}}^{eq}= n_{\widetilde{N}_{-}}^{eq}\equiv
n_{\widetilde{N}}^{eq}$.  Thus, in this approximation, a unique BE for
$Y_{\widetilde{N}_{\rm tot}}\equiv
Y_{\widetilde{N}_{+}}+Y_{\widetilde{N}_{-}}$ suffices to account for the
RHSN densities that, together with the BE for $Y_N$, give two
equations for the out-of-equilibrium heavy neutral states.  
%!!%
Using the SE condition $2 Y_{\ell} = Y_{\widetilde \ell}$ 
%Using the SE condition $2 Y_{\ell}^{eq} = Y_{\widetilde \ell}^{eq} =
%\frac{15}{2\pi^{2}g_{*}}$ 
one can combine the BE for the unflavoured asymmetries 
$Y_{\Delta\ell}$ and $Y_{\Delta\widetilde\ell}$
into a single equation by defining a global density asymmetry in the
$SU(2)_L$ lepton doublets
%!!%
%
\begin{equation}
Y_{\Delta \ell_{\rm tot}}  \equiv  
2 \left( Y_{\Delta\ell}+Y_{\Delta\widetilde\ell} \right),
\label{eq:YLtot} 
\end{equation}
where the factor of 2 comes
from summing over the $SU(2)_L$ degrees of freedom.
We can also define a total CP asymmetry 
\begin{equation}
\label{eq:totalCP}
  \epsilon(T)  =  \epsilon^s(T) + \epsilon^f(T) \equiv 
\bar\epsilon\cdot \Delta_{BF}(T) ,
\end{equation}
where
\be
\label{eq:epsilon-sf}
\epsilon^{s,f}(T) = \pm \sum_{q=S,V,I}\bar\epsilon^q \Delta^{s,f}(T),
\qquad \qquad 
% \epsilon^f(T) = -\sum_{q=S,V,I}\bar\epsilon^q \Delta^f(T), 
\bar\epsilon \equiv \sum_{q=S,V,I}\bar\epsilon^q,
\ee
and the thermal factor $\Delta_{BF}(T)$ is given in
Eq.~\eqref{eq:therm_factor}.  Given that in the one flavour
approximation all lepton flavours are treated on an equal footing, it
is left understood that in the previous equations the various
components of the CP asymmetry have been simply summed over lepton
flavour $ \bar \epsilon^{q} = \sum_\alpha\bar \epsilon_{\alpha}^{q}$.
The relevant parameters that appear in the
CP asymmetries then are $A$, $m_2$, $B$, $M$ and the two CP-violating
phases $\phi_A$ and $\phi_g$. The BE for the unflavoured case read:
%
%%%%%%%%%%%%%%  EQUATIONS REWRITTEN 
%
\begin{eqnarray}
\dot{Y}_{N} \! & = &
- \left(\frac{Y_{N}}{Y_{N}^{eq}}-1  \right)
\left(\gamma_{N}+4\gamma_{t}^{(0)}
+4\gamma_{t}^{(1)}+4\gamma_{t}^{(2)}
+2\gamma_{t}^{(3)}+4\gamma_{t}^{(4)} \right) ,
\label{eq:BEN}\\
\dot{Y}_{\widetilde{N}_{\rm tot}} \!\! & = & \! 
-\! \left( \! \frac{Y_{\widetilde{N}_{\rm tot}}}
{Y_{\widetilde{N}}^{eq}}-2 \! \right)\!\!
\left[\frac{\gamma_{\widetilde{N}}}{2}
\!+\gamma_{\widetilde{N}}^{(3)}+3\gamma_{22}
+2\!\left(\!\gamma_{t}^{(5)}\!
+\gamma_{t}^{(6)}+\gamma_{t}^{(7)}
+\gamma_{t}^{(9)}\right)\!+\!\gamma_{t}^{(8)}\!\right]\!\!,\quad\ \  
\label{eq:BENt}  \\
\dot{Y}_{\Delta \ell_{\rm tot}} & = & 
% \left[
\,\epsilon(T)\,\left(\frac{Y_{\widetilde{N}_{\rm tot}}}
{Y_{\widetilde{N}}^{eq}}-2\right)
% -\frac{Y_{\Delta L_{\rm tot}}} {Y_{c}^{eq}} \right] 
\frac{\gamma_{\widetilde{N}}}{2}
 -W\, \frac{Y_{\Delta \ell_{\rm tot}}} {Y_{\ell_{\rm tot}}^{eq}},
\label{eq:BE_L_tot}
\end{eqnarray}
where the time derivative is defined as $\dot{Y}_X \equiv sHz
\frac{dY_X}{dz}$ with $z \equiv M/T$, 
$Y_{\widetilde N}^{eq} = n_{\widetilde N}^{eq}/s$, 
%is the equilibrium abundance of $\widetilde N$, 
and $Y_{\ell_{\rm tot}}^{eq} \equiv \frac{45}{4\pi^2 g_*}$.  The washout term $W$  
in the equation for ${Y}_{\Delta L_{\rm tot}}$ reads:
\begin{eqnarray}
\nonumber 
W & = & 
\frac{1}{2}\left(\gamma_{\widetilde{N}}+\gamma_{N}\right)
+\gamma_{\widetilde{N}}^{(3)}
+
\frac{Y_{\widetilde{N}_{\rm tot}}}
{Y_{\widetilde{N}}^{eq}}
\left(\gamma_{t}^{(5)}+\frac{1}{2}\gamma_{t}^{(8)}\right) 
 +\frac{Y_{N}}{Y_{N}^{eq}}\left(2 \gamma_{t}^{(0)}+
\gamma_{t}^{(3)} \right) 
\nonumber \\ & + & 
2\left(
\gamma_{t}^{(1)}+\gamma_{t}^{(2)}+\gamma_{t}^{(4)}
+\gamma_{t}^{(6)}+\gamma_{t}^{(7)}+\gamma_{t}^{(9)}
\right)
% \nonumber \\ &  & 
+ \left( 2+\frac{1}{2}
\frac{Y_{\widetilde{N}_{\rm tot}}}{Y_{\widetilde{N}}^{eq}}\right)
\gamma_{22}\,.
\label{eq:BE_W}
\end{eqnarray}
Assuming Maxwell-Boltzmann equilibrium distribution, the RHSN and RHN
equilibrium abundances can be written as:
\be
Y_{\widetilde N}^{eq}
=\frac{45}{4\pi^4 g_*}z^2 \mathcal{K}_2(z),\;\;\;\;\;\;
Y_{N}^{eq}
=\frac{45}{2\pi^4 g_*}z^2 \mathcal{K}_2(z).
\label{eq:RHN_eq_abun}
\ee
(See Ref.~\refcite{Garayoa:2009} for a discussion of the validity of
the use of integrated BE.)  

The derivation of the factorization of the relevant CP asymmetries
including the thermal effects is somewhat lengthy but straightforward
(see \ref{sec:soft_BE}).  The different $\gamma$'s are the thermally
averaged reaction densities for the different processes (they are
defined in \ref{subsec:derivations}).  In all cases a sum over the CP
conjugate final states and lepton flavours is left implicit.

Eqs.~\eqref{eq:BEN}--\eqref{eq:BE_L_tot} include the
$\widetilde{N}_{\pm}$ and $N$ decay and inverse decay processes as
well as all the $\Delta L=1$ scattering processes induced by the
top-quark Yukawa coupling.  $\Delta L=2$ processes involving the
on-shell exchange of $N$ or $\widetilde{N}_{\pm}$ are already
accounted for by the decay and inverse decay processes.  The $\Delta
L=2$ off-shell scatterings involving the pole-subtracted $s$-channel
and the $u$- and $t$-channels, as well as the the $L$-conserving
processes from $N$ and $\widetilde{N}$ pair creation and annihilation,
have not been included. The reaction rates for these processes are
quartic in the neutrino Yukawa couplings and therefore can be safely
neglected as long as these couplings are much smaller than one, as it
is the case for the relevant mass range $M \lesssim 10^9\,{\rm GeV}$
required for successful SL (see next section). The non-resonant
$\Delta L = 2$ processes only become important (strongly suppressing
the final asymmetry) when the neutrino Yukawa couplings become of
order of one which implies $M \gtrsim 10^{14}\,{\rm GeV}$ (see
e.g. Ref.~\refcite{Giudice:2003jh}).  Note that in
Eqs.~\eqref{eq:BEN}--\eqref{eq:BE_L_tot} only the CP asymmetry in the
$\widetilde{N}_{\pm}$ two body decays has been included, while CP
violating effects in three body decays and in
scatterings\cite{Pilaftsis:2004,Pilaftsis:2005b,Abada:2006b,%
  Nardi:2007,Fong:2010bh} have been left out.  Strictly speaking, when
the effects of washout from scatterings are included, for consistency
one should include also the corresponding CP asymmetries.  However, in
the case of standard leptogenesis it has been found that CP
asymmetries in scatterings are important (and dominant) only at high
temperatures $z \lesssim 0.5$.\cite{Nardi:2007} Hence they are only
relevant in the weak washout regime, and in the case of zero initial
RHN abundance.\cite{Abada:2006b,Nardi:2007} In this case, the
inclusion of the scattering CP asymmetries suppresses the final lepton
asymmetry because it results in a balance between the two opposite
sign lepton asymmetries respectively generated during the RHN
production phase and when the RHN eventually decay away, giving rise
to a strong cancellation which, in the limit of vanishing washout, is
actually exact.\cite{Abada:2006b} In SL, however, the inclusion of the
CP asymmetries in scatterings is not straightforward, because
scattering thermal factors constitute a new set of non trivial
quantities.  Nevertheless, it is reasonable to expect that at least
for the strong washout regime the effects of scattering CP asymmetries
are negligible also in SL.  Having said that, a careful quantitative
study in this direction is still lacking.

\subsection{Leptogenesis efficiency}
\label{sec:efficiency}

The effectiveness of leptogenesis for producing a final lepton
asymmetry $Y_{\Delta \ell_{\rm tot}}^\infty\equiv 
Y_{\Delta \ell_{\rm tot}}(z \to \infty)$ (or
$Y_{\Delta_{B-L}}^\infty$ if $L$ violation from sphalerons is accounted for)
% or $Y_{\Delta B}^\infty $ if $Y_{\Delta B-L} \to Y_{\Delta
%   B}$ conversion is also accounted for) 
could be conveniently parametrized in terms of the fractional amount
of the maximum available asymmetry that is eventually converted into
$Y_{\Delta \ell_{\rm tot}}^\infty$.  However, such a parametrization can be
consistently introduced only for standard thermal leptogenesis when it
occurs at temperatures above the onset of flavour effects, and this is
because only in this case the maximum available asymmetry can be
reliably estimated.  In this case $Y_{N}^{eq0}\equiv
Y_{\widetilde N}^{eq}(z\to 0)$ corresponds to the
maximum possible density of decaying RHN, and an amount of
$L$-asymmetry equals to $\epsilon$ is produced in each decay. Then the
maximum available asymmetry is $\epsilon \cdot Y_{N}^{eq0}$, and one
can write
\begin{equation}
\label{eq:efficiency0}
Y_{\Delta \ell_{\rm tot}}^\infty =\eta\cdot \epsilon \,Y_{N}^{eq0}\,,
\end{equation}
where  $\eta$ is a non-negative parameter satisfying $0\leq
\eta\leq 1$ that represents the leptogenesis {\em efficiency}.

However, in other scenarios different from unflavoured thermal
leptogenesis it is more difficult to determine the maximum amount of
available asymmetry.  For example, if the RHN are produced non
thermally,\cite{Giudice:2003jh} it can easily happen that $Y_{N}^{0}$
is much larger than $ Y_{N}^{eq0}$, and in this case, if $Y_{\Delta
  \ell_{\rm tot}}^\infty$ is still expressed in units of $\epsilon \,Y_{N}^{eq0}$,
values of $\eta >1$ will result. Needless to say, this does not
mean that the efficiency of the leptogenesis dynamics is higher than
100\%, but it simply follows from underestimating the maximum
amount of available asymmetry.

In the presence of flavour effects, the available amount of CP
violation is no more described by the total CP asymmetry summed over
lepton flavours $\epsilon = \sum_\alpha \epsilon_\alpha$
($\alpha=e,\mu,\tau$) but rather by the three flavoured CP asymmetries
$\epsilon_\alpha$, and it can easily occur that the absolute value of
some (or even of all) flavoured CP asymmetries are larger than the
absolute value of the total CP asymmetry, with some $\epsilon_\alpha$
having a sign opposite to the one of $\epsilon$.\cite{Nardi:2006a}
Clearly, also in this case $\epsilon \cdot Y_{N}^{eq0}$ does not
account for the maximum available asymmetry, and since particular
flavour configurations can produce $Y_{\Delta \ell_{\rm tot}}^\infty$ with a sign
opposite to the one of the total CP asymmetry $\epsilon$, using
Eq.~\eqref{eq:efficiency0} could even result in {\em negative} values
of the `efficiency' $\eta$.

In SL, estimating the maximum amount of available asymmetry is
basically an impossible task. This is because, besides the effects of
lepton flavours (that is mandatory to include in reliable SL numerical
studies, see Section~\ref{sec:flavor}) the CP asymmetries for RHSN
decays into scalars and fermions $\epsilon^{s,f}(T)$ depend on the
temperature and, as it is depicted in Fig.~\ref{fig:ep_ratio}, the total
CP asymmetry 
obtained from their sum Eq.~\eqref{eq:totalCP} 
% $\epsilon(T) = \bar\epsilon \cdot \Delta_{BF}(T)$
can have different signs, depending on the temperature interval
considered. Nevertheless, it became customary, and it is often
convenient, to express the effectiveness of SL in generating a lepton
asymmetry in terms of the fractional amount $\eta$ of a large {\em
  reference asymmetry} $2\,\bar\epsilon\, Y_{\widetilde N}^{eq0}$,
that is:
\be
\eta =\left| 
%\frac{Y_{\Delta_{L_{\rm tot}}}^\infty}
%!!%
\frac{Y_{\Delta \ell_{\rm tot}}^\infty}
{2\,\bar\epsilon\,  Y_{\widetilde N}^{eq0}}
\right|\,,
% \label{eq:final_lepton_asym}
% \label{eq:yb-l}
\label{eq:eta}
\ee
with a similar definition if $Y_{\Delta_{B-L}}^\infty$ is considered
instead.  In the denominator of the r.h.s. of Eq.~\eqref{eq:eta} the
factor 2 has been included because there are two RHSN states, while
$Y_{\widetilde N}^{eq0}=45\,/(2\pi^4 g_*)$ is defined for one degree
of freedom.  Solving the BE for SL then effectively means finding the
value of $ \eta$ for the specific SL setup. The value of $\eta$ takes
into account the possible inefficiency in the production of RHSN in
the weak washout regime, the erasure of the asymmetry by $L$-violating
washout processes, and the temperature dependence of the CP asymmetry
through the thermal factor $\Delta_{BF}(T)$.  In the more complete
treatment of Section~\ref{sec:flavor}, $\eta$ will also include the
effects of flavours and of spectator processes, and in
Section~\ref{sec:nse} of the non-superequilibration of the particles
and sparticles density asymmetries.

Note that, although as discussed above in general cases $\eta$ does
not correspond to an efficiency, often in comparing different SL
setups with equal initial $Y_{\widetilde N}^{0}$ and equal total (or
flavoured) CP asymmetries, the ratios of the different $\eta$'s do
correspond to the ratios of the corresponding efficiencies, and for
this reason we will follow the general convention of referring to
$\eta$ as to the SL {\sl efficiency}.  Note also that the relative
sign between $\bar\epsilon$ and $Y_{\Delta \ell_{\rm tot}}^\infty$ can
sometimes be important to understand the details of the SL dynamics,
however, as defined in Eq.~\eqref{eq:eta}, $\eta$ is always a positive
quantity. Nevertheless, since the sign of $\bar\epsilon$ is determined
by soft SUSY-breaking phases whose values are presently unknown, and
unlikely to be measured in foreseeable experiments (see
Section~\ref{sec:verifications}), from the practical point of view
no relevant information is lost in characterizing the results
through the `efficiency' $\eta$.

\subsection{Boltzmann equations: qualitative discussion}
\label{sec:qualitative}

Eqs.~\eqref{eq:BEN}--\eqref{eq:BE_L_tot} constitute a rather
nontransparent set of differential equations.  In order to illustrate
their physical content let us discuss an oversimplified example that,
although it refers to $\widetilde N$ decays, it still captures the
most relevant features of the general mechanism of leptogenesis.  Let
us write down simplified BE under the assumption that only the decays
of $\widetilde N$ are relevant to generate the lepton asymmetry
$Y_{\Delta \ell_{\rm tot}}= 2(Y_{\Delta \ell}+ Y_{\Delta \widetilde{\ell}})$ 
%??% 
(where the factor of 2 takes into account the two $SU(2)_L$ degrees of
freedom) and let us describe the evolution of $Y_{\widetilde N}$ and
$Y_{\Delta \ell_{\rm tot}}$ by including only decays and inverse
decays:
\bea
\frac{dY_{\widetilde N}}{dz} & = & 
D \left(Y_{\widetilde N} - Y_{\widetilde N}^{eq}\right), 
\label{eq:sBE_tN}\\
\frac{dY_{\Delta \ell_{\rm tot}}}{dz} & = & 
\epsilon \, D \left(Y_{\widetilde N} - Y_{\widetilde N}^{eq}\right)
-W_{ID} Y_{\Delta \ell_{\rm tot}},
\label{eq:sBE_DL}
\eea
where $\epsilon$ is the CP asymmetry parameter, and the decay and
washout (inverse decay) terms are respectively given by
\be
D  =  K \frac{z \mathcal{K}_1(z)}{\mathcal{K}_2(z)}, 
% \label{eq:D_term}
\qquad \qquad
W_{ID}  =  D \, \frac{Y_{\widetilde N}^{eq}}{Y_{\ell}^{eq}},
\label{eq:W_term}
\ee
with $\mathcal{K}_n$ the modified Bessel function of 
the second kind of order $n$.
From Eq.~\eqref{eq:sBE_DL} we see that in thermal equilibrium, when 
$Y_{\widetilde N} = Y_{\widetilde N}^{eq}$, the source term vanishes
 and no asymmetry can be generated. Let us define the  
decay parameter $K$ as the ratio between the RHSN decay width 
$\,\Gamma\,$ and the Universe expansion rate at $T=M$ 
$H(M) = \sqrt{\frac{4g_*\pi^3}{45}}\frac{M^2}{M_{pl}}$: 
\be
K = \frac{\Gamma}{H(M)} = \frac{m_{\rm eff}}{m^*}.
\label{eq:washout_K}
\ee
In this equation we have introduced the effective neutrino mass 
parameter\cite{Plumacher:1996kc} 
\be
m_{\rm eff} \equiv \frac{1}{M}\displaystyle \sum_\alpha Y_\alpha^2 v_u^2, 
\label{eq:meff} 
\ee
with $v_u=v\, \sin\beta $ (with $v$=174 GeV) the vacuum expectation
value (VEV) of the up-type Higgs doublet, and $\tan\beta \equiv
v_u/v_d$ with $v_d$ the VEV of the down-type Higgs doublet.  Note that
although $m_{\rm eff}$ is related to the light neutrino mass matrix,
it has no direct connection with its eigenvalues, and therefore it is
generally treated as a free parameter.  The {\em equilibrium mass}
appearing in the denominator of the second equality in
Eq.~\eqref{eq:washout_K} is defined as $m^* \equiv \frac{8\pi v_u^2}{9
  M_{pl}}\sqrt{\frac{g_* \pi^3}{45}}$, where $M_{pl}=1.22 \times
10^{19}\,{\rm GeV}$ is the Planck mass. In the MSSM $g_* = 228.75$,
yielding $m^*=7.8 \times 10^{-4}\,{\rm eV}$.

Clearly $m_{\rm eff}$, or equivalently $K$, characterizes the
condition for the RHSN decays to be in equilibrium or out of
equilibrium at $z=1$: the \emph{strong} washout regime corresponds to 
 $K \gg 1$, the \emph{weak} washout regime to $K \ll 1$,
while the \emph{intermediate} washout regime corresponds to $K \sim
1$.  Another factor that concurs to determine the final  result
(in the weak washout regime)  is the assumed 
initial abundance of RHSN $\,Y_{\widetilde N}(z\to 0)$. 
Two possibilities are generally considered: 
\begin{enumerate}
\item[1.] Vanishing initial abundance $Y_{\widetilde N}(z\to 0)
\equiv Y^0_{\widetilde N}  = 0$.  
This case relies on the  assumption that the population of RHSN
  is generated only through neutrino Yukawa interactions in the
  thermal bath.
\item[2.] Thermal initial abundance $Y^0_{\widetilde N}=
  Y_{\widetilde N}^{eq}(z\to 0)\equiv  Y_{\widetilde N}^{eq0}$. 
This possibility can be realized if the RHSN have 
additional interactions with the particles
  in the plasma that at early times are fast enough to generate a
  thermal abundance.
\end{enumerate}

Qualitatively, if $K \gg 1$ decays occur rapidly and quickly generate
a lepton asymmetry. However, inverse decays are also fast and
efficiently erase it.  In this case, irrespectively of the initial
abundance, $Y_{\widetilde N}$ approaches its thermal abundance already
at $z<1$, and any lepton asymmetry generated in the early $\widetilde
N$ production phase, as well as any preexisting asymmetry generated
through some other mechanisms (e.g. from the decays of the heavier
$\widetilde N$) gets washed out completely.  The final lepton
asymmetry can be generated only when $z>1$, that is when the
$\widetilde N$ decays start occurring out of equilibrium
(i.e. $Y_{\widetilde N} > Y_{\widetilde N}^{eq}$ ), and leptogenesis 
proceeds until the last of $\widetilde N$'s have decayed away. In this regime $\eta$,
and hence the final lepton asymmetry, decreases with increasing values
of $K$ because the time at which an asymmetry can be generated is
shifted towards larger values of $z$, where the $\widetilde N$
abundance gets exponentially suppressed by the Boltzmann factor.  In
SL the suppression effect with increasing $K$ is even much larger than
in standard leptogenesis because, as discussed above, the CP asymmetry
quickly decreases with decreasing temperatures.

When $K \ll 1$ the washout of the lepton asymmetry is negligible, and
the initial conditions play an important role. Assuming a thermal
initial abundance $Y^0_{\widetilde N}=Y_{\widetilde N}^{eq0}$ and
taking (just for exemplification) a constant CP asymmetry
$\epsilon(T)=\epsilon$ the final lepton asymmetry saturates to the
maximum possible value $Y^\infty_{\Delta \ell_{\rm tot}}
\approx \epsilon \, Y_{\widetilde N}^{eq0}$ that is $\eta = 1$.  
On the other hand, for zero initial
abundance $Y^0_{\widetilde N}=0$ and $K \ll 1$, basically no
$\widetilde N$'s would decay because none would be produced in first
place, and thus no asymmetry can be generated.  Relaxing the condition
to $K < 1$ a ``wrong'' sign lepton asymmetry is generated as long as
inverse decays keep populating the $\widetilde N$ degree of freedom
(i.e. $Y_{\widetilde N} < Y_{\widetilde N}^{eq}$).\footnote{Notice
  that labeling with ``right'' or ``wrong'' sign of the asymmetry
  is completely arbitrary.}  Since the washout is weak, this asymmetry
only suffers mild washout effects.  Eventually, when at $z>1$ inverse
decays start becoming Boltzmann suppressed and slow down, the out of
equilibrium $\widetilde N$ decays take over (i.e. $Y_{\widetilde N} >
Y_{\widetilde N}^{eq}$) producing a ``right'' sign asymmetry. Because
all washout processes are now Boltzmann suppressed, this asymmetry
suffers an even milder erasure than the ``wrong'' sign one, and the
imperfect cancellation between the two asymmetries of opposite signs
results in a non-vanishing $Y_{\Delta \ell_{\rm tot}}^\infty$.  In this regime the
final asymmetry increases with the value of $K$ because of two
reasons: the first one is that the total $\widetilde N$ population is
created solely through its Yukawa interactions, and thus the larger is
$K$ the larger is the $\widetilde N$ abundance.  The second reason is
that larger $K$ implies stronger washouts processes, and this enhances
the imbalance between the ``wrong'' and ``right'' sign asymmetries.

Finally, for $K \sim 1$ and vanishing $\widetilde N$ initial
abundance a thermal abundance is still reached at $z\sim 1$, while
all washout processes remain as small as possible. This is the
`optimal' regime for thermal leptogenesis, that mediates between the
requirement of generating the largest possible $\widetilde N$
abundance, while at the same time minimizing  washout effects.

\subsection{Quantitative results}
\label{sec:quantitative}

Reliable quantitative results for $Y_{\Delta \ell_{\rm tot}}$ can only be
obtained by solving numerically
Eqs.~\eqref{eq:BEN}--\eqref{eq:BE_L_tot}.  Before embarking in the
details of the analysis, let us remark that $Y_{\Delta \ell_{\rm tot}}$
is not the most convenient quantity for writing the BE for SL. This is
because heavy (s)neutrino decays are not the only source of lepton
number violation: sphaleron transitions, that are the crucial
processes to realize baryogenesis via leptogenesis, also violate
lepton number, and in the temperature regime in which SL can take
place they proceed with in-equilibrium rates violating $L$ at a fast
pace. The quantity that is best suited for numerical studies of
leptogenesis is the density asymmetry $Y_{\Delta_{B-L}}$ (or in the
flavoured case the asymmetries of the flavour charges
$\Delta_\alpha\equiv B/3-L_\alpha$).\cite{Barbieri:2000,Nardi:2006a}
This is because sphalerons conserve $B-L$, and thus $\widetilde N$ and
$N$ related processes are the only ones that can generate such an
asymmetry or change its value.  However, to relate the asymmetry
$Y_{\Delta \ell_{\rm tot}}$ that is generated by $\widetilde N$ and $N$
decays exclusively in the $SU(2)_L$ lepton doublets to
$Y_{\Delta_{B-L}}$ that is given by a sum over the asymmetries of all
the particles with non-vanishing $B-L$, requires also a detailed
knowledge of the network of $B$ and $L$ conserving processes that are
in thermal equilibrium, and this is because through these processes
the asymmetry generated in the decays gets spread among all types of
particle species. We will delay the details of the evaluation of the
$Y_{\Delta_{B-L}} \leftrightarrow Y_{\Delta \ell_{\rm tot}}$ conversion
factors to Section~\ref{sec:flavor} and, as anticipated, here we
will ignore sphalerons as well as all other spectator
effects.\cite{Buchmuller:2001,Nardi:2006a} This boils down to take
simply
\be \frac{dY_{\Delta_{B-L}}}{dz}=-\frac{dY_{\Delta \ell_{\rm tot}}}{dz}\,, 
\label{eq:BE_B-L}
\ee
and the efficiency in producing the $B-L$ asymmetry can 
be expressed in terms of $\eta$  Eq.~\eqref{eq:eta}
with the replacement $Y_{\Delta {\ell_{\rm tot}}}^\infty \to
Y_{\Delta_{B-L}}^\infty$.

After SL is over, the $L$ and $B$ asymmetries keep being converted 
from one into the other by the sphaleron processes until at the EWPT or
slightly after it, $B+L$ violation gets switched off.  How much
$Y_{\Delta B}$ is generated from a certain amount of $Y_{\Delta_{B-L}}$
then depends on the number and types of particles that are present in
the bath with large (thermal) abundances when sphaleron processes drop
out of equilibrium.  Assuming that at the EWPT all supersymmetric
particles already decayed away or have negligible residual densities,
and that the only remaining relativistic degrees of freedom are the SM
states and the up-type and down-type Higgs doublets, the relation
between $Y_{\Delta B}^\infty$ and $Y_{\Delta_{B-L}}^\infty$
is\cite{Harvey:1990qw}
\be
Y_{\Delta B}^\infty = \frac{8}{23}\;Y_{\Delta_{B-L}}^\infty.
\label{eq:yb}
\ee
This relation can change somewhat if the EW sphaleron processes
decouple after the EWPT\cite{Harvey:1990qw,Inui:1993wv} or if
threshold effects for heavy sparticles or particles like the top quark
and Higgs are taken into account.\cite{Inui:1993wv,Chung:2008gv}

\begin{figure}[htb]
\centering
\includegraphics[width=0.7\textwidth]{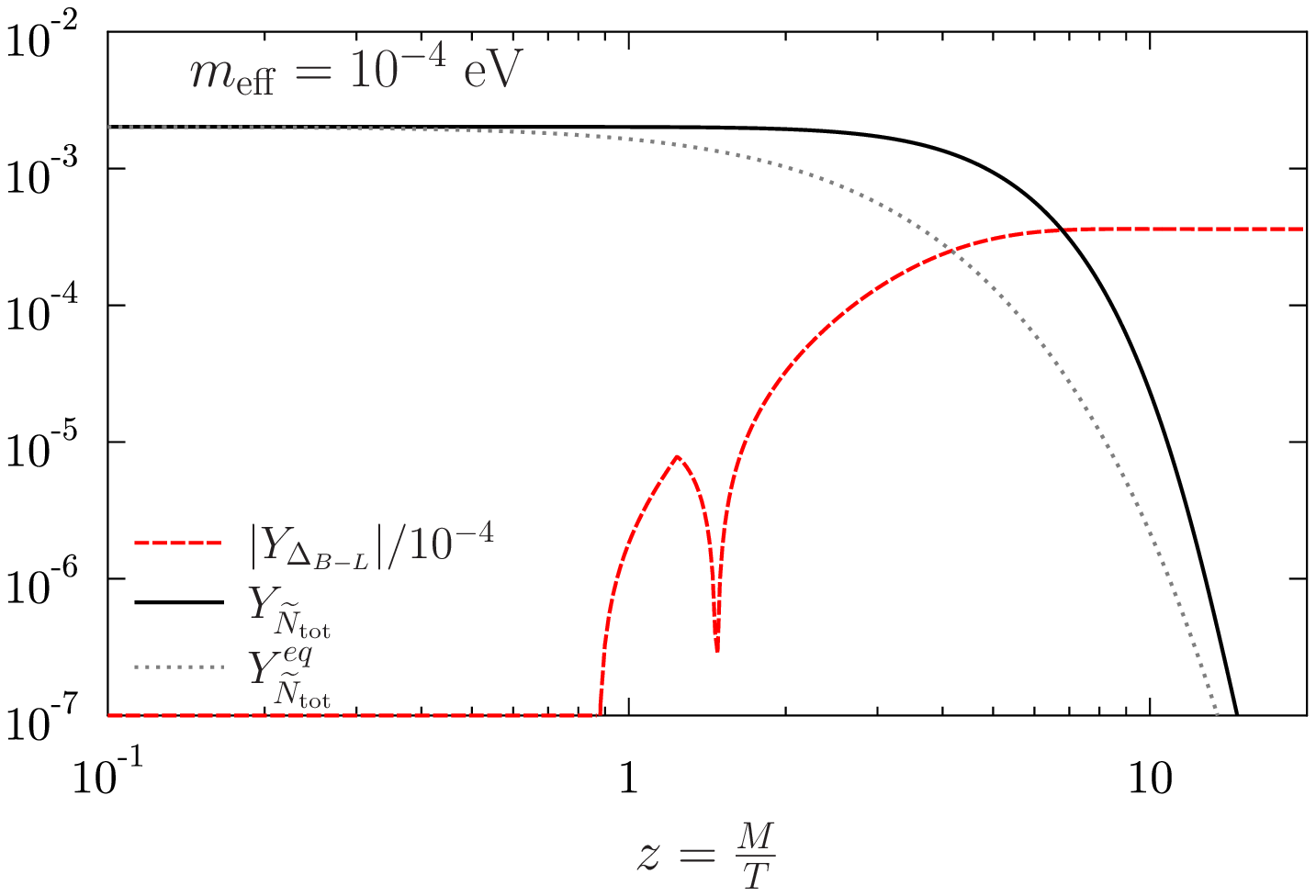}
\includegraphics[width=0.7\textwidth]{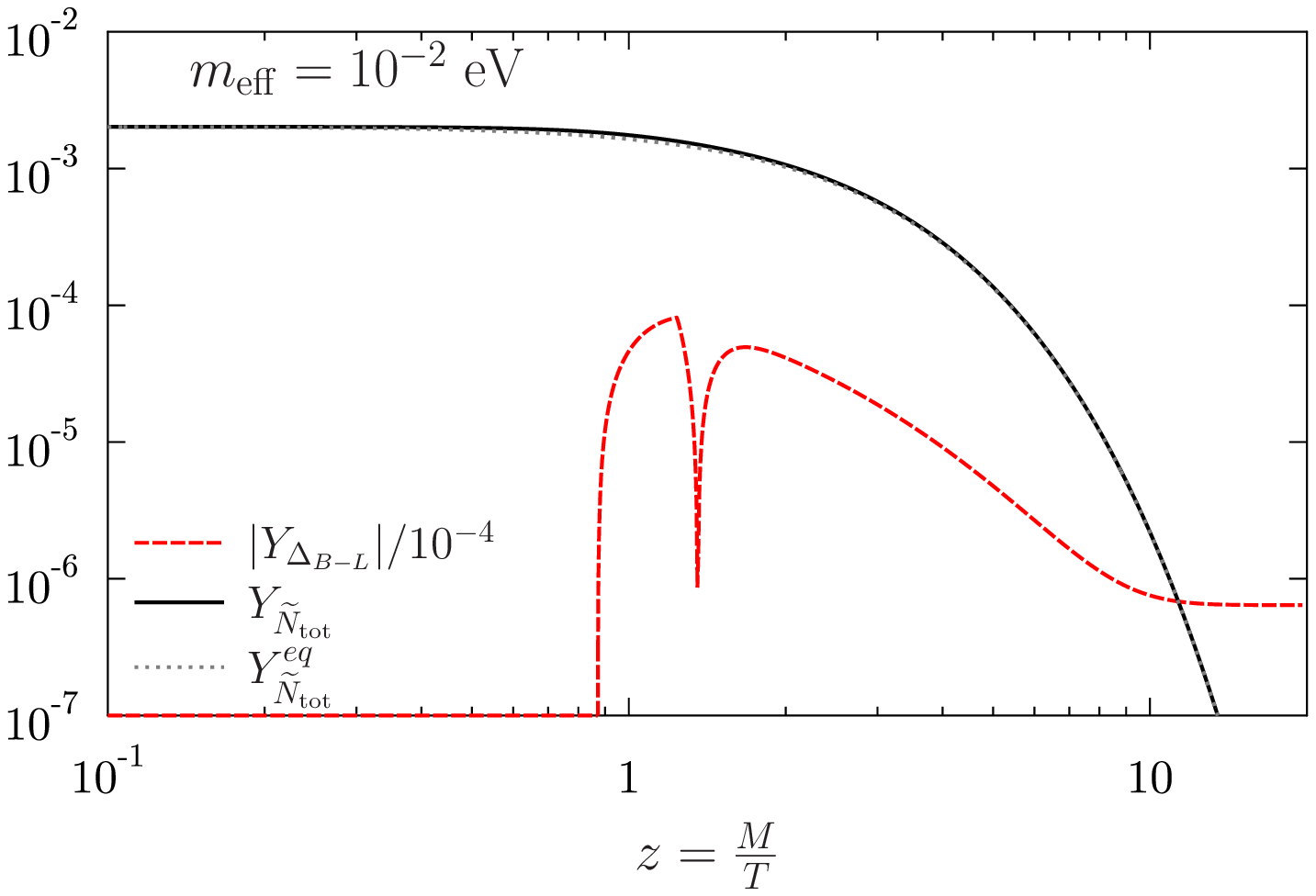}
\caption{Evolution of $Y_{\widetilde N_{\rm tot}}$ (black solid curve) 
and $Y_{\Delta_{B-L}}$ (red dashed curve) assuming an initial thermal 
RHSN abundance $Y^0_{\widetilde N_{\rm tot}} = 2
Y_{\widetilde N}^{eq0}$ for $m_{\rm eff}=10^{-4}\,{\rm eV}$
(top) and $m_{\rm eff}=10^{-2}\,{\rm eV}$ (bottom). The equilibrium
RHSN abundance $Y_{\widetilde N_{\rm tot}}^{eq}$ is given by the gray
dotted curve.}
\label{fig:evolve_BE_th}
\end{figure}

\begin{figure}[htb]
\centering
\includegraphics[width=0.7\textwidth]{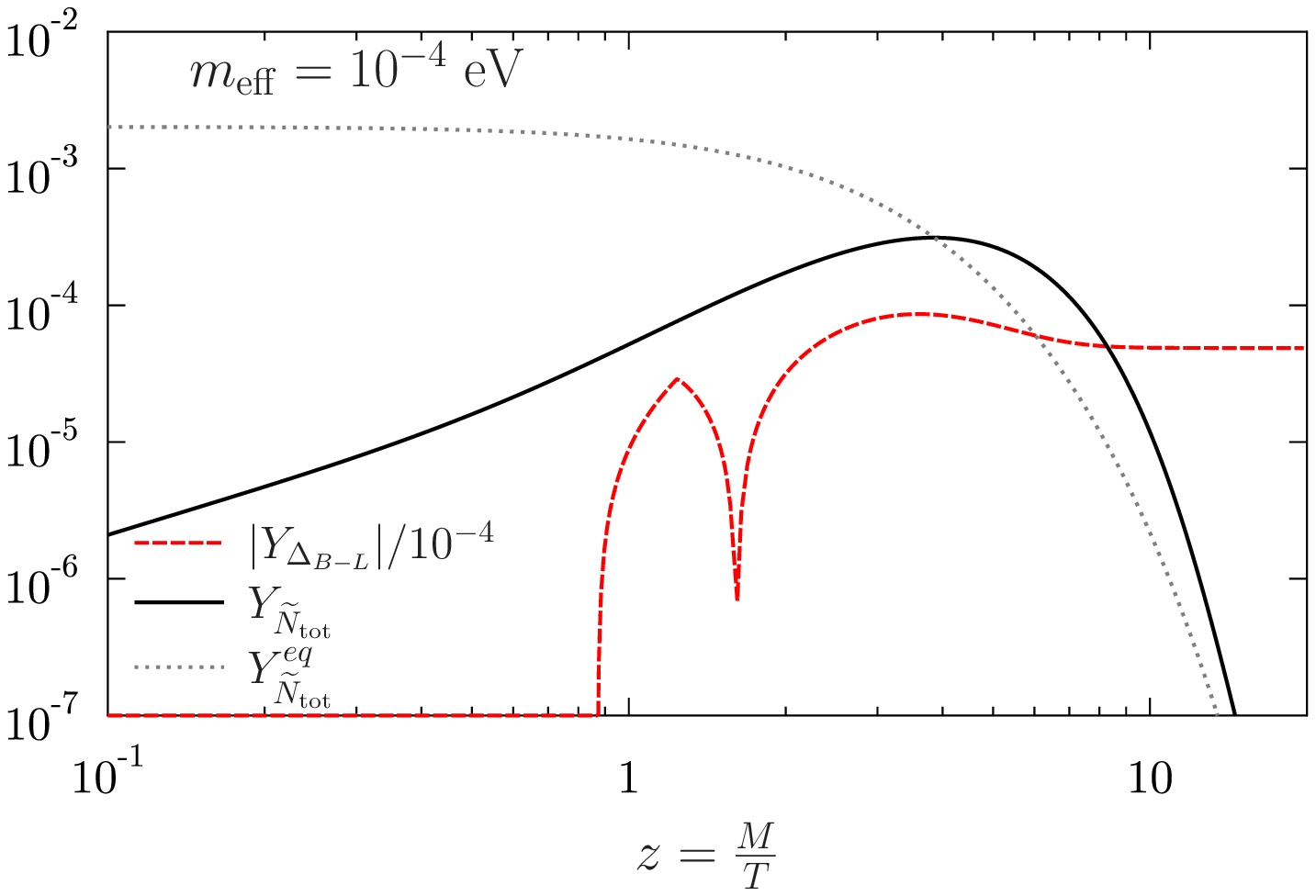}
\includegraphics[width=0.7\textwidth]{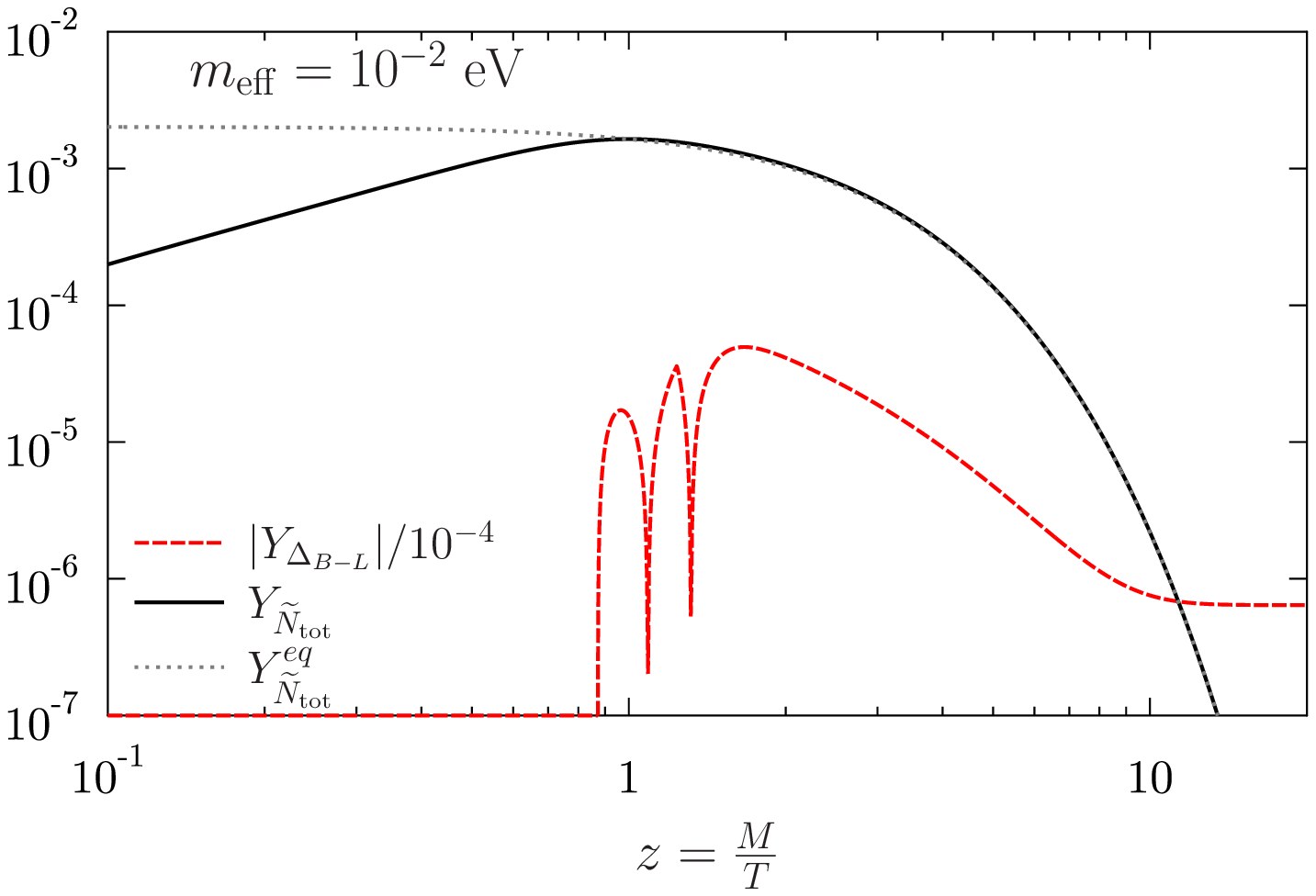}
\caption{ Evolution of $Y_{\widetilde N_{\rm tot}}$ (black solid
  curve) and $Y_{\Delta_{B-L}}$ (red dashed curve) assuming an initial
  vanishing RHSN abundance, for $m_{\rm eff}=10^{-4}\,{\rm eV}$ (top)
  and $m_{\rm eff}=10^{-2}\,{\rm eV}$ (bottom). The equilibrium RHSN
  abundance $Y_{\widetilde N_{\rm tot}}^{eq} $ is given by the gray
  dotted curve.}
\label{fig:evolve_BE}
\end{figure}

\begin{figure}
\centering
\includegraphics[scale=0.60]{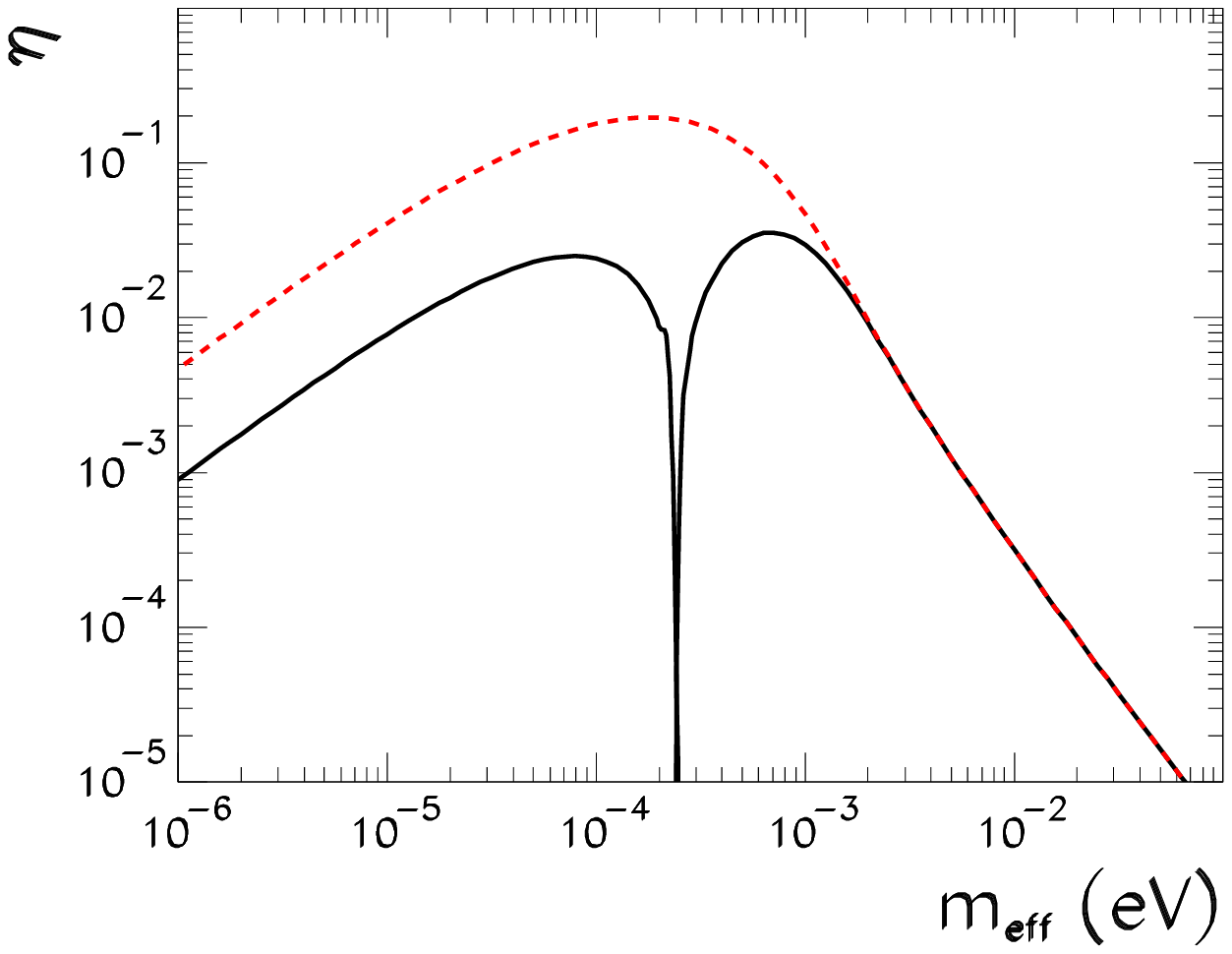}
\caption{Efficiency factor $\eta $ as a function of $m_{\rm eff}$ 
for $M=10^{7}$ GeV and $\tan\beta=30$, and  for  vanishing 
(solid black curve) and thermal initial RHSN abundance 
(dashed red curve). 
\label{fig:etaunf}}
\end{figure}

Solving the BE \eqref{eq:BEN} -- \eqref{eq:BE_L_tot}
one can obtain $\eta$ for different choices of the relevant parameters
$m_{\rm eff}$ and $M$.  Fig.~\ref{fig:etaunf} displays $\eta$ as a
function of $m_{\rm eff}$ for $M=10^7$ GeV and for the two initial
conditions discussed above, and it shows how in the
strong washout regime, the efficiency is independent of the initial
conditions.  This is also illustrated by the evolution of
$Y_{\Delta_{B-L}}$ in the strong regime for both thermal and zero
initial RHSN abundances (bottom panels of Fig.~\ref{fig:evolve_BE_th}
and Fig.  \ref{fig:evolve_BE}).

Notice that with thermal initial RHSN abundance and in the weak
washout regime, the efficiency does not flatten out to a maximum value
as we would have expected if the CP asymmetry were constant,
i.e. temperature independent.  What we observe in Fig.
\ref{fig:etaunf} is that in this case, $\eta$ (dashed red
curve) decreases with decreasing $m_{\rm eff}$ due to the temperature
dependence of the CP asymmetry. As $m_{\rm eff}$ decreases and Yukawa
interactions become correspondingly weaker, the RHSN 
decay at a later time
%attain an overabundant distribution at a later time 
(see the top panel of Fig. \ref{fig:evolve_BE_th}) 
%and hence decay 
when the CP asymmetry is
smaller (see Fig. \ref{fig:ep_ratio}), and this explains the smaller
efficiency.  In the strong washout regime, the efficiency decreases
with increasing $m_{\rm eff}$ due to increasing washout (see the
bottom panel of Fig. \ref{fig:evolve_BE_th}).  If the CP asymmetry
were constant, the efficiency would decrease roughly as $\sim 1/m_{\rm
  eff}$ (see e.g. Ref.~\refcite{Fong:2010a} for a discussion of
leptogenesis in the strong washout regime).  However, larger $m_{\rm
  eff}$ also shifts towards smaller temperatures the moment when the
$B-L$ asymmetry is generated.  Because of the strong temperature
dependence of the CP asymmetry in SL, this implies that the efficiency
decreases faster $\sim 1/m_{\rm eff}$ as can be seen from 
Fig.~\ref{fig:etaunf}.

The solid black curve in Fig.~\ref{fig:etaunf} shows that with zero
initial RHSN abundance, the efficiency $\eta$ quickly drops to zero
somewhere around the intermediate washout regime, to rise again for
larger values of $m_{\rm eff}$. This corresponds to a change of sign
in the ratio $Y_{\Delta_{B-L}}^\infty / \bar \epsilon $ that occurs
for the following reason: during the RHSN production phase
(i.e. $Y_{\widetilde N_{\rm tot}} < 2Y_{\widetilde N}^{eq}$), the
``wrong'' sign lepton asymmetry is generated.  In the weak washout
regime, a large part of ``wrong'' sign asymmetry survives because the
washouts are weak and also because the ``right'' sign asymmetry is
generated at later times when the CP asymmetry is smaller.  As a
result, the ``right'' sign asymmetry cannot overcome the ``wrong''
sign one (see the top panel of Fig. \ref{fig:evolve_BE}).  In the
strong washout regime, the washout of the initial ``wrong'' sign
lepton asymmetry is more efficient and also the RHSN will decay
earlier when the CP asymmetry is larger. The combination of these two
effects results in a final ``right'' sign lepton asymmetry (see the
bottom panel of Fig.  \ref{fig:evolve_BE}).  In the intermediate
regime, a perfect cancellation between the ``wrong'' and ``right''
sign lepton asymmetries occurs in the dip observed in
Fig.~\ref{fig:etaunf} where the efficiency vanishes.

\begin{figure}
\centering
\includegraphics[width=0.88\textwidth]{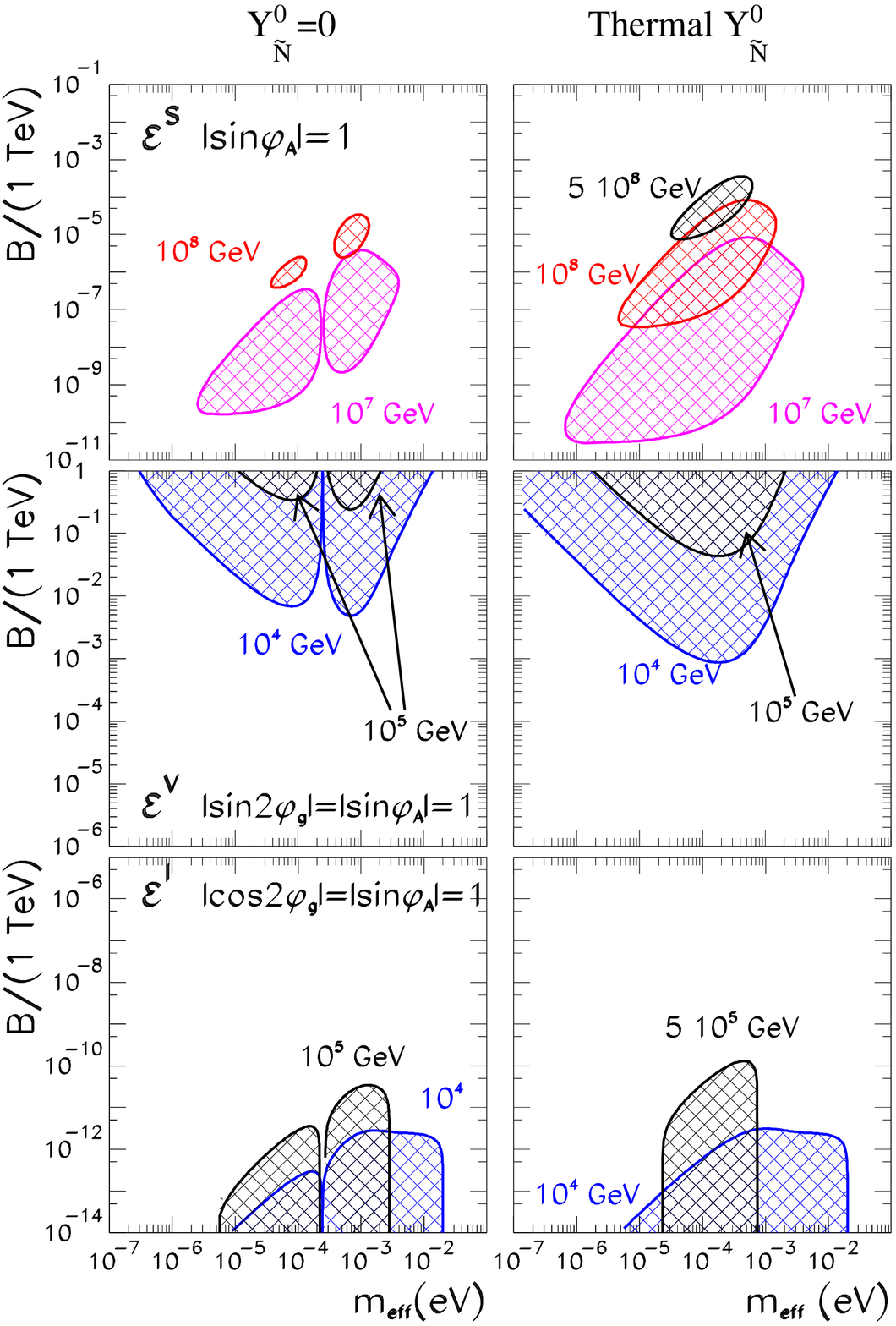}
\caption{ $B, m_{\rm eff}$ regions in which successful SL
   can be achieved (flavour and spectator effects neglected).
   In all cases we take $A=m_2=10^3$ GeV
  and $\tan\beta=30$ and different values of $M$ and $\phi_A$ and
  $\phi_g$ as labeled in the figure (see text for details).  The left
  (right) panels correspond to vanishing (thermal) initial $\widetilde N$
  abundance .
  \label{fig:Bmeffunf}}
\end{figure}

The upper panels in Fig. \ref{fig:Bmeffunf} show the regions of
parameters for which CP violation from pure mixing effects
($\epsilon^S$) can produce the observed asymmetry.  Due to the
resonant nature of this contribution, these effects are sufficiently large 
only when $B\sim {\cal O}(\Gamma)$ which, as discussed in the the
original proposals of SL,\cite{Grossman:2003,DAmbrosio:2003} implies
unconventionally small values of $B$ and an upper bound $M\lesssim
10^{9}\,$GeV.\footnote{Models that can naturally yield small
  values of $B$ are considered in
  Refs.~\refcite{Chun:2004,Chen:2004,Grossman:2005,Chun:2007}.}

The central panels of Fig. \ref{fig:Bmeffunf} give the corresponding
regions for which CP violation from gaugino-induced vertex effects
($\epsilon^V$) can produce the observed baryon asymmetry. Despite
being higher order in $\delta_S$ and proportional to the square of the
gauge couplings ($\alpha_2$), this contribution can be relevant
because it dominates for conventional values of the $B$ parameter.
However, in order to overcome the $\delta_S$ and $\alpha_2$
suppression this contribution can only be sizable if the RHSN
are light $M\lesssim \;10^{6}$ GeV (with the approximation used in
this work: $\delta_S\ll 1$, $A, m_2\sim {\cal O}({\rm TeV})$).

The figure corresponds to values of the parameters 
such that the second term in Eq.\eqref{eq:cp0_V} 
% \eqref{eq:cp_asym_1} 
dominates, so that the allowed region depicts a lower bound on $B$.
Conversely, when the first term in Eq.\eqref{eq:cp_asym_1} dominates,
$\epsilon^V$ becomes independent of $B$. In this case, for a given
value of $M$ and $\delta_S$ the produced baryon asymmetry can be
sizable within the range of $m_{\rm eff}$ values for which $\eta$ 
is large enough. For example for $M= 10^{5}$ GeV, and
$m_2=A=1$ TeV and $|\sin(\phi_A+2\phi_g)=1|$ with vanishing initial
conditions
\begin{equation} 
10^{-5}<\frac{m_{\rm eff}}{\rm eV}<6.5\times 10^{-4} 
\;\;\;\;\;\;\; {\rm or}\;\;\;\;\;\;
8\times 10^{-4}<\frac{m_{\rm eff}}{\rm eV}<3\times 10^{-2}, 
\end{equation}
where each range corresponds to a sign of the CP phase $\sin(\phi_A+2\phi_g)$ 

Finally we show in the lower panels of Fig. \ref{fig:Bmeffunf} the
values of $B$ and $m_{\rm eff}$ for which enough baryon asymmetry can
be generated from the interference of mixing and vertex corrections
$\epsilon^I$ in Eq.\eqref{eq:cp_asym_3}. Generically speaking,
$\epsilon^I$ is subdominant with respect to $\epsilon^S$, since both
involve the same CP phase $\sin(\phi_A)$ but $\epsilon^I$ has
additional $\delta_S$ and $\alpha_2$  suppression:
\begin{equation}
\frac{\bar\epsilon^I}{\bar\epsilon^S}=-\frac{3}{8} \alpha_2 
\frac{m_2}{M}  \ln\frac{m_{2}^{2}}{M^{2}+m_2^{2}} 
\cos(2\phi_g) \frac{\Gamma}{B} \; . 
\end{equation}
Consequently, as it is also shown by the figure, $\epsilon^I$ can
dominate only if $B$ is extremely small ($B\ll \Gamma$), when it
becomes independent of $B$. Note also that for $M\lesssim
10^4\,$GeV and $m_{\rm eff}\gtrsim 10^{-2}\,$eV the baryon asymmetry
generated by this contribution becomes independent of $m_{\rm eff}$.
This is because in this regime of strong washouts the $m_{\rm eff}^2$
dependence from $\Gamma^2$ cancels the approximate $1/m_{\rm eff}^2$
dependence of $\eta$.

%%%%%%%%%%%%%%%%%%%%%%%%%%%%%%%%%%%%%%%%%%%%

\section{The Possible Role of Quantum Effects} 
\label{sec:quantum_role}
Most studies of thermal leptogenesis (both for the standard seesaw
case, as well as for the SL scenario) rely on the classical BE
approach that was described in the previous section.  The possibility
of using quantum BE (QBE) in leptogenesis was first discussed in
Ref.~\refcite{Buchmuller:2000}, and more recently analyzed in greater
detail in
Refs.~\refcite{DeSimone:2007b,Anisimov:2010aq,Anisimov:2010dk,%
Garny:2009rv,Garny:2009qn,Cirigliano:2009yt,Beneke:2010dz}.
In Ref.~\refcite{DeSimone:2007b}, QBE were obtained starting from the
non-equilibrium quantum field theory based on the Closed Time-Path
(CTP) formulation, and differ from the classical BE in that they
contain integrals over the past times.  In the classical kinetic
theory instead the scattering terms do not include any integral over
the past history of the system, which is equivalent to assuming that any
collision in the plasma is independent of the previous ones. In the CTP
formalism, the energy conservation delta functions appearing in the
evaluation of the reaction rates are substituted by retarded time
integrals of time-dependent kernels, and cosine functions whose
arguments are the energy involved in the reactions. In the limit in
which the time range of the kernels is shorter than the relaxation
time of the particle abundances, and the time integrals are taken over
an infinite time (i.e. neglecting memory effects), the standard
time-independent reaction rates are recovered.  Furthermore, the CP
asymmetry also acquires an additional time-dependent piece, that at
any given instant depends upon the previous history of the system.

In Ref.~\refcite{DeSimone:2007b} it was argued that the additional
time dependence of the CP asymmetry is quantitatively the most
relevant effect for leptogenesis.  However, if the time variation of
the CP asymmetry is shorter than the relaxation time of the particles
abundances, the solutions to the quantum and the classical Boltzmann
equations are expected to differ only by terms of the order of the
ratio of the time-scale of the CP asymmetry to the relaxation
time-scale of the distribution. This is typically the case in thermal
leptogenesis with hierarchical RHN. Conversely in the resonant
leptogenesis scenario, $(M_j-M_i)$ is of the order of the decay rate
of the RH neutrinos.  As a consequence the typical time-scale to build
up coherently the time-dependent CP asymmetry, which is of the order
of $(M_j-M_i)^{-1}$, can be larger than the time-scale for the change
of the abundance of the RHN. As shown in
Refs.~\refcite{DeSimone:2007c,Cirigliano:2008}, 
in the case of resonant leptogenesis 
this leads to quantitative
differences between the classical and  quantum approach  
and, in particular, in the weak washout
regime this can enhance the produced asymmetry.
 
Since in SL the CP asymmetry in mixing (Eq.~\eqref{eq:cp_asym_2}) is
produced resonantly, we can expect that this type of effects could be
of some relevance.\cite{Fong:2008b}

%%%%%%%%%%%%%%%%%%%%%%%%%%%%%%%%%%%%%%%%%%%%%%%%%%%%%%%%%%%%%%%%%%

\subsection{Modification to the CP asymmetry and quantification} 
\label{sec:quantum_CP}

We have seen in Section~\ref{sec:cp_asymmetries} that the relevant CP
asymmetry in SL is temperature (i.e. time) dependent already in the
\emph{classical} approximation.  The inclusion of quantum effects
introduces an additional time dependence.  As shown in
Refs.~\refcite{DeSimone:2007b,DeSimone:2007c,Cirigliano:2008} quantum
effects are flavour independent as long as the damping rates of the
leptons are taken to be flavour independent.  Neglecting also the
difference in the width of the two RHSN, one can show that
\begin{equation}
\epsilon^S(T)=\bar\epsilon^S\, \times\,\Delta_{BF}(T)\, \times\, QC(t), 
\end{equation}
where 
$\bar\epsilon^S$ is defined in Eq.~\eqref{eq:cp0_S}  and
\begin{equation} 
QC(t) =2 ~\sin^2\left(\frac{M_+-M_-}{2} t\right)
- \frac{\Gamma}{M_+-M_-}~\sin\left((M_+-M_-)t\right).
\label{eq:qct}
\end{equation}
The factor $QC(t)$ is the one that remains after taking the past time
integral to large time such that only on-shell decay processes
contribute to the CP asymmetry (which is equivalent to neglecting memory
effects in decay processes).  This factor grows for $t \lesssim
1/\Delta M$ and starts oscillating for $t \gtrsim 1/\Delta M$. The
oscillation pattern originates from the CP-violating decays of two
mixed states $N_+$ and $N_-$ analogous to the CP violation in neutral
meson systems.  If the timescale for the decay $\,t \sim 1/\Gamma\,$
is much larger than $1/\Delta M$, the CP asymmetry should average to
the \emph{classical} value.  However, if the decay timescale $t \sim
1/\Gamma$ is shorter than $1/\Delta M$, this additional time dependence on
CP asymmetry may not be negligible.

As usual, it is convenient to change in Eq.~\eqref{eq:qct} from time
$t$ to $z=M/T$.  For a Universe undergoing adiabatic expansion the
entropy per comoving volume is constant, i.e.  $sR^3 =$ constant, and
since $s \propto z^{-3}$ then $R \propto z$.  Thus the Hubble
parameter is given by $H \equiv R^{-1}dR/dt=z^{-1}dz/dt$.  After
integration, one gets
\begin{equation}
t=\frac{1}{H(M)}\frac{z^2-z_0^2}{2},
\label{eq:t_to_z}
\end{equation}
where $z_0$ is the temperature at $t=0$.
Substituting Eq.~\eqref{eq:t_to_z} into Eq.~\eqref{eq:qct}: 
\begin{eqnarray} 
QC(z) &=& 2 ~\sin^2\left(\frac{1}{2} \frac{M_+-M_-}{2 H(M)} z^2\right)
- \frac{\Gamma}{M_+-M_-}
~\sin\left(\frac{M_+-M_-}{2 H(M)}z^2\right),\nonumber \\
&=& 2\sin^2\left(\frac{m_{\rm eff}}{m^*}\, R\, \frac{z^2}{8}\right)-
\frac{2}{R}\sin\left(\frac{m_{\rm eff}}{m^*}\, R\, \frac{z^2}{4}\right), 
\label{eq:qc}
\end{eqnarray}
where $z_0$ has been set equal to $0$ (corresponding to a very high
initial temperature) and $M_+-M_-=B$ has been used (assuming
$\widetilde M \ll M$ (see Eq.~\eqref{eq:mass_eigenvalues}).  Finally, the
degeneracy parameter $R$ has been defined as:
\begin{equation}
R= \frac{2(M_+-M_-)}{\Gamma}=\frac{2B}{\Gamma}.
\end{equation}

In summary, the final CP asymmetry consists of three factors: the
first one is the temperature independent piece $\,\bar\epsilon^S$
which can be rewritten as\footnote{Quantum effects for CP asymmetries
  in decay or interference of decay and mixing can be introduced in
  similar fashion. However, in the interesting parameter space for
  $\bar\epsilon^V$ where $R \gg 1$, these effects are irrelevant,
  while the effects on CP violation in the interference between decay
  and mixing are expected to be of a similar size.}
\begin{equation}
\bar\epsilon^S = \frac{{\rm Im}A}{M} \frac{2R}{R+1}.
\end{equation}
which, as discussed before, it is resonantly enhanced for $R=1$. 
The second one is the thermal factor $\Delta_{BF}(T)$ 
which is non-vanishing only  for $z \gtrsim 0.8$ 
(see Fig. \ref{fig:ep_ratio}).
The third one is the quantum correction factor $QC(z)$.

\begin{figure}
\begin{center}
\hspace*{0.4cm}
\includegraphics[width=\textwidth]{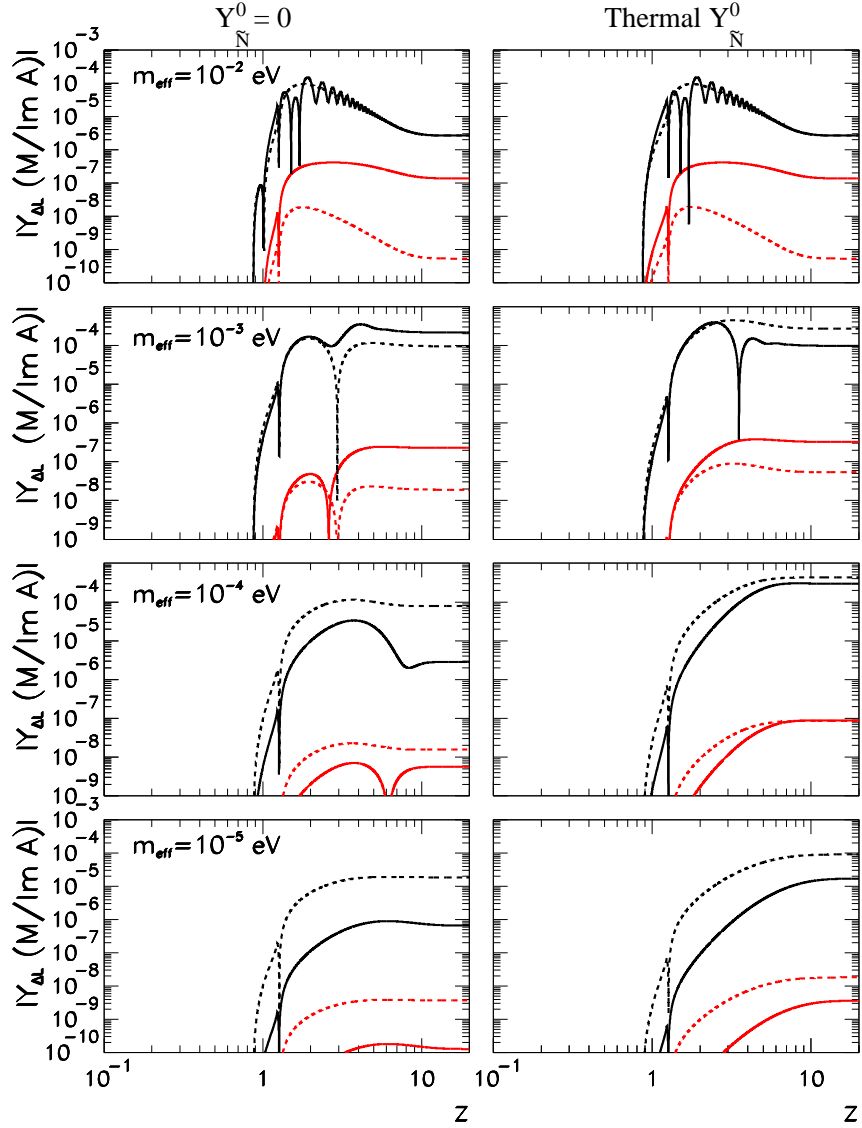}
\caption{Absolute value of the lepton asymmetry with the quantum time
  dependence of the CP asymmetry (solid) and without it (dashed) as a
  function of $z$ for different values of $m_{\rm eff}$ as labeled in
  the figure.  In each panel the two upper curves (black) correspond
  to the resonant case $R=1$ while the lower two curves (red)
  correspond to the very degenerate case $R=2\times 10^{-4}$.  The
  left (right) panels correspond to vanishing (thermal) initial RHSN
  abundance.  The figure corresponds to $M=10^7$ GeV and
  $\tan\beta=30$.
% although, as discussed in the text, the results as
%  normalized in the figure are very weakly dependent on these two
%  parameters.
}
\label{fig:effz}
\end{center}
\end{figure}

The impact of this additional quantum time-dependence of the CP
asymmetry on the final baryon asymmetry can be easily quantified by
introducing   $QC(z)$  in the relevant BE \eqref{eq:BEN}, \eqref{eq:BENt} and
\eqref{eq:BE_L_tot}.  Fig.~\ref{fig:effz} shows the evolution of the
asymmetry with and without the inclusion of the 
quantum correction factor 
for several values of the washout parameter $m_{\rm eff}$ both 
 for the resonant case $R=1$ and for the very degenerate case
$R=2\times 10^{-4}$. The two upper panels correspond to strong and
moderate washout regimes, while the lower two correspond to weak and
very weak washout regimes.

The figure illustrates that, as expected, for strong washouts and
large degeneracy parameter $R$ (see the upper curves in the upper
panels), the quantum effects lead to an oscillation of the produced
asymmetry until it averages out to the {\sl classical} value.
Conversely, for very small values of $R$ and strong washouts, quantum
effects enhance the final asymmetry.  For small enough $R$ the
arguments in the periodic functions in $QC(z)$ are very small for all
relevant values of $z$ and $m_{\rm eff}$, so that the $\sin^2$ term is
negligible. By expanding the $\sin$ term we get
\begin{equation}
QC(z)\simeq - \frac{m_{\rm eff}}{m_*}\frac{z^2}{2},  
\label{eq:qclim}
\end{equation}
which, in the strong washout regime, is always larger than 1.

In the weak washout regime, independently of the initial conditions
and of the value of the degeneracy parameter $R$, the quantum effects
always lead to a suppression of the final asymmetry.  This is
different from what happens in type I seesaw resonant leptogenesis in which
for weak washouts, $R\sim 1$, and vanishing RHN initial abundances,
quantum effects lead to an enhancement of the asymmetry
produced.\cite{DeSimone:2007c}  The origin of the difference is in
the additional time dependence of the CP asymmetry in SL $\Delta_{BF}$.
In order to understand how this works, we must remember that in the
weak washout regime for type I seesaw resonant leptogenesis, the resulting
asymmetry is the one that survives the cancellation between the
opposite sign asymmetry generated when RH neutrinos are initially
produced, and the asymmetry generated when they decay.  Including the
time-dependent quantum corrections spoils this cancellation and as a
consequence a larger asymmetry is obtained.\cite{DeSimone:2007c}

\begin{figure}
\begin{center}
\hspace*{0.5cm}
\includegraphics[width=\textwidth]{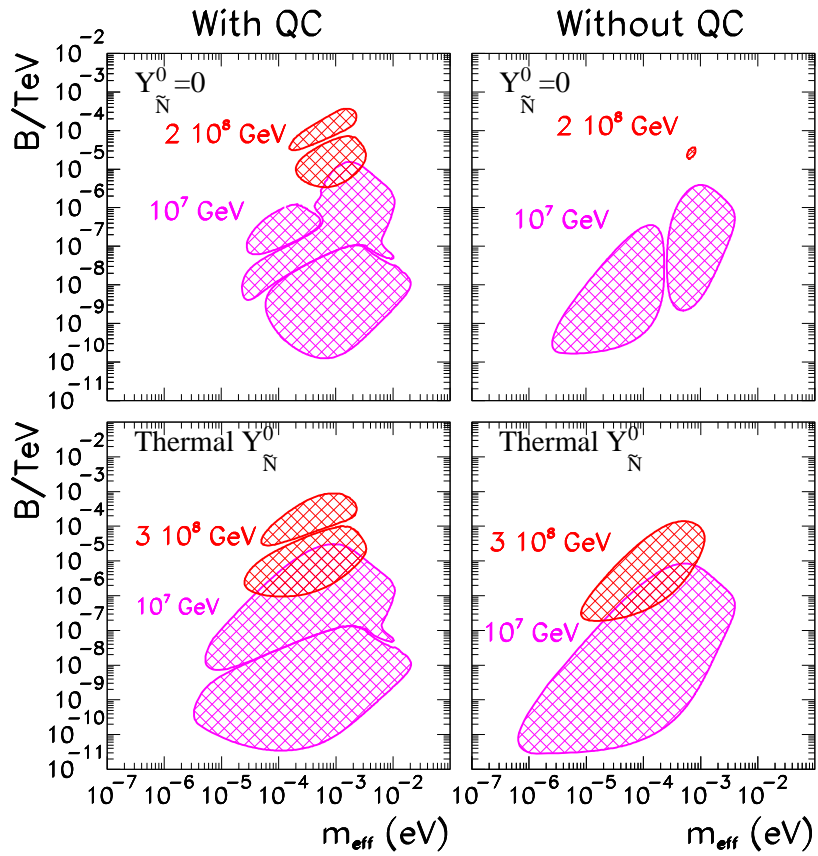}
\caption{ $B, m_{\rm eff}$ regions in which successful SL can be
  achieved with (left panels) and without (right panels) quantum
  effects.  We take $|{\rm Im} A|=10^3$ GeV and $\tan\beta=30$ and
  different values of $M$ as labeled in the figure.  The upper (lower)
  panels correspond to vanishing (thermal) initial RHSN abundance.}
\label{fig:con}
\end{center}
\end{figure}

However, in SL already in the classical approximation the thermal
factor $\Delta_{BF}$ prevents the opposite sign asymmetries
cancellation, and the inclusion of the time dependent quantum effects
only amounts to an additional multiplicative factor which, in this
regime, is smaller than one.  Consequently, for SL, even in the
resonant-regime quantum effects do no lead to major quantitative
differences, at least in the range of parameters for which successful
leptogenesis is possible. This is explicitly shown in
Fig. \ref{fig:con} that depicts the ranges of the parameters $B$ and
$m_{\rm eff}$ for which enough asymmetry is generated ($Y_{\Delta
  B}^\infty\geq 8.35\times 10^{-11}$), with and without the inclusion of
the quantum corrections.  We see that the main effect of including
quantum corrections is that for a given value of $M$ the allowed
regions extend up to larger values of $m_{\rm eff}$.  This is
precisely due to the suppression of the asymmetry in the weak washout
regime just discussed.  Because of the enhancement in the very
degenerate, strong washout regime, for a given value of $M$ the
regions also tend to extend to lower values of $B$ and larger values
of $m_{\rm eff}$.

A qualitative difference obtained from the inclusion of quantum
effects is that, depending on the value of $m_{\rm eff}$, $\eta$ can
take both signs independently of the initial RHSN abundance.  Thus it
is possible to generate the right sign asymmetry with either sign of
${\rm Im} A $ for both thermal and zero initial RHSN abundance.
However, apart from this peculiarity, altogether it can be concluded
that for a given $M$ the values of the $L$-violating soft bilinear
term $B$ required to achieve successful leptogenesis are not
substantially modified.

%% file: flavour.tex
\section{The Role of Lepton Flavours}
\label{sec:flavor}

The role of lepton flavour in the standard thermal leptogenesis
scenario was first discussed in Ref.~\refcite{Barbieri:2000}.
However, until the importance of flavour effects was fully
clarified\cite{Abada:2006a,Nardi:2006b} they had been included in
leptogenesis studies only in a few
cases.\cite{Endoh:2004,Fujihara:2005} Nowadays flavour effects have
been investigated in full 
detail\cite{Abada:2006b,Vives:2009,Pascoli:2007a,Pascoli:2007b,Antusch:2006,%
Antusch:2007a,Branco:2007,Blanchet:2007a,Blanchet:2007b,%
DeSimone:2007b,DeSimone:2007c,Cirigliano:2008,Fong:2010qh} and are 
a mandatory ingredient of any reliable analysis of leptogenesis.

As regards SL, the original works\cite{Grossman:2003,DAmbrosio:2003}
neglected flavour effects and considered the generation of the lepton
asymmetry directly in the $\ell$ and $\widetilde \ell$ lepton states
coupled to the lightest RHSN, a situation  commonly referred to
as the `one flavour approximation'. However, RHSN couple in fact to
certain lepton combinations, in which the different flavours are
weighted by the respective RHN Yukawa couplings: $\ell\propto
\sum_\alpha Y_\alpha \ell_\alpha $ and $\widetilde \ell\propto \sum_\alpha
Y_\alpha \widetilde \ell_\alpha $. Only at very high temperatures $\ell$
and $\widetilde \ell$ remain coherent superpositions, and are the correct
states to describe the dynamics of leptogenesis.  At lower
temperatures scatterings induced by the charged lepton Yukawa
couplings occur at a sufficiently fast pace to distinguish the
different lepton flavour components. In this situation, $\ell$ and
$\widetilde \ell$ cannot be considered anymore as coherent flavour
superpositions, and the dynamics of leptogenesis must be described
instead in terms of the flavour eigenstates $\ell_\alpha$ and $\widetilde
\ell_\alpha$.  The specific temperature when leptogenesis becomes
sensitive to lepton flavour dynamics can be estimated by requiring
that the rates of processes $\Gamma_\alpha$ ($\alpha=e,\mu,\tau$) that
are induced by the charged lepton Yukawa couplings $h_\alpha$ become
faster than the Universe expansion rate $H(T)$. An approximate
relation gives\cite{Campbell:1992jd,Cline:1993bd}
\be
\Gamma_\alpha(T) \simeq 10^{-2} h_\alpha^2 T\,,
\ee
which implies  that 
\be
\Gamma_\alpha(T) > H(T) \qquad {\rm when} \qquad  
T \lesssim  T_\alpha \, (1 + \tan\beta^2)\,, 
\ee
where $T_e \simeq 4\times 10^4\,$GeV, $T_\mu \simeq 2\times 10^9 \,$GeV, and
$T_\tau \simeq 5\times 10^{11}\,$GeV. Notice that to fully distinguish the
three flavours it is sufficient that the $\tau$ and $\mu$ Yukawa
reactions attain thermal equilibrium. 
Therefore,  given that SL can successfully proceed only for 
temperatures below $T_\mu$,  we can conclude that independently 
of the value of  $\tan\beta$ the three flavour regime is always 
the appropriate one to study SL.
The interesting issue is if the region of parameter
space in which SL can be successful is enlarged or reduced when
flavour effects are taken into account, and in particular if the
requirement of unnaturally small values of the soft $B$ term can get
relaxed. These issues are addressed in the following sections.

%%%%%%%%%%%%%%%%%%%%%%%%%%%%%%%%%%%%%%%%%%%%%%%%%%%%%%%%%%%%%%%%%%
%%%%%%%%%%%%%%%%%%%%%%%%%%%%%%%%%%%%%%%%%%%%%%%%%%%%%%%%%%%%%%%%%%

\subsection{Flavored CP asymmetries} 
\label{sec:flavored_CP}

In the flavour regime the CP asymmetries for RHSN decays into the
single lepton flavours become important.  Their flavour structure is
determined by the Yukawa couplings $Y_\alpha$ as well as by the
trilinear couplings $A_\alpha$.  The soft breaking 
Lagrangian of the previous section Eq.~\eqref{eq:soft_terms} 
assumed for simplicity a universal trilinear soft breaking 
scale $A$ and proportionality of the soft breaking trilinear 
couplings to the Yukawa couplings, that is  
$A_\alpha = A Y_\alpha$.  In this section  a more 
general flavour structure  is considered where the $A$-terms have 
the generic flavour structure
\begin{equation}
 -\mathcal{L}_{\rm soft}^{(A)}  =  
A Z_{i\alpha}\,\epsilon_{ab}\widetilde{N}_{i}
\widetilde{\ell}_{\alpha}^{a}H_{u}^{b}+{\rm h.c.}\,. 
\label{eq:soft_terms_gen}
\end{equation}
Considering only the lightest RHSN $\widetilde N= \widetilde N_1$ and
adopting as usual a simplified notation $Z_{\alpha}= Z_{1\alpha}$
etc., with the generalized flavour configuration in
Eq.~\eqref{eq:soft_terms_gen} we have three relevant physical phases:
\begin{equation}
\phi_{A_\alpha} = \arg(AZ_\alpha Y_\alpha^* B^*).
\label{eq:CP_phase1_gen}
\end{equation}
and the CP asymmetries \eqref{eq:cp_asym_2},\eqref{eq:cp_asym_1}
and \eqref{eq:cp_asym_3} are now written as:
\begin{equation}
\label{eq:cp_asym_123-fl}
\epsilon_{\alpha}^{S,V,I}(T)  =  
P_\alpha\,\frac{Z_\alpha}{Y_\alpha}\,\bar\epsilon^{S,V,I}_\alpha\,\Delta_{BF}(T),
\end{equation}
where 
\begin{eqnarray}
\!\! && \bar\epsilon_{\alpha}^{S} % \left(T\right) 
 \equiv  - \frac{A}{M}\sin\left(\phi_{A_\alpha}\right)
\frac{4B\Gamma}{4B^{2}+\Gamma^{2}}
\label{eq:CP_asymres} \\
\!\! && \bar\epsilon_{\alpha}^{V} 
  \equiv  -
\frac{3\alpha_{2}}{4}\frac{m_{2}}{M}
\ln\frac{m_{2}^{2}}{m_{2}^{2}+M^{2}} \left[
\frac{A}{M} \sin\left(\phi_{A_\alpha}+ 2\phi_{g}\right)
-  \frac{Y_\alpha}{ Z_\alpha } \frac{B}{M}\sin\left(2\phi_{g}\right)\right],
\quad \ \ 
 \label{eq:CP_asymver}  \\
\!\! && \bar\epsilon_{\alpha}^{I}   \equiv    
\frac{3\alpha_{2}}{2}\frac{m_2}{M}\frac{A}{M}
\ln\frac{m_{2}^{2}}{m_{2}^{2}+M^{2}}\sin\left(\phi_{A_\alpha}\right)  
\cos\left(2\phi_{g}\right)\frac{\Gamma^{2}}{4B^{2}+\Gamma^{2}}\,. 
 \label{eq:CP_asymint} 
\end{eqnarray}
In these equations the physical complex phases $\phi_{A_\alpha}$ and
$\phi_g$ have been explicitly written, so that all the parameters $A$,
$Z_\alpha$, $Y_\alpha$ etc. are real and positive. Unless
explicitly stated in the text, this convention is adopted also in what
follows.  As regards the flavoured reaction rates, they can be simply
written in terms of the unflavoured rates by means of the flavour
projectors Eq.~\eqref{eq:fla_proj}:
\begin{equation}
\label{eq:flavouredrate}
\gamma_X^{\alpha} =P_\alpha \gamma_X\; .
\end{equation}
%
% where the $\gamma_X$ are defined in eqs.~\eqref{eq:gammas}.

\subsection{Flavour structures}
\label{sec:fla_scenarios}

The flavour structure $Z_\alpha$ of the $A$ terms
Eq.~\eqref{eq:soft_terms_gen} is in principle independent from the
flavour structure of the Yukawa couplings $Y_\alpha$. However, the
study of flavour effects in a completely general flavour configuration
would be rather awkward, because of the very large dimensionality of
the parameter space.  It is thus convenient to define less general
possibilities, but chosen in such a way to render possible a
qualitative extrapolation of the impact of flavour effects to the
general case.  In Refs.~\refcite{Fong:2008a} and~\refcite{Fong:2010zu} 
the following two scenarios were respectively introduced:

1. {\it Universal Trilinear Scenario} (UTS).\cite{Fong:2008a} This
case assumes universal soft terms, that is the soft breaking
Lagrangian of Eq.~\eqref{eq:soft_terms}. It is realized in
supergravity or gauge mediation (when the renormalization group
running of the parameters is neglected) and corresponds to set
\begin{equation}
Z_\alpha=  Y_\alpha. 
\label{eq:uts}
\end{equation}
Thus, in UTS the only flavour structure arises from the Yukawa
couplings and both the flavoured total CP asymmetries
$\epsilon_\alpha=\epsilon^S_\alpha
+\epsilon^V_\alpha+\epsilon^I_\alpha$ and the flavoured washout terms
$W_\alpha$ are proportional to the same flavour projectors $P_\alpha$.
It follows  that $\epsilon_\alpha\propto W_\alpha$, and 
moreover, as seen in Eq.~\eqref{eq:CP_phase1_gen}, 
the trilinear couplings have an unique phase 
$\phi_{A_\alpha}=\phi_A=\arg(AB^*)$.

2. {\it Simplified Misaligned Scenario} (SMS).\cite{Fong:2010zu} To understand 
the possible effects of flavour dynamics, it 
is important to study also a case in which the flavoured 
CP asymmetries $\epsilon_\alpha$ and the washouts $W_\alpha$ are 
misaligned.  This can be done without 
increasing the number of independent 
parameters with respect to  UTS  by imposing the condition\cite{Fong:2010zu} 
\begin{equation}
Z_{\alpha}=\frac{\displaystyle
\sum_\beta |Y_{\beta}|^2}{3 Y^*_{\alpha}}\; , 
\label{eq:sms}
\end{equation}
where we have kept $Z$ and $Y$ explicitly as complex numbers.  With
this condition the CP asymmetries (except for the last term in
Eq.~\eqref{eq:CP_asymver}) are equal for all flavours
$\epsilon_\alpha =\epsilon/3$, while the washouts $W_\alpha$ maintain
their flavour dependence, so that an arbitrary misalignment can be
realized.  From Eq.~\eqref{eq:CP_phase1_gen} we see that there is
again a unique phase $\phi_{A_\alpha}=\phi_A=\arg(A B^*)$.  Note that
both Eq.\eqref{eq:uts} and Eq.~\eqref{eq:sms} yield the same total
asymmetry $\sum_\alpha\epsilon_\alpha=\epsilon$, so that any
difference between the UTS and SMS results can be ascribed directly to
the differences in flavour configuration. Note also that in the case
of flavour equipartition $P_e =P_\mu = P_\tau =1/3$ the two scenarios
are equivalent.

Finally, it should be remarked that in both scenarios it is not
possible to have flavour asymmetries of opposite signs, with
$|\epsilon^\alpha| > |\epsilon|$ for some, or even for all flavours,
as it is also excluded the possibility of having non-zero flavour
asymmetries and a vanishing total CP asymmetry.  This however, is
simply due to the reduction in the number of physical phases and,
similarly to what happens in standard flavoured
leptogenesis,\cite{Abada:2006a,Nardi:2006b,Abada:2006b} in the most
general case asymmetries of opposite signs (and possibly with a
vanishing sum) are an open possibility. Given that these types of
configurations are always characterized by larger CP asymmetries, the
reader should keep in mind that enhancements of the final lepton
asymmetry even larger than the ones found in UTS and SMS are certainly
possible.

%%%%%%%%%%%%%%%%%%%%%%%%%%%%%%%%%%%%%%%%%%%%%%%%%%%%%%%%%%%%

\subsection{Lepton flavour equilibration}
\label{sec:lfe}

The possibility of having large enhancements of the baryon asymmetry
yield of leptogenesis from flavour effects, relies on the fact that
the density asymmetries stored in each flavour are independent from
each other, and if for example a flavour that is weakly coupled to the
washouts has a particularly large  CP asymmetry, the final result 
will be essentially determined by the dynamics of this flavour.

However, in the presence of lepton flavour violating (LFV)
interactions, the density asymmetries in the different flavours are no
more independent, and if LFV rates are sufficiently fast, they can
bring the different flavours into equilibrium, with the result that
the amount of surviving asymmetry will be essentially determined by
the flavour that is more strongly washed out: a potentially
destructive effect.\cite{Aristizabal:2009a}

In SL, LFV interactions are a natural possibility since they are
directly related to off-diagonal entries in the soft mass matrices of
the sleptons. For this reason a reliable study of flavoured SL
must also include an analysis of LFV effects.

In the basis where charged lepton Yukawa couplings are diagonal,
the soft slepton masses read
\begin{eqnarray}
\mathcal{L}_{soft} & \supset & -\widetilde{m}_{\alpha\beta}^{2}
\widetilde{\ell}_{\alpha}^{*}\widetilde{\ell}_{\beta}.
\end{eqnarray}
The off-diagonal soft slepton masses 
$\widetilde{m}_{\alpha\neq\beta}^{2}
\equiv 
\widetilde{m}_{\alpha\beta}^{2}
\,\;(\alpha\neq\beta)$ affect the
flavour composition of the slepton mass eigenstates so generically we
can write  
\begin{equation}
\widetilde{\ell}_{\alpha}^{\left(int\right)}  = 
R_{\alpha\beta}\widetilde{\ell}_{\beta}, 
\end{equation}
where $R_{\alpha\beta}$ is a unitary rotation matrix.  In this basis the
corresponding slepton-gaugino interactions in Eq.~\eqref{eq:mass_basis} 
become
\begin{eqnarray}
-\mathcal{L}_{\widetilde{\lambda},\widetilde \ell} 
& = &g_{2}\left(\sigma_{\pm}\right)_{ab}
\overline{\widetilde{\lambda}_{2}^{\pm}}P_{L}\ell_{\alpha}^{a}
R^*_{\alpha\beta}\widetilde{\ell}_{\beta}^{b*}
+\frac{g_{2}}{\sqrt{2}}\left(\sigma_{3}\right)_{ab}
\overline{\widetilde{\lambda}_{2}^{0}}
P_{L}\ell_{\alpha}^{a}R_{\alpha\beta}^*
\widetilde{\ell}_{\beta}^{b*}\nonumber\\  
& & +\frac{g_{Y}}{\sqrt{2}}\delta_{ab}
\overline{\widetilde{\lambda}_{1}} %\left(
Y_{\ell L}P_{L}   %-Y_{\ell R}P_{R}\right)
\ell_{\alpha}^{a}R^*_{\alpha\beta}
\widetilde{\ell}_{\beta}^{b*} \,+\,{\rm h.c.}\; ,
\label{eq:lgaugino}
\end{eqnarray}
%
%where $Y_{\ell L} = -1$ and $Y_{\ell R=2}$  
%the hypercharges of the left- and right-handed (s)lepton. 
The mixing matrix can be expressed in terms of the off-diagonal
slepton masses as:
\begin{eqnarray}
R_{\alpha\beta} & \sim & \delta_{\alpha\beta}+
\frac{\widetilde{m}_{\alpha\neq\beta}^{2}}{h_{\alpha}^{2}T^{2}}
% \nonumber \\ & = & 
= \delta_{\alpha\beta}+\frac{\widetilde{m}_{\alpha\neq\beta}^{2}
v^2\cos^2\beta}{m_{\alpha}^{2}M^{2}}z^{2},
\end{eqnarray}
where in the first line $h_{\alpha}>h_{\beta}$ is the relevant charged Yukawa
coupling that determines at leading order the thermal mass splittings
of the sleptons, $v$ in the second line is the EW symmetry
breaking VEV with $v^2=v^2_u+v^2_d \simeq 174\,$GeV,
$z\equiv\frac{M}{T}$ 
%where $T$ is the temperature and $M$ the mass of the RHN, 
and $m_\alpha \equiv m_{\ell_\alpha}(T\!=\!0)$ is the zero temperature
mass for the lepton $\ell_\alpha$.  For simplicity we parametrize 
the  $R_{\alpha\neq\beta}$ entries in a way that they are
independent of the particular pair of leptons involved.
Let us define
$\widetilde{m}_{e\tau}=\widetilde{m}_{\mu\tau}
=\widetilde{m}_{od}$ 
and $\widetilde{m}_{e\mu}=\widetilde{m}_{od} \frac{m_\mu}{m_\tau}$, 
where $\widetilde{m}_{od}$ 
is a unique {\it off-diagonal} soft-mass
parameter.  We then obtain for the off-diagonal entries 
$(\alpha\beta)=(e\tau),\,(\mu\tau),\,(e\mu)$
of the matrix $R$:
\begin{eqnarray}
  R_{\alpha\neq\beta} 
  & \sim & \frac{\widetilde{m}_{od}^{2}
    \, v^2\cos^2\beta}{m_{\tau}^{2}M^{2}}\,z^{2},
\label{eq:rij}
\end{eqnarray}
where $m_{\tau}$ is the mass of the tau lepton.

%%%%%%%%%%%%%%%%%%%%%%%%%%%%%%%%%%%%%%%%%%%%%
\begin{figure}[htb]
\includegraphics[width=\textwidth]{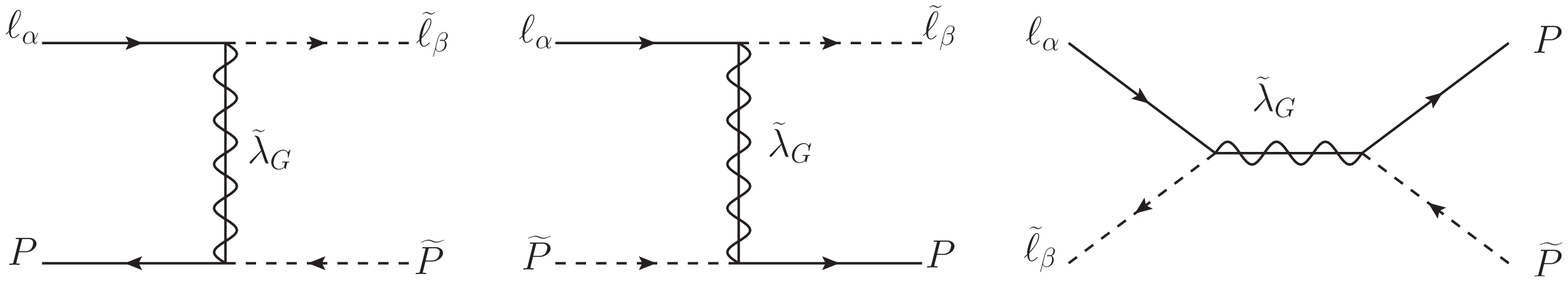}
\caption{LFV lepton-slepton scatterings mediated by 
the  $SU(2)_L$ and $U(1)_Y$ gauginos $\widetilde\lambda_G$.}
\label{fig:lfv_diagrams}
\end{figure} 
%%%%%%%%%%%%%%%%%%%%%%%%%%%%%%%%%%%%%%%%%%%%%

The terms in Eq.~\eqref{eq:lgaugino} induce LFV lepton-slepton
scatterings through the exchange of $SU(2)_L$ and $U(1)_Y$
gauginos.\cite{Aristizabal:2009a} There are two possible t-channel
scatterings
$\,\ell_{\alpha}\overline{P}\leftrightarrow\widetilde{\ell}_{\beta}\widetilde{P}^*$,
$\,\ell_{\alpha}\widetilde{P}\leftrightarrow\widetilde{\ell}_{\beta}P$
and one s-channel scattering
$\,\ell_{\alpha}\widetilde{\ell}_{\beta}^{*}\leftrightarrow
P\widetilde{P}^*$ (we denote $P$ as fermions and $\widetilde{P}$ as
scalars) as shown in Fig. \ref{fig:lfv_diagrams}.  For processes
mediated by $SU(2)_L$ gauginos $P=\ell,Q,\widetilde{H}_{u,d}$, while
when mediated by $U(1)_Y$ gaugino one must include the $SU(2)_L$
singlet states $P=e,u,d$ as well.  The corresponding reduced cross
sections % (defined in Eq.~\eqref{eq:rcs})
read:
\begin{eqnarray}
\hat{\sigma}_{t1,G}^{\alpha\beta}\left(s\right) & = & {\displaystyle \sum_P}
\frac{g_{G}^{4}\left|R_{\alpha\beta}\right|^{2}\Pi_P^{G}}{8\pi}
\left[\left(\frac{2m_{\widetilde{\lambda}_{G}}^{2}}{s}+1\right)
\ln\left|\frac{m_{\widetilde{\lambda}_{G}}^{2}+s}
{m_{\widetilde{\lambda}_{G}}^{2}}\right|-2\right], \nonumber\\ 
\hat{\sigma}^{\alpha\beta}_{t2,G}\left(s\right) 
 & = & {\displaystyle \sum_P}
\frac{g_{G}^{4}\left|R_{\alpha\beta}\right|^{2}\Pi_P^{G}}{8\pi}
\left[\ln\left|\frac{m_{\widetilde{\lambda}_{G}}^{2}+s}
{m_{\widetilde{\lambda}_{G}}^{2}}\right|
-\frac{s}{m_{\widetilde{\lambda}_{G}}^{2}+s}\right], \nonumber\\ 
\hat{\sigma}^{\alpha\beta}_{s,G}\left(s\right) 
 & = & {\displaystyle \sum_P}
\frac{g_{G}^{4}\left|R_{\alpha\beta}\right|^{2}\Pi_P^G}
{16\pi}\left(\frac{s}{s-m_{\widetilde{\lambda}_{G}}^{2}}\right)^{2}  \; ,
\label{eq:sigLFE}
\end{eqnarray}
where $\Pi_{P}^{G}$ counts the numbers of degrees of freedom of the
particle $P$ (isospin, quark flavours and color) involved in the scatterings
mediated by the $SU(2)_L$  ($G=2$) and $U(1)_Y$ ($G=Y$) gauginos respectively. 
In this last case the hypercharges $y_{\ell L}$ and $y_P$ 
are also included in  $\Pi_{P}^{Y}$.
If the flavour changing scatterings in Eq.~\eqref{eq:sigLFE} are fast
enough, they will  lead to \emph{lepton flavour equilibration} 
(LFE).\cite{Aristizabal:2009a}

The values of  $\widetilde{m}_{od}$ for which this occurs can be
estimated by comparing the  LFE scattering rates and
the $\Delta L=1$  washout rates.
Since the dominant $\Delta L=1$  contribution arises from inverse decays, 
the terms to be compared are: 
\begin{equation}
  \label{eq:G-lfe}
  \overline{\Gamma}_{\rm LFE}(T)   
\equiv\frac{\gamma_{\rm LFE}(T)}{n^{c}_L(T)}, \qquad \qquad 
\overline{\Gamma}_{\rm ID}(T)\equiv \frac{\gamma_{\tilde N}(T)}
{n^{c}_L(T)}\,, 
\end{equation}
where  $n^{c}_L=T^3/2$ is the relevant
density factor appearing in the washouts 
(see next section for more details) and 
\begin{eqnarray}
\gamma_{\rm LFE}(T) &=&
{\displaystyle \sum_{G,P}} \Pi^G_P
(\gamma^{\alpha\beta}_{t1,G}
+\gamma^{\alpha\beta}_{t2,G}+\gamma^{\alpha\beta}_{s,G})
 \nonumber \\ &= & 
\frac{T}{64 \pi}  
{\displaystyle \sum_{G}} 
\int ds \sqrt{s} {\cal K}_1\left(\frac{\sqrt{s}}{T}\right)
% \nonumber \\ &  & 
\, \left[\hat \sigma^{\alpha\beta}_{t1,G}(s)
+\hat \sigma^{\alpha\beta}_{t2,G}(s)
+\hat \sigma^{\alpha\beta}_{s,G}(s)\right]\;, 
\label{eq:g-lfe}
\\
\gamma_{\tilde N}(T) & =& n_{\tilde N}^{eq}(T)\,
\frac{{\cal K}_1(z)}{{\cal K}_2(z)}\, \Gamma\;.
\end{eqnarray}
In the first equality in Eq.\eqref{eq:g-lfe}
$\gamma^{\alpha\beta}_{x,G}$ with ($x=t1,t2,s$) is the thermally
averaged LFE reactions for one degree of freedom of the $P$-particle,
$\mathcal{K}_{1,2}(z)$ are the modified Bessel function of the second
kind of order 1 and 2, $\Gamma$ is the zero temperature width
Eq.~\eqref{eq:gamma}, and $n_{\widetilde N}^{eq}(T)$ is the
equilibrium number density for $\widetilde N$ at $T$.

The LFE reaction densities, the Universe expansion rate $H$, and inverse
decay rates all have a different $T$ dependence. In particular
$\overline{\Gamma}_{\rm LFE} \sim T^{-3}$ and $H(T) \sim T^2$, so that
once LFE reactions have attained thermal equilibrium
($\overline{\Gamma}_{\rm LFE} \gsim H$) they will remain in thermal
equilibrium also at lower temperatures. In contrast, for inverse
decays the rate first increases, and then
decreases exponentially $\overline{\Gamma}_{\rm ID}\sim e^{-M/T}$
dropping out of equilibrium at temperatures not much below $T\sim M$.
The relevant temperature where we should compare the rates of these
interactions is precisely when the condition $\overline{\Gamma}_{\rm
  ID} \approx H$ is reached, that is when the lepton asymmetry starts
being generated from the out-of-equilibrium $\widetilde N_\pm$ decays.
Defining $z_{dec}$ through the condition 
$\overline{\Gamma}_{\rm ID}(z_{dec})=H(z_{dec})$,
LFE is expected to be quite relevant for flavoured leptogenesis when
\begin{eqnarray}
\overline{\Gamma}_{\rm LFE}\left(z_{dec}\right) &\geq& 
\overline{\Gamma}_{ID}\left(z_{dec}\right), 
\end{eqnarray}
since when this condition is satisfied, LFE processes are already in
thermal equilibrium at the onset of the out-of-equilibrium decay era.

\begin{figure}[htb]
%\centering
\includegraphics[width=0.49\textwidth]{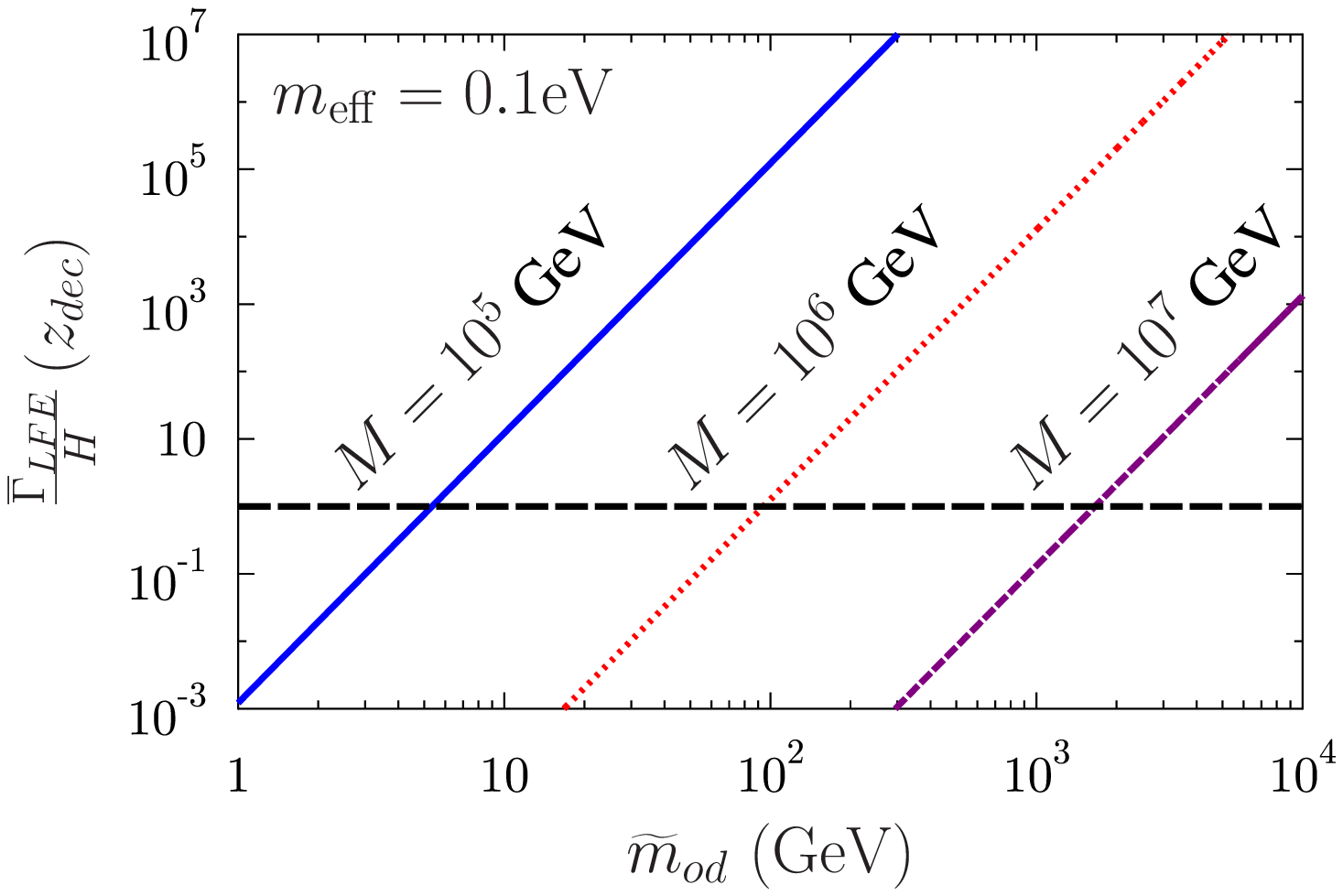}
\includegraphics[width=0.49\textwidth]{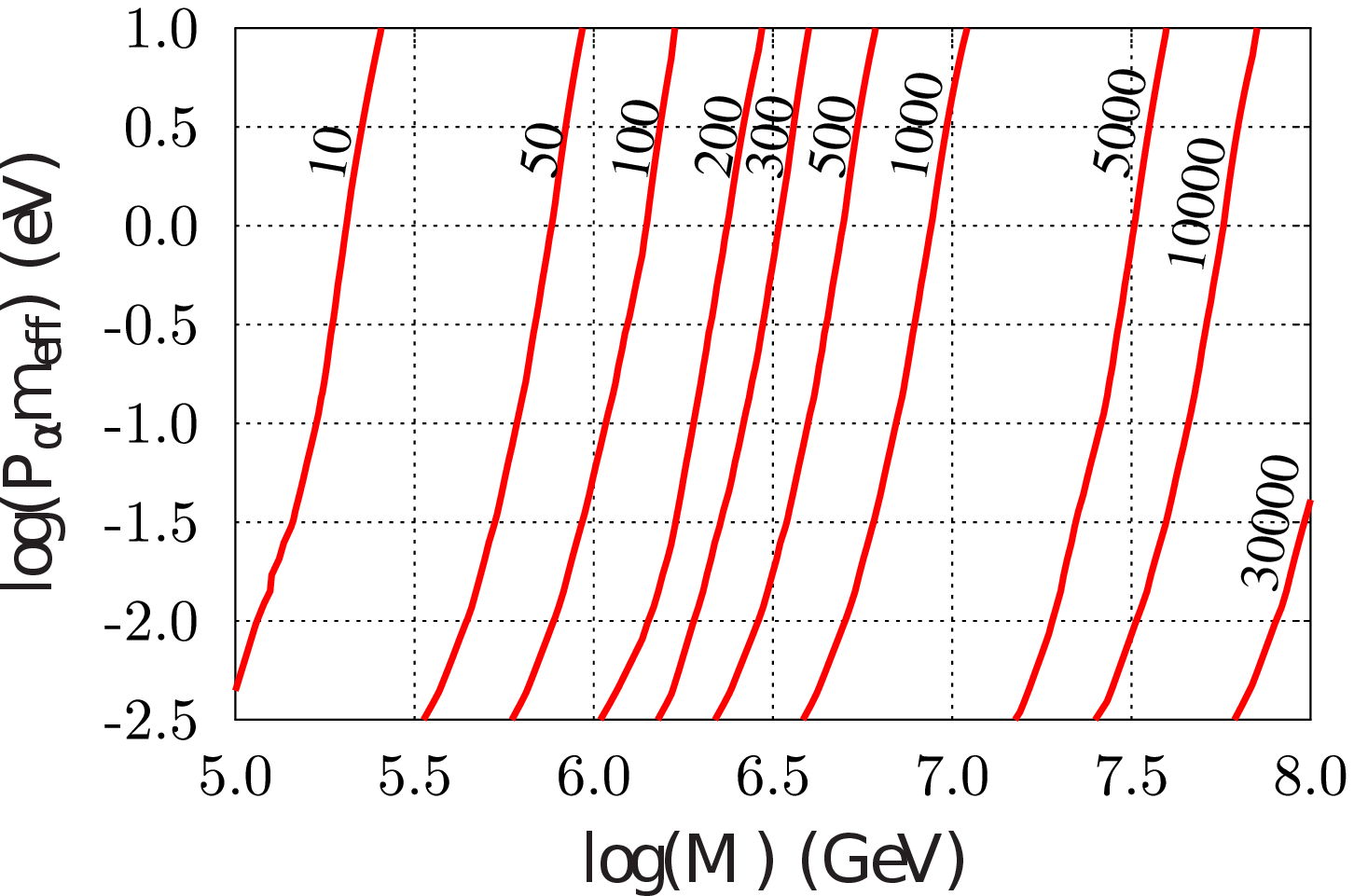}
%\epsfig{file=LFE_mslep,width=0.5\textwidth,angle=0}
%\hspace*{-.2cm} 
%\epsfig{file=M_Pmeff_mod,width=0.5\textwidth,angle=0}
\caption{The left panel shows the ratio $\bar \Gamma_{\rm LFE}/H$ 
% to  the Hubble expansion rate $H$ 
at $z_{dec}$ as a function of
  $\widetilde{m}_{od}$ for $m_{\rm eff}=0.1$ eV, $\tan\beta=30$ and
  three values of $M$.  The right panel shows in the 
($P_\alpha m_{\rm eff}, M$) plane contours of constant values of
  $\widetilde{m}_{od}$ (in GeV) for which
  $\overline{\Gamma}_{\rm LFE} \left(z_{dec}^\alpha\right) \geq
  P_\alpha\overline{\Gamma}_{\rm ID}\left(z_{dec}^\alpha\right)$.}
\label{fig:lfe}
\end{figure}

The left panel of Fig. \ref{fig:lfe} shows the ratio
$\overline{\Gamma}_{\rm LFE}(z_{dec})/H(z_{dec})$ as a function of
$\widetilde{m}_{od}$ for $m_{\rm eff}=0.1$ eV, 
%(defined in Eq.~\eqref{eq:gamma}), 
$\tan\beta=30$, and for different values of $M$.
From the figure we can read the characteristic value of
$\widetilde{m}_{od}$ for which LFE becomes relevant.  Notice that the
dominant dependence on $\tan\beta\sim 1/\cos\beta $ ($\tan\beta \gg
1$) arises from $v\cos\beta=v_d$ in Eq.~\eqref{eq:rij}. Thus the
results from other values of $\tan\beta$ can be easily read from the
figure by rescaling $\widetilde{m}_{od}(\tan\beta)
=\widetilde{m}^{\rm fig}_{od}/(30 \cos\beta)$.

Since we are interested in the dynamics of lepton flavours, to be more
precise about LFE effects we should in fact consider the temperature
at which the inverse decay rate for one specific flavour $\alpha$ goes
out of equilibrium. We then define $z_{dec}^\alpha$ through the
condition $\overline{\Gamma}_{\rm ID}^\alpha(z_{dec}^\alpha) \equiv
P_\alpha\overline{\Gamma}_{\rm ID}(z_{dec}^\alpha) =
H(z_{dec}^\alpha)$. For a hierarchical configuration of flavour
projectors $P_{\gamma}<P_{\beta}<P_{\alpha}$ we will have
$z_{dec}^{\gamma}<z_{dec}^{\beta}<z_{dec}^{\alpha}$, that is the inverse decays
involving the lepton doublet $\ell_{\gamma}$ which is the most weakly
coupled to $\widetilde N_\pm$ will go out of equilibrium first, and
then $\overline\Gamma_{\rm ID}^{\beta}$ and $\overline{\Gamma}_{\rm
  ID}^{\alpha}$ will follow.  Hence, for given values of $m_{\rm eff}$ and
$M$, the minimum value $\widetilde{m}_{od}^{min}$ for which LFE
effects start being important is given by the following condition:
\begin{eqnarray}
\overline{\Gamma}_{\rm LFE}
\left(z_{dec}^{\alpha},\widetilde{m}_{od}^{min}\right) &\simeq& 
\overline{\Gamma}_{\rm ID}^{\alpha}\left(z_{dec}^\alpha \right).
% \;\;\;\Rightarrow 
%   \mbox{\ \ determines\ \  $\widetilde{m}_{od}^{min}$}. 
\label{eq:LFE_condition1}
\end{eqnarray}
For $\widetilde{m}_{od} \ll \widetilde{m}_{od}^{min}$ LFE effects can
be neglected, since they will attain thermal equilibrium only after
leptogenesis is over.

The right panel in Fig. \ref{fig:lfe} shows 
in  the plane of the flavoured effective decay mass 
$P_\alpha m_{\rm eff}$ and of the RHN mass $M$,
various contours corresponding to different values 
of $\widetilde{m}_{od}$  for which 
$\overline{\Gamma}_{\rm LFE}\left(z_{dec}^\alpha\right)
=\Gamma_{\rm ID}^\alpha\left(z_{dec}^\alpha\right)$.
For a given value of $M$ and $m_{\rm eff}$, 
and  for a given set of flavour projections 
$P_\gamma<P_\beta< P_\alpha$, $\widetilde{m}_{od}^{min}$ 
is given by the value of the $\widetilde{m}_{od}$  
curve for  which the vertical line  $x=M$ 
intersects  the corresponding contour at  $y_\alpha=P_\alpha\, m_{\rm eff}$. 
Since $\overline{\Gamma}_{\rm LFE}$ has a rather strong
dependence on $\widetilde{m}_{od}$ ($ \overline{\Gamma}_{\rm LFE}
\propto \widetilde{m}_{od}^4$), one expects that the value
$\widetilde{m}_{od}^{eq}$ for which LFE effects completely equilibrate
the asymmetries in the different lepton flavours will not be much
larger than $\widetilde{m}_{od}^{min}$. Indeed our numerical results
(see Section \ref{sec:num_lfe}) show that $\widetilde{m}_{od}^{eq}
\sim {\cal O}(5-10)\times \widetilde{m}_{od}^{min}$.  Clearly, larger
values $\widetilde{m}_{od}\gg \widetilde{m}_{od}^{eq}$ do not modify
the final results with respect to what is obtained when
$\widetilde{m}_{od} = \widetilde{m}_{od}^{eq}$.

%%%%%%%%%%%%%%%%%%%%%%%%%%%%%%%%%%%%%%%%%%%%%%%%%%%%%%%%%%%%%%%%%%

\subsection{Flavoured Boltzmann equations} 
\label{sec:flavored_BE}

In Refs.~\refcite{Barbieri:2000,Abada:2006a,Nardi:2006b} the relevant
equations including flavour effects associated to the charged lepton
Yukawa couplings were derived in the density operator approach. One
can define a density matrix for the difference of lepton and
antileptons such that $\rho_{\alpha\alpha}=Y_{\Delta\ell_\alpha}$.  As
discussed in Refs.~\refcite{Barbieri:2000,Abada:2006a,Nardi:2006b} as
long as we are in the regime in which a given set of the
charged--lepton Yukawa interactions are out of equilibrium, one can
restrict the general equation for the matrix density $\rho$ to a
subset of equations for the flavour diagonal directions
$\rho_{\alpha\alpha}$.  In the transition regimes in which a given
Yukawa interaction is approaching equilibrium the off-diagonal entries
of the density matrix cannot be
neglected.\cite{Abada:2006a,Abada:2006b,DeSimone:2007a} However, as we
have discussed, SL always occurs in the three flavour regime, and thus
there is no need to worry about this type of effects.  In this Section
we keep working, as in Section~\ref{sec:unflavored}, under the
assumption of superequilibration.  However, we now include the effects
of all the relevant fast reactions of type (i) (see
Section~\ref{sec:effective}) with characteristic time scales much
shorter than $H^{-1}(z=1)$, that is we include both flavour and
spectator effects. This implies that we need to consider  
three flavoured lepton density asymmetries that also include the 
contributions of the $SU(2)_L$ singlet (s)leptons,  defined as:
%!!%
%
\be
\label{eq:YellYe}
Y_{\Delta L_\alpha} 
\equiv 2\left( Y_{\Delta\ell_\alpha} 
+ Y_{\Delta\widetilde\ell_\alpha} \right)
+ Y_{\Delta e_\alpha} + Y_{\Delta\widetilde e_\alpha}, 
\ee
%
%with $\alpha=e,\mu,\tau$.  
where the factor of 2 comes
from summing over the $SU(2)_L$ degrees of freedom.
The contribution of $e_\alpha$ and
$\widetilde e_\alpha$ to the total flavour asymmetries is
non-vanishing because scatterings with the Higgs induced by the
charged lepton Yukawa couplings transfer part of the asymmetry
generated in the $SU(2)_L$ lepton doublets to the singlets.

In the BE for the evolution of the RHN and RHSN densities, a sum over
flavour can be readily taken, and the resulting equations are not
modified with respect to Eqs.~\eqref{eq:BEN} and \eqref{eq:BENt}.  The
lepton charges $\Delta L_\alpha$ are anomalous, and thus are
not only perturbatively violated in RHSN and RHN interactions, but
also in nonperturbative sphalerons transitions.  This type of effects
is however removed by writing evolution equations for the anomaly free
flavoured charges $\Delta_\alpha \equiv B/3-L_\alpha$, with density
asymmetries defined as $Y_{\Delta_{\alpha}} \equiv Y_{\Delta B}/3
-Y_{\Delta L_{\alpha}}$. Including
spectator\cite{Buchmuller:2001,Nardi:2006a} and
LFE\cite{Aristizabal:2009a,Fong:2010zu} effects, 
the corresponding BE is:
%!!%
%
\begin{eqnarray}
-\dot{Y}_{\Delta_{\alpha}}  & = &
\epsilon_{\alpha}\left(z\right)\frac{\gamma_{\widetilde{N}}}{2}
\left(\frac{Y_{\widetilde{N}_{\rm tot}}}{Y_{\widetilde{N}}^{eq}}-2\right)
\nonumber \\
& & - \left[\frac{\gamma^\alpha_{\widetilde{N}}}{2}
+ \frac{\gamma^\alpha_{N}}{2}+
\gamma_{\widetilde{N}}^{\left(3\right)\alpha}+
\left(\frac{1}{2}\frac{Y_{\widetilde{N}_{\rm tot}}}
{Y_{\widetilde{N}}^{eq}}
+2\right)\gamma^\alpha_{22}\right] 
\left(
{\cal Y}_{\Delta\ell_{\alpha}}
+
{\cal Y}_{\Delta \widetilde H_u}
\right)
\nonumber \\
&& -2\left(\gamma_{t}^{\left(1\right)\alpha}
+\gamma_{t}^{\left(2\right)\alpha}
+\gamma_{t}^{\left(4\right)\alpha}+\gamma_{t}^{\left(6\right)\alpha}
+\gamma_{t}^{\left(7\right)\alpha}+\gamma_{t}^{\left(9\right)\alpha}\right)
{\cal Y}_{\Delta\ell_{\alpha}}
 \nonumber \\ &  & 
-\left[\left(2\gamma_{t}^{\left(0\right)}
+\gamma_{t}^{\left(3\right)\alpha}\right)\frac{Y_{N}}{Y_{N}^{eq}}
+\left(\gamma_{t}^{\left(5\right)\alpha}
+\frac{1}{2}\gamma_{t}^{\left(8\right)\alpha}\right)
\frac{Y_{\widetilde{N}_{\rm tot}}}{Y_{\widetilde{N}}^{eq}}\right]
{\cal Y}_{\Delta\ell_{\alpha}}
 \nonumber \\  &  & 
 \hspace{-1.5cm}
-\left(2\gamma_{t}^{\left(0\right)\alpha}
+\gamma_{t}^{\left(1\right)\alpha}
+\gamma_{t}^{\left(3\right)\alpha}+\gamma_{t}^{\left(4\right)\alpha}
+2\gamma_{t}^{\left(5\right)\alpha}\right.
% \nonumber \\ && 
\left.+\gamma_{t}^{\left(6\right)\alpha}
+\gamma_{t}^{\left(7\right)\alpha}+\gamma_{t}^{\left(8\right)\alpha}
+\gamma_{t}^{\left(9\right)\alpha}\right)
{\cal Y}_{\Delta \widetilde H_u}
 \nonumber \\
 &  & \hspace{-1.5cm}-\left[\left(\gamma_{t}^{\left(1\right)\alpha}
+\gamma_{t}^{\left(2\right)\alpha}
+\gamma_{t}^{\left(4\right)\alpha}\right)\frac{Y_{N}}{Y_{N}^{eq}}
+\frac{1}{2}\left(\gamma_{t}^{\left(6\right)\alpha}
+\gamma_{t}^{\left(7\right)\alpha}+\gamma_{t}^{\left(9\right)\alpha}\right)
\frac{Y_{\widetilde{N}_{\rm tot}}}{Y_{\widetilde{N}}^{eq}}\right]
{\cal Y}_{\Delta \widetilde H_u} 
\nonumber \\ &  & 
\hspace{-1.5cm}- \sum_{\beta \neq \alpha}\left[
 84 \left(\gamma^{\beta\alpha}_{t1,2}+\gamma^{\beta\alpha}_{t2,2}
 +\gamma^{\beta\alpha}_{s,2}\right)
% \frac{Y_{\Delta L^\alpha_{\rm tot}}-Y_{\Delta L^\beta_{\rm tot}}}{Y_c^{eq}}
% \nonumber \\ & - & 
+ 76 % \sum_{\beta\neq \alpha}
\left(\gamma^{\beta\alpha}_{t1,Y}
+\gamma^{\beta\alpha}_{t2,Y}+\gamma^{\beta\alpha}_{s,Y}\right)
\right]
\left( {\cal Y}_{\Delta \ell_\alpha}
-{\cal Y}_{\Delta \ell_\beta} \right), 
\quad 
\label{eq:BE_L_tot_fla}
\end{eqnarray}
%
%!!%
where ${\cal Y}_{\ell_\alpha} \equiv 
Y_{\Delta \ell_\alpha}/Y_{\ell}^{eq}$
and ${\cal Y}_{\widetilde H_u} \equiv 
Y_{\Delta \widetilde H_u}/Y_{\widetilde H_u}^{eq}$
with $Y_{\ell}^{eq} = Y_{\widetilde H_u}^{eq} = \frac{15}{8\pi^2 g_*}$.
Notice that SE implies
$2 Y_{\Delta\ell_\alpha} = Y_{\Delta\widetilde\ell_\alpha}$
and $2 Y_{\Delta \widetilde H_u} = Y_{\Delta H_u}$ and
we chose to express the asymmetries in term of the fermionic ones. 
The higgsino asymmetries
%where $Y_{\Delta H_{{\rm tot}}}$ is the total Higgs asymmetry for
%$H_u$ and $\widetilde H_u$, 
enter in two ways:
directly, when the Higgs(inos) are involved in the relevant process as
external particles, as in RHSN decays and inverse decays; indirectly,
when the top and stop quarks are involved in the relevant scatterings,
and their chemical potentials are rewritten in terms of 
$Y_{\Delta \widetilde H_u}$ (see \ref{sec:BE_SE}).  The last line in
Eq.~\eqref{eq:BE_L_tot_fla} includes the reaction densities for the
LFE processes in Eq.~\eqref{eq:sigLFE}, that as discussed above play
the role of controlling the effectiveness of flavour effects.

To bring Eq.~\eqref{eq:BE_L_tot_fla} in the form of a closed
differential equations for the charge density asymmetries
$Y_{\Delta_{\alpha}}$, the three asymmetries $Y_{\Delta \ell_\alpha}$ 
as well as the higgsino asymmetry $Y_{\Delta \widetilde H_u}$
must be expressed in terms of $Y_{\Delta_{\alpha}}$ according to:
\begin{equation}
\label{eq:leptonHiggs}
Y_{\Delta \ell_{\alpha}} =  \sum_{\beta}A_{\alpha\beta}
Y_{\Delta_{\beta}},\qquad\qquad 
Y_{\Delta \widetilde H_u} = \sum_{\beta}C_{\beta}Y_{\Delta_{\beta}}.
\end{equation}
The matrix $A$ was first introduced in Ref.~\refcite{Barbieri:2000}
and the vector $C$ in Ref.~\refcite{Nardi:2006a}.  The values of their
entries are determined by which interactions are in chemical
equilibrium, and thus eventually by the specific range of temperature
when leptogenesis takes place (see \ref{app:se_new}).

%%%%%%%%%%%%%%%%%%%%%%%%%%%%%%%%%%%%%%%%%%%%%%%%%%%%%%%%%%%%%%%%%%

\subsection{Results}
\label{sec:flavored_results}

Unless differently stated, our results are obtained for $M=10^6\,$GeV
and $\tan\beta=30$, although as long as LFE effects are negligible
they are practically independent of $M$.
The dependence on $\tan\beta$ is also rather mild since it mainly
arises from $m_{\rm eff}$ as given in Eq.~(\ref{eq:meff})
through $v_u^2 \simeq v^2 (1+1/\tan\beta^2)^{-1}$.  For  
 $\tan\beta=30$ the $d$-quark and electron Yukawa couplings 
are sufficiently large to ensure that around 
$T\sim 10^6\,$GeV the corresponding interactions are,  
like all other Yukawa interactions, in thermal equilibrium.
In this regime the $A$ and $C$ matrices are\cite{Fong:2010qh}
%
%!!%
\begin{eqnarray}
A=\frac{1}{9 \times 237}\left(\begin{array}{ccc}
-221 & 16 & 16\\
16 & -221 & 16\\
16 & 16 & -221
\end{array}\right), &\;\;\;\;\;& C=-\frac{4}{237}\left(1\;\;1\;\;1\right).
\label{eq:AC_Matrix}
\end{eqnarray}
%
% \begin{eqnarray}
% A=\frac{2}{711}\left(\begin{array}{ccc}
% -221 & 16 & 16\\
% 16 & -221 & 16\\
% 16 & 16 & -221
% \end{array}\right), &\;\;\;\;\;& C=-\frac{8}{79}\left(1\;\;1\;\;1\right).
% \label{eq:AC_Matrix}
% \end{eqnarray}
%

We anticipate that the impact of flavour and spectator corrections
on the results will be sizable only in the strong washout regime. 
This can be easily understood by adding the equations for the three 
flavour asymmetries, Eq.~\eqref{eq:BE_L_tot_fla}. This results 
in an equation for  
$Y_{\Delta_{B-L}}$ that can be recast in the form 
%??% Y_{\ell}^{eq}
%
\begin{equation}
-\dot{Y}_{\Delta_{B-L}} = 
\epsilon(z)
\left(\frac{Y_{\widetilde{N}_{\rm tot}}}{Y_{\widetilde{N}}^{eq}}-2\right) 
\frac{\gamma_{\widetilde{N}}}{2} -  
W \sum_{\alpha\beta} P_\alpha \,A_{\alpha\beta}
\frac{Y_{\Delta_{\beta}}}{Y_{\ell}^{eq}}
- W_H \sum_{\alpha} P_\alpha\, C_{\alpha}\
\frac{Y_{\Delta_{\alpha}}}{Y_{\ell}^{eq}} \,,
\label{eq:Be_fla_sum}
\end{equation} 
where $\epsilon(z) =\sum_\alpha\epsilon_\alpha(z)$, $W$ is the washout
term related to the lepton density asymmetries given in
Eq.~\eqref{eq:BE_W}, and $W_H$ is an additional washout term related
to the Higgs density asymmetry, whose expression can be easily worked
out from Eq.~\eqref{eq:BE_L_tot_fla}. Of course,
Eq.~\eqref{eq:Be_fla_sum} cannot be solved alone, since at least other
two equations for two density asymmetries $Y_{\Delta_{\alpha}}$ are
needed to get a closed system.  However, since flavour and spectator
effects are encoded in the $\sum_{\alpha\beta}$ and $\sum_\alpha$
terms, this equation clearly shows that if $W\,,W_H \to 0$ none of
these effects will be important.

\begin{figure}[htb]
\centering
\includegraphics[width=0.7\textwidth]{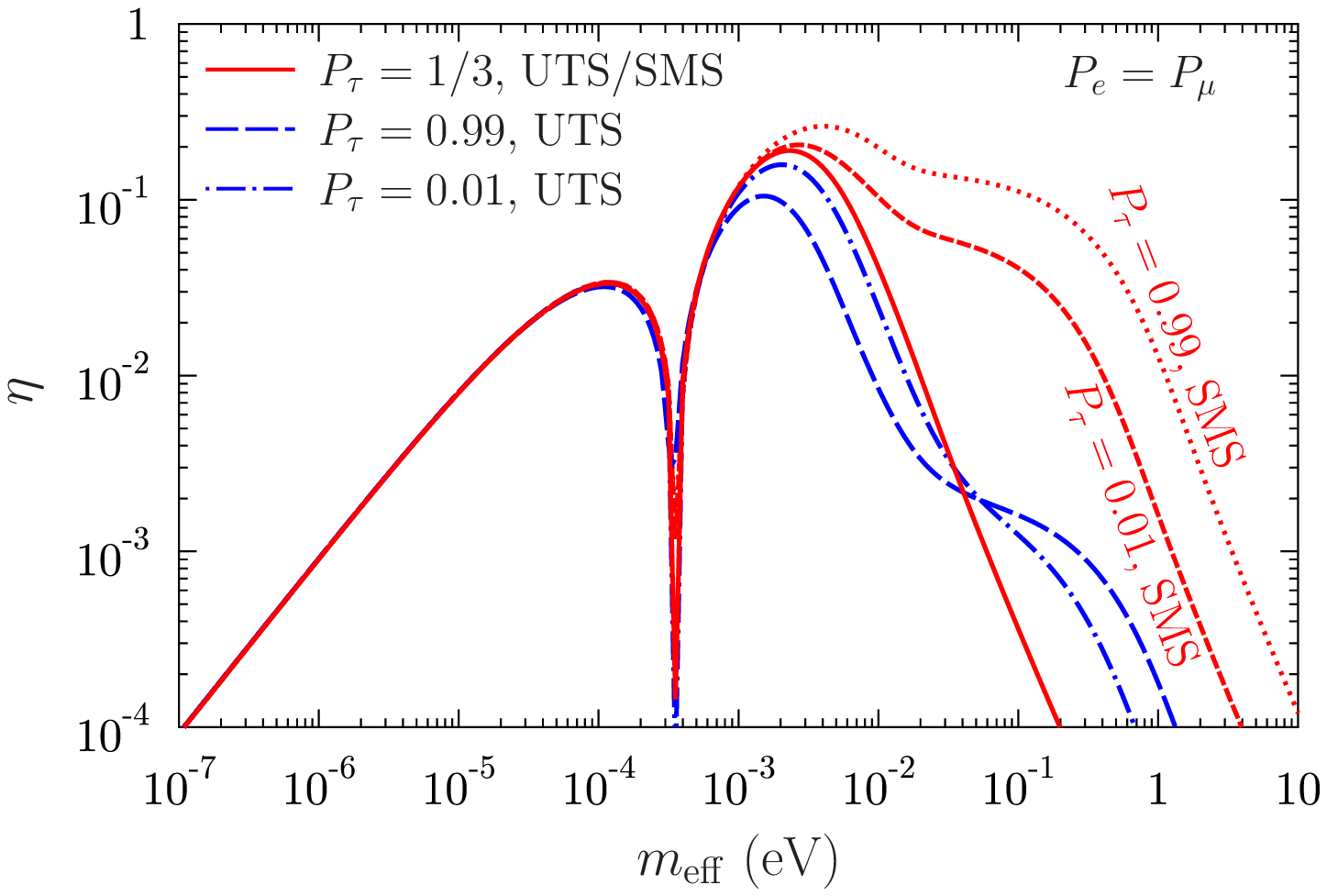}
\includegraphics[width=0.7\textwidth]{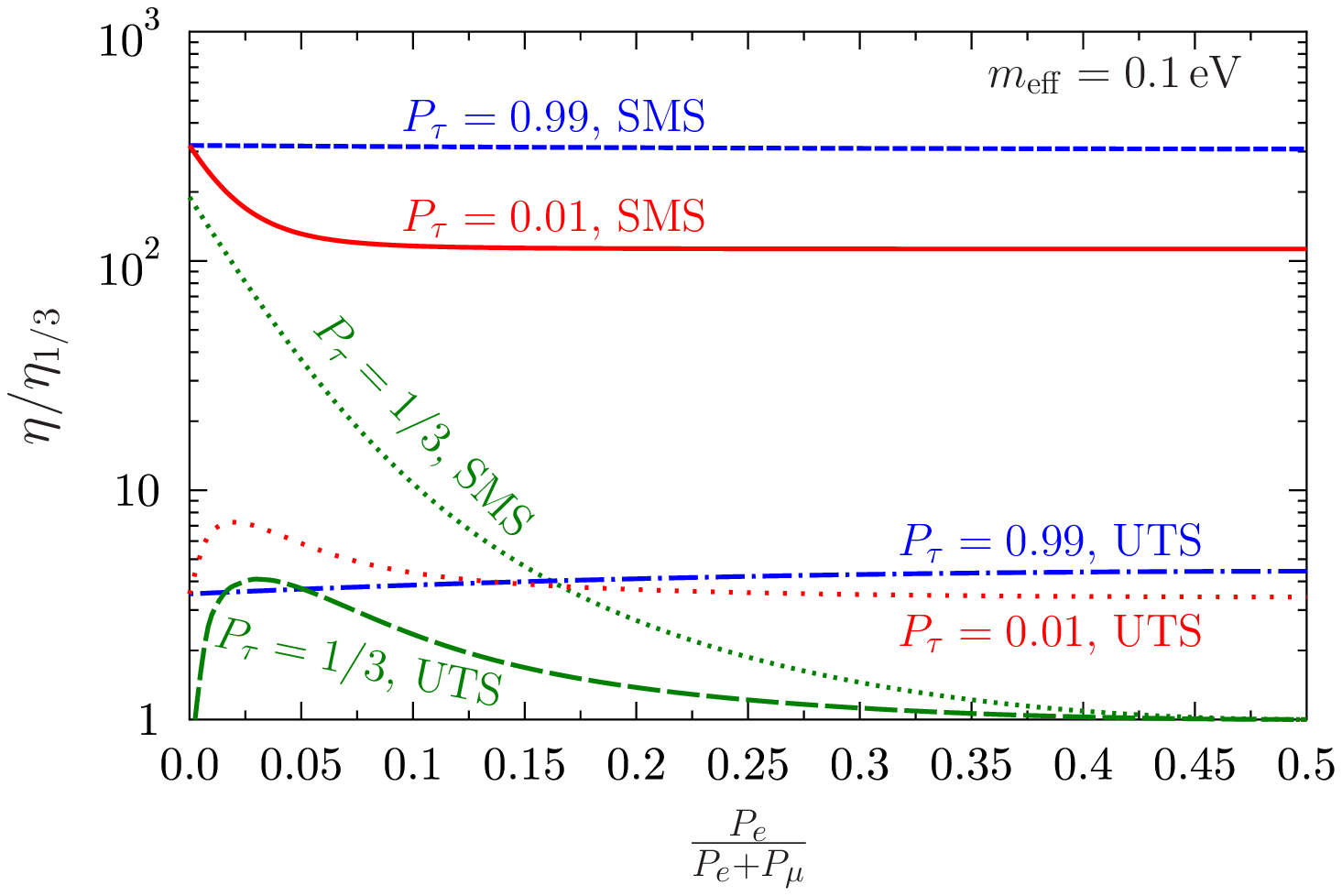}
\caption{
Top panel: the dependence of  $\eta$ on the washout
parameter $m_{\rm eff}$ for $P_e=P_\mu$ and different values of $P_\tau$.
Bottom panel: the dependence of $\eta$ for different
flavour configurations $P_\alpha$ normalized to $\eta_{1/3}$ that
corresponds to the flavour equipartition case $P_\alpha=1/3$.
}
\label{fig:eff_fla}
\end{figure} 

The dependence of the efficiency factor on the flavour projectors
$P_\alpha$ and on $m_{\rm eff}$ is shown in Fig. \ref{fig:eff_fla}.
The top panel shows the dependence of the efficiency on $m_{\rm eff}$
in the flavour equipartition case and for two other sets of flavour
projections.  As already mentioned, flavour effects become more
relevant when the washouts get stronger.  This is confirmed in this
picture where it is seen that for the SMS scenario the possible
enhancements quickly grow with $m_{\rm eff}$.  Note that in SL this
dependence is even stronger than in standard leptogenesis. This is due
to the fact that the flavoured washout parameters $P_\alpha m_{\rm
  eff}$ also determine the value of $z^\alpha_{dec}$ when the lepton
asymmetry in the $\alpha$ flavour starts being generated, and since
the CP asymmetry has a strong dependence on $z$, different values of
$P_e,\,P_\mu,\,$ and $P_\tau$ imply that the corresponding flavour
asymmetries are generated with different values of the CP asymmetry
even when, as in the SMS, the fundamental quantity $\bar \epsilon$ is
flavour independent.  In summary, what happens is that the flavour
that suffers the weakest washout is also the one for which
inverse-decays go out of equilibrium earlier, and thus also the one
for which the lepton asymmetry starts being generated when
$\bar\epsilon\times\Delta_{BF}$ is larger. This realizes a very
efficient scheme in which the flavour that is more weakly washed out
has effectively the largest CP asymmetry, and this explains
qualitatively the origin of the large enhancements that we have
found. Furthermore, when $P_\alpha m_{\rm eff}\ll m_*$ so that the
inverse decay of flavour $\alpha$ never reaches equilibrium and the
washout of the asymmetry $Y_{\Delta_\alpha}$ is negligible, the
maximum efficiency is reached.

The bottom panel in Fig. \ref{fig:eff_fla} further illustrates 
how the departure from the equipartition flavour case results in an 
enhancement of the efficiency, and that particularly large enhancements 
are possible for the SMS scenario.  Note that the top line in the top panel of
Fig. \ref{fig:eff_fla} labeled $P_\tau=0.99$ represents the maximum
enhancement that can be obtained in the SMS (relaxing the constraint
in Eq.~\eqref{eq:sms} that defines our SMS, larger enhancements are
however possible).  This is because for $P_\tau=0.99$ both the
asymmetries $Y_{\Delta_e}$ and $Y_{\Delta_\mu}$ are
generated in the weak washout regime, that is, approximately within
the same temperature range, and in the SMS this implies $\epsilon_e
(T_e) \approx \epsilon_\mu (T_\mu)$. The related combined efficiency is
then simply determined by $(P_e+P_\mu)\,m_{\rm eff}\simeq m_*$ and is
thus always maximal, independently of the individual values of $P_e$
and $P_\mu$, as it is apparent from the figure.

We should however spend a word of caution for the reader about
interpreting these results in the weak washout regime and, for
the SMS, also in the limit of extreme flavour hierarchies 
($P_\alpha \to 0$).  
At high temperatures $(z < 1)$ the Higgs bosons (higgsinos)
develop a sufficiently large thermal mass to decay into sleptons
(leptons) and sneutrinos.  The new CP asymmetries associated with these decays
could be particularly large,\cite{Giudice:2003jh} and thus sizable lepton
flavour asymmetries could be generated at high temperatures. 
This type of thermal effects are not included in the  analysis here 
described.  

Concerning the flavour decoupling limit
within the SMS, clearly when $P_\alpha \to 0$ no asymmetry can be generated
in the flavour $\alpha$.  However, in our SMS flavour asymmetries are
defined to be independent of the projectors $P$ and thus survive in
the $P\to 0$ limit. On physical grounds, one would expect for example
that when one decay branching ratio is suppressed, say, as 
$P < 10^{-5}$, the associated CP asymmetry will be at most of 
$ {\cal O}(10^{-7})$ and thus irrelevant for leptogenesis.
This means that for extreme flavour hierarchies, the SMS breaks down
as a possible physical realization of SL, and thus in
what follows we will restrict our considerations to a range of
hierarchies $P\gsim 10^{-3}$.

As a result one finds that for the SMS scenario with hierarchical
Yukawa couplings, successful leptogenesis is possible even for $m_{\rm
  eff}\gg {\cal O} ({\rm eV})$.  For example, as it is shown in the right
panel of Fig. \ref{fig:eff_fla}, for $P_e=P_\mu=5 \times 10^{-3}$ and
$m_{\rm eff}\sim 5$ eV, we obtain $\eta \sim 10^{-3}$, that yields the
estimate
\begin{equation}
Y_{\Delta B}^0 ({\rm SMS}, P_e=P_\mu=5 \times 10^{-3},
m_{\rm eff}=5\,{\rm  eV})\sim 10^{-6} \times 
\overline\epsilon \;. 
\end{equation}
Thus assuming a large, but still acceptable value of 
$\bar\epsilon \sim 10^{-4}$, SL can successfully generate the observed
baryon asymmetry 
for values of  $m_{\rm eff}$ that are about two orders of magnitude larger 
than what is found in the unflavoured standard  leptogenesis scenario.

%%%%%%%%%%%%%%%%%%%%%%%%%%%%%%%%%%%%%%%%%%%%%%%%%%%%

\subsection{Natural values of B}
\label{sec:natural_B}

We next describe the impact that flavour enhancements can have in
relaxing the constraints on the values of $B$ (and $M$). 
Here we include only the leading CP asymmetry contribution
from mixing Eq.~\eqref{eq:CP_asymres}.  From
Eqs.~\eqref{eq:yb},~\eqref{eq:eta}, and~\eqref{eq:cp0_S}, it is easy
to derive  the maximum value of $B$ for given values of $M$ and
$m_{\rm eff}$:
\begin{eqnarray}
B  & \leq & \frac{\Gamma\left(m_{\rm eff},M\right)}{2}
\frac{\left|\mbox{Im}A\right|}{M}\frac{y_{\widetilde{N}}\,\eta(m_{\rm eff})}
{Y_{\Delta B}^{CMB}} 
\left[1+\sqrt{1-\left(\frac{M}
{\left|\mbox{Im}A\right|}\frac{Y_{\Delta B}^{CMB}}
{ y_{\widetilde{N}}\,\eta(m_{\rm eff})}\right)^{2}}\right],
\label{eq:B_pm}
\end{eqnarray}
where $y_{\widetilde{N}}  =\frac{16}{23}Y_{\widetilde{N}}^{eq\,0}$,
$\Gamma(m_{\rm eff},M)$ is given in Eq.~\eqref{eq:gamma},  
$\mbox{Im}A= A\sin\phi_A$ and $Y_{\Delta B}^{CMB}$ is the observed 
baryon asymmetry Eq.\eqref{eq:YB_CMB}. 
Consequently  we obtain
\begin{eqnarray}
\label{eq:max_M}
M & \leq & 
\frac{\left|\mbox{Im}A\right|y_{\widetilde{N}}\,\eta(m_{\rm eff})
}{Y_{B_{obs}}}\; , \\
B & \leq & \frac{3\sqrt{3} m_{\rm eff}}{32\pi v^2} \left(
\frac{\left|\mbox{Im}A\right|y_{\widetilde{N}}\,\eta(m_{\rm eff})}
{Y_{\Delta B}^{CMB}}
\right)^2\; , 
\end{eqnarray}
where $\eta(m_{\rm eff})\equiv\eta(m_{\rm eff},P_\alpha,Z_\alpha)$ 
and all residual dependence of $\eta$ on $M$ has been neglected.  
As seen in the upper panel of Fig. \ref{fig:eff_fla}, 
assuming the SMS and for
sufficiently hierarchical $P_\alpha$, $\eta(m_{\rm eff})$ decreases first
very mildly with $m_{\rm eff}$ and -- once all the flavours have
reached the strong washout regime-- it decreases roughly as 
$\sim m_{\rm eff}^{-2}$.  Thus the product $m_{\rm eff}\times 
\eta(m_{\rm eff})^2$ first grows with $m_{\rm eff}$ till it reaches a maximum
and then for sufficiently large $m_{\rm eff}$ it decreases $\sim
m_{\rm eff}^{-3}$.  Therefore, for a fixed value of the projectors, 
the upper bound on $B$ does not corresponds simply to 
the maximum allowed value of $m_{\rm eff}$, but
it has a more complicated dependence.

\begin{figure}[htb]
\includegraphics[width=0.49\textwidth]{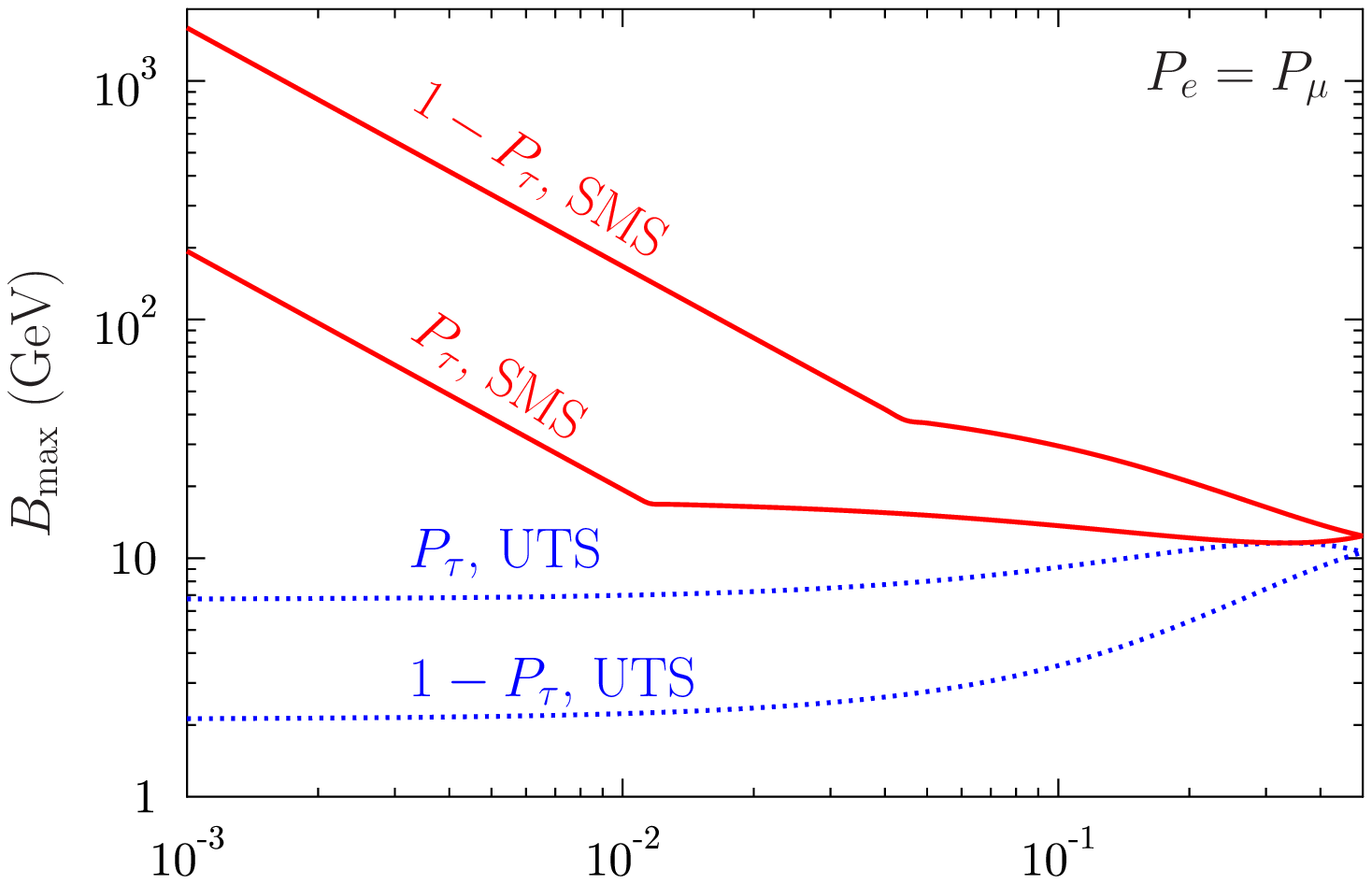}
\includegraphics[width=0.49\textwidth]{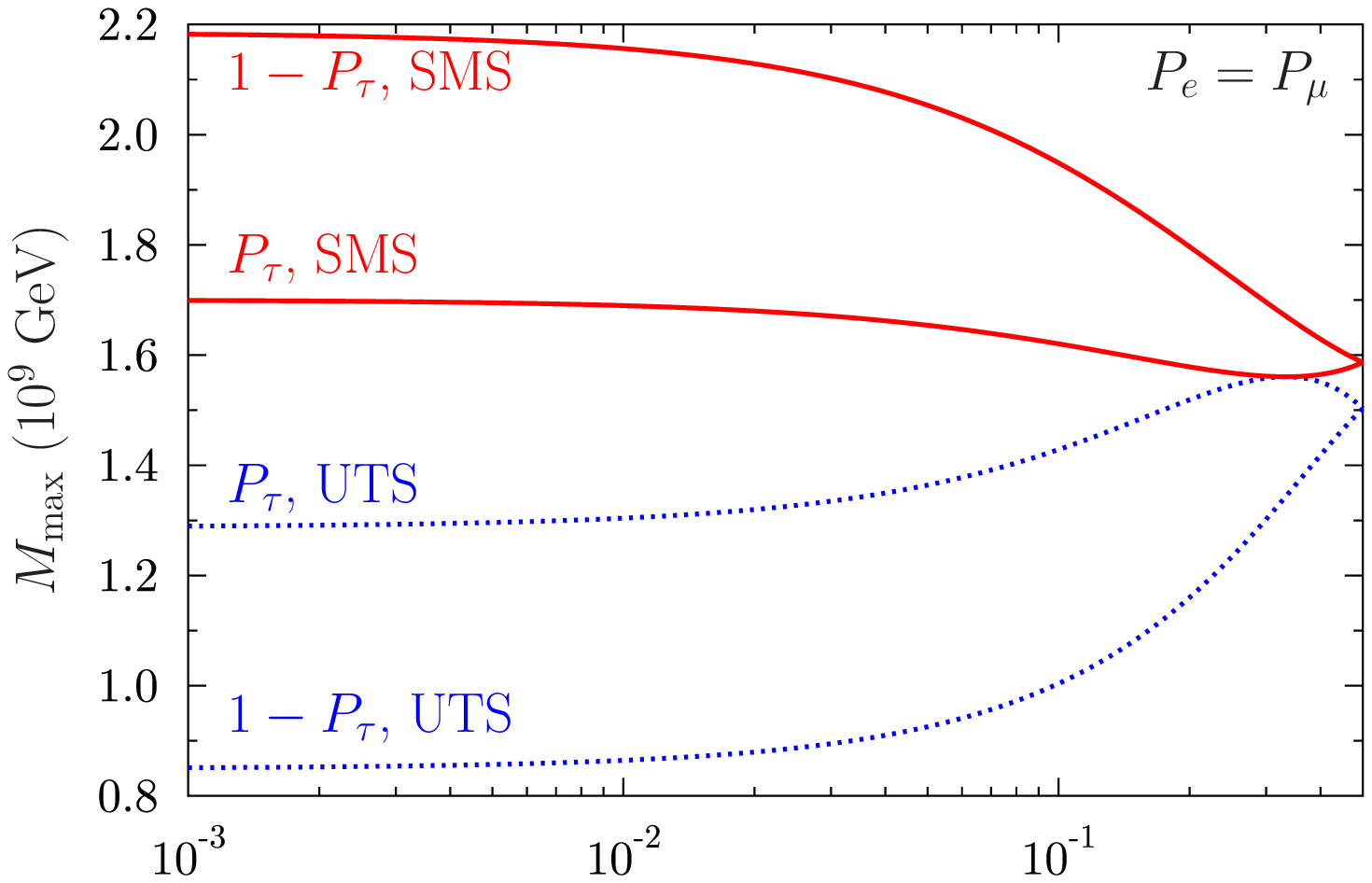}
\caption{Maximum values of $B$ and $M$ as a function of the flavour
  projections.  To highlight the effects when $P_\tau$ is very large or
   very small, the curves are given as a function of $P_\tau$ and
of  $1-P_\tau$.  
% The figure corresponds to $A\sin\phi_A=1\,$TeV.
%and $\tan\beta=30$. 
}
\label{fig:MmaxPBmaxP}
\end{figure}

Fig. \ref{fig:MmaxPBmaxP} shows the maximum values of $B$ and $M$
obtained for both the UTS and SMS cases as a function of the flavour
projections. In order to have better resolution when either $P_\tau$ 
or $1-P_\tau$ is very small, we plot them both as a function of $P_\tau$ or
$1-P_\tau$.  In the figure we set $\mbox{Im}A=1\,$TeV.  The figure
illustrates that within the UTS, the parameter space for successful
leptogenesis is very little modified by departing from the flavour
equipartition case (that corresponds to the point where the UTS and SMS
curves join). On the contrary, in the SMS case with
hierarchical flavour projections $1-P_\tau \sim {\rm few}\,\times
10^{-3}$ successful SL is allowed also with 
$B\sim{\cal O}({\rm TeV})$, that is for quite natural values 
of the bilinear term.  
As mentioned above, even for hierarchical projections the maximum allowed
values of $B$ and $M$ do not correspond to the maximum allowed value
of $m_{\rm eff}$.  In particular, for the range of flavour projections
shown in the figure we obtain that the maximum values of $B$ and $M$
correspond to  $m_{\rm eff}\lsim 2\,$eV.

%%%%%%%%%%%%%%%%%%%%%%%%%%%%%%%%%%%%%%%%%%%%%%%%%%%%%%%

\subsection{Lepton flavour equilibration and low energy constraints}
\label{sec:num_lfe}

\begin{figure}[htb]
\centering
\includegraphics[width=0.75\textwidth]{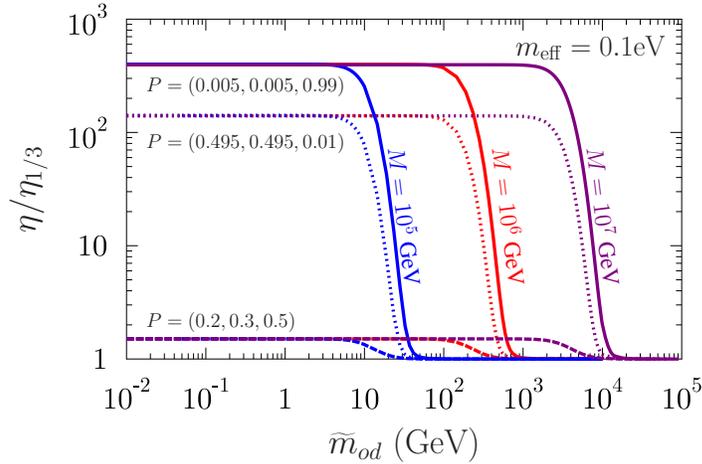}
\caption{The dependence of the efficiency (normalized to the flavour
  equipartition case $P_\alpha=1/3$) on the off-diagonal soft
  slepton mass parameter $\widetilde{m}_{od}$, for different values of
$M$ and of the flavour projections (see text for details).}
\label{fig:eff_fci}
\end{figure}

We finish this section by discussing the impact that the presence of
the LFE scatterings discussed in Section \ref{sec:lfe} can have on the
enhancement of the efficiency due to flavour effects.  We plot in
Fig. \ref{fig:eff_fci} the dependence of the flavour enhancement of
the efficiency as a function of the off-diagonal slepton mass
parameter $\widetilde{m}_{od}$. As can be seen in the figure (and as
it was expected from the discussion in the previous section) for any
given value of $M$, LFE quickly becomes efficient damping completely
the lepton flavours enhancements of the efficiency within a very
narrow range of values $\widetilde{m}_{od}^{min} \leq
\widetilde{m}_{od} \leq \widetilde{m}_{od}^{max}$.  The figure
corresponds to $\tan\beta=30$ however, as already said, the dependence
on $\tan\beta$ arises from $v_d=v\cos\beta$ in Eq.~\eqref{eq:rij} and
is rather mild.  Results from other values of $\tan\beta$ can be
easily read from the figure by rescaling
$\widetilde{m}^\beta_{od}=\widetilde{m}^{\rm fig}_{od}/(30
\cos\beta)$.

Notice that while neglecting LFE the efficiency for a given value of
$m_{\rm eff}$ is practically insensitive to the particular value of
$M$, this is not the case when the  efficiency 
is evaluated by accounting for LFE effects. In fact, given the
different scaling with the temperature of the $\overline{\Gamma}_{\rm
  LFE}$ and $\overline{\Gamma}_{\rm ID}$ rates, the precise
temperature at which leptogenesis occurs is crucial.  For example, we
see from Fig. \ref{fig:eff_fci} that for reasonable values $
\widetilde{m}_{od} \lsim 200\,$GeV and for $M\gsim 10^6\,$GeV, LFE is
not effective, and the large enhancements of the efficiency due to
flavour effects can survive, while for $M\lsim 10^5\,$GeV all flavour
enhancements disappear.

\begin{figure}[t!]
\centering
\includegraphics[width=0.75\textwidth]{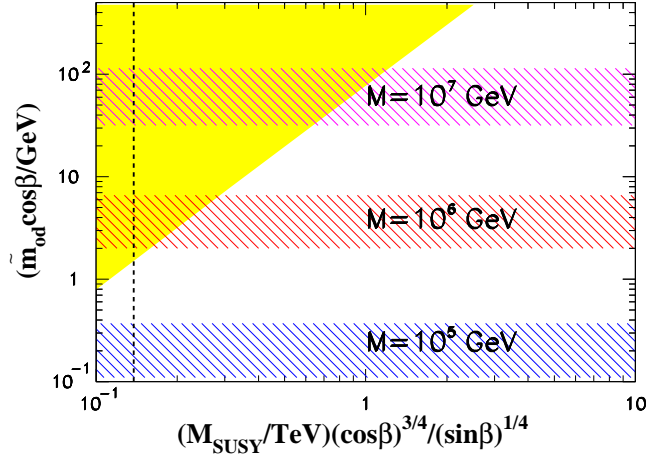}
\caption{ Shaded in yellow: the region of
  $\widetilde{m}_{od}\cos\beta$ versus
  $m_{SUSY}(\cos\beta)^\frac{3}{4}/(\sin\beta)^\frac{1}{4}$ excluded
  by the present bound $BR(\mu\rightarrow e\gamma)\leq 1.2 \times
  10^{-11}$.  The three bands corresponding to $M=10^5\,$GeV,
  $M=10^6\,$GeV and $M=10^7\,$GeV show the minimum value of
  $\widetilde{m}_{od}{\cos\beta}$ for which LFE effects start damping
  flavour effects.  The width of the bands corresponds to $P_\alpha
  m_{\rm eff}$ varying in the range $0.003\,$eV$-10\,$eV, with
  $P_\alpha$ the largest flavour projection. The vertical dashed line
  corresponds to the value of $m_{SUSY}/(\tan\beta)^\frac{1}{2}$
  required to explain the discrepancy between the SM prediction and
  the measured value of $a_\mu$, assuming
  $\tan\beta=1$.\protect\cite{Davidson:2008pf}}
\label{fig:susylfv}
\end{figure}
%\nocite{Davidson:2008pf}%[129]\cite{Davidson:2008pf}

It is interesting notice that the presence of a sizable 
$\widetilde{m}_{od}$ would induce various LFV decays, 
like for example $l_\alpha \rightarrow l_\beta\gamma$
with rate
\bea
&& \!\!\!\!\!\!\!\!\!\!\!\!\!\!\!\!\!\!\!\!\!\!\!\!\!  
\frac{BR(l_\alpha \rightarrow l_\beta\gamma)}
{BR(l_\alpha\rightarrow l_\beta \nu_\alpha\overline\nu_\beta)}
 \sim  \frac{\alpha^3}{G_F^2}\frac{\tan^2\beta}{m^8_{SUSY}}
\widetilde{m}_{od}^4 \nonumber \\
 && \simeq  2.9\times 10^{-19} 
\frac{\sin^2\beta}{\cos^6\beta}\left(\frac{\rm TeV}{m_{SUSY}}\right)^8
\left(\cos^2\beta\frac{\widetilde{m}_{od}^2}{\rm GeV^2}\right)^2,
\eea
where $m_{SUSY}$ is a generic SUSY scale for the gauginos and sleptons
masses running in the LFV loop. So it is possible 
to compare the values of $\widetilde{m}_{od}$ 
for which LFE occurs with  the existing bounds imposed from non-observation 
of such flavour violation in leptonic decays. 
The result of such comparison in presented in  Fig. \ref{fig:susylfv}.
The  yellow shade gives  the excluded region of 
$\widetilde{m}_{od}\,{\cos\beta}$ versus
$m_{SUSY}(\cos\beta)^\frac{3}{4}/(\sin\beta)^\frac{1}{4}$ arising from
the present bound BR$(\mu\rightarrow e\gamma)\leq 1.2 \times 10^{-11}$, 
together with the minimum value of $\widetilde{m}_{od}\,
{\cos\beta}$ for which LFE effects start damping out flavour
enhancements in SL.  Three bands are shown respectively
for $M=10^5\,$GeV, $M=10^6\,$GeV and $M=10^7\,$GeV.  
The width of the bands represents the range associated 
with variations of the effective flavoured decay parameter 
$P_\alpha m_{\rm eff}$ in the range $0.003\,$eV$-10\,$eV, 
where $P_\alpha$ is the largest of the three flavour
projections. For illustration we also show in the figure the 
characteristic SUSY scale that allows to explain the small discrepancy
between the SM prediction and the measured value of
the muon anomalous magnetic moment, $a_\mu$.  This values is
$m_{SUSY}/(\tan\beta)^\frac{1}{2}=141$ GeV,\cite{Davidson:2008pf} 
and the vertical dashed line in the picture corresponds to $\tan\beta= 1$.  
As seen in the figure, in this case the off-diagonal slepton masses are
bound to be small enough to allow for flavour enhancements in SL 
for $M$ as low as $10^6$ GeV. For larger values of
$\tan\beta$, even lower values of $M$ are allowed.

In brief, LFE effects induced by off-diagonal soft slepton masses,
when constrained with the bounds imposed from the non-observation of
flavour violation in leptonic decays, are ineffective for damping the
flavour enhancements.

%% file: nonseq.tex
\section{Soft Leptogenesis without Superequilibration: {\it R}-genesis}
\label{sec:nse}

As mentioned in the previous Sections, early works on leptogenesis
were carried out from the start within the unflavoured effective
theory. Quite likely this happened because the corresponding
Lagrangian is much simpler, given that the number of relevant
parameters is reduced to a few.  The main virtue of subsequent studies
on lepton flavour effects was that of recognizing that for $T \lesssim
10^{12}\,$GeV the unflavoured theory breaks down, and the new theory
brings in new fundamental parameters, which can give genuinely
different answers for the amount of baryon asymmetry that is
generated.

\begin{figure}[t]
\includegraphics[width=\textwidth]{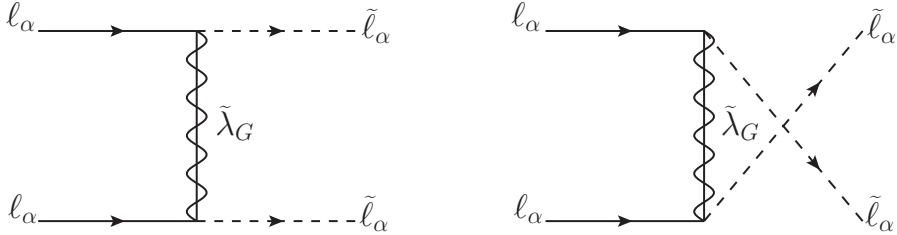}
\caption{Feynman diagrams for lepton-slepton scatterings induced 
by soft gaugino masses
  $m_{\widetilde g}$ through the exchange of $SU(2)_L$ and $U(1)_Y$
  gauginos $\widetilde\lambda_G$.  The squared amplitudes of these
  processes are proportional to $m_{\widetilde g}^2$ and vanish in the
  $m_{\widetilde g}\to 0$ limit.}
\label{fig:seq_diagrams}
\end{figure}

In supersymmetric leptogenesis the opposite happened: the effective
theory that was generally used assumed fast particle-sparticle
equilibration reaction.  But this assumption is only appropriate for
temperatures much lower than the typical temperatures $T\gg 10^8\,$GeV
in which leptogenesis can be successful. In fact, only quite recently
Ref.~\refcite{Fong:2010qh} clarified that in the relevant temperature
range, a completely different effective theory holds instead. More
specifically, in supersymmetric leptogenesis studies prior to
Ref.~\refcite{Fong:2010qh} it was always assumed (often implicitly)
that lepton-slepton reactions like e.g. $\ell\ell \leftrightarrow
\widetilde\ell\widetilde\ell$ (see Fig. \ref{fig:seq_diagrams}) that
are induced by soft gaugino masses
\be
\mathcal{L}_{\widetilde \lambda} = -\frac{1}{2}\left(
%m_{3}\overline{\widetilde{\lambda}_{3}^{a}}
%P_{L}\widetilde{\lambda}_{3}^{a}+
m_{2}\overline{\widetilde{\lambda}_{2}^{\pm,0}}
P_{L}\widetilde{\lambda}_{2}^{\pm,0}
%+m_{2}^{*}\overline{\widetilde{\lambda}_{2}^{a}}P_{R}\widetilde{\lambda}_{2}^{a}
+m_{1}\overline{\widetilde{\lambda}_{1}}P_{L}\widetilde{\lambda}_{1}
%+m_{1}^{*}\overline{\widetilde{\lambda}_{1}}P_{R}\widetilde{\lambda}_{1}
+{\rm h.c.}\right),
\ee
as well as  higgsino mixing transitions $\widetilde H_u 
\leftrightarrow \overline{\widetilde H_d}$, that
are induced by the superpotential term
\be W_H = \muh \hat H_u \hat H_d, \ee are in thermal equilibrium.
This implies that the lepton and slepton chemical potentials
equilibrate.  However, in general, in supersymmetric leptogenesis {\sl
  superequilibration} (SE) between particles and sparticles chemical
potentials does not occur. In fact, the rates of interactions induced
by SUSY-breaking scale ($\Lambda_{susy}$) parameters, like soft
gaugino masses $m_{\widetilde g}$ or the higgsino mixing parameter
$\muh$, are slower than the Universe expansion rate when 
\begin{equation}
\label{eq:Tgmu}
  \frac{\Lambda_{susy}^2}{T} \lsim 25\; \frac{T^2}{M_{pl}}
\qquad 
 \Rightarrow \qquad  
T> T_{\rm SE} 
\sim 5\cdot 10^7 
\left(\frac{\Lambda_{susy}}{500\,{\rm GeV}}\right)^{2/3}\;{\rm GeV}.
\end{equation}
Thus, when this condition is realized, these reactions should be
classified as reactions of type (ii) (see Section~\ref{sec:effective})
and handled accordingly.  Since leptogenesis occurs when the
temperature is of the order of the heavy neutrino mass, in
terms of $M$ the assumption of SE breaks down when
\begin{equation}
 M \gsim 5\cdot 10^7
 \left(\frac{\Lambda_{susy}}{500\,{\rm GeV}}\right)^{2/3}\,{\rm GeV}. 
  \label{eq:NSE}
\end{equation}
Following the discussion in Section~\ref{sec:effective}, the effective
theory appropriate for studying supersymmetric leptogenesis, in which
the heavy Majorana masses certainly satisfy the bound~\Eqn{eq:NSE}, is
thus obtained by setting $m_{\widetilde g},\,\muh \to
0$.\cite{Ibanez:1992aj,Fong:2010qh} In this limit, the supersymmetric
Lagrangian acquires two additional anomalous global symmetries
(respectively a $R$-symmetry and a $PQ$-like symmetry) under which,
besides the $SU(2)_L$ and $SU(3)_c$ SM fermions, also the gauginos and
higgsinos transform non-trivially.  As a consequence, the EW and QCD
sphaleron equilibrium conditions are modified with respect to the
usual ones: winos, binos and higgsinos are now coupled to the
$SU(2)_L$ sphalerons, while gluinos get coupled to the QCD
sphalerons.\cite{Ibanez:1992aj} Therefore, besides the occurrence of
non-superequilibration (NSE) effects, also the pattern of sphaleron
induced lepton-flavour mixing is different from what is obtained with
a naive supersymmetrization of the SM case.\cite{Fong:2010qh} Besides
this, a new anomaly-free $R$-symmetry can be defined and the
corresponding charge, being exactly conserved, provides a constraint
on the particle density asymmetries that is not present in the
SM.\cite{Fong:2010qh} Nevertheless, in spite of all these
modifications, in Ref.~\refcite{Fong:2010qh} it was found that
eventually the resulting baryon asymmetry would not differ much from
what would be obtained with the (incorrect) assumption of SE.
Basically, the reason for this is that by dropping the SE assumption
and accounting for all the new effects only modifies spectator
processes, while the overall amount of CP asymmetry that drives
leptogenesis remains the same.

In SL however, when $M\gsim 10^7\,$GeV the appropriate NSE effective
theory not only implies profound theoretical modifications, but also
results in very large quantitative differences. The reason for this is
twofold:

(I)\ \ In the NSE regime the leptonic density asymmetries for scalar and
fermion evolve independently, and this implies that the corresponding
efficiencies $\eta_{s,f}$ are different.  When these different
`weights' are taken into account, the strong cancellation between the
scalar and fermion contributions to the baryon asymmetry, that is 
characteristic of SL, gets spoiled, and a non-vanishing result is
obtained even without the inclusion of  thermal corrections.\cite{Fong:2010bv}

(II)\ \ While the new symmetries $R$ and $PQ$ that arise in the high
temperature effective theory after setting $m_{\widetilde g},\mu\to 0$
are anomalous, two anomaly free combinations involving $R$ and $PQ$
can be defined.  These combinations, that have been denoted in
Ref.~\refcite{Fong:2010bv} as ${R_B}$ and ${R_\chi}$, are conserved in
sphaleron transitions, and are only (slowly) violated by RHSN
dynamics. Thus their evolutions must be followed by means of two new BE
and, since all density asymmetries get mixed by EW sphalerons, these
equations couple to the BE that control the evolution of ${B-L}$.
What is important is that the sources for ${R_B}$ and ${R_\chi}$ are
respectively the CP-violating asymmetries $\epsilon_s$ and
$\epsilon_s-\epsilon_f$, which are not suppressed by any kind of
cancellation.  Thus the corresponding density asymmetries $Y_{\Delta
  R_B}$ and $Y_{\Delta R_\chi}$ remain large during leptogenesis, and
act as source terms for the $\Delta_\alpha = B/3-L_\alpha$ density asymmetries
$Y_{\Delta_\alpha}$ that are thus driven to comparably large values.

As regards the final values of ${R_B}$ and ${R_\chi}$ at the end of
leptogenesis, they are instead irrelevant for the computation of the
baryon asymmetry since, well before the temperature when the EW
sphalerons are switched off, soft SUSY-breaking effects attain
in-equilibrium rates, implying that $R$ and $PQ$ cease to be good
symmetries also at the perturbative level implying that $Y_{\Delta
  R_B},\,Y_{\Delta R_\chi}\to 0$.  The baryon asymmetry is then
determined only by the amount of $B-L$ asymmetry at freeze-out of the
EW sphalerons according to the usual relation $B=\frac{8}{23}\,
(B-L)$.

%In this section, for simplicity, we will only consider 
%a universal trilinear soft term $A_\alpha = A Y_\alpha$ 
%(the UTS scenario discussed in Section \ref{sec:fla_scenarios}).

%%%%%%%%%%%%%%%%%%%%%%%%%%%%%%%%%%%%%%%%%%%%%%%%%%%%%%%%%%%%%%%%%%

\subsection{Anomalous and non-anomalous symmetries above 
$T_{\rm SE}$ } 
\label{sec:symmetries}

As discussed above, when the temperature of the thermal bath satisfies
the condition~\Eqn{eq:Tgmu} the appropriate effective supersymmetric
Lagrangian is obtained by setting $m_{\widetilde g}, \muh \to 0$,
which results in the two new $U(1)$ symmetries $R$ and
$PQ$.\cite{Ibanez:1992aj}
\begin{table}[t!]
% \begin{center} 
% \begin{tabular}{|c|c|c|c|c|c|c|c|c|c|}
\tbl{$B$, $L$, $PQ$ and $R$ charges for the particle supermultiplets. 
  The labels in the top row refer to the supermultiplets  L-handed fermion
  components.  
  The $R$ charges for bosons are determined by $R(b)=R(f)+1$.}
{\begin{tabular}{|cc|c|c|c|c|c|c|c|c|c|c|c|c|} \hline & &\quad $\tilde
    g\ $ &\quad $Q\ $ &\quad$u^c \ $ &\quad $d^c \ $ &\quad $\ell \ $
    &\quad $e^c \ $ &\quad $\tilde H_d\phantom{\Big|}$ &\quad$\tilde
    H_u\ $ &\quad$N^c\ $
    \\
    \hline \multicolumn{2}{|c|}{$B\phantom{\Big|}$} &$ 0$&$
    \frac{1}{3}$&$-\frac{1}{3}$&$-\frac{1}{3}$&$ 0$&$ 0$&$ 0$&$ 0$& 0
    \\ \hline \multicolumn{2}{|c|}{$L\phantom{\Big|}$} &$ 0$&$ 0$ &$
    0$ &$ 0$ &$ 1$&$-1$&$ 0$&$ 0$& 0 \\ \hline
    \multicolumn{2}{|c|}{$PQ\phantom{\Big|}$}&$ 0$&$ 0$ &$-2$ &$ 1$
    &$-1$&$ 2$&$-1$&$ 2$& 0 \\ \hline
    \multirow{2}{*}{$R$}&$\phantom{\!\!\!\!\Big|}f$ &$ 1$&$-1$ &$-3$
    &$ 1$ &$-1$&$ 1$&$-1$&$3$&$-1$ \\ \cline{3-11}
    &$\phantom{\!\!\!\!\Big|}b$ &$ 2$&$ 0$ &$-2$ &$ 2$ &$ 0$&$ 2$&$
    0$&$ 4$& 0 \\ \hline
\end{tabular}\label{tab:1}}
% \end{center}
\end{table}
The charges of the various states under these symmetries, together
with the values of the other two global symmetries $B$ and $L$ are
given in Table~\ref{tab:1}.  Note that to facilitate the evaluation of
the anomalies in the table we list the charges of the L-handed chiral
multiplets, and in particular those of  $u^c,\,d^c,\,e^c$ that are
opposite with respect to the charges of the R-handed states $u,\,d,\,e$
whose chemical potentials will enter the chemical equilibrium
equations below.  

Like $L$, also $R$ and $PQ$ are not symmetries of the seesaw
superpotential terms $M \hat N^c \hat N^c + \lambda \hat N^c \hat\ell
\hat H_u$, since it is not possible to find any charge assignment that
would leave both terms invariant.  In Table~\ref{tab:1} the charges of
the heavy $N^c$ supermultiplets have been fixes so that RHSN do
not carry any charge.  This has the advantage of ensuring that all the
RHSN bilinear terms, corresponding to the mass parameters
$M,\,\widetilde{M},\,B$, are invariant, and thus RHSN mixing does
not break any internal symmetry.  However, since $R(\hat N^c \hat
N^c)=0$, it follows that the mass term for the heavy RHN
breaks $R$ by two units.\footnote{Under $R$-symmetry the superspace
  Grassman parameter transforms as $\theta \to e^{i\alpha}\theta$
  . Invariance of $\int d\theta\, \theta =1$ then requires
  $R(d\theta)=-1$. Then the chiral superspace integral of the
  superpotential $\int d\theta^2\, W $ is invariant if $R(W)=2$. By
  expanding a chiral supermultiplet in powers of $\theta$ it follows
  that the supermultiplet $R$ charge equals the charge of the bosonic
  scalar component $R(b)=R(f)+1$, and thus for the fermion bilinear
  term $R\left(\overline{N^c_R}N^c_L\right)=-2$.}

All the four global symmetries $B$, $L$, $PQ$ and $R$ have mixed gauge
anomalies with $SU(2)_L$, and $R$ and $PQ$ have also mixed gauge
anomalies with $SU(3)_c$.  Two linear combinations of $R$ and $PQ$,
having respectively only $SU(2)_L$ and $SU(3)_c$ mixed anomalies, have been
identified in Ref.~\refcite{Ibanez:1992aj}.  They are:\footnote{ With
  respect to Ref.~\refcite{Ibanez:1992aj}, for definiteness we restrict
  ourselves to the case of three generations $N_g=3$ and one pair of
  Higgs doublets $N_h=1$, and we also normalize $R_{2,3}$ in such a
  way that $R_{2,3}(b) =R_{2,3}(f)+1$.}
\bea
\label{eq:R2}
R_2 &=& R-2\,PQ \\
\label{eq:R3}
R_3 &=&R-3\,PQ\,. 
\eea
The values of $R_{2,3}$ for the different states are given in
Table~\ref{tab:2}.  The authors of Ref.~\refcite{Ibanez:1992aj} have also
constructed the effective multi-fermions operators generated by the
mixed anomalies: 
\bea
\label{eq:tO-EW}
\tilde O_{EW} &=&
\Pi_\alpha \left(QQQ\ell_\alpha\right)\; \tilde H_u\tilde H_d\;\tilde W^4\,,\\
\label{eq:tO-QCD}
\tilde O_{QCD} &=& \Pi_i \left(QQu^c d^c\right)_{i}\; \tilde g^6 \,.
\eea
Given that the three charges $R_2$, $B$ and $L$ all have mixed
$SU(2)_L$ anomalies, two anomaly free combinations can be defined.
The most convenient are $B-L$ and\cite{Fong:2010bv}
\be
\label{eq:RB}
 {R_B}=\frac{2}{3}B+R_2,  
\ee
whose values are also given in Table~\ref{tab:2}.  The fact that $R_B$
does not contain any $B-L$ fragment, ensures that it will not enter in
the final computation of the baryon asymmetry, and the fact that it is
independent of $L$ renders easier writing the BE for its evolution.
The values of $R_B$ in Table~\ref{tab:2} imply that the superpotential
term $N^c\,\ell H_u$ has charge $R_B=2$ and thus is invariant. It
follows that RHSN decays into fermions conserve $R_B$.  In
contrast, the soft $A$ term in~\Eqn{eq:soft_terms} responsible for
RHSN decays into scalars violates $R_B$ by 2 units, and more
precisely for $\widetilde{N}_\pm \to H_u\tilde\ell$ we have $\Delta
R_B=+2$, while for $\widetilde{N}_\pm \to H_u^*\tilde\ell^*$ we have
$\Delta R_B=-2$.  As regards the heavy neutrinos, their mass term
violates $R_B$ by two units. Note that this is precisely like the case
when one chooses to assign a lepton number $-1$ to the singlet
neutrinos $N$. Accordingly, the decays of the heavy Majorana neutrino
violate $R_B$ by one unit: for $N\to \ell H_u,\, \tilde\ell\tilde H_u$
we have $\Delta R_B=+1$ while for the CP conjugate final states
$\Delta R_B=-1$.  All $R_B$ violating reactions have, by assumption,
rates that are comparable to the Universe expansion rate, and then a
specific BE is needed to track the evolution of $Y_{\Delta R_B}$.

 \begin{table}[t!]
\tbl{Charges for the fermionic and bosonic components of the SUSY
  multiplets under the $R$-symmetries defined in Eqs.\protect
  \eqref{eq:R2}, \protect \eqref{eq:R3} and \protect \eqref{eq:RB}.
  Supermultiplets are labeled in the top row by their L-handed fermion
  component.}
% \begin{center}
{\begin{tabular}{|cc|c|c|c|c|c|c|c|c|c|c|c|c|}
 \hline
&
&\quad  $\tilde g\ $
&\quad $Q\ $
&\quad$u^c \ $
&\quad $d^c \ $
&\quad $\ell \ $
&\quad $e^c \ $
&\quad $\tilde H_d\phantom{\Big|}$
&\quad$\tilde H_u\ $
&\quad$N^c\  $
\\
\hline
\multirow{2}{*}{$R_2$} &$\phantom{\!\!\!\!\Big|}f$     
&$ 1$&$-1$&$ 1$&$-1$&$ 1$&$ -3$&$1$&$-1$& $-1$\\ \cline{3-11}
                     &$\phantom{\!\!\!\!\Big|}b$       
&$ 2$&$ 0$&$ 2$&$ 0$&$ 2$&$-2$&$ 2$&$ 0$& 0 \\ \hline
\multirow{2}{*}{$R_3$} &$\phantom{\!\!\!\!\Big|}f$     
&$ 1$&$-1$&$ 3$&$-2$&$ 2$&$ -5$&$2$&$-3$& $-1$\\ \cline{3-11}
                     &$\phantom{\!\!\!\!\Big|}b$       
&$ 2$&$ 0$&$ 4$&$ -1$&$ 3$&$-4$&$ 3$&$-2$& 0 \\ \hline
\hline
\multirow{2}{*}{$R_B$}&$\phantom{\!\!\!\!\Big|}f$ 
&$ 1$&$-\frac{7}{9}$&$\frac{7}{9}$&$-\frac{11}{9}$&$1$&$ -3$&$
1$&$-1$& $-1$ \\
 \cline{3-11}   &$\phantom{\!\!\!\!\Big|}b$       
&$ 2$&$\frac{2}{9}$&$\frac{16}{9}$&$-\frac{2}{9}$&$ 2$&$-2$&$ 2$&$ 0$& 0 \\ 
\hline
\end{tabular}\label{tab:2}}
% \end{center}
\end{table}

At temperatures satisfying condition~\Eqn{eq:Tgmu} there is at least
one other anomalous global symmetry, that in
Ref.~\refcite{Fong:2010bv} has been denoted by $\chi$.
It corresponds to $U(1)$ phase rotations of the $u^c$ chiral
multiplet that, for its fermionic component, can be readily identified
with chiral symmetry for the right-handed up-quark. In fact, above
$T\sim 2\times 10^{6}\,$GeV, reactions mediated by $h_u$ do not occur
and the condition $h_u\to 0$ must be imposed, resulting in a new
anomalous `chiral' symmetry.  In the $SU(3)_c$ sector we then have two
anomalous symmetries $R_3$ and $\chi$, and one anomaly free
combination can be constructed. Assigning to the $L$-handed $u^c_L$
supermultiplet a chiral charge $\chi=-1$ this combination has the
form\cite{Fong:2010qh}
\begin{equation}
\label{eq:chiralup}
R_\chi= \raise 2pt \hbox{$\chi$}_{u^c_L}+\kappa_{u^c_L}\,R_3, 
\end{equation}
where $\kappa_{u^c_L}=1/3$. When the additional condition $h_d\to 0$
is imposed, a chiral symmetry arises also for the $d^c$
supermultiplet.  A second anomaly free $R_\chi$ symmetry can then be
defined in a way completely analogous to \Eqn{eq:chiralup}, with
$\kappa_{d^c_L}=\kappa_{u^c_L}=1/3$.\cite{Fong:2010qh}  As regards
perturbative violation of $R_\chi$, this charge inherits the same
violation $R_3$ suffers.  The soft $A$ term in~\Eqn{eq:soft_terms}
violates $R_3$ by one unit, and so do RHSN decays into scalars.
Moreover, since $N^c\,\ell H_u$ has an overall charge $R_3=1$, a
violation by one unit occurs also for RHSN decays into
fermions. Correspondingly, we have $\Delta R_3=+1$ for the decays
$\tilde N,\,\tilde N^* \to H_u\tilde\ell,\; \overline{\widetilde{H}}_u
\overline\ell $ and $\Delta R_3=-1$ for $\tilde N,\,\tilde N^* \to
\widetilde{H}_u \ell,\; H_u^*\tilde\ell^*$. Of course, similarly to
$R_B$, also the evolution of $R_\chi$ needs to be tracked by means of
one BE.

%%%%%%%%%%%%%%%%%%%%%%%%%%%%%%%%%%%%%%%%%%%%%%%%%%%%%%

\subsection{Chemical equilibrium conditions and conservation laws}
\label{sec:chem}

Because of the network of fast particle reactions occurring in the
thermal bath, asymmetries generated in RHSN decays spread around
among the various particle species, and this can affect directly or
indirectly leptogenesis processes.  In principle there is one
asymmetry for each particle degree of freedom. There are however
several conditions and constraints that reduce the number of
independent asymmetries to a few. 
\begin{itemize}
\item[(i)] Constraints imposed by reactions whose rates are much  
faster than the Universe expansion have to be formulated in
terms of chemical equilibrium conditions for the chemical potentials of
incoming $\mu_I$ and  final state particles $\mu_F$:
\begin{equation}
  \label{eq:muimuf}
  \sum_I \mu_I = \sum_F \mu_F.
\end{equation}

\item[(ii)] Conservation laws that  arise when all the reactions 
that violate some specific charge are much slower than the 
the Universe expansion have to be formulated in terms of particle 
number densities $\Delta n = n - \bar n$ and, for a generic charge
$Q$, read: 
\begin{equation}
  \label{eq:Qtot}
  Q=\sum_i   Q_i \Delta n_i = {\rm const}, 
\end{equation}
where $Q_i$ is the charge of the $i$-particle species. 
We will always assume as initial conditions for leptogenesis 
that all particle asymmetries vanish, and thus we will put the 
constant value of \Eqn{eq:Qtot} equals to zero.

\item[(iii)] Reactions with rates comparable with the Universe
  expansion  have to be treated by means of appropriate dynamical
  equations. In this case, in order to reabsorb the dilution effects due to the
  Universe expansion, it is convenient to introduce as basic variables
  the number densities of particles normalized 
 to the entropy density $s$ so we define the density asymmetries per 
 degree of freedom $g_i$:
%(also known as \emph{abundances})
%
\begin{equation}
  \label{eq:Y}
  Y_{\Delta_i} = \frac{1}{g_i}\frac{\Delta n_i}{s}\,. 
\end{equation}
\end{itemize}

Clearly, $\mu_i$, $\Delta n_i$ and $Y_{\Delta_i}$ are all related to
particle asymmetries. In particular, the number density asymmetry
of a particle for which a chemical potential can be defined is
directly related with this chemical potential. For both bosons ($b$)
and fermions ($f$) this relation acquires a particularly simple form
in the relativistic limit $m_{b,f}\ll T$, and at first order in
$\mu_{b,f}/T\ll 1$:
\be
\label{eq:Dnmu}
\Delta n_{b}=\frac{g_b}{3}
  T^2\mu_b, \qquad\quad
\Delta n_{f}=\frac{g_f}{6}T^2\mu_f\,.
\ee
Eventually, to solve for the large set of conditions in a closed form
one needs to use a single set of variables. Here we will take this to
be the set $\left\{Y_{\Delta_i}\right\}$, leaving understood that
the solutions to the constraining conditions are obtained after
expressing $\mu_i$ and $\Delta n_i$ in terms of this set through
\Eqn{eq:Dnmu} and \Eqn{eq:Y}.

%%%%%%%%%%%%%%%%%%%%%%%%%%%%%%%%%%%%%%%%%%%%%%%%%%%%%%%%%%%%%

\subsubsection{General constraints}
\label{sec:general}

We first list in items 1, 2 and 3 below the conditions that 
hold in the whole temperature range $M_W \ll T \lsim 10^{14}\,$GeV. 
Conversely, some of the Yukawa coupling conditions given in items 4 and 5  
will have to be dropped as the temperature is increased and the corresponding
reactions go out of equilibrium.  For simplicity of notations, in the
following we denote the chemical potentials with the same notation
that labels the corresponding field: $\phi \equiv \mu_\phi$.

%%%%%%%%%%%%%%%%%%%%%%%%%%%%%%%%%%%%%%%%%%%%%%%%%%%%%%%%%%%%%

\begin{enumerate}

\item[(1)] At scales much higher than $M_W$, gauge fields have vanishing
  chemical potential $W=B=g=0$.\cite{Harvey:1990qw} This also implies that all
  the particles belonging to the same $SU(2)_L$ or $SU(3)_c$ multiplets 
  have the same chemical potential. For example
  $\phi(I_3=+\frac{1}{2})=\phi(I_3=-\frac{1}{2})$ for a field $\phi$
  that is a doublet of weak isospin $\vec I$, and similarly for color.

\item[(2)] Denoting by $\tilde W_R$, $\tilde B_R$ and $\tilde g_R$ the
  right-handed winos, binos and gluinos chemical potentials, and by
  $\ell,\,Q$ ($\tilde\ell,\,\tilde Q$) the chemical potentials of the
  (s)lepton and (s)quarks left-handed doublets, the following
  reactions: $\tilde Q +\tilde g_R \to Q$,\ $\tilde Q +\tilde W_R \to
  Q$,\ $\tilde \ell +\tilde W_R \to \ell $,\ $\tilde \ell +\tilde B_R
  \to \ell $,\ 
  imply that all gauginos have the same chemical potential:
\begin{equation}
\label{eq:g}
-\tilde g = Q-\tilde Q=
-\tilde W= \ell-\tilde \ell=-\tilde B,
\end{equation}
where $\tilde W$, $\tilde B$ and $\tilde g$
denote the chemical potentials of {\it left-handed} gauginos.
It follows that the chemical potentials of the SM particles are related 
to those of their respective superpartners as 
\
\begin{eqnarray}
  \label{eq:tQtell}
   \tilde{Q},\tilde \ell &=&    Q,\ell+  \tilde g \\
  \label{eq:HuHd}
   H_{u,d} &=&   \tilde H_{u,d}+  \tilde g \\
  \label{eq:tutdte}
   \tilde u,\tilde d,\tilde e  &=&   u,d,e-  \tilde g. 
\end{eqnarray}
The last relation, in which $u,d,e\equiv u_R,d_R,e_R$ denote the
$R$-handed $SU(2)_L$ singlets, follows e.g. from $ \tilde u^c_L= u^c_L+
\tilde g$ for the corresponding $L$-handed fields, together with
$u^c_L=-u_R$, and from the analogous relation for the $SU(2)_L$ singlet
squarks. 

\end{enumerate}

\smallskip

Eqs.~\eqref{eq:tQtell}--\eqref{eq:tutdte} together with the vanishing
of the chemical potentials of the gauge fields and the equality of the
chemical potentials for all the gauginos, imply that we are left
with 18 chemical potentials (or number density asymmetries) that we
choose to be the ones of the fermionic states. They are 15 for the SM
quarks and leptons, 2 for the up-type and down-type higgsinos, and 1
for the gauginos. These 18 quantities are further constrained by
additional conditions.

\smallskip

\begin{enumerate}

\item[(3)] Before EW symmetry breaking hypercharge is an exactly conserved
  quantity, and we can assume a vanishing  total hypercharge for the Universe: 
  \begin{equation}
    \label{eq:Ytot}
    {y}_{\,\rm tot} = \sum_{b} \Delta n_b\, y_b +   \sum_{f} \Delta
    n_f\,y_f =0, 
  \end{equation}
  where $y_{b,f}$ denotes the hypercharge of the $b$-bosons or
  $f$-fermions.  It is useful to rewrite explicitly 
this condition in terms of the rescaled density asymmetries per
  degree of freedom $\left\{Y_{\Delta_i}\right\}$ defined in \Eqn{eq:Y}:
 \begin{equation}
    \label{eq:YtotY}
\sum_i\left(Y_{\Delta Q_i}+2Y_{\Delta u_i}-Y_{\Delta d_i}\right)
-\sum_\alpha\left(Y_{\Delta \ell_\alpha}+Y_{\Delta e_\alpha}\right)+
Y_{\Delta \tilde{H}_u}-Y_{\Delta \tilde{H}_d}= 0. 
\end{equation}

\item[(4)] When the reactions mediated by the lepton Yukawa couplings are
  faster than the Universe expansion rate, the following chemical
  equilibrium conditions are enforced:
\be 
\label{eq:leptons}
\ell_\alpha - e_\alpha + \tilde H_d + \tilde g =0, \qquad
(\alpha=e,\,\mu,\,\tau).  
\ee
For $\alpha=e$ the corresponding Yukawa condition holds only as long
as
\be
\label{eq:Te}
T\lsim 10^5(1+\tan^2\beta)\,{\rm GeV,} 
\nonumber
\ee
when Yukawa reactions between the first generation left-handed $SU(2)_L$
lepton doublet $\ell_e$ and the right-handed singlet $e$ are faster
than the expansion.\cite{Cline:1993vv,Cline:1993bd}  Note also that, as 
discussed in Section~\ref{sec:num_lfe}, if the
temperature is not too low lepton flavour equilibration induced by
off-diagonal slepton soft masses will not occur. We assume that this
is the case, and thus we take the three $\ell_\alpha$ to be 
independent quantities.

\item[(5)]  Reactions mediated by the quark Yukawa couplings enforce the
  following six  chemical equilibrium conditions:
\bea 
\label{eq:upquarks}
Q_i - u_i + \tilde H_u + \tilde g &=&0, \qquad
(u_i=u,\,c,\,t),\\  
\label{eq:downquarks}
Q_i - d_i + \tilde H_d + \tilde g &=&0, \qquad
(d_i=d,\,s,\,b)\,.  
\eea
The up-quark Yukawa coupling maintains chemical equilibrium between
the left- and right-handed up-type quarks up to $T\sim 2\cdot
10^6\,$GeV.  Note that when the Yukawa reactions of at least two
families of quarks are in equilibrium, the mass basis is fixed for all
the quarks and squarks.  Intergeneration mixing then implies that
family-changing charged-current transitions are also in equilibrium:
$b_L \to c_L$ and $t_L \to s_L$ imply $Q_2 = Q_3$; $s_L \to u_L$ and
$c_L \to d_L$ imply $Q_1 = Q_2$. Thus, up to temperatures $T\lsim
10^{11}\,$GeV, that are of the order of the equilibration temperature
for the charm Yukawa coupling, the three quark doublets have the same
chemical potential:
  \begin{equation}
    \label{eq:Q}
    Q\equiv Q_3=Q_2=Q_1. 
  \end{equation}
  At higher temperatures, when only the third family is in
  equilibrium, we have instead $Q\equiv Q_3=Q_2\neq Q_1$.  Above
  $T\sim 10^{13}$ when (for moderate values of $\tan\beta$) also the
  $\tau$ and $b$-quark $SU(2)_L$ singlets decouple from their Yukawa
  reactions, all intergeneration mixing becomes negligible and
  $Q_3\neq Q_2\neq Q_1$.

\end{enumerate}

\subsubsection{Above the superequilibration temperature}
\label{sec:nse_regime}

%%%%%%%%%%%%%%%%%%%%%%%%%%%%%%%%%%%%%%%%%%%%%%%%%%%%%%%%%%%%%

We now discuss the condition specific for ranges of temperatures
satisfying Eq.~\eqref{eq:Tgmu}, for which the chemical potentials of
particle $\phi$ and of its superpartner $\widetilde \phi$ are not
equal (NSE) but are related through a (non-vanishing) gaugino chemical
potential $\widetilde g$, as in
Eqs.~\eqref{eq:tQtell}--\eqref{eq:tutdte}.  For definiteness, we fix
the relevant temperature around $T\sim 10^{8}\,$GeV, and to emphasize
that this condition applies only to the NSE regime we put the
subscript `NSE' on the numbering.

\begin{enumerate}

\item[(6$_{\rm NSE}$)] Fast reactions induced by the generalized QCD
  and EW sphaleron multi-fermion operators~\Eqn{eq:tO-EW}
  and~\Eqn{eq:tO-QCD} imply\cite{Ibanez:1992aj}\footnote{These
    equations should be compared with the SE sphaleron conditions
    \eqref{eq:EW} and \eqref{eq:QCD}.}
\bea
\label{eq:tEWmu}
&&3\sum_i Q_i+\sum_\alpha\ell_\alpha
+\tilde H_u+\tilde H_d+4\,\tilde g=0,\\
\label{eq:tQCDmu}
&&2\sum_i Q_i-\sum_i\left(u_i+d_i\right)+6\,\tilde g=0.
\eea

\end{enumerate}

At $T \sim 10^{8}\,$GeV, Yukawa equilibrium for the $u$ quark is never
realized.  For $\alpha=e$ and for the $d$-quark Yukawa, equilibrium
holds as long as $ T\lsim 10^5(1+\tan^2\beta)\,{\rm
GeV}$\cite{Cline:1993vv,Cline:1993bd} and $ T\lsim 4\cdot
10^6(1+\tan^2\beta)\,{\rm GeV}$ respectively.  Then, for $T \sim
10^{8}\,$GeV both condition hold only if $\tan\beta \gsim 35$, while
they both do not hold if $\tan\beta \lsim 5$. As we will discuss
below, in the latter case the Yukawa equilibrium conditions get
replaced by other two conditions, and thus the overall number of
constraints does not change.  Later in Section~\ref{sec:yuknse} 
we will present results for the large and small
$\tan\beta$ cases, and since they do not differ much, we omit the
corresponding results for the intermediate case $5\lsim \tan\beta\lsim
35$.

Counting the number of additional conditions listed in items 3--5 
and 6$_{\rm NSE}$, we have 1 from global hypercharge neutrality, 8
from Yukawa equilibrium plus 2 from intergeneration quark mixing,
and 2 from the EW and QCD sphaleron equilibrium. This adds to a total
of 13 constraints for the initial 18 variables, meaning that 5
quantities must be determined from dynamical evolution equations.
These quantities can be chosen, for example, as the
density asymmetries of the three fermionic lepton flavours
$Y_{\Delta\ell_\alpha}$, of the up-type higgsinos $Y_{\Delta \tilde
  H_u}$ and of the gauginos $Y_{\Delta\tilde g}$, given that the last
one allows to relate $Y_{\Delta\ell_\alpha}$ and $Y_{\Delta \tilde
  H_u}$ to the densities asymmetries of the corresponding
superpartners.  This choice would be a natural one since these are the
density asymmetries that `weight' the various interactions entering
the BE for SL. However, the EW and QCD sphaleron
reactions~\Eqn{eq:tO-EW} and \Eqn{eq:tO-QCD} imply fast changes of
these asymmetries. A much more convenient choice will be to use 
appropriate linear combinations of the various asymmetries
corresponding to anomaly free and quasi-conserved charges, where with
`quasi-conserved' we refer to charges that are not conserved only by
the `slow' RHSN-related reactions.  These quantities can be
identified with the three flavoured charges $\Delta_\alpha$ and
with the two charges $R_B$ and $R_\chi$ discussed in the previous
section. In terms of the rescaled density asymmetries (asymmetry
abundances) per degree of freedom they read:
\begin{eqnarray} 
\label{eq:YDeltaAlpha}
Y_{\Delta_\alpha} &=& 6\,Y_{\Delta Q}+ \sum_i\left(Y_{\Delta
    u_i}+Y_{\Delta d_i}\right)- 3\,(2Y_{\Delta \ell_\alpha}+Y_{\Delta
  e_\alpha})-2\,Y_{\Delta\tilde g}\,,   \\
Y_{\Delta R_B}&=&
- 6 Y_{\Delta Q} - 
\sum_i \left(13\,Y_{\Delta u_i}-5\,Y_{\Delta d_i}\right)
\nonumber \\
&& 
+ \sum_\alpha\left(10\,Y_{\Delta \ell_\alpha}+7\,Y_{\Delta e_\alpha}\right)
+68\,Y_{\Delta\tilde g} 
+10\,Y_{\Delta \tilde H_d}-2\,Y_{\Delta \tilde H_u},  
\label{eq:YDeltaRB}
\end{eqnarray}
and
\begin{eqnarray} 
Y_{\Delta R_\chi}&=&  3\,\sum_i \left(3\,Y_{\Delta u_i}-2\,Y_{\Delta \tilde
  g}\right)+\frac{1}{3}\, Y_{\Delta R_3},  
\label{eq:YDeltaRchi}
\end{eqnarray}
where,  in this last expression, 
\bea
\nonumber 
Y_{\Delta R_3}&=&
- 18 Y_{\Delta Q} - 
3\, \sum_i \left(11\,Y_{\Delta u_i}-4\,Y_{\Delta d_i}\right)
\nonumber \\
&& \hspace{-0.5cm}
+ \sum_\alpha\left(16\,Y_{\Delta \ell_\alpha}+13\,Y_{\Delta e_\alpha}\right)
+82\,Y_{\Delta\tilde g} 
+16\,Y_{\Delta \tilde H_d}-14\,Y_{\Delta \tilde H_u}.   
\label{eq:YDeltaR3}
\eea
% 
% In Eq.~\eqref{eq:YDeltaRchi} we have left in clear some numerical
% factors: the overall factor of 3 adds the contributions of scalars
% (that is twice that of fermions), the factor of 2 in front of the
% $Y_{\Delta Q_i}$ and $Y_{\Delta \ell_\alpha}$ accounts for the $SU(2)_L$
% gauge multiplicity, while the color factor compensates against the the
% quark baryon number $B=1/3$.  
The asymmetry abundances of the five charges in
\Eqns{eq:YDeltaAlpha}{eq:YDeltaRchi}  define the basis
$Y_{\Delta_a}=\left\{Y_{\Delta_\alpha},Y_{\Delta R_B},Y_{\Delta
    R_\chi} \right\}$ 
in terms of which the five fermionic asymmetry abundances 
$Y_{\Delta\psi_a}=
\{Y_{\Delta\ell_\alpha},\,Y_{\Delta\tilde g},\,Y_{\Delta \tilde
    H_u}\}$, 
that are the relevant ones for the SL
processes,  have to be expressed. We will do this by introducing a
$5\times 5$ $A$-matrix defined according to:
\be
\label{eq:A5x5}
Y_{\Delta\psi_a} = A_{ab}\,  Y_{\Delta_b}\,, 
\ee
where the numerical values of $A_{ab}$ are obtained from
\Eqns{eq:YDeltaAlpha}{eq:YDeltaRchi} subjected to the constraining
conditions listed in items 3--5 and 6$_{\rm NSE}$.  Let us note at this point
that the $3\times 5$ submatrix $A_{\ell_\alpha b}$ for the lepton asymmetry abundances
represents the generalization of the $A$ matrix introduced
in Ref.~\refcite{Barbieri:2000}, $A_{\tilde H_u b}$ generalizes the Higgs
$C$-vector first introduced in Ref.~\refcite{Nardi:2006a}, and $A_{\tilde g b}$
generalizes the $C$-vector for the gauginos first introduced
in Ref.~\refcite{Fong:2010qh}. As regards the asymmetry abundances for the bosonic
partners of $\ell_\alpha$ and of $\widetilde H_u$, they are simply given
by: $A_{\widetilde{\ell}_\alpha\, b}=2 \left(A_{\ell_\alpha\, b} +
  A_{\widetilde{g} b}\right)$ and $A_{H_{u}\, b} = 
2\left(A_{\widetilde{H}_u}+A_{\widetilde{g}\, b}\right)$.

%%%%%%%%%%%%%%%%%%%%%%%%%%%%%%%%%%%%%%%%%% 
\medskip
%%%%%%%%%%%%%%%%%%%%%%%%%%%%%%%%%%%%%%%%%%

\subsubsection{Additional conditions from Yukawa reactions}
\label{sec:yuknse}

\begin{enumerate}
%\item[(7$_{\rm NSE}$-I)] 
\item[(7-I)] 
Case I: {\sl Electron and down-quark  Yukawa reactions in equilibrium.}\\
If the down-type Higgs VEV is relatively small $v_d\ll v$, the values
of the electron and down-quark masses are obtained for correspondingly
large values of the $h_d$ and $h_e$ Yukawa couplings.  For
$v_u/v_d=\tan\beta \gsim 35$ we have a regime in which at $T \sim
10^8\,$GeV, that is well above the NSE threshold Eq.~\eqref{eq:Tgmu},
both $h_d$ and $h_e$ related reactions are in equilibrium.  
%!!%
Since $u$-quark Yukawa reactions are never in equilibrium, in this
case, only eight Yukawa conditions in \Eqns{eq:leptons}{eq:downquarks}
hold. Solving for the density asymmetries $Y_{\Delta\psi_a}=
\{Y_{\Delta\ell_\alpha},\,Y_{\Delta\tilde g},\,Y_{\Delta \tilde
  H_u}\}$ in terms of the charge-asymmetries
$Y_{\Delta_a}=\left\{Y_{\Delta_\alpha},Y_{\Delta R_B},Y_{\Delta
    R_\chi} \right\}$ subject to the constraints in items  3 to
5 and 6$_{\rm NSE}$, yields 
\begin{equation}
A=\frac{1}{9\times 827466}\left(
\begin{array}{rrrrr}
 -788776 &  38690 &  38690 & -56295 &  41931  \\
   38690 &-788776 &  38690 & -56295 &  41931  \\
   38690 &  38690 &-788776 & -56295 &  41931  \\
   41913 &  41913 &  41913 & 124281 &  12798  \\
 -102411 &-102411 &-102411 & 108108 &-335907
\end{array}
\right).  
\label{eq:Aedin}
\end{equation}
%
%\item[(7$_{\rm NSE}$-II)] 
\item[(7-II)] 
Case II: {\sl Electron and down-quark Yukawa reactions out of equilibrium}\\
If $v_d$ is not much smaller than $v_u$, resulting in $\tan\beta \lsim
5$, then both $h_e$ and $h_d$ are sufficiently small that at $T \sim
10^8\,$GeV the related Yukawa reactions do not occur.  In this case we
have to set $h_d,\,h_e\to 0$ and the corresponding two Yukawa
equilibrium conditions in~\Eqns{eq:leptons}{eq:downquarks} do not hold
%!!%
(on top of $u$-quark Yukawa reactions which are never in equilibrium). 
However, two conservation laws replace these conditions.
$h_e\to 0$ implies that we gain a `chiral' symmetry for the
right-handed fermion and scalar electrons, ensuring that the total
number density asymmetry $\Delta n_{e}+\Delta n_{\tilde e}$ is
conserved. As usual, we assume that the constant value of this
quantity vanishes, which in terms of the rescaled density asymmetries
per degree of freedom implies:
\begin{equation}
  \label{eq:YeR}
  Y_{\Delta e}-\frac{2}{3}\, Y_{\Delta \tilde g}=0\,.
\end{equation}
For the right-handed down quark we could define an anomaly-free charge 
completely equivalent to $Y_{\Delta R_\chi}$ in~\Eqn{eq:YDeltaRchi}  
but, given that in this regime all the dynamical equations  
are symmetric  under the exchange $u \leftrightarrow d$, it is
equivalent, and much more simple, to impose the condition
\begin{equation}
  \label{eq:YdR}
  Y_{\Delta d}=  Y_{\Delta u}\,.
\end{equation}
The net result is that, with respect to the previous case, the total
number of constraints is not changed, and again five quantities
suffice to express the rescaled density asymmetries for all the
fields. For the $5\times 5$ $A$ matrix defined in~\Eqn{eq:A5x5}
we obtain:
\begin{equation}
A=\frac{1}{9\times 162332}\left(
\begin{array}{rrrrr}
-210531 &  21573 &  21573 & -12414 & 12483 \\
   8676 &-165529 &  -3197 & -17958 & 29709 \\
   8678 &  -3197 &-165529 & -17958 & 29709 \\
   7497 &   7299 &   7299 &  23634 &  4833 \\
 -11322 & -18477 & -18477 &  23940 &-74385
\end{array}
\right). 
\label{eq:Aedout}
\end{equation}
\end{enumerate}
%%%%%%%%%%%%%%%%%%%%%%%%%%%%%%%%%%%%
%%%%%%%%%%%%%%%%%%%%%%%%%%%%%%%%%%%%%%%%%%%%%%%

\subsection{Boltzmann equations for {\it R}-genesis}
\label{sec:nse_BE}
In order to render clear the role played by the new charges $\Delta
R_B$ and $\Delta R_\chi$ and by NSE effects, in this section we
introduce a simplified set of BE including only decays and inverse
decays of RHN and RHSN.   In this approximation 
the evolutions of the number density of the heavy states normalized to
the entropy density $s$ is  obtained from 
Eqs.~\eqref{eq:BEN} and \eqref{eq:BENt}
by retaining only the two 
reactions rates  $\gamma_{\widetilde{N}}$ 
and $\gamma_{{N}}$.  
In writing down the evolution equations for the five charges
$Y_{\Delta_\alpha},Y_{\Delta R_B},Y_{\Delta R_\chi}$ it is convenient
to introduce a special notation for the scalar and fermionic asymmetry
abundances (per degree of freedom) normalized to the respective
equilibrium abundances $Y^{eq}_s = 2 Y^{eq}_f= \frac{15}{4\pi^2 g_*}$:
\begin{equation}
  \label{eq:calY}
{\cal Y}_{\Delta s,\Delta f} \equiv \frac{Y_{\Delta s,\Delta f}}{Y^{eq}_{s,f}}.  
\end{equation}
Using Eqs.  \eqref{eq:Y} and \eqref{eq:Dnmu}  together   
with Eqs. \eqref{eq:tQtell} and \eqref{eq:HuHd} it is then easy to 
verify that
\begin{equation}
  \label{eq:Ytg}
  {\cal Y}_{\Delta \tilde \ell,\Delta H_u}=  
{\cal Y}_{\Delta \ell,\Delta \tilde H_u}
+ {\cal Y}_{\Delta \tilde g}\,. 
\end{equation}
Retaining  only decays and inverse decays, the BE for
the flavour density asymmetries read:
\begin{eqnarray}
\dot Y_{\Delta_\alpha}&=& 
- \epsilon^f_\alpha\left(z\right)
\left(\frac{Y_{\widetilde{N}}}{Y_{\widetilde{N}_+}^{eq}}-2\right)
\frac{\gamma_{\widetilde{N}}}{2}
+\left({\cal Y}_{\Delta \ell_\alpha}
+{\cal Y}_{\Delta \widetilde H_u}\right)
\frac{\gamma_{\widetilde{N}}^{f,\alpha}}{2}
+\Big({\cal Y}_{\Delta \ell_\alpha}+{\cal Y}_{\Delta H_u}\Big)
\frac{\gamma_{N}^\alpha}{4}  
\nonumber \\
&& \hspace{-13mm}-\epsilon^s_\alpha\left(z\right)
\left(\frac{Y_{\widetilde{N}}}{Y_{\widetilde{N}_+}^{eq}}-2\right)
\frac{\gamma_{\widetilde{N}}}{2}
+\left({\cal Y}_{\Delta \widetilde \ell_\alpha}
+{\cal Y}_{\Delta H_u}\right)\frac{\gamma_{\widetilde{N}}^{s,\alpha}}{2}
+\left({\cal Y}_{\Delta \widetilde \ell_\alpha}
+{\cal Y}_{\Delta \widetilde H_u}\right)\frac{\gamma_{N}^\alpha}{4}, 
\end{eqnarray}
where $\gamma_{\widetilde{N}}^{s(f),\alpha}$ denotes the rate of RHSN
decays into scalars (fermions) of flavour $\alpha$, while quantities
without a flavour index are understood to be summed over all flavours.
To an excellent approximation we have
$\gamma_{\widetilde{N}}^{s,\alpha}=
\gamma_{\widetilde{N}}^{f,\alpha}$, and furthermore the density
asymmetries of the scalars can be expressed in terms of the ones of
the fermions by means of~\Eqn{eq:Ytg}. This yields:
\begin{eqnarray}
  \dot Y_{\Delta_\alpha}\! &=& \!
  - \epsilon_\alpha\left(z\right)
  \left(\frac{Y_{\widetilde{N}}}{Y_{\widetilde{N}_+}^{eq}}-2\right)
  \frac{\gamma_{\widetilde{N}}}{2}+
  \left({\cal Y}_{\Delta \ell_\alpha}
    +{\cal Y}_{\Delta \widetilde H_u}+{\cal Y}_{\Delta \widetilde g}
  \right)\!\!
  \frac{\gamma_{N}^\alpha+\gamma_{\widetilde{N}}^\alpha}{2},   
  \label{eq:DeltaAlpha}
\\
\dot Y_{\Delta R_B}\!&=&\!
 \epsilon^s\left(z\right) \!\!
\left(\frac{Y_{\widetilde{N}}}{Y_{\widetilde{N}_+}^{eq}}-2\right)
\!\gamma_{\widetilde{N}}
% \nonumber \\ && 
-\!\sum_\alpha 
\left(
{\cal Y}_{\Delta \ell_\alpha}+
{\cal Y}_{\Delta \widetilde H_u}+{\cal Y}_{\Delta \widetilde g}
\right)
\!\!\frac{\gamma_{N}^\alpha+\gamma_{\widetilde{N}}^\alpha}{2}
- 
{\cal Y}_{\Delta \widetilde g}
\frac{\gamma_{\widetilde{N}}}{2}
, \ \quad \ \ 
  \label{eq:DeltaRB}
\\
\dot Y_{\Delta R_\chi}&=& 
\left[\epsilon^s\left(z\right)- 
\epsilon^f\left(z\right)\right]
\left(\frac{Y_{\widetilde{N}}}{Y_{\widetilde{N}_+}^{eq}}-2\right)
\frac{\gamma_{\widetilde{N}}}{6}
% \frac{\gamma_{\widetilde{N}}}{2}-\frac{1}{3}\, 
- {\cal Y}_{\Delta \widetilde g}\,
\frac{\gamma_{\widetilde{N}}}{6}\,.  
  \label{eq:DeltaRchi}
\end{eqnarray}

It is possible, and formally straightforward, to add to these
equations the appropriate terms that allow to extend their validity
also in the SE regime, when the RHSN masses are below the
bound~\Eqn{eq:NSE}. In order to do this, one must add a
$\gamma_{\tilde g}^{\rm eff}$ term characterizing the set of
gaugino-mediated reactions with chirality flip on the gaugino line
that are responsible for processes that equilibrate particle-sparticle
chemical potentials.\footnote{Ref.~\refcite{Plumacher:1997ru} included a
  similar term $\gamma_{\rm MSSM}$ in the BE for supersymmetric
  leptogenesis, corresponding to the thermally averaged cross section
  for the photino mediated process $e+e\leftrightarrow
  \widetilde{e}+\widetilde{e}$ that was computed in
  Ref~\refcite{Keung:1983nq}. However, the  only contributions 
that do not vanish in the $m_{\tilde \gamma}\to 0$
  limit are those that, like e.g.  $e^-_L+e^-_R \leftrightarrow
  \widetilde{e}_L+\widetilde{e}_R$, do not enforce  SE. 
%SE reactions like $e^-_L+e^-_L \leftrightarrow\widetilde{e}_L
%+\widetilde{e}_L$ all vanish in the $m_{\tilde \gamma}\to 0$ limit.
} 
Equivalently $\gamma^{\rm eff}_{\submuh}$
characterizes the set of reactions induced by the higgsino mixing
parameter $\mu$ that enforce the chemical equilibrium condition
$\tilde H_u+\tilde H_d=0$. The thermally averaged rates for these
reactions can be written in an approximated form as:
\begin{equation}
  \label{eq:approxSE}
\frac{\gamma_{\tilde g}^{\rm eff}}{n^{eq}_f} = \frac{m^2_{\widetilde g}}{T}, 
\qquad\qquad 
\frac{\gamma_{\submuh}^{\rm eff}}{n^{eq}_f} = \frac{\muh^2}{T}, 
\end{equation}
where $n_f^{eq}$ is the equilibrium number density for one fermionic
degree of freedom, while $m_{\widetilde g }$ and $\muh$ have to be
understood as effective mass parameters in which all coupling
constants as well as reaction multiplicities are reabsorbed.
Extension of the validity of \Eqns{eq:DeltaAlpha}{eq:DeltaRchi} to the
SE domain is then achieved by adding the following terms to the
equations for $R_B$ and $R_\chi$:
\begin{eqnarray}
  \label{eq:DeltaRBSE}
\dot Y_{\Delta R_B}^{SE}&=&\left\{\dot Y_{\Delta R_B}\right\}  
- {\cal Y}_{\Delta \widetilde g}\>\gamma_{\tilde g}^{\rm eff}\,,
\\ \label{eq:DeltaRchiSE}
\dot Y_{\Delta R_\chi}^{SE}&=&
\left\{\dot Y_{\Delta R_\chi}\right\}  
- \frac{1}{3}{\cal Y}_{\Delta \widetilde g}\> \gamma_{\tilde g}^{\rm eff}
+ \frac{1}{3}\left({\cal Y}_{\Delta \widetilde H_u}+ 
{\cal Y}_{\Delta \widetilde H_d}\right)\,\gamma_{\submuh}^{\rm eff}\,,
\end{eqnarray}
where the $\big\{\dot Y_{\Delta R}\big\}$ above stand for the r.h.s of
the corresponding Eqs. \eqref{eq:DeltaRB} and
\eqref{eq:DeltaRchi}. Note that since the $R_B$ charge of the $\muh$
term is $R_B(H_uH_d)=2$, $\muh$ conserves $R_B$ and accordingly there
is no term proportional to $\gamma_{\submuh}^{\rm eff}$ in
\Eqn{eq:DeltaRBSE}.  Since higgsino equilibration involves also the
density asymmetry ${\cal Y}_{\Delta \widetilde H_d}$ we give below the
$C^{\tilde H_d}$ vectors for the two cases:
\begin{eqnarray}
  \label{eq:C_I}
{\rm Case\ I:} \; C^{\tilde H_d} &=&
\frac{1}{827466}\left(14237,\;
14237,\;14237,\;1260,\,-3915\right)\,,
\\  \label{eq:C_II}
 {\rm Case\ II:} \; C^{\tilde H_d} &=& 
\frac{1}{3\times 162332  }\left(12469,\;
16768,\;16768,\;7056,\,-21924\right).
\end{eqnarray}
It can be shown that the results of numerically solving 
\Eqn{eq:DeltaAlpha} and \Eqns{eq:DeltaRBSE}{eq:DeltaRchiSE}
with  increasing values of $m_{\widetilde g }$ and $\muh$ 
converge to the solutions of the usual BE for the SE regime
(see \ref{sec:BE_SE}).

\medskip

\subsection{NSE regime: {\it R}-genesis in a simple case} 
\label{sec:simple}

The role played by the asymmetries of the two $R$
charges is easy to understand in a simple scenario, in which lepton flavour
effects play basically no role and thus do not shadow the new effects.
This scenario is defined by the following two conditions:
\begin{itemize}
\item We assume equal branching fractions for the decays of $N$ and of
  $\widetilde{N}_\pm$ into the three lepton flavours, that is the
  $P_\alpha$ defined in \Eqn{eq:fla_proj} are all equal to
  $\frac{1}{3}$ implying $\epsilon_\alpha = \frac{1}{3}\epsilon$ 
  %!!%
  \footnote{Here we assume the UTS scenario 
discussed in Section \ref{sec:fla_scenarios}.}
  and $\gamma^\alpha_{N,\widetilde{N}} =\frac{1}{3} \gamma_{N,\widetilde{N}}$.
\item We assume the regime described in Case I in which the 
Yukawa equilibrium condition for the electron holds, 
and thus the three lepton flavours are 
all treated on equal footing (see the $3\times 3$ 
upper-left corner in the $A$-matrix 
\Eqn{eq:Aedin}). Given the previous 
condition, it is then useful to define  a `flavour averaged' 
lepton asymmetry as: 
\begin{equation}
  \label{eq:Yell}
{\cal Y}_{\Delta \ell} =\frac{1}{3} \sum_\alpha
{\cal Y}_{\Delta \ell_\alpha}  
\end{equation}
\end{itemize}

With these conditions, the three equations for the flavour
charges~\Eqn{eq:DeltaAlpha} can be resummed in closed form into a
single equation for the $B-L$ asymmetry:
\begin{equation}
\label{eq:DeltaB-LS}
\dot Y_{\Delta_{B-L}}=
  - \epsilon\left(z\right)
  \left(\frac{Y_{\widetilde{N}}}{Y_{\widetilde{N}_+}^{eq}}-2\right)
  \frac{\gamma_{\widetilde{N}}}{2}+
\left({\cal Y}_{\Delta \ell}
    +{\cal Y}_{\Delta \widetilde H_u}+{\cal Y}_{\Delta \widetilde g}
  \right)
  \frac{\gamma_{N}+\gamma_{\widetilde{N}}}{2}\,,    
\end{equation}
yielding a reduced set of just 3 BE.
The $3\times 3$ matrix relating
$\{Y_{\Delta\ell},\,Y_{\Delta\tilde g},\,Y_{\Delta \tilde H_u}\}$ to
the three charge-asymmetries $\left\{Y_{\Delta_{B-L}},Y_{\Delta
    R_B},Y_{\Delta R_\chi} \right\}$
can be readily evaluated
from \Eqn{eq:Aedin}:
\begin{equation}
A=\frac{1}{827466}\left(
\begin{array}{rrrrr}
  -26348 & -6255 &  4659  \\
    4657 & 13809 &  1422  \\
  -11379 & 12012 &-37323
\end{array}
\right).  
\label{eq:Aedin3x3}
\end{equation}

It is now easy to see that in the NSE regime we can rewrite the BE  as
\begin{eqnarray}
  \dot Y_{\Delta_{B-L}}
\! &=&\!3\,\dot Y_{\Delta R_\chi}-\dot Y_{\Delta R_B}\,,   
  \label{eq:DeltaB-LnoS}
\\
\dot Y_{\Delta R_B}
&=& \epsilon^s(z)\!\!
\left(\frac{Y_{\widetilde{N}}}{Y_{\widetilde{N}_+}^{eq}}-2\right)
{\gamma_{\widetilde{N}}}
% \nonumber \\&&
-\left(
{\cal Y}_{\Delta \ell}+
{\cal Y}_{\Delta \widetilde H_u}+{\cal Y}_{\Delta \widetilde g}
\right)
\frac{\gamma_{N}+\gamma_{\widetilde{N}}}{2}
- 
{\cal Y}_{\Delta \widetilde g}
\frac{\gamma_{\widetilde{N}}}{2},\quad  
\label{eq:DeltaRBS} 
\\
\dot Y_{\Delta R_\chi}\! &=& \!
\left[\epsilon^s\left(z\right)- 
\epsilon^f\left(z\right)\right]
\left(\frac{Y_{\widetilde{N}}}{Y_{\widetilde{N}_+}^{eq}}-2\right)
\frac{\gamma_{\widetilde{N}}}{6}
% \frac{\gamma_{\widetilde{N}}}{2}-\frac{1}{3}\, 
- {\cal Y}_{\Delta \widetilde g}
\frac{\gamma_{\widetilde{N}}}{6} \,, 
\label{eq:DeltaRchiS}
\end{eqnarray}
since the difference in the r.h.s. of \Eqn{eq:DeltaB-LnoS} gives
precisely \Eqn{eq:DeltaB-LS}.  \Eqn{eq:DeltaB-LnoS} makes apparent how
$Y_{\Delta R_\chi}$ and $Y_{\Delta R_B}$, that in the $T\to 0$ limit
keep having non vanishing CP asymmetries, are sources of the $B-L$
asymmetry.  Note that the only role of the two conditions listed above
is simply that of allowing to collapse the three equations for
$\Delta_\alpha$ into a single one for $\Delta_{B-L}$, while
maintaining the BE in closed form. Therefore the above result is
completely general, and in particular it holds also when scattering
processes are included, and is independent of the particular NSE
temperature regime (e.g. Case I and Case II) and flavour
configuration. In short, in the NSE regime the evolution of
$\Delta_{B-L}$ can be always obtained from the evolution of
$3\Delta_{R_\chi}- \Delta_{R_B}$, and the final value of
$Y_{\Delta_{B-L}}$ can be equally well obtained from summing the
values of the flavour charges asymmetries $\sum_\alpha
Y_{\Delta_\alpha}$ or from the final value of $3Y_{\Delta R_\chi}-
Y_{\Delta R_B}$.  The reason why this happens is simple: by using the
definitions~\Eqns{eq:RB}{eq:chiralup} together with
\Eqns{eq:R2}{eq:R3} one obtains that $3R_\chi- R_B= \raise 2pt
\hbox{$\chi$}_{u^c_L} -\frac{2}{3}B-PQ $. Of course, only the $PQ$
fragment of this charge is violated in RHSN interactions, and
from Table~\ref{tab:1} we see that this violation is precisely the
same as for $B-L$ (e.g. for $\widetilde N \to \ell \tilde H_u$ we
have $\Delta(B-L)=-\Delta L=-\Delta (PQ)=-1$). Thus, regardless of the
fact that $B-L$, $R_B$ and $R_\chi$ are all independent charges, in
the NSE regime the BE for $3Y_{\Delta R_\chi}- Y_{\Delta R_B}$ 
always coincides with the BE for $Y_{\Delta_{B-L}} =\sum_\alpha
Y_{\Delta_\alpha}$.

In this particularly simple case one can take a further step and
rewrite the density asymmetry ${\cal Y}_{\Delta \widetilde g}$ and the
combination $({\cal Y}_{\Delta \ell}+ {\cal Y}_{\Delta \widetilde
  H_u}+{\cal Y}_{\Delta \widetilde g})$ in the r.h.s of
\Eqns{eq:DeltaRBS}{eq:DeltaRchiS} in terms of
$Y_{\Delta_{B-L}},\,Y_{\Delta R_B},\,Y_{\Delta R_\chi}$ by means of
the $A$ matrix \Eqn{eq:Aedin3x3}. 
Replacing $Y_{\Delta_{B-L}}\to 3 Y_{\Delta R_\chi} - Y_{\Delta R_B}$ and  
 using $\gamma_N =\gamma_{\widetilde N}$ one obtains:
\begin{eqnarray}
\label{eq:DeltaRchiS2}
3 \dot Y_{\Delta R_\chi}&=& 
\left[\epsilon^s\left(z\right)- 
\epsilon^f\left(z\right)\right]
\left(\frac{Y_{\widetilde{N}}}{Y_{\widetilde{N}_+}^{eq}}-2\right)
\frac{\gamma_{\widetilde{N}}}{2}
% \nonumber \\ &&
-\frac{
9152\,Y_{\Delta R_B} 
+15393\, Y_{\Delta R_\chi}
}{827466}
\frac{\gamma_{\widetilde{N}}}{2}
\,,\qquad 
\\
\dot Y_{\Delta R_B}
&=& 2\,\epsilon^s(z)\,
\left(\frac{Y_{\widetilde{N}}}{Y_{\widetilde{N}_+}^{eq}}-2\right)
\frac{\gamma_{\widetilde{N}}}{2}\
% \nonumber \\ &&
-\ \frac{114424\,Y_{\Delta R_B} 
-245511\, Y_{\Delta R_\chi}}{827466}
\frac{\gamma_{\widetilde{N}}}{2}
\,. \ \ 
\label{eq:DeltaRBS2}
\end{eqnarray}
These two equations show that although $3 R_\chi$ and $R_B$
have the same  $T=0$ source term so that the difference of their
asymmetries tends to cancel, their respective washouts are quite
different, and such a cancellation will never occur.  With a 
general flavour configuration 
the set of BE cannot be collapsed to just
two equations, but still the same mechanism is at work: because of the
different washouts, the difference between $3 Y_{\Delta R_\chi}$ and
$Y_{\Delta R_B}$ becomes of the same order of these density asymmetries, and
so does $Y_{\Delta_{B-L}}$.  Consequently, one expects that by
increasing the washouts from weak strengths up to (not too)
large strengths, the final value of $B-L$ will increase.  The
numerical results in the next section confirm this picture.

In the SE regime instead, things proceed in a different way.
\Eqns{eq:DeltaRBSE}{eq:DeltaRchiSE} show that the BE for $Y_{\Delta R_\chi}$
and $Y_{\Delta R_B}$ acquire new washout terms, that are proportional to the
SE rates, while on the contrary no analogous terms enter the BE
\Eqn{eq:DeltaAlpha} for $Y_{\Delta_\alpha}$ or \Eqn{eq:DeltaB-LS} for
$Y_{\Delta_{B-L}}$.  Thus, in the SE regime, \Eqn{eq:DeltaB-LnoS} does
not hold. One can argue instead that, because of the SE washouts, the
roles of $\Delta_{B-L}$ and of $3\Delta R_\chi-\Delta R_B$ get
reversed, since now we have
\begin{equation}
3\,\dot Y_{\Delta R_\chi}-\dot Y_{\Delta R_B}=
\dot Y_{\Delta_{B-L}}+
\left({\cal Y}_{\Delta \widetilde H_u}+ 
{\cal Y}_{\Delta \widetilde H_d}\right)
\,\gamma_{\submuh}^{\rm eff}\,.
\end{equation}
In other words, since SE reactions conserve $B-L$ but violate the $R$
and $PQ$ charges, the only source of asymmetry surviving SE is the
$Y_{\Delta_{B-L}}$ asymmetry generated by thermal corrections. Given
that $\Delta{R_\chi}$ and $\Delta{R_B}$ both contain `fragments' that
carry $B$ number, they do not vanish in the SE limit, but are driven
to values that are proportional to $\Delta_{B-L}$.  The constants of
proportionality are determined by the chemical equilibrium and
conservation law conditions appropriate for the specific regime and,
for example, in Case I are given by $Y_{\Delta
  {R_B}}=-\frac{1}{3}Y_{\Delta_{B-L}}$ and $Y_{\Delta
  {R_\chi}}=-\frac{3}{79} \,Y_{\Delta_{B-L}}$.

%%%%%%%%%%%%%%%%%%%%%%%%%%%%%%%%%%%%%%%%%%%%%%%%%%%%%%%%%%%%%%%%%%

\subsection{Numerical analysis  of {\it R}-genesis}
\label{sec:nse_results}

We summarize here some of the numerical results for SL in the NSE
regime.  They are obtained by integration of the BE given in
\ref{Appendix-B} that also include the various scattering
processes.  The comparative results for the SE case can be obtained in
two formally different, but physically equivalent, ways.  A first
possibility is that of taking the limit $m_{\widetilde g},\muh \to
\infty$ in the complete BE (given, for example, in their basic form
in~\Eqns{eq:DeltaRBSE}{eq:DeltaRchiSE}).  A second possibility, that
corresponds to usual treatments, is to solve only the three equations
for the flavour charge density asymmetries $Y_{\Delta_\alpha}$ with the
corresponding $A$ matrix and $C$ vectors obtained under the assumption
of SE.  For the two cases described in Section~\ref{sec:yuknse} the
corresponding matrices assuming SE are given in ~\ref{app:se_new}
in Eqs.~\eqref{eq:ACT1} and \eqref{eq:ACT3SE}.  

To single out the new NSE effects, for all the results a flavour
equipartition configuration, with equal flavour branching
fractions~\Eqn{eq:fla_proj} $P_\alpha = \frac{1}{3}$ is assumed, so
that flavour effects are basically switched off.  In all cases, the
heavy RHSN mass is held fixed at $M=10^8\,$GeV, that is above the
temperature threshold for the validity of the effective
theory~\Eqn{eq:Tgmu}. The values of the other relevant parameters are:
$A=1\,$TeV, $\phi_A=\frac{\pi}{2}$ and
$\bar\epsilon=\frac{A}{M}=10^{-5}$ that correspond to a resonantly
enhanced CP asymmetry in mixing $\bar\epsilon^S$ Eq.~\eqref{eq:cp0_S}.
This is obtained for $2\,B\sim \Gamma\sim 2.6\, \left(\frac{m_{\rm
      eff}}{0.1\,{\rm eV}}\right)\,$GeV.  As regards gaugino mass
dependent contributions to the CP asymmetries from vertex corrections:
$\bar\epsilon^V$ Eq.~\eqref{eq:cp0_V} and $\bar\epsilon^I$
Eq.~\eqref{eq:cp0_I}, they are suppressed by additional powers of
$\Lambda_{susy}/M$ and thus have been neglected.  Given the large
value of $M$, they remain irrelevant even in the cases labeled as the
``$m_{\tilde g}\to \infty$ limit'', since in practice $m_{\tilde
  g}\approx 10\,$TeV is more than sufficient to enforce SE.
The results are presented for Case I because, as shown 
Ref.~\refcite{Fong:2010bv}, 
the differences  between the situations in which the $h_{e,d}$ Yukawa 
reactions are in equilibrium and when they are out of equilibrium are 
rather mild.

%
%%%%%%%%%%%%%%%%%%%%%%%%%%%%%%%%%%%%%%%%%%%%%%%%%%%%%%%%%%%%%%%%%%%%%%%
% PANEL
\begin{figure}[t!]
\includegraphics[width=\textwidth]{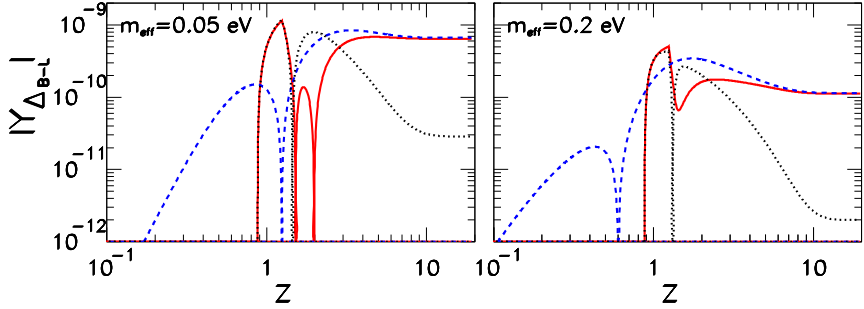}
%
% \caption[]{\baselineskip 12pt
%
\caption{Evolution of $Y_{\Delta_{B-L}}$. The solid continuous (red)
  lines depict the complete results in the $m_{\widetilde g}=\muh\to
  0\,$ limit.  The dashed (blue) lines correspond to the same limit
  when thermal corrections to the $CP$ asymmetries are neglected.
  The dotted (black) lines gives $Y_{\Delta_{B-L}}$ with thermal
  effects when SE is assumed. The picture on the left is for $m_{\rm
    eff}=0.05\,$eV and that on the right for $m_{\rm eff}=0.20\,$eV.}
\label{fig:evolutions}
\end{figure} 
%
%%%%%%%%%%%%%%%%%%%%%%%%%%%%%%%%%%%%%%%%%%%%%%%%%%%%%%%%%%%%%%%%%%%%%%%

Fig.~\ref{fig:evolutions} displays the evolution of
$Y_{\Delta_{B-L}}$ with increasing $z=M/T$.  The solid (red) lines
correspond to the full results obtained in the $m_{\widetilde
  g},\,\muh\to 0\,$GeV limit, that is when particle-sparticle
superequilibrating processes are completely switched off.  The dashed
(blue) lines give the results obtained in the same limit, but when all
thermal corrections to the CP asymmetries are neglected, and
$\epsilon^s=-\epsilon^f=\bar\epsilon/2$. Both pictures display clearly
that in the NSE regime neglecting thermal corrections in evaluating
the final values of $Y_{\Delta_{B-L}}$ is an excellent approximation.
The dotted (black) lines give $Y_{\Delta_{B-L}}$ with thermal
corrections included and under the assumption of SE, that in the
BE~\eqref{eq:DeltaRBSE}-\eqref{eq:DeltaRchiSE} corresponds to taking
the limit $m_{\widetilde g},\muh \to \infty$.  The two panels are for
two different washout strengths $m_{\rm eff}=0.05\,$eV (left) and
$m_{\rm eff}=0.20\,$eV (right) and, as anticipated, they show that 
stronger washouts result in larger  gain in the efficiency.

%%%%%%%%%%%%%%%%%%%%%%%%%%%%%%%%%%%%%%%%%%%%%%%%%%%%%%%%%%%
\begin{figure}[ht]
\begin{center}
\includegraphics[width=0.7\textwidth]{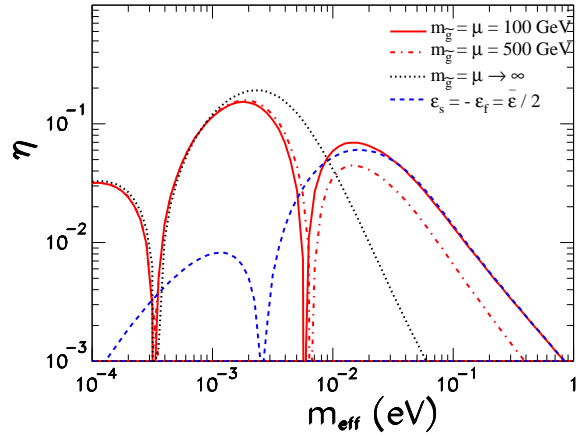}
\end{center}
\caption{Efficiency factor $\eta$ as a function of the washout
  parameter $m_{\rm eff}$ for Case I ($h_{e,d}$ Yukawa equilibrium)
  and different values of $m_{\widetilde g}=\muh$.  The red continuous
  line is for the NSE regime with $m_{\widetilde g}=\muh=100\,$GeV.
   The dashed blue line give the result for the same regime when
  thermal corrections are neglected.  The red dash-dotted line
  corresponds to $m_{\widetilde g}=\muh= 500\,$GeV, and the black
  dotted line to SE with $m_{\widetilde g},\,\muh\to \infty$.}
\label{fig:etameff}
\end{figure}
%%%%%%%%%%%%%%%%%%%%%%%%%%%%%%%%%%%%%%%%%%%%%%%%%%%%%%%%%%%

Fig.~\ref{fig:etameff} shows the efficiency $\eta$ as a function of
the washout parameter $m_{\rm eff}$.  The red continuous line
corresponds to $m_{\widetilde g}=\muh=100\,$GeV, and since it is
practically indistinguishable from the $m_{\widetilde g}=\muh\to 0\,$
case, the evolution occurs in the NSE regime in agreement with
\Eqn{eq:Tgmu}.  The red dash-dotted line corresponds to $m_{\widetilde
  g}=\muh=500\,$ GeV. In this case we see that SE rates start
suppressing the efficiency even without attaining full thermal
equilibrium. The black dotted line corresponds to the $m_{\widetilde
  g},\,\muh\to \infty\,$ limit of complete SE.  The figure illustrates
that at $T\gsim 10^7\,$GeV the leptogenesis efficiency could be
significantly underestimated if SE is incorrectly assumed.  The size
of this underestimation is a fast increasing function of the washouts,
and for particularly large values of $m_{\rm eff}$ can reach the two
orders of magnitude level.  Also, for $m_{\rm eff}\gsim 6\times
10^{-3}\,$eV, the assumption of SE results in a baryon asymmetry of
the wrong sign. Graphically, one can see this from the fact that at
small values of $m_{\rm eff}$ the black dotted and red dash-dotted and
continuous lines approximately overlap, and there is a change of sign
in $Y_{\Delta B-L}^\infty/\bar \epsilon$ around $m_{\rm eff}\sim
3\times 10^{-4}\,$eV. But around $m_{\rm eff}\sim 6\times 10^{-3}\,$eV
for the red dash-dotted and continuous lines there is another sign
change. This marks the onset of $R$-genesis domination; therefore,
from this point onward, baryogenesis does not proceed through
leptogenesis, but rather through $R$-genesis.  In the same figure the
dashed blue continuous line shows the NSE results in the approximation
of neglecting all thermal corrections to the $CP$ asymmetries.  By
comparing with the full results (red continuous line) we see that for
$m_{\rm eff}\gsim {\rm few}\,\times 10^{-2}\,$eV thermal corrections
give negligible effects. Thus, for $R$-genesis, the zero temperature
approximation yields quite reliable results.

%%%%%%%%%%%%%%%%%%%%%%%%%%%%%%%%%%%%%%%%%%%%%%%%%%%%%%%%%%%
\begin{figure}[t!]
\begin{center}
\includegraphics[width=0.7\textwidth]{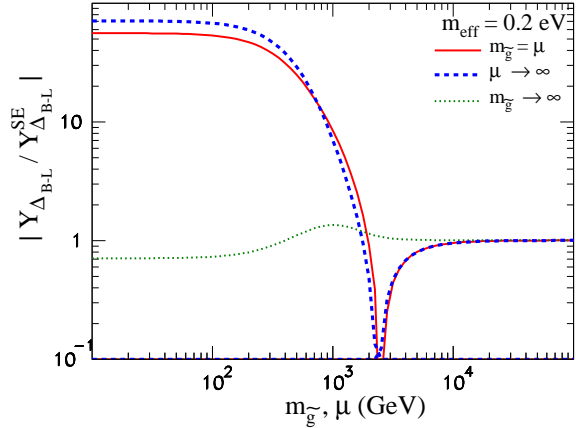}
\end{center}
\caption{The final value of $Y_{\Delta_{B-L}}$ normalized to the SE
  result $Y^{SE}_{\Delta_{B-L}}$ as a function of $m_{\widetilde g}$
  and $\muh$ for Case I ($h_{e,d}$ Yukawa equilibrium) and $m_{\rm
    eff}=0.20\,$eV.  The red continuous line corresponds to varying
  simultaneously both parameters holding $m_{\widetilde g}=\muh$.  The
  blue dashed line corresponds to varying only $m_{\widetilde g}$ with 
$\muh \to \infty$.  The green dotted line corresponds to
  varying only $\muh$ with  $m_{\widetilde g}\to \infty$.}
\label{fig:mg-eta}
\end{figure}

Fig.~\ref{fig:mg-eta} displays the value of $Y_{\Delta_{B-L}}^\infty$
(labeled just as $Y_{\Delta_{B-L}}$ for simplicity) as a function of
different values of $m_{\widetilde g}$ and $\muh$, normalized to
$Y^{SE}_{\Delta_{B-L}}$ that is the final value of the asymmetry
obtained assuming SE.  In order to enhance the impact of the new
effects, the washout parameter has been fixed to a rather large value
$m_{\rm eff}=0.20\,$eV.  The red continuous line corresponds to
varying simultaneously both SE parameters keeping their values equal:
$m_{\widetilde g}=\muh$.  We see that for $m_{\widetilde g}=\muh\lsim
1\,$TeV the amount of $B-L$ asymmetry produced by SL can be up to two
orders of magnitude larger (and of the opposite sign) with respect to
what would be obtained in the usual approach with SE.  SE effects
start suppressing the asymmetry around $m_{\widetilde g}=\muh\sim
1\,$TeV.  The asymmetry then changes sign around $3\,$TeV, that marks
the transition from the $R$-genesis to the leptogenesis regime, and
eventually around $5\,$TeV SE reactions attain complete thermal
equilibrium and
$Y_{\Delta_{B-L}}/Y^{SE}_{\Delta_{B-L}}\to 1$.
%
%%%%%%%%%%%%%%%%%%%%%%%%%%%%%%%%%%%%%%%%%%%%%%%%%%%%%%%%%%%%%%%
\begin{figure}[t!]
\begin{center}
\includegraphics[width=0.7\textwidth]{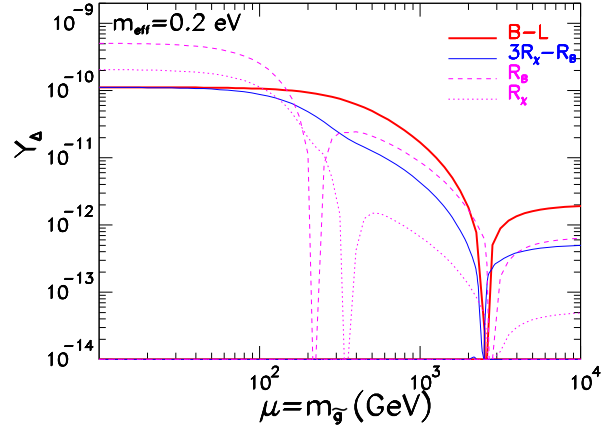}
\end{center}
\caption{Final values of the charge density asymmetries as a function
  of $m_{\widetilde g}=\muh$ for Case I ($h_{e,d}$ Yukawa equilibrium)
  and $m_{\rm eff}=0.20\,$eV.  Thick red line: $Y_{\Delta_{B-L}}$;
  thick blue line: $3Y_{\Delta R_\chi}-Y_{\Delta R_B}$; thin dashed
  purple line: $Y_{\Delta {R_B}}$; thin dotted purple line: $Y_{\Delta
    {R_\chi}}$.  }
\label{fig:rcharges}
\end{figure}
%%%%%%%%%%%%%%%%%%%%%%%%%%%%%%%%%%%%%%%%%%%%%%%%%%%%%%%%%%%%%%%
%

The BE~\eqref{eq:DeltaRBSE}-\eqref{eq:DeltaRchiSE}  are general enough 
to allow to study what would happen if only one of the two anomalous
symmetries $U(1)_R$ or $U(1)_{PQ}$ were present. The corresponding
results are also depicted in Fig.~\ref{fig:mg-eta}.  The blue dashed
line corresponds to the $U(1)_R$-theory where $m_{\widetilde g}$ is
varied while $U(1)_{PQ}$ is broken.\footnote{Note that since $\muh$
  breaks both symmetries, the case of the $U(1)_R$-theory is somewhat
  academic. We include it to put in evidence the fundamental role of
  $U(1)_R$ in enhancing the baryon asymmetry.}  The green dotted line
corresponds to the alternative $U(1)_{PQ}$-theory in which
$m_{\widetilde g} \to \infty$ and only $\muh$ is varied.
These results clearly show that the real responsible of the large
effects is the $R$-symmetry, while the effects of the $PQ$ symmetry
remains qualitatively more at the level of typical spectator effects.
A theoretical justification of this behavior is not difficult to find,
and we will discuss it in the following  section.

Some important aspects of the transition from $R$-genesis (NSE regime)
to leptogenesis (SE regime) are highlighted in
Fig.~\ref{fig:rcharges} which displays the final value of the
relevant charge density asymmetries as a function of $m_{\widetilde
g}=\muh$, assuming Case~I and $m_{\rm eff}=0.20\,$eV.  The thick solid
red line corresponds to $Y_{\Delta_{B-L}}$, while the thin solid blue
line corresponds to $3Y_{\Delta R_\chi}-Y_{\Delta R_B}$.  The thin
dashed and dotted purple lines display respectively $Y_{\Delta {R_B}}$
and $Y_{\Delta {R_\chi}}$.  We see that up to $m_{\widetilde
g}=\muh\sim 100\,$GeV we have $Y_{\Delta_{B-L}}\simeq 3Y_{\Delta
R_\chi}-Y_{\Delta R_B}$ that is in agreement
with~\Eqn{eq:DeltaB-LnoS}, and thus implies that baryogenesis occurs
almost only via $R$-genesis.  As the soft SUSY-breaking parameters are
increased, SE reactions begin to wash out efficiently $Y_{\Delta
{R_B}}$ and $Y_{\Delta {R_\chi}}$ but the difference $3Y_{\Delta
{R_\chi}}-Y_{\Delta {R_B}}$ still remains of the order of
$Y_{\Delta_{B-L}}$, and $R$-genesis still gives the dominant
contribution to baryogenesis.
Around $m_{\widetilde g}=\muh\sim 3\,$TeV all the charge asymmetries
change simultaneously their sign. This is the benchmark of the onset
of the regime in which leptogenesis dominates.  The only relevant
source for generating the density-asymmetries is now the
(opposite-sign) thermally induced $B-L$ asymmetry, that is not
affected by SE washouts, and that is feeding (small) asymmetries into
all the other charges.  In this regime $Y_{\Delta_{R_B}}$ and
$Y_{\Delta_{R_\chi}}$ do not have anymore an independent dynamics, and
can be simply computed in terms of $Y_{\Delta_{B-L}}$ yielding
$Y_{\Delta {R_B}}=-\frac{1}{3}Y_{\Delta_{B-L}}$ and $Y_{\Delta
{R_\chi}}=-\frac{3}{79} \,Y_{\Delta_{B-L}}$.

%%%%%%%%%%%%%%%%%%%%%%%%%%%%%%%%%%%%%%%%%%%%%%%%%%%%%%%%%%

\subsection{Discussion}
\label{sec:nonseq_conclusions}
In the temperature regime quantified by \Eqn{eq:Tgmu} 
all reactions that depend on the
soft gaugino masses do not occur. In this regime 
the early Universe effective theory
includes a new $R$-symmetry. In SL, this $R$-symmetry
is violated in the out of equilibrium interactions of RHN and
RHSN. In particular, $R$-number CP asymmetries in heavy
RHSN decays can be defined, and constitute  important
quantities.  In fact, given that $R$-symmetries do not commute with
SUSY transformations, it is hardly surprising that 
no cancellation occurs between the $R$-number CP asymmetries for scalars
and fermions.  For this reason, a sizable density asymmetry for the
$R$ charge can develop in the thermal bath, and this asymmetry turns
out to play the main role for the generation of the baryon
asymmetry.

To keep higgsinos sufficiently light, in SUSY one needs to
assume $\muh \sim m_{\tilde g}$, and thus when the gaugino masses are
set to zero, one must set $\muh \to 0$ as well.  In this limit the
effective theory acquires another quasi-conserved global symmetry,
that is a $U(1)_{PQ}$ symmetry of the Peccei-Quinn type. $PQ$ is also
violated in RHSN interactions and thus it also has an associated
CP asymmetry.  However, since $U(1)_{PQ}$ is a bosonic symmetry that
commutes with SUSY, the same cancellation between
fermion/boson $CP$ asymmetries occurring for lepton number also occurs
for $PQ$. Accordingly, $PQ$ does not play an equivalently important
role in the generation of the baryon asymmetry.

In order to make more understandable the previous two remarks, let us
start from the beginning, by listing the relevant global symmetries of
the effective theory. For simplicity we concentrate on Case~I
($h_{e,d}$ Yukawa equilibrium).  Neglecting lepton flavour, that is
irrelevant for the present discussion, these symmetries are:
$L,\,R,\,PQ,\,B$ and $\raise 2pt\hbox{$\chi$}_{u^c_L}$.  The first
three $L,\,R,\,PQ$ are violated perturbatively in the interactions of
the RHSN, and all five symmetries are violated by
non-perturbative sphaleron processes. In this review, in carrying out
our analysis, we have first identified the anomaly free combinations
of the five charges, that are $B-L$, $R_B$ and $R_\chi$, and then we
have written down the BE to describe their evolutions. Here, we want to
sketch a different procedure. We first write  a set of evolution
equations for the five anomalous charges, that  have the form:
\begin{equation}
  \label{eq:calQ}
  \dot Y_{\Delta_{\cal Q}} = {\cal S}_{\Delta_{\cal Q}} +
 {\cal G}_{\Delta_{\cal Q}} + {\cal G}^{NP}_{\Delta_{\cal Q}}\,.
\end{equation}
In this equation ${\cal S}$ represents the source term for $
Y_{\Delta}$, ${\cal G}$ is the (s)neutrino-related washouts with all
density-asymmetries and signs absorbed, and ${\cal G}^{NP}$ represents
the non-perturbative EW and/or QCD sphaleron reactions that violate
$\Delta_{\cal Q}$.  The latter are reactions of type (i) discussed in
the introduction in Section~\ref{sec:effective} and in this Section
in~\ref{sec:chem}, that is fast processes, that eventually will be
convenient to eliminate in favour of chemical equilibrium conditions.
Now, given that $B$ and $\raise 2pt\hbox{$\chi$}_{u^c_L}$ are good
symmetries at the perturbative level, they have no CP-violating source
terms: ${\cal S}_{\Delta_{B}},\, {\cal S}_{\Delta_{\chi}}=0$ (they
also do not have perturbative washouts, and ${\cal G}_{\Delta_{B}}$
${\cal G}_{\Delta_{\chi}}=0$ too). The only source terms thus are
${\cal S}_{\Delta_{L}},\, {\cal S}_{\Delta_{PQ}}$ and ${\cal
  S}_{\Delta_{R}}$. However, as we already know, in the $T\to 0$
limit, for ${\cal S}_{\Delta_{L}}$ we have a cancellation between the
fermion and scalar contributions: ${\cal S}^f_{\Delta_{L}}+ {\cal
  S}^s_{\Delta_{L}} \to 0$.  This straightforwardly implies that
${\cal S}^f_{\Delta_{PQ}}+ {\cal S}^s_{\Delta_{PQ}} \to 0$ too, since
the RHSN processes contributing to the CP asymmetry for $PQ$ are
the same as for $L$: they are simply multiplied by the appropriate
$PQ$ charge that is, however, the same for fermion and scalar final
states. For the $R$ charge we have instead ${\cal S}_{\Delta_{R}}
\propto R_f\cdot {\cal S}^f_{\Delta_{L}}+ R_s\cdot {\cal
  S}^s_{\Delta_{L}}$, where $R_{f,s}$ are respectively the overall
$R$-charges of the fermion and boson {\it two particle} final states,
and thus satisfy $R_{s}=R_f+2$.  We then straightforwardly obtain that
in the $T\to 0$ limit the $R$-charge source term does not vanish, and
is given by ${\cal S}_{\Delta_{R}}\to 2\, {\cal S}^s_{\Delta_{L}}$.
Fast in-equilibrium sphaleron processes enforce equilibrium conditions
between particle densities carrying $R$ charge, and those carrying a
$B$ and $L$ numbers and, as a result, baryon and lepton
asymmetries roughly of the same order than the $R$ charge-asymmetry
develop.  Eventually, with the decreasing of the temperature, gaugino
mass related reactions will start occurring with in-equilibrium rates
erasing any asymmetry in the $R$ charge.  It is important to notice
that when the $R$-symmetry gets explicitly broken, generalized EW
sphalerons reduce to the standard EW sphalerons, and sphaleron induced
multi-fermion operators decouple from gauginos, and reduce to their
standard $B+L$ violating form.\footnote{Here we concentrate on the
  role and fate of the $R$-symmetry.  However, given that eventually
  also the $PQ$ symmetry gets explicitly broken, higgsinos decouple
  from sphalerons as well.}  Since gaugino mass reactions as well as
all other MSSM processes conserve $B-L$, the asymmetry initially
generated through $R$-genesis will remain unaffected.

Now that we have identified where the large density asymmetries come
from, we can complete our  procedure by constructing
suitable linear combinations of the five equations \eqref{eq:calQ} for
which the sphaleron terms ${\cal G}^{NP}$ cancel out. Since there are
only two such terms, ${\cal G}^{NP}_{EW}$ and ${\cal G}^{NP}_{QCD}$,
we can construct three linear combinations in which only processes of
type (iii) enter.  These are the BE for the three anomaly
free charges that have been discussed at length in
Section~\ref{sec:symmetries}.  The equilibrium conditions enforced by
${\cal G}^{NP}_{EW}$ and ${\cal G}^{NP}_{QCD}$ have to be imposed on
the system, and to obtain the BE in closed form, the various
density-asymmetries appearing in the washout terms ${\cal G}$ must be
rotated into the densities of the anomaly free charges by means of an
appropriate $A$ matrix.

%% file: conclusions.tex
\section{Soft Leptogenesis Testability and Variations}
%{Variations and experimental verifications}
\label{sec:verifications}

As is well-known, leptogenesis models are plagued with the undesirable
feature that their experimental verification is very difficult, and in
minimal scenarios it appears to be impossible, at least in the
light of foreseeable experimental tests.  This is because, to
establish leptogenesis experimentally, we would need to produce the
heavy states responsible for the generation of the lepton asymmetry,
and measure the CP asymmetry in their decays.  With the dawn of the
LHC, this issue has been more pressing than
ever, since new states with masses of the order of the TeV could
become for the first time accessible.  However, in the most natural
scenarios, the new states relevant for leptogenesis lie at a scale
that is several orders of magnitude above the TeV. With some severe
fine tuning, masses light enough to fall within the energy range
accessible at the LHC could be accommodated. However, in this case to
keep the light neutrino mass scale within the experimental limits the
Yukawa couplings of the heavy states must be extremely tiny,
preventing again any possibility of direct production.  The
possibility of indirect verifications of leptogenesis, for example by
pinning down the whole set of the eighteen parameters of the seesaw, 
and from this deriving a prediction for the baryon asymmetry, is also
not viable. This is because only half of the seesaw parameters are (in
principle) accessible at low energy, while the values of some of the
remaining (unmeasurable) high energy parameters are crucial for
leptogenesis predictions.

As regards SL, it is clear that the discovery of SUSY at the
LHC can be regarded as a basic condition to keep considering this
scenario as a possible explanation of the cosmic baryon asymmetry.
However, in spite of an energy scale that is intrinsically much lower
than that of standard leptogenesis, and that could even fall within
the range of energies accessible at the LHC, with respect to the
possibility of direct experimental verifications SL is in no better
shape than standard (or supersymmetric) leptogenesis.  Even if the
RHSN mass is low enough, direct production of RHSN faces the same
no-go issue of extremely suppressed couplings.  Indirect evidences
could in principle come from measurements unrelated to the neutrino
sector because, as it has been thoroughly discussed in
Section~\ref{sec:SL}, SL depends also on soft SUSY-breaking
parameters that are not directly related with the seesaw.  These
parameters could in principle be measurable through their effects on
other low energy observables like LFV lepton decays, or the electric
dipole moments of the charged leptons, that receive contributions from
the complex phases of the soft SUSY-breaking terms.  However,
Ref.~\refcite{Kashti:2004vj} that addressed specifically this issue,
found that all the related effects are much smaller than other MSSM
contributions, and therefore unobservable.  One can thus conclude
that, much alike standard leptogenesis and supersymmetric
leptogenesis, the simplest SL scenario based on the supersymmetric
type I seesaw, and on the related soft SUSY-breaking terms,
also escapes the possibility of experimental verification.

\subsection{Variations of soft leptogenesis}
\label{sec:variations}

The exceedingly strong suppression of the production rates of
relatively light RHSN states is clearly the direct consequence of
their gauge-singlet nature, that leaves the tiny Yukawa interactions
as the only mechanism for their production.
%\footnote{The main difficulties that one has to face in constructing
%  models of leptogenesis at TeV scale were discussed in length in
%  Ref.~\refcite{Hambye:2001eu}.}
In order to obviate this problem one can assume that these states
are non-singlet under $SU(2)_L$, or under some new gauge symmetry, so
that they could be produced at colliders through the corresponding
gauge interactions.  However, in following this approach, one has to
be very careful because fast gauge scatterings could potentially keep
the RHSN in complete thermal equilibrium, and/or RHSN annihilation
through gauge boson channels could leave a too small fraction of
out-of-equilibrium decaying states.

In Refs.~\refcite{D'Ambrosio:2004fz,Chun:2005ms} the MSSM with the
addition of an $SU(2)_L$ triplet of scalars with non-zero hypercharge
(as required for the type II seesaw) was considered, and the
possibility of SL at a low scale was investigated.  Fast annihilations
of the scalar triplets through gauge interactions keep the triplet
abundance very close to its equilibrium value, and strongly suppress
the final lepton asymmetry.  However, it is argued that for a triplet
mass scale $10^3-10^4\,{\rm GeV}$ leptogenesis could still be
successful.\cite{Chun:2005ms} Due to the small neutrino Yukawa
coupling $\sim 10^{-6}$ all low-energy LFV processes like $\mu \to e
\gamma$ or $\mu \to 3e$ remain strongly
suppressed.\cite{Chun:2003ej,Akeroyd:2009nu} On the other hand,
Tevatron and LHC have the potential to produce these triplets, and a
marked signature will be the decay of the doubly charged component of
the triplet to lepton and Higgsino
pairs.\cite{Akeroyd:2005gt,Akeroyd:2011ir}

A supersymmetric seesaw model associated with an extra $U(1)'$ gauge
symmetry spontaneously broken at the TeV scale has been studied in
Ref.~\refcite{Chun:2005tg}, and shown to be a viable option for the
generation of the cosmological baryon asymmetry via the SL mechanism.
Such a scenario leads to testable predictions in colliders, through
the production of $Z'$ and their subsequent decays into RHSN. The RHSN
will further decay to final states with pairs of same-sign leptons
(sleptons) and  charginos (charged Higgses) through $\widetilde
N-\widetilde N^*$ mixing and CP violation.

SL in the inverse seesaw model was considered in
Ref.~\refcite{Garayoa:2006xs}. This is another interesting possibility
with the potential of being verified experimentally.  In this scenario
the lightness of the active neutrinos is not associated with tiny
neutrino Yukawa couplings, but with a small dimensional parameter that
breaks $U(1)_L$.  The unsuppressed Yukawa couplings together with a
low mass scale for the RHSN can result in a relatively large mixing
with the SM leptons, and through this mixing direct production and
detection of the RHSN at the LHC become possible.

Ref.~\refcite{Medina:2006hi} put forth the more speculative idea of
implementing SL in a warped five dimensional scenario. It was shown
that, within the context of extra dimensions, the condition of
out-of-equilibrium decay and the phenomenological constraints on the
neutrino mass can be both satisfied in a natural way, and that all
necessary elements needed for SL to predict a correct value for the
baryon asymmetry can be obtained.  While the specific 
SL mechanism of this model does not seem to be easily
verifiable experimentally, the general idea of extra dimensions could
potentially be probed at LHC through the production of the
Kaluza-Klein excitations. An experimental confirmation of this
scenario would certainly increase the phenomenological interest of SL
in the context of extra dimensions.

Other interesting alternative models which utilize soft-SUSY
breaking terms to realize leptogenesis at a low scale have been
considered in Refs.~\refcite{Allahverdi:2004ix,Boubekeur:2004ez,%
Ellis:2005uk}. It remains to be seen if any of these models can
yield some clear experimental signature.

\section{Final Remarks and Conclusions}
\label{sec:conclusions}

The matter-antimatter asymmetry of the Universe and the experimental
confirmation of tiny but non-vanishing neutrino masses are two among
the very few evidences of physics beyond the Standard Model.  The type
I seesaw can elegantly explain the strong suppression of the neutrino
mass scale, and through the leptogenesis mechanism can also provide a
natural explanation of the cosmic matter-antimatter asymmetry.
Leptogenesis can be quantitatively successful without any fine-tuning
of the seesaw parameters, yet, in the non-supersymmetric seesaw
framework, a fine-tuning problem arises due to the large corrections
to the mass-squared parameter of the Higgs potential, that are
proportional to the heavy Majorana neutrino masses.  The
supersymmetric version of the seesaw has the virtue of stabilizing the
Higgs mass-squared parameter under radiative corrections, but at the 
same time also introduces a serious tension between the lower limit on 
the seesaw scale that follows from the requirement of successful 
baryogenesis, and the upper bound on the reheating temperature that 
must be satisfied to avoid an overproduction of gravitinos.

However, supersymmetry (SUSY) has to be broken, and this yields the
possibility that leptogenesis could proceed through right-handed sneutrino decays,
thanks to the new sources of CP violation from complex phases in the
SUSY-breaking sector.  In this scenario the leptogenesis
scale is naturally lowered and successful baryogenesis can be obtained
anywhere in the temperature range $10^4\,{\rm GeV} \lesssim T \lesssim
10^9\,{\rm GeV}$.  Accordingly, the tension with the gravitino problem
gets generically relaxed and, in the lower temperature window, is
completely avoided.  This scenario, termed soft
leptogenesis\cite{Grossman:2003,DAmbrosio:2003} (SL) has been the
subject of this review.

As discussed in Section~\ref{sec:SL}, SL is plagued by
the problem of a congenital low efficiency, that is related to the
cancellation between the asymmetries produced in fermions and bosons
carrying lepton number.  It should be stressed that it is the fact
that lepton number commutes with supersymmetric transformations (that
is that scalar and fermionic members of the lepton supermultiplets
have the same lepton charge) that plays the crucial role in enforcing
this cancellation.  Finite temperature corrections break SUSY
spoiling the cancellation between the scalar and fermionic CP
asymmetries, and  can eventually rescue SL from a
complete failure.

The basic mechanism of SL was reviewed in
Section~\ref{sec:unflavored}.  To highlight the role of thermal
factors and of the different types of CP asymmetries, in this section
the simplifying assumptions of a single lepton flavour and of equal
density asymmetries for the lepton and slepton states
(superequilibration) were adopted.

When the SL CP asymmetries are dominated by the contribution from
mixing, the RHSN have to be highly degenerated, and in this situation
quantum effects can become important. These issues have been reviewed
in Section~\ref{sec:quantum_role}.  In the strong washout regime
quantum effects can enhance the absolute value of the final asymmetry
and also induce a change of its sign. However, altogether, the allowed
parameter space for SL to be successful is not modified substantially.

Issues related with lepton flavour effects have been addressed in
Section~\ref{sec:flavor}.  Given that SL can only occur at
temperatures low enough that all the three lepton flavours are
resolved by their fast charged lepton Yukawa interactions, the
inclusion of flavour effects in SL studies is mandatory.  We have seen
that enhancements of the produced asymmetry up to factors $\sim 10^3$
are possible when flavour effects are accounted for, and this is
sufficient to avoid the need of additional enhancements from resonant
conditions. Thus, the RHSN do not need to be highly degenerated, and a
natural scale for the sneutrino mixing parameter $B \sim m_{SUSY}$ is
allowed.

In Section~\ref{sec:nse} we discussed the recently discovered
possibility that baryogenesis could proceed through
$R$-genesis.\cite{Fong:2010bv} If the RHSN mass lies above $M\sim
10^7\,$GeV, a new scenario different from SL must be
considered.  In this scenario the asymmetry is first generated in a
new $R$ charge that is conserved in the effective theory when all
gaugino-related reactions can be neglected, which is the case when
$T\gsim 10^7\,$GeV.  This asymmetry is then transferred in part to
baryons via generalized electroweak sphaleron interactions. Given that
$R$-symmetries do not commute with SUSY transformations, the
scalar and fermionic members of the lepton supermultiplets have
different $R$-charges, and no cancellation between the $R$-asymmetries
produced in fermions and bosons occurs. Thus, in the high temperature
window where SL is replaced by $R$-genesis, a sizable
baryon asymmetry can be generated regardless of thermal effects.

In conclusion, SUSY allows for baryogenesis via leptogenesis
to occur at any temperature from somewhat below the GUT scale down to
the TeV.  Above $T \sim 10^9\,{\rm GeV}$ baryogenesis can occur
through the supersymmetric version of standard leptogenesis, although
this possibility is disfavoured by considerations related to gravitino
overproduction, whose decays would affect Big Bang nucleosynthesis.
In the intermediate temperature range $10^7\,{\rm GeV} \lesssim T
\lesssim 10^9\,{\rm GeV}$, where the gravitino constraint gets
relaxed, baryogenesis can occur through $R$-genesis, which is a highly
efficient mechanism in which the production of asymmetries does not
rely on thermal effects.  In the lower temperature window $T \lesssim
10^7\,{\rm GeV}$ where the gravitino problem is usually evaded, the usual SL
mechanism can take place, with an efficiency that is suppressed by the
scalar-boson asymmetries cancellation, but with possible large
enhancements from flavour effects. Finally, carefully constructed
variations of SL can allow for a scale as low as the
TeV and, as was briefly reviewed in Section~\ref{sec:variations}, in
some of these cases experimental verifications are also possible.

%% file: boltzmanneqs.tex
\section{Boltzmann Equations} 
\label{app:boltzmanneqs_new}

\subsection{General Boltzmann equations}\label{sec:gen_BE}
Our Universe is very well described by a spatially homogeneous and isotropic
metric known as Robertson-Walker (RW) metric
\be
ds^2 = dt^2 - R^2(t)\left[\frac{dr^2}{1-kr^2} + r^2d\theta^2
+r^2 \sin^2\theta d\phi^2\right],
\label{eq:RW_metric}
\ee
where $(t,r,\theta,\phi)$ are comoving coordinates, $R(t)$ is the 
cosmic scale factor, and $k=0,+1,-1$ describe spaces of zero, positive, 
or negative spatial curvature, respectively. 
For a general process $a+b+...\leftrightarrow i+j+....$ in the RW space, 
the Boltzmann equation (BE) for the phase-space distribution of the particle
species $a$ can be written as:
\begin{eqnarray}
\frac{\partial f_{a}}{\partial t}
-H\,|\vec p_a|\,\frac{\partial f_{a}}{\partial |\vec p_a|} & = & 
- \frac{1}{2E_a}C[f_a],
\label{eq:general_BE0}
\end{eqnarray}
where
\bea
C[f_a] 
& \equiv & \frac{1}{g_a}\sum_{b,...i,j,...}\Lambda_{b...}^{ij...}\left[\left|\mathcal{M}
\left(ab...\rightarrow ij...\right)\right|^{2}f_{a}f_{b}...
\left(1+\eta_{i}f_{i}\right)\left(1+\eta_{j}f_{j}\right)...\right.\nonumber \\
 &  & \left.-\left|\mathcal{M}\left(ij...\rightarrow ab...\right)
\right|^{2}f_{i}f_{j}...\left(1+\eta_{a}f_{a}\right)
\left(1+\eta_{b}f_{b}\right)...\right].
\label{eq:collision}
\eea
In the above $g_a$ is the number of spin degrees of freedom of particle $a$ and
\bea
\Lambda_{b...}^{ij...} \! & \equiv & \!\! \int d\Pi_{b}...
d\Pi_{i}d\Pi_{j}...\left(2\pi\right)^{4}\delta^{\left(4\right)}
\left(p_{a}+p_{b}+...-p_{i}-p_{j}-...\right), \nonumber \\
d\Pi_{x} & \equiv & \frac{d^{3}p_{x}}{\left(2\pi\right)^{3}2E_{x}}.
\label{eq:gen_def0}
\eea
In Eq.~\eqref{eq:collision}, $\left|\mathcal{M}_{ab...\rightarrow ij...}\right|^{2}$
is the squared amplitude summed over initial and final quantum numbers
%!!% 
(spin states and gauge multiplicity) and 
$f_x$ is the distribution function of $x$ with 
$\eta_{x}=\pm$ if $x$ is a boson or fermion respectively. The 
factors $(1\pm f_x)$ are known as Pauli-blocking (for $x$ being fermion)
and Bose-enhancement or stimulated emission 
(for $x$ being boson) factors, respectively. 
In Eq.~\ref{eq:general_BE0}, the Hubble expansion rate of the Universe $H$
in the radiation-dominated era is given by
\begin{eqnarray}
H & \equiv & \frac{\dot R}{R}
=\frac{2}{3}\sqrt{\frac{g_{*}\pi^{3}}{5}}\frac{T^{2}}{M_{pl}},
\end{eqnarray}
where $M_{pl}=1.22 \times 10^{19}\,$GeV is the Planck mass, $g_*$ is
the total number of relativistic degrees of freedom ($g_*=228.75$ for
MSSM).

Using the definition of the number density in terms of the phase space
distribution
\be
n_a = g_a \int \frac{d^3 p}{(2\pi)^3} f_a,
\label{eq:num_den}
\ee
and upon integration by parts, the BE \eqref{eq:general_BE0} can be 
rewritten in the form
\begin{eqnarray}
\frac{dn_{a}}{dt}+3Hn_{a} & = & -\sum_{b,...i,j,...}
\left[ab...\leftrightarrow ij...\right],
\label{eq:general_BE}
\end{eqnarray}
where
\begin{eqnarray}
\left[ab...\leftrightarrow ij...\right] 
& \equiv & \Lambda_{ab...}^{ij...}\left[\left|\mathcal{M}
\left(ab...\rightarrow ij...\right)\right|^{2}f_{a}f_{b}...
\left(1+\eta_{i}f_{i}\right)\left(1+\eta_{j}f_{j}\right)...\right. 
\nonumber \\
 &  & \hspace{-5mm}\left.-\left|\mathcal{M}\left(ij...\rightarrow ab...\right)
\right|^{2}f_{i}f_{j}...\left(1+\eta_{a}f_{a}\right)
\left(1+\eta_{b}f_{b}\right)...\right].
\label{eq:gen_def}
\end{eqnarray}
Notice that in writing the BE \eqref{eq:general_BE}, we have
implicitly assumed that the right hand side of this equation can also
be written in term of $n_a$.  However, without certain approximations,
this cannot be done in general and we have to resort to the BE
\eqref{eq:general_BE0} and solve it in term of phase space
distribution.  In \ref{subsec:derivations} we list the 
approximations that  allow us
to write the right hand side of Eq.~\eqref{eq:general_BE} in terms of
number densities, and then to use the BE \eqref{eq:general_BE}.

In order to scale out the effect of the expansion of the Universe.
one defines the particle \emph{abundance} i.e. the particle density
$n_a$ normalized to the entropy density $s$ as:
\be
Y_a \equiv \frac{n_a}{s},
\label{eq:abundance}
\ee
where the entropy density in the radiation dominated era is given by
\be
s=\frac{2\pi^{2}}{45}g_{*}T^{3}.
\label{eq:entropy_density}
\ee
Using the conservation of entropy per comoving volume 
(i.e. $s\,R^3 = $ constant), 
replacing the time $t$
with the temperature $T$ (in the radiation dominated era 
$t=\frac{1}{2H}\sim T^{-2}$) and defining the 
dimensionless parameter 
\be
z \equiv \frac{M}{T},
\label{eq:z_MT}
\ee
where $M$ is any convenient mass scale, the left hand side of 
Eq.~\eqref{eq:general_BE} becomes:
\be
\frac{dn_{a}}{dt}+3Hn_{a} = 
s\frac{dY_{a}}{dt}  =  sHz\frac{dY_{a}}{dz}.
\ee
Regarding the distribution functions in Eqs.~\eqref{eq:gen_def}, for
particles for which the elastic scatterings are much faster than the
inelastic scatterings, one can assume that they are in kinetic
equilibrium and have either Fermi-Dirac distribution (for fermions $f$) or
Bose-Einstein distribution (for scalars $s$) given respectively by
\begin{equation}
f_{f,\overline{f}}  =  \frac{1}{e^{\left(E_{f}\mp\mu_{f}\right)/T}+1},
\;\;\;\;\;
f_{s,s^{*}}  =  \frac{1}{e^{\left(E_{s}\mp\mu_{s}\right)/T}-1},
\label{eq:FD_BE_dist}
\end{equation}
where $\mu$'s are the chemical potentials and the ``bar'' or ``star''
refers to the corresponding antiparticles. 
The equilibrium distributions $f^{eq}_x$ are defined as those with  $\mu=0$:
\begin{equation}
f_{f,\overline{f}}^{eq}  =  \frac{1}{e^{E_{f}/T}+1}, \;\;\;\;\;
f_{s,s^{*}}^{eq}  =  \frac{1}{e^{E_{s}/T}-1}.
\label{eq:FD_BE_dist_eq}
\end{equation}
So for $\frac{\mu}{T}\ll1$ one can expand the kinetic equilibrium
distribution function in $\frac{\mu}{T}$ as
\begin{eqnarray}
f_{f,\overline{f}} & = & f_{f}^{eq}\pm 
f_{f}^{eq,2}e^{E_{f}/T}\frac{\mu_{f}}{T}
+\mathcal{O}\left[\left(\frac{\mu_{f}}{T}\right)^{2}\right],\nonumber \\
f_{s,s^{*}} & = & f_{s}^{eq}\pm f_{s}^{eq,2}e^{E_{s}/T}
\frac{\mu_{s}}{T}+\mathcal{O}\left[\left(\frac{\mu_{s}}{T}\right)^{2}\right].
\label{eq:chem_pot_expand}
\end{eqnarray}
It follows that
\begin{eqnarray}
f_{f}-f_{\overline{f}} 
 & = & 2\left(1-f_{f}^{eq}\right)f_{f}^{eq}\frac{\mu_{f}}{T}
+\mathcal{O}\left[\left(\frac{\mu_{f}}{T}\right)^{3}\right],
\nonumber \\
f_{s}-f_{s^*} 
 & = & 2\left(1+f_{s}^{eq}\right)f_{s}^{eq}\frac{\mu_{s}}{T}+
\mathcal{O}\left[\left(\frac{\mu_{s}}{T}\right)^{3}\right],
\label{eq:diff_dist_chem_pot}
\end{eqnarray}
where we have used the identities
\begin{equation}
1-f_{f,\overline{f}} = e^{(E_{f}\mp\mu_{f})/T}f_{f,\overline{f}},
\;\;\;\;\;
1+f_{s,s^{*}} = e^{(E_{s}\mp\mu_{s})/T}f_{s,s^{*}}.
\label{eq:dist_identities}
\end{equation}
Using Eq.~\eqref{eq:num_den} one gets that the difference
between number densities of massless particles and antiparticles 
at leading order in chemical potentials is
\begin{equation}
n_{\Delta f}  \equiv  n_{f}-n_{\overline{f}}
=\frac{g_{f}}{6}T^{3}\frac{\mu_{f}}{T},
\;\;\;\;\;
n_{\Delta s}  \equiv  n_{s}-n_{\bar{s}} 
=\frac{g_{s}}{3}T^{3}\frac{\mu_{s}}{T}.
\end{equation}
Defining the {\sl density asymmetries} 
per degree of freedom
$Y_{\Delta f,s} \equiv n_{\Delta f,s}/(g_{f,s}s)$ as the the number
density asymmetries per degree of freedom 
$n_{\Delta f,s}/g_{f,s}$ normalized to the entropy density
$s$, we can rewrite the chemical potentials for massless fermions and
bosons as:
\begin{equation}
2\frac{\mu_{f}}{T}  =  
%\frac{9\zeta(3)}{2\pi^{2}}\frac{Y_{\Delta f}}{Y_{f}^{eq}}=
\frac{8\pi^{2}g_{*}}{15}
Y_{\Delta f}\equiv \frac{Y_{\Delta f}}{Y_{f}^{eq}},
\;\;\;\;\;
2\frac{\mu_{s}}{T}  =  
%\frac{3\zeta(3)}{\pi^{2}}\frac{Y_{\Delta s}}{Y_{s}^{eq}}=
\frac{4\pi^{2}g_{*}}{15}
Y_{\Delta s}\equiv\frac{Y_{\Delta s}}{Y_{s}^{eq}},
\label{eq:chem_pot_asym}
\end{equation}
where $Y_{f}^{eq}\equiv\frac{15}{8\pi^{2}g_{*}}$ and  
$Y_{s}^{eq}\equiv\frac{15}{4\pi^{2}g_{*}}$.

Let us introduce the following shorthand notation: 
\begin{eqnarray}
F_{ab...ij...}\left(...\right) & \equiv & \Lambda_{ab...}^{ij...}
\left|\mathcal{M}\left(ab...\rightarrow ij...\right)\right|^{2}
\left(...\right),\nonumber \\
\overline{F_{ab...ij...}}\left(...\right) & \equiv & 
\Lambda_{ab...}^{ij...}\left|\mathcal{M}
\left(ij...\rightarrow ab...\right)\right|^{2}\left(...\right), 
\label{eq:def_Fabij}
\end{eqnarray}
where $(...)$ denotes some function to be integrated over. 
Note that CPT invariance implies 
$\overline{F_{ab...ij...}}\left(...\right)
=F_{\overline{ab...ij...}}\left(...\right)$.
The thermally averaged reaction densities can be defined as 
\bea
\gamma\left(ab...\rightarrow ij...\right) 
\!& \equiv & \!
F_{ab...ij...}  
\times f_{a}^{eq} f_{b}^{eq}...
\left(1+\eta_{i}f_{i}^{eq}\right)\left(1+\eta_{j}f_{j}^{eq}\right)...,\qquad 
\label{eq:therm_ave_rate}
\eea
where we have used the equilibrium distribution functions with 
vanishing chemical potentials Eqs.~\eqref{eq:FD_BE_dist_eq}.

Neglecting  Pauli-blocking and Bose-enhancement factors 
and assuming that all the particles follow Maxwell-Boltzmann 
distribution $f=e^{-E/T}$, 
and that $\left|\mathcal{M}\left(ab...\rightarrow ij...\right)\right|^{2}$
does not depend on the relative motion of particles with respect 
to the plasma, 
Eq.~\eqref{eq:therm_ave_rate} for the decay $N \to ij...$ reduces to 
\be
\gamma(a \to ij...) = \gamma(ij... \to a) 
= n_a^{eq}\frac{\mathcal{K}_1(z)}{\mathcal{K}_2(z)}\Gamma_a,
\label{eq:decay_rea}
\ee
where $\Gamma_a$ is the decay width in the rest frame of $a$,
$\mathcal{K}_q$ is the modified Bessel function of the second 
kind of order $q$,
and $n_a^{eq}$ is the equilibrium number density of $a$:
\be
n_a^{eq} = g_a\int \frac{d^3p_a}{(2\pi)^3} e^{-E_a/T} = \frac{g_a T^3}{\pi^2}.
\ee
For a two-body scattering $ab\to ij$, 
Eq.~\eqref{eq:therm_ave_rate} reduces to 
\be
\gamma(ab \to ij) = \frac{T}{64\pi^4}\int^\infty_{s_{\rm min}}
ds \,\sqrt{s} \,\hat\sigma(s)\, \mathcal{K}_1\left(\frac{\sqrt{s}}{T}\right),
\label{eq:scatt_rea}
\ee
where $s$ is the center of mass energy squared, 
$s_{\rm min} = \max[(m_a+m_b)^2,(m_i+m_j)^2]$ and
$\hat\sigma(s)$ is the \emph{reduced cross section} which is related
to the total cross section $\sigma(s)$ (summing over initial and final 
spin states) by
\bea
\hat\sigma(s) &\equiv& \frac{2 \lambda^2(s,m_a^2,m_b^2)}{2} \sigma(s) 
= \frac{1}{8\pi s}\int^{t+}_{t-} dt \,
\left|\mathcal{M}\left(ab\rightarrow ij\right)\right|^{2},
\label{eq:rcs}
\eea
with 
\bea
\lambda(a,b,c) &\equiv& \sqrt{(a-b-c)^2-4bc}\,, \\ 
t_{\pm} &=& \frac{m_a^2-m_b^2-m_i^2+m_j^2}{4s} 
 \nonumber \\ && \hspace{-14mm}
-\left[\sqrt{\frac{(s+m_a^2-m_b^2)^2}{4s}-m_a^2}\mp
\sqrt{\frac{(s+m_i^2-m_j^2)^2}{4s}-m_i^2}\right]^2.
\eea
Because of the thermal-statistical nature of the CP asymmetry in SL, a
rigorous treatment would require the use of the BE for the particle
distribution functions Eq.~\eqref{eq:general_BE0} rather than the
integrated BE for the number densities. We will describe in the next
section the derivation of the BE for SL and the approximations
required to write them in integrated form, while keeping the relevant
thermal statistical factors.

\subsection{Boltzmann equations for soft leptogenesis}
\label{sec:soft_BE}

%%%%%%%%%%%%%%%%%%%%%%%%%%%%%%%%%%%%%%%%%%%%5

%\subsubsection{Definitions} 
%\label{subsec:def}

\subsubsection{Unflavoured Boltzmann equations} 
\label{subsec:derivations}

In the rest of this section, unless otherwise stated, we will ignore
thermal masses. The BE for the RHN abundance $Y_N$ can be written
down as:
\begin{eqnarray}
{\dot Y}_{N} & = & -\left[N\leftrightarrow
\widetilde{H}_u\widetilde{\ell}\right]_+
-\left[N\leftrightarrow H_u\ell\right]_+
-\left[N\widetilde{\ell}\leftrightarrow Q\widetilde{u}^*\right]_{+}
-\left[N\widetilde{\ell}\leftrightarrow\widetilde{Q}\overline{u}\right]_{+}
\nonumber \\ &  & 
\hspace{-0.0cm}
-\left[N\overline{Q}\leftrightarrow\widetilde{\ell}^{*}\widetilde{u}^*\right]_{+} 
% \nonumber \\ &  & 
-\left[Nu\leftrightarrow\widetilde{\ell}^{*}\widetilde{Q}\right]_{+}
-\left[N\widetilde{u}\leftrightarrow\widetilde{\ell}^{*}Q\right]_{+}
-\left[N\widetilde{Q}^{*}\leftrightarrow\widetilde{\ell}^{*}\overline{u}\right]_{+} 
\nonumber \\ &  & -\left[N\ell\leftrightarrow Q\overline{u}\right]_{+} 
-\left[Nu\leftrightarrow\overline{\ell}Q\right]_{+}
-\left[N\overline{Q}\leftrightarrow\overline{\ell}\overline{u}\right]_{+}
 \hspace{2cm}
\nonumber \\ & = &
2\widetilde{F}_{N}\left(\frac{f_{N}}{f_{N}^{eq}}-
\frac{1-f_{N}}{1-f_{N}^{eq}}\right)+2F_{N}\left(\frac{f_{N}}{f_{N}^{eq}}-
\frac{1-f_{N}}{1-f_{N}^{eq}}\right)
\nonumber \\ &  &  
\hspace{-0.0cm} 
+\left(4F_t^{\left(0\right)}+4F^{\left(1\right)}+4F_t^{\left(2\right)}
+2F_t^{\left(3\right)}+4F_t^{\left(4\right)}\right) 
% \nonumber \\ &  & \times 
\left(\frac{f_{N}}{f_{N}^{eq}}
-\frac{1-f_{N}}{1-f_{N}^{eq}}\right),\quad \label{eq:N_BE}
\end{eqnarray}
where the time derivative is 
defined as $\dot{Y} \equiv sHz\frac{dY}{dz}$,
% \begin{eqnarray}
$ \left[ab\leftrightarrow ij\right]_{+} 
% & \equiv & 
 \equiv  \left[ab\leftrightarrow ij\right]
+\left[\overline{a}\overline{b}\leftrightarrow
\overline{i}\overline{j}\right]$,  
and  
% ,\nonumber \\
$ \left[ab\leftrightarrow ij\right]_{-} 
% & \equiv & 
\equiv  \left[ab\leftrightarrow ij\right]
-\left[\overline{a}\overline{b}\leftrightarrow\overline{i}\overline{j}\right]
$  
% .\end{eqnarray}
and we have further defined the following shorthand notations:
\begin{eqnarray}
\widetilde{F}_{N}\left(...\right) & \equiv & 
F_{N\widetilde{H}_u\widetilde{\ell}}f_{N}^{eq}
\left(1+f_{\widetilde{\ell}}^{eq}\right)
\left(1-f_{\widetilde{H}_u}^{eq}\right)
\left(...\right),
\nonumber \\
F_{N}\left(...\right) & \equiv & F_{NH_u\ell}f_{N}^{eq}
\left(1-f_{\ell}^{eq}\right)\left(1+f_{H_u}^{eq}\right)\left(...\right)\,, 
\nonumber \\
%\end{eqnarray}
%
% \begin{eqnarray}
F_t^{\left(0\right)}\left(...\right) & \equiv & 
F_{N\widetilde{\ell}Q\widetilde{u}^*}f_{N}^{eq}f_{\widetilde{\ell}}^{eq}
\left(1-f_{Q}^{eq}\right)\left(1+f_{\widetilde{u}}^{eq}\right)\left(...\right)
\nonumber \\
& = & F_{N\widetilde{\ell}\widetilde{Q}\overline{u}}f_{N}^{eq}
f_{\widetilde{\ell}}^{eq}\left(1+f_{\widetilde{Q}}^{eq}\right)
\left(1-f_{u}^{eq}\right)\left(...\right),
\nonumber \\
F_t^{\left(1\right)}\left(...\right) & \equiv & 
F_{N\overline{Q}\widetilde{\ell}^{*}\widetilde{u}^*}f_{N}^{eq}f_{Q}^{eq}
\left(1+f_{\widetilde{\ell}}^{eq}\right)\left(1+f_{\widetilde{u}}^{eq}\right)
\left(...\right) \nonumber \\
& = & F_{Nu\widetilde{\ell}^{*}\widetilde{Q}}f_{N}^{eq}f_{u}^{eq}
\left(1+f_{\widetilde{\ell}}^{eq}\right)
\left(1+f_{\widetilde{Q}}^{eq}\right)
\left(...\right),
\nonumber \\
F_t^{\left(2\right)}\left(...\right) & \equiv & 
F_{N\widetilde{u}\widetilde{\ell}^{*}Q}f_{N}^{eq}f_{\widetilde{u}}^{eq}
\left(1+f_{\widetilde{\ell}}^{eq}\right)\left(1-f_{Q}^{eq}\right)
\left(...\right)
\nonumber \\
& = & F_{N\widetilde{Q}^{*}\widetilde{\ell}^{*}\overline{u}}
f_{N}^{eq}f_{\widetilde{Q}}^{eq}\left(1+f_{\widetilde{\ell}}^{eq}\right)
\left(1-f_{u}^{eq}\right)\left(...\right),
\nonumber \\
F_t^{\left(3\right)}\left(...\right) & \equiv & F_{N\ell Q\overline{u}}
f_{N}^{eq}f_{\ell}^{eq}\left(1-f_{Q}^{eq}\right)
\left(1-f_{u}^{eq}\right)\left(...\right),
\nonumber \\
F_t^{\left(4\right)}\left(...\right) & \equiv & F_{Nu\overline{\ell}Q}
f_{N}^{eq}f_{u}^{eq}\left(1-f_{\ell}^{eq}\right)
\left(1-f_{Q}^{eq}\right)\left(...\right)
\nonumber \\
& = & F_{N\overline{Q}\overline{\ell}\overline{u}}f_{N}^{eq}
f_{Q}^{eq}\left(1-f_{\ell}^{eq}\right)
\left(1-f_{u}^{eq}\right)\left(...\right).
\end{eqnarray}
The BE for the RHSN abundances $Y_{\widetilde{N}_{\pm}}$ are:
\begin{eqnarray}
\dot{Y}_{\widetilde{N}_{\pm}} 
& = & - \! \left[\widetilde{N}_{\pm}
\leftrightarrow\widetilde{H}_u\ell\right]_+
- \! \left[\widetilde{N}_{\pm}\leftrightarrow H_u\widetilde{\ell}\right]_+ 
- \! \left[\widetilde{N}_{\pm}\leftrightarrow
\widetilde{\ell}\widetilde{u}\widetilde{Q}^{*}\right]_+
- \! \left[\widetilde{N}_{\pm}\widetilde{\ell}
\leftrightarrow\widetilde{u}^*\widetilde{Q}\right]_+
\nonumber \\ &  & 
- \! \left[\widetilde{N}_{\pm}\widetilde{Q}
\leftrightarrow\widetilde{\ell}\widetilde{u}\right]_+ 
- \! \left[\widetilde{N}_{\pm}\widetilde{u}
\leftrightarrow\widetilde{\ell}^{*}\widetilde{Q}\right]_+
- \! \left[\widetilde{N}_{\pm}\ell
\leftrightarrow Q\widetilde{u}^*\right]_{+} 
- \! \left[\widetilde{N}_{\pm}\ell\leftrightarrow
\widetilde{Q}\overline{u}\right]_{+}
\nonumber \\ &  & 
- \! \left[\widetilde{N}_{\pm}\widetilde{u}
\leftrightarrow\overline{\ell}Q\right]_{+}
- \! \left[\widetilde{N}_{\pm}\widetilde{Q}^{*}
\leftrightarrow\overline{\ell}\overline{u}\right]_{+} 
- \! \left[\widetilde{N}_{\pm}\overline{Q}
\leftrightarrow\overline{\ell}\widetilde{u}^*\right]_{+}
- \! \left[\widetilde{N}_{\pm}u\leftrightarrow
\overline{\ell}\widetilde{Q}\right]_{+}
\nonumber \\
&  & - \! \left[\widetilde{N}_{\pm}\widetilde{\ell}^{*}
\leftrightarrow\overline{Q}u\right]_{+}
-\! \left[\widetilde{N}_{\pm}Q\leftrightarrow\widetilde{\ell}u\right]_{+} 
-\! \left[\widetilde{N}_{\pm}\overline{u}
\leftrightarrow\widetilde{\ell}\overline{Q}\right]_{+} 
 \nonumber \\
& = & -\left(F_{\widetilde{N}_{\pm}}^f+F_{\widetilde{N}_{\pm}}^s+
2F_{\widetilde{N}_{\pm}}^{\left(3\right)}+6F_{22_{\pm}}\right)
\left(\frac{f_{\widetilde{N}_{\pm}}}{f_{\widetilde{N}_{\pm}}^{eq}}-
\frac{1+f_{\widetilde{N}_{\pm}}}{1+f_{\widetilde{N}_{\pm}}^{eq}}\right)
\nonumber \\ &  & 
-2\left(2F_{t\pm}^{\left(5\right)}+2F_{t\pm}^{\left(6\right)}
+2F_{t\pm}^{\left(7\right)}+F_{t\pm}^{\left(8\right)}
+2F_{t\pm}^{\left(9\right)}\right) 
% \nonumber \\ &  & 
 \left(\frac{f_{N_{\pm}}}{f_{N_{\pm}}^{eq}}
-\frac{1+f_{\widetilde{N}_{\pm}}}
{1+f_{\widetilde{N}_{\pm}}^{eq}}\right),
\label{eq:Nplus_BE}
\end{eqnarray}
where terms of order 
$\mathcal{O}\left(\epsilon\frac{\mu}{T}\right)$, 
have been dropped. The shorthand notations are 
\begin{eqnarray}
F_{\widetilde{N}_{\pm}}^f \left(...\right) & \equiv & 
\left(F_{\widetilde{N}_{\pm}\widetilde{H}_u\ell}+
F_{\widetilde{N}_{\pm}\overline{\widetilde{H}_u}\overline{\ell}}\right)
f_{\widetilde{N}_{\pm}}^{eq}\left(1-f_{\ell}^{eq}\right)
\left(1-f_{\widetilde{H}_u}^{eq}\right)\left(...\right),
\nonumber \\
F_{\widetilde{N}_{\pm}}^s \left(...\right) & \equiv & 
\left(F_{\widetilde{N}_{\pm}H_u\widetilde{\ell}}
+F_{\widetilde{N}_{\pm}H_u^{*}
\widetilde{\ell}^{*}}\right)f_{\widetilde{N}_{\pm}}^{eq}
\left(1+f_{\widetilde{\ell}}^{eq}\right)
\left(1+f_{H_u}^{eq}\right)\left(...\right),
\nonumber \\
F_{\widetilde{N}_{\pm}}^{\left(3\right)}\left(...\right) 
& \equiv & F_{\widetilde{N}_{\pm}\widetilde{\ell}\widetilde{u}
\widetilde{Q}^{*}}f_{\widetilde{N}_{\pm}}^{eq}
\left(1+f_{\widetilde{\ell}}^{eq}\right)\left(1+f_{\widetilde{u}}^{eq}\right)
\left(1+f_{\widetilde{Q}}^{eq}\right)\left(...\right), \nonumber 
%\\
\end{eqnarray}
\begin{eqnarray}
F_{22_{\pm}}\left(...\right) 
& \equiv & F_{\widetilde{N}_{\pm}
\widetilde{\ell}\widetilde{u}^*\widetilde{Q}}f_{\widetilde{N}_{\pm}}^{eq}
f_{\widetilde{\ell}}^{eq}\left(1+f_{\widetilde{u}}^{eq}\right)
\left(1+f_{\widetilde{Q}}^{eq}\right)\left(...\right)
\nonumber \\
& = & F_{\widetilde{N}_{\pm}\widetilde{Q}\widetilde{\ell}\widetilde{u}}
f_{\widetilde{N}_{\pm}}^{eq}f_{\widetilde{Q}}^{eq}
\left(1+f_{\widetilde{u}}^{eq}\right)\left(1+f_{\widetilde{\ell}}^{eq}\right)
\left(...\right) \nonumber \\
& = & F_{\widetilde{N}_{\pm}\widetilde{u}^*\widetilde{\ell}\widetilde{Q}^{*}}
f_{\widetilde{N}_{\pm}}^{eq}f_{\widetilde{u}}^{eq}
\left(1+f_{\widetilde{\ell}}^{eq}\right)
\left(1+f_{\widetilde{Q}}^{eq}\right)\left(...\right),
\nonumber \\
F_{t\pm}^{\left(5\right)}\left(...\right) & \equiv &
F_{\widetilde{N}_{\pm}\ell
Q\widetilde{u}^*}f_{\widetilde{N}_{\pm}}^{eq}f_{\ell}^{eq}
\left(1-f_{Q}^{eq}\right)\left(1+f_{\widetilde{u}}^{eq}\right)
\left(...\right) 
\nonumber \\
& = & F_{\widetilde{N}_{\pm}\ell\widetilde{Q}\overline{u}}
f_{\widetilde{N}_{\pm}}^{eq}f_{\ell}^{eq}\left(1+f_{\widetilde{Q}}^{eq}\right)
\left(1-f_{u}^{eq}\right)\left(...\right),
\nonumber\\ 
F_{t\pm}^{\left(6\right)}\left(...\right) & \equiv &
F_{\widetilde{N}_{\pm}\widetilde{u}^*\overline{\ell}Q}
f_{\widetilde{N}_{\pm}}^{eq}f_{\widetilde{u}}^{eq}\left(1-f_{\ell}^{eq}\right)
\left(1-f_{Q}^{eq}\right)\left(...\right) 
\nonumber \\
& = & F_{\widetilde{N}_{\pm}
\widetilde{Q}^{*}\overline{\ell}\overline{u}}f_{\widetilde{N}_{\pm}}^{eq}
f_{\widetilde{Q}}^{eq}\left(1-f_{\ell}^{eq}\right)
\left(1-f_{u}^{eq}\right)\left(...\right),
\nonumber\\ 
F_{t\pm}^{\left(7\right)}\left(...\right) & \equiv & 
F_{N_{\pm}\ell Q\overline{u}}f_{N_{\pm}}^{eq}f_{\ell}^{eq}
\left(1-f_{Q}^{eq}\right)\left(1-f_{u}^{eq}\right)\left(...\right),
\nonumber\\ 
F_{t\pm}^{\left(8\right)}\left(...\right) & \equiv &
F_{N_{\pm}u\overline{\ell}Q}f_{N_{\pm}}^{eq}f_{u}^{eq}\left(1-f_{\ell}^{eq}\right)
\left(1-f_{Q}^{eq}\right)\left(...\right) 
\nonumber \\
& = & F_{N_{\pm}\overline{Q}\overline{\ell}
\overline{u}}f_{N_{\pm}}^{eq}f_{Q}^{eq}\left(1-f_{\ell}^{eq}\right)
\left(1-f_{u}^{eq}\right)\left(...\right),
\nonumber\\ 
F_{t\pm}^{\left(9\right)}\left(...\right) & \equiv &
F_{N_{\pm}u\overline{\ell}Q}f_{N_{\pm}}^{eq}f_{u}^{eq}\left(1-f_{\ell}^{eq}\right)
\left(1-f_{Q}^{eq}\right)\left(...\right) 
\nonumber \\
& = & F_{N_{\pm}\overline{Q}\overline{\ell}
\overline{u}}f_{N_{\pm}}^{eq}f_{Q}^{eq}\left(1-f_{\ell}^{eq}\right)
\left(1-f_{u}^{eq}\right)\left(...\right).
\end{eqnarray}
The BE for the asymmetry in the lepton doublets 
%!!%
$Y_{\Delta\ell} \equiv \left( Y_{\ell}-Y_{\overline{\ell}} \right) /2$ 
is: \footnote{Here $Y_{\Delta\ell}$ refers to lepton
asymmetry abundance in single $SU(2)_L$ gauge degree of freedom. 
However, since the amplitude on the r.h.s is summed over gauge multiplicity, 
we have to multiply by a factor of two in the l.h.s. of the BE.}
\begin{eqnarray}
2 \dot{Y}_{\Delta\ell} & = &
\sum_{i=\pm} \left[\widetilde{N}_{i}
\leftrightarrow\widetilde{H}_u\ell\right]_-
-\sum_{ij} \left[\widetilde{H}_u
\ell\leftrightarrow ij\right]_{-}^{\rm sub}
%\nonumber \\
%&& 
+\left[N\leftrightarrow H_u\ell\right]_-
-\left[\ell\ell\leftrightarrow
\widetilde{\ell}\widetilde{\ell}\right]_-
\nonumber \\
 &  & -\left[N\ell\leftrightarrow Q\overline{u}\right]_{-}
-\left[Nu\leftrightarrow\overline{\ell}Q\right]_{-}-
\left[N\overline{Q}\leftrightarrow\overline{\ell}\overline{u}\right]_{-} 
\nonumber \\
&  & -\sum_{i=\pm}\left(\left[\widetilde{N}_{i}\ell
\leftrightarrow Q\widetilde{u}^*
\right]_{-}+\left[\widetilde{N}_{i}\ell
\leftrightarrow\widetilde{Q}
\overline{u}\right]_{-}
+\left[\widetilde{N}_{i}\widetilde{u}
\leftrightarrow\overline{\ell}Q\right]_{-}\right.
\nonumber \\
&  & \left. +\left[\widetilde{N}_{i}\widetilde{Q}^{*}
\leftrightarrow\overline{\ell}\overline{u}\right]_{-}
+\left[\widetilde{N}_{i}\overline{Q}
\leftrightarrow\overline{\ell}\widetilde{u}^*\right]_{-}
+\left[\widetilde{N}_{i}u\leftrightarrow\overline{\ell}
\widetilde{Q}\right]_{-}\right)
\nonumber \\
& = & 
\sum_{i=\pm} \left\{ \left(F_{\widetilde{N}_{i}
\widetilde{H}_u\ell}-F_{\widetilde{N}_{i}
\overline{\widetilde{H}_u}\overline{\ell}}\right)
f_{\widetilde{N}_{i}}^{eq}
\left(1-f_{\ell}^{eq}\right)\left(1-f_{\widetilde{H}_u}^{eq}\right) 
\left(\frac{f_{\widetilde{N}_{i}}}{f_{\widetilde{N}_{i}}^{eq}}
-\frac{1+f_{\widetilde{N}_{i}}}{1+f_{\widetilde{N}_{i}}^{eq}}\right)\right.
\nonumber \\  &  &
-F_{\widetilde{N}_{i}} \!\! \left[\frac{f_{\widetilde{N}_{i}}}
{f_{\widetilde{N}_{i}}^{eq}}f_{\ell}^{eq}+\frac{1+f_{\widetilde{N}_{i}}}
{1+f_{\widetilde{N}_{i}}^{eq}}
\left(1-f_{\ell}^{eq}\right)\right]\frac{\mu_{\ell}}{T}
-(f_\ell^{eq} \to f_{\widetilde H_u}^{eq}, \mu_\ell \to \mu_{\widetilde H_u}) 
\nonumber \\ &  &
-2F_{N} \!\! \left[\frac{f_{N}}{f_{N}^{eq}}f_{\ell}^{eq}+
\frac{1-f_{N}}{1-f_{N}^{eq}}\left(1-f_{\ell}^{eq}\right)\right]
\frac{\mu_{\ell}}{T}
-(f_\ell^{eq} \to - f_{H_u}^{eq}, \mu_\ell \to \mu_{H_u}) 
\nonumber \\
 &  & +4F_{\ell\ell\widetilde{\ell}\widetilde{\ell}}
\left(\frac{\mu_{\widetilde{\ell}}}{T}-\frac{\mu_{\ell}}{T}\right)
f_{\ell}^{eq,2}\left(1+f_{\widetilde{\ell}}^{eq}\right)^{2}
+ S_{t} + W_{\Delta L= 2},
\label{eq:lepton_BE}
\end{eqnarray}
where $(f_\ell^{eq} \to f_{\widetilde H_u}^{eq},  
\mu_\ell \to \mu_{\widetilde H_u})$ etc. 
refers to the term obtained by replacing the corresponding
$f$ and $\mu$ in the preceding term.
In Eq.~\eqref{eq:lepton_BE}, $\sum_{ij}\left[\widetilde{H}_u
\ell\leftrightarrow ij\right]_{-}^{\rm sub}$ refers to the sum
of all possible $\Delta L =2$ scatterings 
$\widetilde{H}_u \ell \leftrightarrow ij$ and,  
if $\widetilde N_\pm$ exchange  in the s-channel
is involved, the on-shell contributions 
are subtracted out to avoid double counting. 
The $\Delta L =2$ scatterings $\widetilde{H}_u \ell \leftrightarrow ij$ 
with t- and u-channel exchange of $\widetilde N_\pm$, and the leftover 
off-shell contribution for s-channel exchange of $\widetilde N_\pm$ 
are all collected in $W_{\Delta L= 2}$. 
The details of the subtraction procedure is given 
in \ref{subsec:22sub}. In the numerical calculation, 
we neglect $W_{\Delta L =2}$ since it is subdominant in the SL 
temperature range $T \lesssim 10^9$ GeV.
The top and stop scattering term $S_t$ in Eq.~\eqref{eq:lepton_BE} 
is given by
\begin{eqnarray}
S_{t} \! & = & \! -2F_t^{\left(3\right)} \!\! \left[\frac{f_{N}}{f_{N}^{eq}}
\left(1-f_{\ell}^{eq}\right)+\! \frac{1-f_{N}}{1-f_{N}^{eq}}f_{\ell}^{eq}\right]
\!\! \frac{\mu_{\ell}}{T} 
%  \nonumber \\ &  & 
-4F_t^{\left(4\right)} \!\! \left[\frac{f_{N}}{f_{N}^{eq}}
f_{\ell}^{eq}+\! \frac{1-f_{N}}{1-f_{N}^{eq}}\left(1-f_{\ell}^{eq}\right)\right]
\!\!\frac{\mu_{\ell}}{T}
\nonumber \\
 &  & + 2\left(F_t^{\left(3\right)}+F_t^{\left(4\right)}\right)
\left[\frac{f_{N}}{f_{N}^{eq}}f_{Q}^{eq}+\frac{1-f_{N}}{1-f_{N}^{eq}}
\left(1-f_{Q}^{eq}\right)\right]\frac{\mu_{Q}}{T} %-\left(Q\to u\right)
\nonumber \\
 &  & + 2F_t^{\left(4\right)}\left[\frac{f_{N}}{f_{N}^{eq}}
\left(1-f_{Q}^{eq}\right)+\frac{1-f_{N}}{1-f_{N}^{eq}}f_{Q}^{eq}\right]
\frac{\mu_{Q}}{T} %-\left(Q\to u\right) 
\nonumber \\ &  & 
-\sum_{i=\pm} \left\{ 4F_{ti}^{\left(5\right)}\left[
\frac{f_{\widetilde{N}_{i}}}{f_{\widetilde{N}_{i}}^{eq}}
\left(1-f_{\ell}^{eq}\right)
+\frac{1+f_{\widetilde{N}_{i}}}{1+f_{\widetilde{N}_{i}}^{eq}}
f_{\ell}^{eq}\right]\frac{\mu_{\ell}}{T} \right.
\nonumber \\
&  & + 4\left(F_{ti}^{\left(6\right)}+F_{ti}^{\left(7\right)}\right)
\left[\frac{f_{\widetilde{N}_{i}}}{f_{\widetilde{N}_{i}}^{eq}}
f_{\ell}^{eq}+\frac{1+f_{\widetilde{N}_{i}}}{1+f_{\widetilde{N}_{i}}^{eq}}
\left(1-f_{\ell}^{eq}\right)\right]\frac{\mu_{\ell}}{T}
\nonumber \\
 &  & - 2\left(F_{ti}^{\left(5\right)}
+F_{ti}^{\left(6\right)}\right)
\left[\frac{f_{\widetilde{N}_{i}}}{f_{\widetilde{N}_{i}}^{eq}}f_{Q}^{eq}
+\frac{1+f_{\widetilde{N}_{i}}}{1+f_{\widetilde{N}_{i}}^{eq}}
\left(1-f_{Q}^{eq}\right)\right]\frac{\mu_{Q}}{T} %-\left(Q\to u\right)
\nonumber \\
&  & -  2F_{ti}^{\left(7\right)}
\left[\frac{f_{\widetilde{N}_{i}}}{f_{\widetilde{N}_{i}}^{eq}}
\left(1-f_{Q}^{eq}\right)
+\frac{1+f_{\widetilde{N}_{i}}}{1+f_{\widetilde{N}_{i}}^{eq}}f_{Q}^{eq}\right]
\frac{\mu_{Q}}{T} %-\left(Q\to u\right)
\nonumber \\
&  & -  2\left(F_{ti}^{\left(5\right)}
+F_{ti}^{\left(7\right)}\right)  
\left[-\frac{f_{\widetilde{N}_{i}}}{f_{\widetilde{N}_{i}}^{eq}}
f_{\widetilde{Q}}^{eq}
+\frac{1+f_{\widetilde{N}_{i}}}{1+f_{\widetilde{N}_{i}}^{eq}}
\left(1+f_{\widetilde{Q}}^{eq}\right)\right]\frac{\mu_{\widetilde{Q}}}{T}
%-\!\!\left(\widetilde{Q}\to\widetilde{u}\right)
\nonumber \\
&  & - \left. 2F_{ti}^{\left(6\right)}
\left[\frac{f_{\widetilde{N}_{i}}}{f_{\widetilde{N}_{i}}^{eq}}
\left(1+f_{\widetilde{Q}}^{eq}\right)-
\frac{1+f_{\widetilde{N}_{i}}}{1+f_{\widetilde{N}_{i}}^{eq}}
f_{\widetilde{Q}}^{eq}\right]\frac{\mu_{\widetilde{Q}}}{T} \right\}
% \nonumber \\ && 
-\left(Q\to u\right)
-\left(\widetilde{Q}\to\widetilde{u}\right),\qquad \quad 
\label{eq:St}
\end{eqnarray}
where in the last line $\left(Q\to u\right)$
and $(\widetilde{Q}\to\widetilde{u})$
denote respectively terms in which $Q$ is replaced by $u$ and
$\widetilde Q$ is replaced by $\widetilde u$.

The BE for the slepton asymmetry 
%!!%
$Y_{\Delta\widetilde{\ell}} \equiv 
\left( Y_{\widetilde{\ell}} -Y_{\widetilde{\ell}^{*}} \right)/2$ 
can be written as: 
\begin{eqnarray}
2 \dot{Y}_{\Delta\widetilde{\ell}} 
& = &  \sum_{i=\pm}  \left[\widetilde{N}_{+}
\leftrightarrow H_u\widetilde{\ell}\right]_-
\!\!-\!\!\sum_{ij}\left[H_u\widetilde{\ell}
\leftrightarrow ij\right]_{-}^{\rm sub}
%\nonumber \\
%&  & 
+\left[N\leftrightarrow\widetilde{H}_u
\widetilde{\ell}\right]_-
+\left[\ell\ell\leftrightarrow\widetilde{\ell}\widetilde{\ell}\right]_-
\nonumber \\ &  & 
\hspace{-1.0cm} 
+\sum_{i}\left(\left[\widetilde{N}_{i}\leftrightarrow
\widetilde{\ell}\widetilde{u}\widetilde{Q}^{*}\right]_-
+\left[\widetilde{N}_{i}\widetilde{\ell}^*\leftrightarrow
\widetilde{u}\widetilde{Q}^*\right]_- \right.
%\nonumber \\
%&  & 
\left.+\left[\widetilde{N}_{i}\widetilde{Q}
\leftrightarrow\widetilde{\ell}\widetilde{u}\right]_-
+\left[\widetilde{N}_{i}\widetilde{u}^*
\leftrightarrow\widetilde{\ell}\widetilde{Q}^{*}\right]_-\right)
\nonumber \\ &  & 
\hspace{-1.0cm} 
-\! \left[N\widetilde{\ell}\leftrightarrow Q\widetilde{u}^*\right]_{-}
\!\!-\! \left[N\widetilde{\ell}\leftrightarrow\widetilde{Q}\overline{u}\right]_{-}
\!\!-\! \left[N\overline{Q}\leftrightarrow\widetilde{\ell}^{*}\widetilde{u}^*\right]_{-}
% \nonumber \\ &  & 
\!\!-\! \left[Nu\leftrightarrow\widetilde{\ell}^{*}\widetilde{Q}\right]_{-}
\!\!-\! \left[N\widetilde{u}\leftrightarrow\widetilde{\ell}^{*}Q\right]_{-}
\nonumber \\  &  & 
\hspace{-1.0cm} 
-\left[N\widetilde{Q}^{*}\leftrightarrow\widetilde{\ell}^{*}
\overline{u}\right]_{-} 
+\sum_{i}\left(\left[\widetilde{N}_{i}\widetilde{\ell}^{*}
\leftrightarrow\overline{Q}u\right]_{-}
+\left[\widetilde{N}_{i}Q\leftrightarrow\widetilde{\ell}u\right]_{-}
+\left[\widetilde{N}_{i}\overline{u}
\leftrightarrow\widetilde{\ell}\overline{Q}\right]_{-}\right)
\nonumber \\
& = & \sum_{i=\pm} 
\left\{ \left(F_{\widetilde{N}_{i}H_u\widetilde{\ell}}-
F_{\widetilde{N}_{i}H_u^{*}\widetilde{\ell}^{*}}\right)
f_{\widetilde{N}_{i}}^{eq}\left(1+f_{\widetilde{\ell}}^{eq}\right)
\left(1+f_{H_u}^{eq}\right)
\left(\frac{f_{\widetilde{N}_{i}}}{f_{\widetilde{N}_{i}}^{eq}}
-\frac{1+f_{\widetilde{N}_{i}}}{1+f_{\widetilde{N}_{i}}^{eq}}\right) \right.
\nonumber \\ &  & 
- \widetilde{F}_{\widetilde{N}_{i}} \!\!
\left[-\frac{f_{\widetilde{N}_{i}}}{f_{\widetilde{N}_{i}}^{eq}}
f_{\widetilde{\ell}}^{eq}+\frac{1+f_{\widetilde{N}_{i}}}
{1+f_{\widetilde{N}_{i}}^{eq}}
\left(1+f_{\widetilde{\ell}}^{eq}\right)\right]
\frac{\mu_{\widetilde{\ell}}}{T} 
-(f_{\widetilde\ell}^{eq} \to f_{H_u}^{eq}, 
\mu_{\widetilde\ell} \to \mu_{H_u}) 
\nonumber \\ &  & 
-2F_{N} \!\! \left[-\frac{f_{N}}{f_{N}^{eq}}
f_{\widetilde{\ell}}^{eq}+\frac{1-f_{N}}{1-f_{N}^{eq}}
\left(1+f_{\widetilde{\ell}}^{eq}\right)\right]
\frac{\mu_{\widetilde{\ell}}}{T} 
-(f_{\widetilde \ell}^{eq} \to - f_{\widetilde H_u}^{eq}, 
\mu_{\widetilde \ell} \to \mu_{\widetilde H_u}) 
\nonumber \\  &  & 
-4F_{\ell\ell\widetilde{\ell}\widetilde{\ell}}
\left(\frac{\mu_{\widetilde{\ell}}}{T}-\frac{\mu_{\ell}}{T}\right)
f_{\ell}^{eq,2}\left(1+f_{\widetilde{\ell}}^{eq}\right)^{2}
+\widetilde{S}_{t} +S_{22} +\widetilde{W}_{\Delta L=2},
\label{eq:slepton_BE}
\end{eqnarray}
where the $\Delta L =2$ scatterings $H_u \widetilde\ell
\leftrightarrow ij$ with t- and u-channel exchange of $\widetilde
N_\pm$ and the off-shell contribution for s-channel exchange of
$\widetilde N_\pm$ are all collected in $\widetilde W_{\Delta L= 2}$
that, as already said, can be neglected in the SL temperature range.

The term $S_{22}$ in Eq.~\ref{eq:slepton_BE} from scalar
potential terms is given by
\begin{eqnarray}
S_{22} 
& = & \sum_{i}\left\{ 2\widetilde{F}_{\widetilde{N}_{i}}^{\left(3\right)}
\left[\frac{f_{\widetilde{N}_{i}}}{f_{\widetilde{N}_{i}}^{eq}}
f_{\widetilde{\ell}}^{eq}-\frac{1+f_{\widetilde{N}_{i}}}
{1+f_{\widetilde{N}_{i}}^{eq}}\left(1+f_{\widetilde{\ell}}^{eq}\right)\right]
\frac{\mu_{\widetilde{\ell}}}{T}\right. \nonumber \\
&  & -2F_{22_{i}}\left[\frac{f_{\widetilde{N}_{i}}}{f_{\widetilde{N}_{i}}^{eq}}
\left(1-f_{\widetilde{\ell}}^{eq}\right)
+\frac{1+f_{\widetilde{N}_{i}}}{1+f_{\widetilde{N}_{i}}^{eq}}
\left(2+f_{\widetilde{\ell}}^{eq}\right) \right]
\frac{\mu_{\widetilde{\ell}}}{T}
\nonumber \\
&  & -2\widetilde{F}_{\widetilde{N}_{i}}^{\left(3\right)}
\left[\frac{f_{\widetilde{N}_{\pm}}}{f_{\widetilde{N}_{\pm}}^{eq}}
f_{\widetilde{Q}}^{eq}-
\frac{1+f_{\widetilde{N}_{\pm}}}{1+f_{\widetilde{N}_{\pm}}^{eq}}
\left(1+f_{\widetilde{Q}}^{eq}\right)\right]\frac{\mu_{\widetilde{Q}}}{T} 
\nonumber \\
&  & \left. +2F_{22_{i}}\left[\frac{f_{\widetilde{N}_{i}}}{f_{\widetilde{N}_{i}}^{eq}}
\left(1-f_{\widetilde{Q}}^{eq}\right)
+\frac{1+f_{\widetilde{N}_{i}}}{1+f_{\widetilde{N}_{i}}^{eq}}
\left(2+f_{\widetilde{Q}}^{eq}\right) \right]
\frac{\mu_{\widetilde{Q}}}{T} \right\}
% \nonumber \\ && 
-\left(\widetilde Q \to \widetilde u \right), \qquad 
\end{eqnarray}
while the top and stop scatterings term $\widetilde{S}_{t}$ reads:
\begin{eqnarray}
\widetilde{S}_{t} 
& = & -4F^{\left(0\right)}\left[\frac{f_{N}}{f_{N}^{eq}}
\left(1+f_{\widetilde{\ell}}^{eq}\right)-\frac{1-f_{N}}{1-f_{N}^{eq}}
f_{\widetilde{\ell}}^{eq}\right]\frac{\mu_{\widetilde{\ell}}}{T} 
\nonumber \\
&  & -4\left(F^{\left(1\right)}+F^{\left(2\right)}\right)
\left[-\frac{f_{N}}{f_{N}^{eq}}f_{\widetilde{\ell}}^{eq}
+\frac{1-f_{N}}{1-f_{N}^{eq}}\left(1+f_{\widetilde{\ell}}^{eq}\right)\right]
\frac{\mu_{\widetilde{\ell}}}{T}
\nonumber \\
&  & + 2F^{\left(0\right)}\left[\frac{f_{N}}{f_{N}^{eq}}f_{Q}^{eq}
+\frac{1-f_{N}}{1-f_{N}^{eq}}\left(1-f_{Q}^{eq}\right)\right]\frac{\mu_{Q}}{T}
%-\left(Q\to u\right)
\nonumber \\
&  & + 2F^{\left(1\right)}\left[\frac{f_{N}}{f_{N}^{eq}}
\left(1-f_{Q}^{eq}\right)+\frac{1-f_{N}}{1-f_{N}^{eq}}f_{Q}^{eq}\right]
\frac{\mu_{Q}}{T}%-\left(Q\to u\right) 
\nonumber \\
&  & + 2\left(F^{\left(0\right)}+F^{\left(1\right)}\right)
\left[-\frac{f_{N}}{f_{N}^{eq}}f_{\widetilde{Q}}^{eq}+
\frac{1-f_{N}}{1-f_{N}^{eq}}\left(1+f_{\widetilde{Q}}^{eq}\right)\right]
\frac{\mu_{\widetilde{Q}}}{T} %-\left(\widetilde{Q}\to\widetilde{u}\right)
\nonumber \\
 &  & + 2F^{\left(2\right)}\left[\frac{f_{N}}{f_{N}^{eq}}
\left(1+f_{\widetilde{Q}}^{eq}\right)-\frac{1-f_{N}}{1-f_{N}^{eq}}
f_{\widetilde{Q}}^{eq}\right]\frac{\mu_{\widetilde{Q}}}{T}
%-\left(\widetilde{Q}\to\widetilde{u}\right)
\nonumber %\\
\end{eqnarray}
\begin{eqnarray} 
\hspace{0.5cm}
&  & -\sum_{i=\pm} \left\{ 2F_{i}^{\left(8\right)}
\left[\frac{f_{\widetilde{N}_{i}}}{f_{\widetilde{N}_{i}}^{eq}}
\left(1+f_{\widetilde{\ell}}^{eq}\right)-
\frac{1+f_{\widetilde{N}_{i}}}{1+f_{\widetilde{N}_{i}}^{eq}}
f_{\widetilde{\ell}}^{eq}\right]\frac{\mu_{\widetilde{\ell}}}{T} \right.
\nonumber \\
&  & + 4F_{i}^{\left(9\right)}\left[-
\frac{f_{\widetilde{N}_{i}}}{f_{\widetilde{N}_{i}}^{eq}}
f_{\widetilde{\ell}}^{eq}+
\frac{1+f_{\widetilde{N}_{i}}}{1+f_{\widetilde{N}_{i}}^{eq}}
\left(1+f_{\widetilde{\ell}}^{eq}\right)\right]
\frac{\mu_{\widetilde{\ell}}}{T}
\nonumber \\
 &  & - 2 F_{i}^{\left(8\right)} 
\left[\frac{f_{\widetilde{N}_{i}}}
{f_{\widetilde{N}_{i}}^{eq}}f_{Q}^{eq}
+\frac{1+f_{\widetilde{N}_{i}}}{1+f_{\widetilde{N}_{i}}^{eq}}
\left(1-f_{Q}^{eq}\right)\right]\frac{\mu_{Q}}{T}
% -\left(Q\to u\right) 
\nonumber \\
 &  & \left. - 2F_{i}^{\left(9\right)}
\left[\frac{f_{\widetilde{N}_{i}}}{f_{\widetilde{N}_{i}}^{eq}}
+\frac{1+f_{\widetilde{N}_{i}}}
{1+f_{\widetilde{N}_{i}}^{eq}}\right]\frac{\mu_{Q}}{T} \right\}
%-\left(Q\to u\right) 
-\left(Q\to u\right)
-\left(\widetilde{Q}\to\widetilde{u}\right).
\end{eqnarray}
%

%%%%%%%%%%%%%%%%%%%%%%%%%%%%%

\subsubsection{Approximations: integrated Boltzmann equations}
\label{app:approx_int_BE}

In order to write the BE in the integrated form of equations for the
number densities, the following approximations are needed:
\begin{eqnarray}
\frac{1+\eta_{a}f_{a}}{1+\eta_{a}f_{a}^{eq}}  \to 1,
&\;\;\;\;\;&
\eta_{i}f_{i}^{eq}\frac{\mu_{i}}{T} \to  0,
\label{eq:approx}
\end{eqnarray}
where $a$ refers to $N$ or $\widetilde{N}_{\pm}$ and $i$ for
any other particles. 
The approximations above are equivalent to neglect the 
chemical potentials in the Fermi-blocking and Bose-enhancement factors.
In addition, we need to assume that $N$ and $\widetilde{N}_{\pm}$
are in kinetic equilibrium, that is 
\begin{eqnarray}
\frac{f_{N}}{f_{N}^{eq}} 
=  \frac{Y_{N}}{Y_{N}^{eq}}, & \;\;\;\;\;\;&
\frac{f_{\widetilde{N}_{\pm}}}{f_{\widetilde{N}_{\pm}}^{eq}} 
=  \frac{Y_{\widetilde{N}_{\pm}}}{Y_{\widetilde{N}_{\pm}}^{eq}}.
\label{eq:N_kinetic_equil_approx}
\end{eqnarray}
The BE in their non-integrated form have been considered in
Refs.~\refcite{basboll:2006,HahnWoernle:2009,Garayoa:2009}, and it was
found that in the strong washout regime the distributions of $N$ and
$\widetilde{N}_{\pm}$ are close to kinetic equilibrium, and thus
Eqs.~\eqref{eq:N_kinetic_equil_approx} represent a very good
approximation.  However, it should be mentioned that, as
discussed in Ref.~\refcite{Garayoa:2009}, in the weak washout regime
numerical differences when using the integrated form of the BE can be
up to one order of magnitude.
In any case, by means of the approximations \eqref{eq:approx} 
and \eqref{eq:N_kinetic_equil_approx} on can write a set of 
integrated BE as
\begin{eqnarray}
\dot{Y}_{N}
& = & -\left(\frac{Y_{N}}{Y_{N}^{eq}}-1\right) 
%\nonumber \\
%&  & \times 
\left(\gamma_{N}+4\gamma_{t}^{(0)}+4\gamma_{t}^{(1)}+
4\gamma_{t}^{(2)}+2\gamma_{t}^{(3)}+4\gamma_{t}^{(4)}\right),\qquad 
\label{eq:int_BE_RHN}\\
\dot{Y}_{\widetilde{N}_{\pm}}
& = &  -\left(\gamma_{\widetilde{N}_{\pm}}^{f}
+\gamma_{\widetilde{N}_{\pm}}^{s}\right)
\left(\frac{Y_{\widetilde{N}_{\pm}}}{Y_{\widetilde{N}_{\pm}}^{eq}}-1\right)
-2\left(\gamma_{\widetilde{N}_{\pm}}^{\left(3\right)}+3\gamma_{22_{\pm}}\right)
\left(\frac{Y_{\widetilde{N}_{\pm}}}{Y_{\widetilde{N}_{\pm}}^{eq}}-1\right)
\nonumber \\
&  & -2\left(2\gamma_{t\pm}^{\left(5\right)}+2\gamma_{t\pm}^{\left(6\right)}
+2\gamma_{t\pm}^{\left(7\right)}+\gamma_{t+}^{\left(8\right)}
+2\gamma_{t\pm}^{\left(9\right)}\right)
\left(\frac{Y_{\widetilde{N}_{\pm}}}{Y_{\widetilde{N}_{\pm}}^{eq}}-1\right), 
\label{eq:int_BE_RHSN_pm}
\end{eqnarray}
where we have defined the reaction densities for RHN decays as
$\gamma_N \equiv F_N(1) + \widetilde F_N(1)$, RHSN decays as
$\gamma_{\widetilde N_\pm}^{f,s} \equiv F_{\widetilde N_\pm}^{f,s}(1)$,
interactions from scalar potential as
$\gamma_{\widetilde N_\pm}^{(3)} \equiv F_{\widetilde N_\pm}^{(3)}(1)$,
$\gamma_{22\pm} \equiv F_{22\pm}(1)$,
and top/stop scatterings as
$\gamma_{t}^{(n)} \equiv F_{t}^{(n)}(1)$
$\gamma_{t}^{(n)} \equiv F_{t}^{(n)}(1)$ and 
$\gamma_{t\pm}^{(n)} \equiv F_{t\pm}^{(n)}(1)$.

By using the following approximations 
\begin{eqnarray}
\nonumber
&& \!\!\!\!\!\! 
Y_{\widetilde{N}_{+}}^{eq}  \approx  
Y_{\widetilde{N}_{-}}^{eq}\equiv Y_{\widetilde{N}}^{eq},
% \;\; &
\qquad \quad 
Y_{\widetilde{N}_{+}}  \approx  Y_{\widetilde{N}_{-}}
\equiv\frac{1}{2}Y_{\widetilde{N}_{\rm tot}}, \\
% \;\;&
&& \!\!\!\!\!\! 
\gamma_{\widetilde{N}_{+}}^{f}+\gamma_{\widetilde{N}_{+}}^{s} 
 \approx  \gamma_{\widetilde{N}_{-}}^{f}+\gamma_{\widetilde{N}_{-}}^{s}
\approx\frac{\gamma_{\widetilde{N}}}{2}, 
\nonumber
\end{eqnarray}
that are justified by the fact that the $\widetilde{N}_{\pm}$ mass
splitting is small $B \ll M$, we can sum up the BE for
$\widetilde{N}_{+}$ and $\widetilde{N}_{-}$ \eqref{eq:int_BE_RHSN_pm}
by defining $Y_{\widetilde N_{\rm tot}} \equiv 
Y_{\widetilde N_+} + Y_{\widetilde N_-}$ and we end up with:
\begin{eqnarray}
\dot{Y}_{\widetilde{N}_{\rm tot}}
& = & 
-\left[\frac{\gamma_{\widetilde{N}}}{2}+
\gamma_{\widetilde{N}}^{\left(3\right)}+
3\gamma_{22}+\gamma_{t}^{\left(8\right)}+
2\left(
\gamma_{t}^{\left(5\right)}+
\gamma_{t}^{\left(6\right)}+
\gamma_{t}^{\left(7\right)}+
\gamma_{t}^{\left(9\right)}\right)\right] 
 \nonumber \\ &  &  \times 
\left(\frac{Y_{\widetilde{N}_{\rm tot}}}{Y_{\widetilde{N}}^{eq}}-2\right).
\label{eq:int_BE_RHSN}
\end{eqnarray}
%
%!!%
With the same approximations we can write
% and also rewrite 
the BE for 
$Y_{\Delta\ell}$ and $Y_{\Delta\widetilde{\ell}}$ as follows:
\begin{eqnarray}
2\dot{Y}_{\Delta\ell} 
& = & \epsilon^{f}\left(T\right)
\frac{\gamma_{\widetilde{N}}}{2}
\left(\frac{Y_{\widetilde{N}_{\rm tot}}}{Y_{\widetilde{N}}^{eq}}-2\right)
- %\frac{
\gamma_{\widetilde{N}}^{f}
% }{2}
\left(\frac{\mu_{\ell}}{T} +\frac{\mu_{\widetilde{H}_u}}{T}\right)
% \nonumber \\ &  & 
- % \frac{1}{2}
\gamma_{N}^{f}\left(\frac{\mu_{\ell}}{T}
+\frac{\mu_{H_u}}{T}\right)
\nonumber \\
 &  & -\left(\gamma_{t}^{\left(3\right)}\frac{Y_{N}}{Y_{N}^{eq}}+
2\gamma_{t}^{\left(4\right)}+2\gamma_{t}^{\left(6\right)}+
2\gamma_{t}^{\left(7\right)}+\gamma_{t}^{\left(5\right)}
\frac{Y_{\widetilde{N}_{\rm tot}}}{Y_{\widetilde{N}}^{eq}}\right)
\frac{2\mu_{\ell}}{T}
\nonumber \\
&  & + \left[\!\gamma_{t}^{\left(3\right)}
+
\gamma_{t}^{\left(4\right)}\left(1+
% +\gamma_{t}^{\left(4\right)}
\frac{Y_{N}}{Y_{N}^{eq}}\right)+
\gamma_{t}^{\left(5\right)}+\gamma_{t}^{\left(6\right)}+
\frac{1}{2}\gamma_{t}^{\left(7\right)}
\frac{Y_{\widetilde{N}_{\rm tot}}}{Y_{\widetilde{N}}^{eq}}\!
\right] \frac{2\left(\mu_{Q}-\mu_{u}\right)}{T}
\nonumber \\
 &  & +\left(\gamma_{t}^{\left(5\right)}+\gamma_{t}^{\left(7\right)}
+\frac{1}{2}\gamma_{t}^{\left(6\right)}
\frac{Y_{\widetilde{N}_{\rm tot}}}{Y_{\widetilde{N}}^{eq}}\right)
\frac{2\left(\mu_{\widetilde{Q}} -\mu_{\widetilde{u}}\right)}{T}
\nonumber \\
&  & + 4
% \gamma_{\ell\ell\widetilde{\ell}\widetilde{\ell}}
\gamma_{\widetilde g}^{{\rm eff}}
\left(\frac{\mu_{\widetilde{\ell}}}{T}
-\frac{\mu_{\ell}}{T}\right)+ W_{\Delta L=2}\, ,
\label{eq:BE_deltal}
\end{eqnarray}
\begin{eqnarray}
2\dot{Y}_{\Delta\widetilde{\ell}}
& = & \epsilon^{s}\left(T\right)\frac{\gamma_{\widetilde{N}}}{2}
\left(\frac{Y_{\widetilde{N}_{\rm tot}}}{Y_{\widetilde{N}}^{eq}}-2\right)
-
% \frac{
\gamma_{\widetilde{N}}^{s}
% }{2}
\left(\frac{\mu_{\widetilde{\ell}}}{T}+\frac{\mu_{H_u}}{T}\right)
-
% \frac{1}{2}
\gamma_{N}^{s}\left(\frac{\mu_{\widetilde{\ell}}}{T}
+\frac{\mu_{\widetilde{H}_u}}{T}\right)
\nonumber \\ &  & 
-2 \gamma_{\widetilde{N}}^{\left(3\right)}
\frac{\mu_{\widetilde{\ell}}-\mu_{\widetilde{Q}}+\mu_{\widetilde{u}}}{T}
%  \nonumber \\&  & 
-\gamma_{22}\left(
% \frac{1}{2}
\frac{Y_{\widetilde{N}_{\rm tot}}}{Y_{\widetilde{N}}^{eq}} +4 \right)
\frac{\mu_{\widetilde{\ell}}-\mu_{\widetilde{Q}}+\mu_{\widetilde{u}}}{T}
\nonumber \\ &  & 
-\left(2\gamma_{t}^{\left(0\right)}\frac{Y_{N}}{Y_{N}^{eq}}
+2\gamma_{t}^{\left(1\right)}+2\gamma_{t}^{\left(2\right)}
+\frac{1}{2}\gamma_{t}^{\left(8\right)}
\frac{Y_{\widetilde{N}_{\rm tot}}}{Y_{\widetilde{N}}^{eq}}
+2\gamma_{t}^{\left(9\right)}\right)\frac{2\mu_{\widetilde{\ell}}}{T}
\nonumber \\ &  & 
+\left[\gamma_{t}^{\left(0\right)}+\gamma_{t}^{\left(1\right)}
\frac{Y_{N}}{Y_{N}^{eq}}+\gamma_{t}^{\left(8\right)}
+\gamma_{t}^{\left(9\right)} \left(1
+\frac{1}{2}
% \gamma_{t}^{\left(9\right)}
\frac{Y_{\widetilde{N}_{\rm tot}}}{Y_{\widetilde{N}}^{eq}}\right)
\right]
\frac{2\left(\mu_{Q}-\mu_{u}\right)}{T}
\nonumber \\ &  & 
+\left(\gamma_{t}^{\left(0\right)}+\gamma_{t}^{\left(1\right)}
+\gamma_{t}^{\left(2\right)}\frac{Y_{N}}{Y_{N}^{eq}}
\right)
\frac{2\left(\mu_{\widetilde{Q}} -\mu_{\widetilde{u}}\right)}{T}
\nonumber \\ &  & 
-4
% \gamma_{\ell\ell\widetilde{\ell}\widetilde{\ell}}
\gamma_{\widetilde g}^{{\rm eff}}
\left(\frac{\mu_{\widetilde{\ell}}}{T}-\frac{\mu_{\ell}}{T}\right)
+\widetilde{W}_{\Delta L=2}, 
\label{eq:BE_deltatl}
\end{eqnarray}
where we define the CP asymmetries for $\widetilde N_{\pm}$ decays as follows: 
\begin{eqnarray}
\epsilon^{f}\left(T\right) & \equiv & \sum_{i=\pm}
\frac{\left(F_{\widetilde{N}_{i}{\widetilde{H}_u}\ell}
-F_{\widetilde{N}_{i}\overline{\widetilde{H_u}}\overline{\ell}}\right)
f_{\widetilde{N}_{i}}^{eq}\left(1-f_{\ell}^{eq}\right)
\left(1-f_{\widetilde{H}_u}^{eq}\right)}{\gamma_{\widetilde{N}}}, \\
\epsilon^{s}\left(T\right) & \equiv & \sum_{i=\pm} 
\frac{\left(F_{\widetilde{N}_{i}H_u\widetilde{\ell}}-
F_{\widetilde{N}_{i}H_u^{*}\widetilde{\ell}^{*}}\right)
f_{\widetilde{N}_{i}}^{eq}
\left(1+f_{\widetilde{\ell}}^{eq}\right)\left(1+f_{H_u}^{eq}\right)}
{\gamma_{\widetilde{N}}}.
\label{eq:asyms}
\end{eqnarray}
%%%%%%%%%%%%%%%%%%%%%%%%%%%%%%%%%%%%%%%%%
In order to write the BE \eqref{eq:BE_deltal} and
\eqref{eq:BE_deltatl} in a closed form, all chemical potentials
% $\mu_{Q,\widetilde Q, u, \widetilde u, H_u, \widetilde H_u}$ 
have to be expressed in terms of $\mu_{\ell,\widetilde\ell}$, and then
these quantities must be rewritten in terms of the lepton and slepton
density asymmetries by means of Eqs.\eqref{eq:chem_pot_asym}.

%%%%%%%%%%%%%%%%%%%%%%%%%%%%%%%%%%%%%%%%%%%%%%%%%%%%%%%%%%%%%%%%%

\subsubsection{Subtracted 2 $\leftrightarrow$ 2 scatterings}
\label{subsec:22sub}

Although $2\leftrightarrow 2$ scatterings are of higher order 
$\mathcal{O}(Y^4)$ with respect to decays and inverse decays  which 
are $\mathcal{O}(Y^2)$, the CP asymmetries of the 
 $2\leftrightarrow 2$ subtracted rates 
are of the same order than that of the decays,\cite{Kolb:1980,Fry:1980}
and hence cannot be ignored.
The term  $\left[\widetilde{H}_u\ell\leftrightarrow ij\right]_-^{\rm sub}$
in the BE \eqref{eq:lepton_BE} consists of the following two terms
%and \eqref{eq:slepton_BE}
%
\begin{eqnarray}
&& \left[\widetilde{H}_u\ell\leftrightarrow ij\right]^{\rm sub} 
% \nonumber \\ & = & 
= \left(F_{\widetilde{H}_u\ell ij}
-\overline{F_{\widetilde{H}_u\ell ij}}\right)^{\rm sub}
\left(1+f_{\ell}^{eq}e^{\frac{E_{\ell}}{T}}\frac{\mu_{\ell}}{T}
+f_{\widetilde{H_u}}^{eq}e^{\frac{E_{\widetilde{H}_u}}{T}}
\frac{\mu_{\widetilde{H}_u}}{T}\right) 
\nonumber \\
&  & \times f_{\ell}^{eq}f_{\widetilde{H_u}}^{eq}
\left(1+\eta_{i}f_{i}^{eq}\right)\left(1+\eta_{j}f_{j}^{eq}\right) 
\nonumber \\ &  & 
\hspace{-0.3cm} 
-\overline{F_{\widetilde{H}_u\ell ij}}^{\rm sub}
\left(\frac{\mu_{i}+\mu_{j}}{T}-\frac{\mu_{\ell}
+\mu_{\widetilde{H}_u}}{T}\right)
f_{\ell}^{eq}f_{\widetilde{H}_u}^{eq}\left(1+\eta_{i}f_{i}^{eq}\right)
\left(1+\eta_{j}f_{j}^{eq}\right)
\nonumber \\ &  & 
\hspace{-0.3cm} 
+\left(F_{\widetilde{H}_u\ell ij}
\!-\overline{F_{\widetilde{H}_u\ell ij}}\right)^{\rm sub}
\!\!\! f_{\ell}^{eq}f_{\widetilde{H}_u}^{eq}
\left(1\!+\eta_{i}f_{i}^{eq}\right)
\left(1\!+\eta_{j}f_{j}^{eq}\right) 
% \nonumber \\ &  & \times 
\left(\eta_{i}f_{i}^{eq}
\frac{\mu_{i}}{T}+\eta_{j}f_{j}^{eq}\frac{\mu_{j}}{T}\right) ,\qquad \quad
\label{eq:sub1}
\end{eqnarray}
and, for the CP conjugate states
\begin{eqnarray}
&& \left[\overline{\widetilde{H}_u}\overline{\ell}
\leftrightarrow ij\right]^{\rm sub} 
% \nonumber \\ & = & 
= \left(F_{\overline{\widetilde{H}_u}\overline{\ell}ij}
-\overline{F_{\overline{\widetilde{H}_u}
\overline{\ell}ij}}\right)^{\rm sub}
\left(1 - f_{\ell}^{eq}e^{\frac{E_{\ell}}{T}}\frac{\mu_{\ell}}{T}
- f_{\widetilde{H}_u}^{eq}e^{\frac{E_{\widetilde{H}_u}}{T}}
\frac{\mu_{\widetilde{H}_u}}{T}\right)
\nonumber \\ &  & 
\times f_{\ell}^{eq}f_{\widetilde{H}_u}^{eq}
\left(1+\eta_{i}f_{i}^{eq}\right)\left(1+\eta_{j}f_{j}^{eq}\right)
\nonumber \\ &  & 
\hspace{-0.3cm} 
-\overline{F_{\overline{\widetilde{H}_u}\overline{\ell}ij}}^{\rm sub}
\left(\frac{\mu_{i}+\mu_{j}}{T}
+\frac{\mu_{\ell}+\mu_{\widetilde{H}_u}}{T}\right)
f_{\ell}^{eq}f_{\widetilde{H}_u}^{eq}\left(1+\eta_{i}f_{i}^{eq}\right)
\left(1+\eta_{j}f_{j}^{eq}\right)
\nonumber \\ &  &
\hspace{-0.3cm} 
+\left(F_{\overline{\widetilde{H}_u}\overline{\ell}ij}
-\overline{F_{\overline{\widetilde{H}_u}\overline{\ell}ij}}\right)^{\rm sub}
\!\!\!f_{\ell}^{eq}f_{\widetilde{H}_u}^{eq}
\left(1\!+\eta_{i}f_{i}^{eq}\right)
\left(1\!+\eta_{j}f_{j}^{eq}\right) 
% \nonumber \\ &  & \times
\left(\eta_{i}f_{i}^{eq}\frac{\mu_{i}}{T}
+\eta_{j}f_{j}^{eq}\frac{\mu_{j}}{T}\right),\qquad \quad
\label{eq:sub2} 
\end{eqnarray}
where $\eta_i = +1,-1$ for boson and fermion respectively.
From Eqs.~\eqref{eq:sub1} and \eqref{eq:sub2} and using CPT invariance, 
we obtain 
\bea
&& 
\sum_{ij} % \left\{ 
\left[\widetilde{H}_u\ell
\leftrightarrow ij\right]_-^{\rm sub} 
% -\left[\overline{\widetilde{H}_u}\overline{\ell}
% \leftrightarrow ij\right]^{\rm sub} 
% \right\}
% \nonumber \\ & = & 
= \sum_{ij} \left\{ 2\left(F_{\widetilde{H}_u\ell ij}
-\overline{F_{\widetilde{H}_u\ell ij}}\right)^{\rm sub}
f_{\ell}^{eq}f_{\widetilde{H_u}}^{eq}
\left(1+\eta_{i}f_{i}^{eq}\right)\left(1+\eta_{j}f_{j}^{eq}\right) \right.
\nonumber \\ &  & 
\hspace{-0.3cm}
+\left(\overline{F_{\widetilde{H}_u\ell ij}}
-\overline{F_{\overline{\widetilde{H}_u}\overline{\ell} ij}}\right)^{\rm sub}
\frac{\mu_{i}+\mu_{j}}{T}
f_{\ell}^{eq}f_{\widetilde{H}_u}^{eq}\left(1+\eta_{i}f_{i}^{eq}\right)
\left(1+\eta_{j}f_{j}^{eq}\right)
\nonumber \\ &  & 
\hspace{-0.3cm}
+\left(\overline{F_{\widetilde{H}_u\ell ij}}
+\overline{F_{\overline{\widetilde{H}_u}\overline{\ell} ij}}\right)^{\rm sub}
\frac{\mu_{\ell}+\mu_{\widetilde{H}_u}}{T}
f_{\ell}^{eq}f_{\widetilde{H}_u}^{eq}\left(1+\eta_{i}f_{i}^{eq}\right)
\left(1+\eta_{j}f_{j}^{eq}\right)
\nonumber \\ &  & 
\hspace{-0.3cm}
+2\left(F_{\widetilde{H}_u\ell ij}
\! -\overline{F_{\widetilde{H}_u\ell ij}}\right)^{\rm sub}
\!\!\!f_{\ell}^{eq}f_{\widetilde{H}_u}^{eq}
\left(1\!+\eta_{i}f_{i}^{eq}\right)
\left(1\!+\eta_{j}f_{j}^{eq}\right) 
% \nonumber \\ &  &  \times 
\left. 
\!\! \left(\eta_{i}f_{i}^{eq}
\frac{\mu_{i}}{T}+\eta_{j}f_{j}^{eq}\frac{\mu_{j}}{T}\right) \! \right\}.
\qquad
\label{eq:subrate}
\eea
For  $\widetilde H_u \ell \leftrightarrow ij$ 
with $\widetilde N_\pm$ exchanged in 
s-channel we have 
\begin{eqnarray}
&  & 
 \hspace{-.5cm}
\sum_{ij}\left(F_{\widetilde{H}_u\ell ij}^s
-\overline{F_{\widetilde{H}_u\ell ij}^s}\right)^{\rm sub}
\left(1+\eta_{i}f_{i}^{eq}\right)
\left(1+\eta_{j}f_{j}^{eq}\right)
 \nonumber \\ & = & 
\sum_{ij,k}\left[F_{\widetilde{H}_u\ell ij}^s
-F_{\widetilde{H}_u\ell\widetilde{N}_{k}}
\left(1+f_{\widetilde{N}_{k}}\right)\mbox{Br}
\left(\widetilde{N}_{k}\to ij\right)\right. 
\nonumber \\
&  & \left. -\overline{F_{\widetilde{H}_u\ell ij}^s}
+\overline{F_{\widetilde{H}_u\ell\widetilde{N}_{k}}}
\left(1+f_{\widetilde{N}_{k}}\right)\mbox{Br}
\left(\widetilde{N}_{k}\to ij\right)\right]
%\nonumber \\
% &  & \times 
\left(1+\eta_{i}f_{i}^{eq}\right)
\left(1+\eta_{j}f_{j}^{eq}\right) 
\nonumber \\
& = &
\sum_{k}\left(F_{\widetilde{N}_{k}\widetilde{H}_u\ell}
-F_{\widetilde{N}_{k}\overline{\widetilde{H}_u}\overline{\ell}}\right)
\left(1+f_{\widetilde{N}_{k}}\right),
\label{eq:ssub}
\end{eqnarray}
where $\mbox{Br}\left(\widetilde{N}_{k}\to ij\right)$ is the branching
ratio for the corresponding process, that satisfies the unitarity condition
\begin{eqnarray}
\sum_{ij}\mbox{Br}\left(\widetilde{N}_{k}\to ij\right)
\left(1+\eta_{i}f_{i}^{eq}\right)\left(1+\eta_{j}f_{j}^{eq}\right) & = & 1.
\nonumber 
\end{eqnarray}
In the last equality in Eq.~\eqref{eq:ssub}
we have neglected terms proportional to 
the CP asymmetry $\left(F_{\widetilde{H}_u\ell ij}^s
-F_{\overline{\widetilde{H}_u}\overline{\ell} ij}^s\right)$ 
which are of higher order  $\sim \mathcal{O}(Y^6)$. 
Substituting Eq.~\eqref{eq:ssub} into Eq.~\eqref{eq:subrate} and
ignoring the term of order $\mathcal{O}(\epsilon \frac{\mu}{T})$,
we have
\begin{eqnarray}
\sum_{ij}\left[\widetilde{H}_u\ell\leftrightarrow ij\right]_-^{\rm sub} 
& = & 2\sum_{k}\left(F_{\widetilde{N}_{k}\widetilde{H}_u\ell}
-F_{\widetilde{N}_{k}\overline{\widetilde{H}_u}\overline{\ell}}\right)
\frac{1+f_{\widetilde{N}_{k}}}{1+f_{\widetilde{N}_{k}}^{eq}}
  \nonumber \\ &  & \times 
f_{\widetilde{N}_{k}}^{eq}\left(1-f_{\ell}^{eq}\right)
\left(1-f_{\widetilde{H}_u}^{eq}\right) - W_{\Delta L=2}\, ,
\end{eqnarray}
where we have used the identity $f_{\ell}^{eq}f_{\widetilde{H}_u}^{eq}
=\frac{f_{\widetilde{N}_{k}}^{eq}}
{1+f_{\widetilde{N}_{k}}^{eq}}\left(1-f_{\ell}^{eq}\right)
\left(1-f_{\widetilde{H}_u}^{eq}\right)$ and
\bea
W_{\Delta L =2 } = - \sum_{ij} \! \left( \! F_{\widetilde{H}_u\ell ij}
\!+F_{\overline{\widetilde{H}_u}\overline{\ell}ij}\right)^{\rm sub}
\!\!\!
f_{\ell}^{eq}f_{\widetilde{H}_u}^{eq}\frac{\mu_{\ell}+\mu_{\widetilde{H}_u}}{T}
% \nonumber \\&  & \times 
\left(1\!+\eta_{i}f_{i}^{eq}\right)\!\left(1\!+\eta_{j}f_{j}^{eq}\right) \!.
\qquad \ 
\label{eq:off22}
\eea
Following the same procedure we also obtain
\begin{eqnarray}
\sum_{ij}\left[H_u\widetilde{\ell}\leftrightarrow ij\right]_{-}^{\rm sub}
& = & 2\sum_{k}\left(F_{\widetilde{N}_{k}H_u\widetilde\ell}
-F_{\widetilde{N}_{k}{H}_u^*\widetilde{\ell}^*}\right)
\frac{1+f_{\widetilde{N}_{k}}}{1+f_{\widetilde{N}_{k}}^{eq}} 
 \nonumber \\ &  & \times 
f_{\widetilde{N}_{k}}^{eq}\left(1+f_{\widetilde{\ell}}^{eq}\right)
\left(1+f_{H_u}^{eq}\right) - \widetilde{W}_{\Delta L=2}\, ,
\end{eqnarray}
where 
\begin{eqnarray}
\widetilde{W}_{\Delta L =2} = 
% & \equiv & 
-\sum_{ij} \! \left( \! F_{H_u\widetilde{\ell}ij}
\! +F_{H_u^{*}\widetilde{\ell}^{*}ij}\right)^{\rm sub}
\!\!\!
f_{\widetilde{\ell}}^{eq}f_{H_u}^{eq}
\frac{\mu_{\widetilde\ell}+\mu_{H_u}}{T}
% \nonumber \\ &  & \times 
\left(1\!+\eta_{i}f_{i}^{eq}\right)\!\left(1\!+\eta_{j}f_{j}^{eq}\right) \!.
\qquad\ 
\label{eq:off22t}
\end{eqnarray}

%%%%%%%%%%%%%%%%%%%%%%%%%%%%%%%%%%%%%%%%%%%%

\subsection{Lepton flavours and lepton flavour equilibration} 
\label{sec:flavor_lfe}

The unflavoured BE \eqref{eq:BE_deltal} and \eqref{eq:BE_deltatl} can
be easily generalized to the flavoured case. By denoting (s)lepton
flavours by $\alpha=e,\mu,\tau$ we simply have to replace
$Y_{\Delta\ell}\to Y_{\Delta\ell_\alpha}$, $\epsilon\to
\epsilon_\alpha$, $\gamma\to \gamma_\alpha$ (except for the
normalization factor of $\epsilon_\alpha$ that is still 
$\gamma_{\widetilde{N}}$), $ W_{\Delta L=2}\to W_{\Delta L=2}^\alpha $
etc. However, before writing the flavoured BE we now take some
further step.  Chemical equilibrium enforced by the top Yukawa
interactions implies
\be
-\mu_{Q}+\mu_{u} = \mu_{H_u},
\;\;\;\;
-\mu_{Q} + \mu_{\widetilde{u}} = \mu_{\widetilde{H}_u},
\;\;\;\;
-\mu_{\widetilde{Q}}+\mu_{u}  =  \mu_{\widetilde{H}_u}, 
\label{eq:top_equilibrium}
\ee
which also yields
\be
-\mu_{\widetilde{Q}}+\mu_{\widetilde u}  =  
2 \mu_{\widetilde{H}_u} -\mu_{H_u}.
\label{eq:top_equilibrium_deriv}
\ee
This equation and the first relation in \Eqn{eq:top_equilibrium} allow
to eliminate in the BE the chemical potentials of the (s)quarks for
those of the Higgs(inos).  It is also convenient to trade chemical
potentials $\mu_\phi$ for the corresponding density asymmetries
$Y_{\Delta\phi}$.  This can be done by using
Eqs.~\eqref{eq:chem_pot_asym}. For $f=\ell_\alpha,\widetilde{H}_u$ and
$s=\widetilde{\ell}_\alpha, H_u$ we have 
%!!%
%
\be
\frac{2\mu_{f}}{T} = \frac{Y_{\Delta f}}{Y_f^{eq}}
%\frac{2\mu_{f}}{T} = \frac{Y_{\Delta f}}{2Y_f^{eq}}
\equiv {\cal Y}_{\Delta f},
\;\;\;\;\;\quad
\frac{2\mu_{s}}{T} = \frac{Y_{\Delta s}}{Y_{s}^{eq}}
%\frac{2\mu_{s}}{T} = \frac{Y_{\Delta s}}{2Y_{s}^{eq}}
\equiv {\cal Y}_{\Delta s}, 
\label{eq:mu_to_Y}
\ee
where $2Y_f^{eq}=Y_{s}^{eq}=\frac{15}{4\pi^2 g_*}$.  
With these conventions the flavoured BE can be written as:
%
% \begin{equation}
\bea
2\dot{Y}_{\Delta\ell_\alpha} 
 &=&  E_\alpha + W_{\Delta L=2}^\alpha,
\nonumber \\
% \qquad\quad
2\dot{Y}_{\Delta \widetilde{\ell}_\alpha} 
 &=&  \widetilde{E}_\alpha +\widetilde{W}_{\Delta L=2}^\alpha,
\label{eq:BE_deltal_fla}
\eea
% \end{equation}
%
where:
\begin{eqnarray}
E_\alpha & = & \epsilon^f_\alpha\left(z\right)
\frac{\gamma_{\widetilde{N}}}{2}
\left(\frac{Y_{\widetilde{N}_{\rm tot}}}{Y_{\widetilde{N}}^{eq}}-2\right)
-\frac{\gamma_{\widetilde{N}}^{f,\alpha}}{2}
\left({\cal Y}_{\Delta\ell_\alpha}+{\cal Y}_{\Delta \widetilde H_u}\right)
-\frac{1}{4}\gamma_{N}^\alpha\left({\cal Y}_{\Delta\ell_\alpha}+
{\cal Y}_{\Delta H_u}
\right)
\nonumber \\
 &  & -\left(\gamma_{t}^{\left(3\right)\alpha}\frac{Y_{N}}{Y_{N}^{eq}}+ %!%
2\gamma_{t}^{\left(4\right)\alpha}+2\gamma_{t}^{\left(6\right)\alpha}+
2\gamma_{t}^{\left(7\right)\alpha}+\gamma_{t}^{\left(5\right)\alpha}
\frac{Y_{\widetilde{N}_{\rm tot}}}{Y_{\widetilde{N}}^{eq}}\right)
{\cal Y}_{\Delta\ell_\alpha}
\nonumber \\
 &  & -\left(\gamma_{t}^{\left(3\right)\alpha}
% +\gamma_{t}^{\left(4\right)\alpha} 
+\gamma_{t}^{\left(4\right)\alpha}
\left(1+\frac{Y_{N}}{Y_{N}^{eq}}\right)+
\gamma_{t}^{\left(5\right)\alpha}+\gamma_{t}^{\left(6\right)\alpha}+
\frac{1}{2}\gamma_{t}^{\left(7\right)\alpha}
\frac{Y_{\widetilde{N}_{\rm tot}}}{Y_{\widetilde{N}}^{eq}}\right)
{\cal Y}_{\Delta H_u}
\nonumber \\
 &  & -\left(\gamma_{t}^{\left(5\right)\alpha}+\gamma_{t}^{\left(7\right)\alpha}+
\frac{1}{2}\gamma_{t}^{\left(6\right)\alpha}
\frac{Y_{\widetilde{N}_{\rm tot}}}{Y_{\widetilde{N}}^{eq}}\right)
\left(2{\cal Y}_{\Delta \widetilde H_u}-{\cal Y}_{\Delta H_u}\right)
 \nonumber \\ &  & 
+ 2\gamma_{\tilde g}^{\rm eff}
\left({\cal Y}_{\Delta \widetilde \ell_\alpha}
-{\cal Y}_{\Delta\ell_\alpha}\right),
\label{eq:BE_Delta_lep}
\end{eqnarray}
and 
\begin{eqnarray}
\widetilde E_\alpha & = & 
\epsilon^s_\alpha\left(z\right)\frac{\gamma_{\widetilde{N}}}{2}
\left(\frac{Y_{\widetilde{N}_{\rm tot}}}{Y_{\widetilde{N}}^{eq}}-2\right)
-\frac{\gamma_{\widetilde{N}}^{s,\alpha}}{2}
\left({\cal Y}_{\Delta \widetilde \ell_\alpha}+{\cal Y}_{\Delta H_u}\right)
-\frac{1}{4}\gamma_{N}^\alpha\left({\cal Y}_{\Delta \widetilde \ell_\alpha}
+{\cal Y}_{\Delta \widetilde H_u}\right)
\nonumber \\
&  & -\left(\gamma_{\widetilde{N}}^{\left(3\right)\alpha}+
\frac{1}{2}\gamma_{22}^{\alpha}\frac{Y_{\widetilde{N}_{\rm tot}}}{Y_{\widetilde{N}}^{eq}}
+2\gamma_{22}^{\alpha}\right)\left({\cal Y}_{\Delta\widetilde{\ell}_\alpha}
+2{\cal Y}_{\Delta\widetilde{H}_u}-{\cal Y}_{\Delta H_u}
\right)\nonumber \\
&  & -\left(2\gamma_{t}^{\left(0\right)\alpha}\frac{Y_{N}}{Y_{N}^{eq}}+
2\gamma_{t}^{\left(1\right)\alpha}+2\gamma_{t}^{\left(2\right)\alpha}+
\frac{1}{2}\gamma_{t}^{\left(8\right)\alpha}
\frac{Y_{\widetilde{N}_{\rm tot}}}{Y_{\widetilde{N}}^{eq}}+
2\gamma_{t}^{\left(9\right)k}\right){\cal Y}_{\Delta \widetilde \ell_\alpha}
\nonumber \\
 &  & -\left(\gamma_{t}^{\left(0\right)\alpha}+\gamma_{t}^{\left(1\right)\alpha}
\frac{Y_{N}}{Y_{N}^{eq}}+\gamma_{t}^{\left(8\right)\alpha}
+\gamma_{t}^{\left(9\right)\alpha}
\left(1+\frac{1}{2} % \gamma_{t}^{\left(9\right)\alpha}
\frac{Y_{\widetilde{N}_{\rm tot}}}{Y_{\widetilde{N}}^{eq}}
\right)
\right){\cal Y}_{\Delta H_u}
\nonumber \\
 &  & -\left(\gamma_{t}^{\left(0\right)\alpha}+\gamma_{t}^{\left(1\right)\alpha}+
\gamma_{t}^{\left(2\right)\alpha}\frac{Y_{N}}{Y_{N}^{eq}}\right)
\left(2{\cal Y}_{\Delta \widetilde H_u}-
{\cal Y}_{\Delta H_u}
\right)
\nonumber \\
 &  & -2 \gamma_{\tilde g}^{\rm eff}
\left({\cal Y}_{\Delta \widetilde \ell_\alpha}
-{\cal Y}_{\Delta\ell_\alpha}\right). 
\label{eq:BE_Delta_slep}
\end{eqnarray}

\subsubsection{Lepton flavour equilibrating  interactions} 

The off-diagonal soft slepton masses induce lepton flavour violating 
(LFV) interactions through the exchange of $SU(2)_L$ gauginos
$\widetilde{\lambda}_{2}^{a}$ and $U(1)_Y$ gaugino
$\widetilde{\lambda}_{1}$ (see the Lagrangian
\eqref{eq:lgaugino}), and this can result in lepton flavour equilibration
(LFE).  There are two t-channel scatterings
$\ell_{\alpha}\overline{P}\leftrightarrow\widetilde{\ell}_{\beta}\widetilde{P}^*$,
$\ell_{\alpha}\widetilde{P}\leftrightarrow\widetilde{\ell}_{\beta}P$
and one s-channel scattering
$\ell_{\alpha}\widetilde{\ell}_{\beta}^{*} \leftrightarrow
P\widetilde{P}$ as shown in Fig. \ref{fig:lfv_diagrams}.  
We denote fermions as $P$ and scalars as $\widetilde{P}$.
For processes mediated by $SU(2)_L$ gauginos
$P=\ell,Q,\widetilde{H}_{u,d}$, while for $U(1)_Y$
gaugino one must include the $SU(2)_L$ singlet states $P=e,u,d$ as
well.  We have in general
\begin{eqnarray}
\left[\ell_{\alpha}P\leftrightarrow
\widetilde{\ell}_{\beta}\widetilde{P}\right]_{-} 
& = & -2\overline{F_{\ell_{\alpha}P\widetilde{\ell}_{\beta}\widetilde{P}}}
\left|R_{\alpha\beta}\right|^{2}
\left(\frac{\mu_{\widetilde{\ell}_{\beta}}+\mu_{\widetilde{P}}}{T}
-\frac{\mu_{\ell_{\alpha}}+\mu_{P}}{T}\right) 
\nonumber \\
&  & f_{\ell_{\alpha}}^{eq}f_{P}^{eq}\left(1+f_{\widetilde{\ell}_{\beta}}^{eq}\right)
\left(1+f_{\widetilde{P}}^{eq}\right),
\nonumber \\
\left[\ell_{\alpha}\widetilde{P}\leftrightarrow
\widetilde{\ell}_{\beta}P\right]_{-} 
& = & -2\overline{F_{\ell_{\alpha}\widetilde{P}\widetilde{\ell}_{\beta}P}}
\left|R_{\alpha\beta}\right|^{2}
\left(\frac{\mu_{\widetilde{\ell}_{\beta}}
+\mu_{P}}{T}-\frac{\mu_{\ell_{\alpha}}+\mu_{\widetilde{P}}}{T}\right) 
\nonumber \\
&  & f_{\ell_{\alpha}}^{eq}f_{\widetilde{P}}^{eq}
\left(1+f_{\widetilde{\ell}_{\beta}}^{eq}\right)\left(1-f_{P}^{eq}\right),
\nonumber \\
\left[\ell_{\alpha}\widetilde{\ell}_{\beta}^{*}
\leftrightarrow P\widetilde{P}\right]_{-} 
& = & -2\overline{F_{\ell_{\alpha}\widetilde{\ell}_{\beta}^{*}P
\widetilde{P}}}\left|R_{\alpha\beta}\right|^{2}
\left(\frac{\mu_{P}+\mu_{\widetilde{P}}}{T}
-\frac{\mu_{\ell_{\alpha}}-\mu_{\widetilde{\ell}_{\beta}}}{T}\right) 
\nonumber \\
&  & f_{\ell_{\alpha}}^{eq}f_{\widetilde{\ell}_{\beta}}^{eq}\left(1-f_{P}^{eq}\right)
\left(1+f_{\widetilde{P}}^{eq}\right),
\label{eq:gen_LFV}
\end{eqnarray}
where the factors of two come from the CP conjugate processes.
% (CP violation is irrelevant in here and is neglected). 
Each $\ell_{\alpha}\widetilde{\ell}_{\beta}-\mbox{gaugino}$ vertex
involves one element $R_{\alpha\beta}$ of a unitary matrix.  
In Eq.~\eqref{eq:gen_LFV} we have explicitly factored out the entries
$\left|R_{\alpha\beta}\right|^{2}$ and hence, if we ignore the zero
temperature lepton and slepton masses,
$\overline{F_{\ell_{\alpha}P\widetilde{\ell}_{\beta}\widetilde{P}}}\left(...\right)$,
$\overline{F_{\ell_{\alpha}\widetilde{P}\widetilde{\ell}_{\beta}P}}\left(...\right)$
and
$\overline{F_{\ell_{\alpha}\widetilde{\ell}_{\beta}^{*}P\widetilde{P}}}\left(...\right)$
are flavour independent.
With the same approximation the distributions $f_{\ell_\alpha}^{eq}$
and $f_{\widetilde\ell_\alpha}^{eq}$ are also flavour independent thus, 
from now on, we will drop the flavour index whenever possible.
For simplicity, we only keep the thermal masses
$m_{\widetilde{\lambda}_{2}}$ and $m_{\widetilde{\lambda}_{Y}}$ of the
$SU(2)_L$ and $U(1)_Y$ gauginos. With this approximations, we can
define the flavour independent LFE reaction
densities % (where we factor out e.g. $|R_{\alpha\beta}|^2$) 
as follows
\begin{eqnarray}
\gamma_{t1,G} & \equiv & 
\overline{F_{\ell P \widetilde{\ell}\widetilde{P}}}(g_G)
f_{\ell}^{eq}f_{P}^{eq}
\left(1+f_{\widetilde{\ell}}^{eq}\right)
\left(1+f_{\widetilde{P}}^{eq}\right),
\nonumber \\
\gamma_{t2,G} & \equiv & 
\overline{F_{\ell \widetilde P \widetilde{\ell}P}}(g_G)
f_{\ell}^{eq}f_{\widetilde P}^{eq}
\left(1+f_{\widetilde{\ell}}^{eq}\right)
\left(1-f_{P}^{eq}\right),
\nonumber \\
\gamma_{s,G} & \equiv & 
\overline{F_{\ell \widetilde\ell^* P\widetilde{P}}}(g_G)
f_{\ell}^{eq}f_{\widetilde\ell}^{eq}
\left(1-f_{P}^{eq}\right)
\left(1+f_{\widetilde{P}}^{eq}\right),
\end{eqnarray}
where $G=2,Y$ for the scatterings mediated respectively by the
$SU(2)_L$ and $U(1)_Y$ gauginos and correspondingly $g_G=g_2$ or
$g_Y$. For example, to rack the evolution of the abundance
of $\ell_{\alpha}$  for $P=\ell$ $P=\ell$ we need the following terms:
\begin{eqnarray}
&&
\hspace{-.8cm}
\sum_{g_{\ell},\zeta,\beta,\eta}\left[\ell_{\alpha}
\overline{\ell}_{\zeta}\leftrightarrow\widetilde{\ell}_{\beta}
\widetilde{\ell}_{\eta}^{*}\right]_{-}\!\!= 
-2\Pi_{\ell}\,\gamma_{t1}^{G} 
%\nonumber \\
%&  & 
%\hspace{-12mm}  \times
\left(\!3\sum_{\beta}\left|R_{\alpha\beta}\right|^{2}
\frac{\mu_{\widetilde{\ell}_{\beta}}-\mu_{\ell_{\alpha}}}{T}
+\sum_{\zeta}\frac{\mu_{\ell_{\zeta}}
-\mu_{\widetilde{\ell}_{\zeta}}}{T}\right)\!\!,\qquad 
\label{eq:LFV_ll1} 
\end{eqnarray}
\begin{eqnarray}
&&
\hspace{-.8cm}
\sum_{g_{\ell},\zeta,\beta,\eta}\left[\ell_{\alpha}\widetilde{\ell}_{\zeta}
\leftrightarrow\widetilde{\ell}_{\beta}\ell_{\eta}\right]_{-} 
\!\!= 
-2\Pi_{\ell}\,\gamma_{t2}^{G} 
%% \times
\left(\!3\sum_{\beta}\left|R_{\alpha\beta}\right|^{2}
\frac{\mu_{\widetilde{\ell}_{\beta}}-\mu_{\ell_{\alpha}}}{T}
+\sum_{\zeta}\frac{\mu_{\ell_{\zeta}}
-\mu_{\widetilde{\ell}_{\zeta}}}{T}\right)\!\!,\qquad 
\label{eq:LFV_ll2} 
\end{eqnarray}
\begin{eqnarray}
&&
\hspace{-.8cm}
\sum_{g_{\ell},\zeta,\beta,\eta}
\left[\ell_{\alpha}\widetilde{\ell}_{\beta}^{*}
\leftrightarrow\widetilde{\ell}_{\zeta}^{*}\ell_{\eta}\right]_{-} 
 \!\! =  
-2\Pi_{\ell}\,\gamma_{s}^{G} 
%\nonumber \\
%&  & \hspace{-15mm}\times
\left(\!3\sum_{\beta}\left|R_{\alpha\beta}\right|^{2}
\frac{\mu_{\widetilde{\ell}_{\beta}}-\mu_{\ell_{\alpha}}}{T}
+\sum_{\zeta}\frac{\mu_{\ell_{\zeta}}
-\mu_{\widetilde{\ell}_{\zeta}}}{T}\right)\!\!.\qquad
\label{eq:LFV_ll3}
\end{eqnarray}
where we have used the following properties of unitary matrices: 
\begin{eqnarray}
\sum_{\beta}\left|R_{\alpha\beta}\right|^{2} 
& = & \delta_{\alpha\alpha},\;\;\;\mbox{(no sum over $\alpha$)},  
% \nonumber \\
\qquad 
\sum_{\alpha,\beta}\left|R_{\alpha\beta}\right|^{2} = 3.
% & = & 3.
\end{eqnarray}
In Eqs.~\eqref{eq:LFV_ll1}-\eqref{eq:LFV_ll3} $\Pi_{\ell}$ is a factor
arising from summing over isospin degrees of freedom of leptons and
sleptons, and for example for the scatterings mediated by $\widetilde
\lambda_2^a$ we have $\Pi_{\ell}=3$.
Note that 
since $\widetilde\ell_\alpha - \ell_\alpha = \mu_{\widetilde g}$, for
$\beta=\alpha$ the sum of chemical potentials always cancel (soft
slepton masses can only induce LFV interactions but not
superequilibration).  Hence
Eqs.~\eqref{eq:LFV_ll1}--\eqref{eq:LFV_ll3} simply become
\begin{eqnarray}
\sum_{g_{\ell},\zeta,\beta,\eta}\left[\ell_{\alpha}
\overline{\ell}_{\zeta}\leftrightarrow\widetilde{\ell}_{\beta}
\widetilde{\ell}_{\eta}^{*}\right]_{-} 
& = & -6\Pi_{\ell}\,\gamma_{t1}^{G} 
\sum_{\beta\neq\alpha}\left|R_{\alpha\beta}\right|^{2}
\frac{\mu_{\widetilde{\ell}_{\beta}}-\mu_{\ell_{\alpha}}}{T},
\label{eq:LFV_ll1s} \\
\sum_{g_{\ell},\zeta,\beta,\eta}\left[\ell_{\alpha}\widetilde{\ell}_{\zeta}
\leftrightarrow\widetilde{\ell}_{\beta}\ell_{\eta}\right]_{-} 
& = & -6\Pi_{\ell}\,\gamma_{t2}^{G} 
\sum_{\beta \neq\alpha}\left|R_{\alpha\beta}\right|^{2}
\frac{\mu_{\widetilde{\ell}_{\beta}}-\mu_{\ell_{\alpha}}}{T},
\label{eq:LFV_ll2s} \\
\sum_{g_{\ell},\zeta,\beta,\eta}\left[\ell_{\alpha}\widetilde{\ell}_{\beta}^{*}
\leftrightarrow\widetilde{\ell}_{\zeta}^{*}\ell_{\eta}\right]_{-} 
& = & -6\Pi_{\ell}\,\gamma_{s}^{G} 
\sum_{\beta\neq\alpha}\left|R_{\alpha\beta}\right|^{2}
\frac{\mu_{\widetilde{\ell}_{\beta}}-\mu_{\ell_{\alpha}}}{T}.
\label{eq:LFV_ll3s}
\end{eqnarray}

Similarly, for processes mediated by $SU(2)_L$ gauginos we have the
scatterings with $P=Q,\widetilde{H}_{u,d}$ for processes mediated by
$U(1)_Y$ gauginos $P=Q,\widetilde{H}_{u,d},e,u,d$.  The corresponding
term that has to be added to the BE for $Y_{\Delta \ell_\alpha}$
is
\begin{eqnarray}
% && 
\left(\dot{Y}_{\Delta \ell_\alpha}\right)_{\!\!LFE}  
\!\!\! = -\!\!\!
\sum_{G=2,Y}
\! n_G\left(\gamma_{t1}^{G}+\gamma_{t2}^{G}
+\gamma_{s}^{G}\right)
 \sum_{\beta\neq\alpha}
 \left|R_{\alpha\beta}\right|^{2}
\left(
{\cal Y}_{\Delta\ell_{\alpha}}- 
{\cal Y}_{\Delta\widetilde\ell_{\beta}} 
\right),\qquad 
\label{eq:ellLFE}
\end{eqnarray}
with $n_2=42$ and $n_Y=38$.  The analogous term for
$\dot{Y}_{\Delta\widetilde{\ell}_{\alpha}}$ can be obtained from
\Eqn{eq:ellLFE} by exchanging the asymmetry labels
${\ell_{\alpha}}\to {\widetilde\ell_{\alpha}}$, $
{\widetilde\ell_{\beta}} \to {\ell_{\beta}}$.  Given that the LFV
factor $\left|R_{\alpha\beta}\right|^{2}$ is explicitly accounted for
in the terms above, the reduced cross sections for the LFE
interactions can be defined in a flavour independent way:
\begin{eqnarray}
\hat{\sigma}_{t1}^{G}\left(s\right) 
& = & \frac{g_{G}^{4}}{8\pi}\left[
\left(\frac{2m_{\widetilde{\lambda}_{G}}^{2}}{s}+1\right)
\ln\left|\frac{m_{\widetilde{\lambda}_{G}}^{2}+s}
{m_{\widetilde{\lambda}_{G}}^{2}}\right|-2\right],
\nonumber \\
\hat{\sigma}_{t2}^{G}\left(s\right) 
& = & \frac{g_{G}^{4}}{8\pi}\left[
\ln\left|\frac{m_{\widetilde{\lambda}_{G}}^{2}+s}
{m_{\widetilde{\lambda}_{G}}^{2}}\right|
-\frac{s}{m_{\widetilde{\lambda}_{G}}^{2}+s}\right],
\nonumber \\
\hat{\sigma}_{s}^{G}\left(s\right) 
& = & \frac{g_{G}^{4}}{16\pi}\left(\frac{s}
{s-m_{\widetilde{\lambda}_{G}}^{2}}\right)^{2},
\end{eqnarray}
where the gaugino thermal mass is
$m_{\widetilde{\lambda}_{G}}^{2}=\left(9/2\right)g_{G}^{2}T^{2}$,  and
quantum statistical factors have been neglected.

%%%%%%%%%%%%%%%%%%%%%%%%%%%%%%%%%%%%%%%%%%%%%%%%%%%%

\subsection{The superequilibration regime} 
\label{sec:BE_SE}

Superequilibration (SE)  is defined by chemical potentials condition 
\begin{eqnarray}
\mu_{\phi} & = & \mu_{\widetilde \phi}.
\label{eq:superequilibration}
\end{eqnarray}
where $(\phi,\widetilde\phi)$ are the fermion and boson components of
a generic supermultiplet.  In the SE regime, the
BE for the RHN and RHSN abundances are still given respectively by
Eqs.~\eqref{eq:int_BE_RHN} and \eqref{eq:int_BE_RHSN}.  However, 
Eqs.~\eqref{eq:top_equilibrium} combined with
Eq.~\eqref{eq:superequilibration} yields 
\begin{eqnarray}
\mu_{Q}-\mu_{u} = \mu_{\widetilde{Q}}-\mu_{\widetilde{u}} & = & 
-\mu_{\widetilde{H}_u} = -\mu_{H_u},
% \nonumber \\
% \mu_{Q}-\mu_{u} & = & -\mu_{\widetilde{H}_u}=-\mu_{H_u},
\end{eqnarray}
and, as we will now see, this allows us to sum up the two BE
\eqref{eq:BE_deltal_fla}
% and \eqref{eq:BE_deltatl_fla} 
into a single BE since 
with SE the relation between scalars ($s$) and fermions ($f$)
density asymmetries is $Y_{\Delta s} = 2Y_{\Delta f}$ which implies
$Y_{\Delta \widetilde\ell_\alpha} = 2Y_{\Delta \ell_\alpha}$
and $Y_{\Delta \widetilde H_u} = Y_{\Delta H_u}/2$.

Summing up the two equations  \eqref{eq:BE_deltal_fla} 
and including LFE effects 
we obtain:
\begin{eqnarray}
\dot{Y}_{\Delta\ell_{\rm tot}^\alpha} & = & 
\left(E_\alpha+\widetilde E_\alpha\right)_{SE}
%+ W_{\Delta L=2}^\alpha + \widetilde{W}_{\Delta L=2}^\alpha 
+ \left(\dot{Y}_{\Delta\ell_{\rm tot}^\alpha}\right)_{LFE}, \quad  
\label{eq:BE_L_tot_LFE}
\end{eqnarray}
where $Y_{\Delta \ell_{\rm tot}^\alpha} = 2 \left( Y_{\Delta \ell_\alpha} 
+ Y_{\Delta \widetilde\ell_\alpha} \right)$ 
and we have dropped the $\Delta L=2$ off-shell scattering terms
Eqs.~\eqref{eq:off22} and \eqref{eq:off22t} which 
in  SL are always negligible. 
The LFE term is obtained by summing \Eqn{eq:ellLFE} to the analogous term  
$(\dot{Y}_{\Delta\widetilde{\ell}_\alpha})_{LFE}$ which gives: 
\begin{eqnarray}
\left(\dot{Y}_{\Delta\ell_{\rm tot}^\alpha}\right)_{\!\!LFE} 
= -2\sum_{G=2,Y}
\! n_G\left(\gamma_{t1}^{G}+\gamma_{t2}^{G}
+\gamma_{s}^{G}\right)
 \sum_{\beta\neq\alpha}
 \left|R_{\alpha\beta}\right|^{2}
\left(
{\cal Y}_{\Delta\ell_{\alpha}}
- 
{\cal Y}_{\Delta \ell_{\beta}} 
\right),\qquad 
\label{eq:totellLFE}
\end{eqnarray}
while summing up \Eqn{eq:BE_Delta_lep} and \eqref{eq:BE_Delta_slep}
under the assumption of SE yields: 
\begin{eqnarray}
\left(E_\alpha+\widetilde E_\alpha\right)_{SE} 
& = &
\epsilon_{\alpha}\left(z\right)\frac{\gamma_{\widetilde{N}}}{2}
\left(\frac{Y_{\widetilde{N}_{\rm tot}}}{Y_{\widetilde{N}}^{eq}}-2\right)
\nonumber \\
& & - \left[\frac{\gamma^\alpha_{\widetilde{N}}}{2}
+ \frac{\gamma^\alpha_{N}}{2}+
\gamma_{\widetilde{N}}^{\left(3\right)\alpha}+
\left(\frac{1}{2}\frac{Y_{\widetilde{N}_{\rm tot}}}
{Y_{\widetilde{N}}^{eq}}
+2\right)\gamma^\alpha_{22}\right] 
\left(
{\cal Y}_{\Delta\ell_{\alpha}}
+
{\cal Y}_{\Delta \widetilde H_u}
\right)
\nonumber \\
&& -2\left(\gamma_{t}^{\left(1\right)\alpha}
+\gamma_{t}^{\left(2\right)\alpha}
+\gamma_{t}^{\left(4\right)\alpha}+\gamma_{t}^{\left(6\right)\alpha}
+\gamma_{t}^{\left(7\right)\alpha}+\gamma_{t}^{\left(9\right)\alpha}\right)
{\cal Y}_{\Delta\ell_{\alpha}}
 \nonumber \\ &  & 
-\left[\left(2\gamma_{t}^{\left(0\right)}
+\gamma_{t}^{\left(3\right)\alpha}\right)\frac{Y_{N}}{Y_{N}^{eq}}
+\left(\gamma_{t}^{\left(5\right)\alpha}
+\frac{1}{2}\gamma_{t}^{\left(8\right)\alpha}\right)
\frac{Y_{\widetilde{N}_{\rm tot}}}{Y_{\widetilde{N}}^{eq}}\right]
{\cal Y}_{\Delta\ell_{\alpha}}
 \nonumber \\  &  & 
 \hspace{-25mm}
-\left(2\gamma_{t}^{\left(0\right)\alpha}
+\gamma_{t}^{\left(1\right)\alpha}
+\gamma_{t}^{\left(3\right)\alpha}+\gamma_{t}^{\left(4\right)\alpha}
+2\gamma_{t}^{\left(5\right)\alpha}\right.
% \nonumber \\ && 
\left.+\gamma_{t}^{\left(6\right)\alpha}
+\gamma_{t}^{\left(7\right)\alpha}+\gamma_{t}^{\left(8\right)\alpha}
+\gamma_{t}^{\left(9\right)\alpha}\right)
{\cal Y}_{\Delta \widetilde H_u}
 \nonumber \\
 &  & \hspace{-25mm}-\left[\left(\gamma_{t}^{\left(1\right)\alpha}
+\gamma_{t}^{\left(2\right)\alpha}
+\gamma_{t}^{\left(4\right)\alpha}\right)\frac{Y_{N}}{Y_{N}^{eq}}
+\frac{1}{2}\left(\gamma_{t}^{\left(6\right)\alpha}
+\gamma_{t}^{\left(7\right)\alpha}+\gamma_{t}^{\left(9\right)\alpha}\right)
\frac{Y_{\widetilde{N}_{\rm tot}}}{Y_{\widetilde{N}}^{eq}}\right]
{\cal Y}_{\Delta \widetilde H_u}\,.  
\label{eq:BE_L_tot_fla2}
\end{eqnarray}
%

%!!%
We note at this point that Eq.~\eqref{eq:BE_L_tot_LFE} is in fact
incomplete, since ${\Delta\ell_\alpha}$ is also violated by EW
sphalerons, however, no term accounting for these processes has been
included.  This can be corrected by writing instead BE for the
flavour charge asymmetries $Y_{\Delta_{\alpha}} \equiv Y_{\Delta B}/3 
-Y_{\Delta L_{\alpha}}$ that are violated only by the RHN and RHSN
interactions appearing in the r.h.s of Eq.~\eqref{eq:BE_L_tot_fla2}.
Here
$
Y_{\Delta L_\alpha} 
\equiv 2\left( Y_{\Delta\ell_\alpha} 
+ Y_{\Delta\widetilde\ell_\alpha} \right)
+ Y_{\Delta e_\alpha} + Y_{\Delta\widetilde e_\alpha}
$
and the contributions of $e_\alpha$ and
$\widetilde e_\alpha$ to the total flavour asymmetries
have to be included because scatterings with the Higgs induced by the
charged lepton Yukawa couplings can transfer part of the asymmetry
generated in the $SU(2)_L$ lepton doublets to the singlets.
To take into account the EW sphalerons, 
we write down the complete BE for 
$Y_{\Delta L_\alpha}$ and $Y_{\Delta B}$:
\bea
\label{eq:L_full}
\dot{Y}_{\Delta L_\alpha} & = & 
\left(\dot{Y}_{\Delta L_\alpha}\right)_{\rm pert}
+ \left( \dot{Y}_{\Delta L_\alpha} \right)_{\rm non-pert}, 
\\
\label{eq:B_full}
\dot{Y}_{\Delta B} & = & \left( \dot{Y}_{\Delta B} \right)_{\rm
  non-pert}, \eea where `pert' refers to the violation of $\Delta L$
from perturbative interactions, i.e.  the r.h.s. of
Eq.~\eqref{eq:BE_L_tot_LFE}, while `non-pert' refers to the violation
of $\Delta L$ and $\Delta B$ from the non-perturbative EW sphaleron
processes.  Since the EW sphalerons conserve $B/3-L_\alpha$ we have
\be 
\frac{1}{3}\left( \dot{Y}_{\Delta B} \right)_{\rm non-pert}
-\left( \dot{Y}_{\Delta L_\alpha} \right)_{\rm non-pert} = 0, 
\ee 
and then by taking the difference between Eqs.~\eqref{eq:L_full} and
\eqref{eq:B_full} with the proper factor of 1/3, we arrive at \be
\dot{Y}_{\Delta_\alpha} = - \left( \dot{Y}_{\Delta L_\alpha}
\right)_{\rm pert}.
\label{eq:BE_Delta}
\ee
To get the BE for $Y_{\Delta_\alpha}$ in closed form, the
density asymmetries $Y_{\Delta \ell_\alpha}$ and $Y_{\Delta \widetilde H_u}$
multiplying the washout reactions on the r.h.s.  of
\Eqn{eq:BE_Delta} must then be expressed in terms of
$Y_{\Delta_\alpha}$ according to
\begin{equation}
\label{eq:leptonHiggs2}
{\cal Y}_{\Delta\ell_{\alpha}}
=  
\sum_{\beta}
A^\ell_{\alpha\beta}
{\cal Y}_{\Delta_{\beta}},\qquad\qquad 
{\cal Y}_{\Delta H_u}   
= \sum_{\beta}C^{\widetilde H_{u}}_{\beta}
{\cal Y}_{\Delta_\beta}\,.  
\nonumber 
\end{equation}
The values of the entries in the matrices $A^\ell$ and $C^{\widetilde
  H_{u}}$ depend on the particular set of reactions that are in
equilibrium when leptogenesis is taking place, and are given
in~\ref{app:se_new}.

\subsection{The non-superequilibration regime}
\label{Appendix-B}

The BE describing the RHN and RHSN abundances in the
non-superequilibration (NSE) regime (when $\mu_\phi\neq
\mu_{\widetilde\phi}$) are still given by Eqs.~\eqref{eq:int_BE_RHN}
and \eqref{eq:int_BE_RHSN}.  The evolution of the flavour charges
$Y_{\Delta_\alpha}$ is given by:
\begin{eqnarray}
\dot{Y}_{\Delta_\alpha} &=
& -\left(E_\alpha+\widetilde E_\alpha\right),
\label{eq:BE_Delta_alpha} 
\end{eqnarray}
where $E_\alpha$ and $\widetilde E_\alpha$ are respectively  given in
 \Eqns{eq:BE_Delta_lep}{eq:BE_Delta_slep}, 
and the $\Delta L=2$
$W$-terms as well as LFE effects (that are irrelevant at $T\gsim 10^7\,$GeV) 
have been neglected.
To derive the BE for the evolution of $R_B$ and $R_\chi$ defined
 in~\Eqns{eq:RB}{eq:chiralup} we need to know by which amount these
 charges are violated in the different scattering processes.  This
 information is collected in the following table:
\begin{displaymath}
\begin{array}{|l|l|l|}
\hline
\rm Reaction & \Delta R_B & \Delta R_3 \\ \hline
\gamma_{22}^{\alpha} 
\! \equiv \! \gamma \!\left(\!\widetilde{N}_{\pm}\widetilde{\ell}_{\alpha}\leftrightarrow\widetilde{Q}\widetilde{u}^*\right)
\!=\!\gamma \!\left(\!\widetilde{N}_{\pm}\widetilde{Q}^{*}\leftrightarrow\widetilde{\ell}_{\alpha}^{*}\widetilde{u}^*\right) 
\!=\!\gamma\!\left(\!\widetilde{N}_{\pm}\widetilde{u}\leftrightarrow\widetilde{\ell}_{\alpha}^{*}\widetilde{Q}\right)\quad\!\!
& 0 & 1 \\
\gamma_{\widetilde{N}}^{\left(3\right)\alpha} \! \equiv \!
\gamma\left(\!\widetilde{N}_{\pm}\leftrightarrow \widetilde{u}^*\widetilde{\ell}_{\alpha}^{*}\widetilde{Q}\right)
& 0 & 1 \\ %!%
\gamma_{t}^{(0)\alpha} 
 \equiv  \gamma\left(N\widetilde{\ell}_{\alpha}\leftrightarrow Q\widetilde{u}^*\right)
=\gamma\left(N\widetilde{\ell}_{\alpha}\leftrightarrow\widetilde{Q}\overline{u}\right)
& -1 & 0 \\
\gamma_{t}^{(1)\alpha} 
\equiv  \gamma\left(N\overline{Q}\leftrightarrow\widetilde{\ell}_{\alpha}^{*}\widetilde{u}^*\right)
=\gamma\left(Nu\leftrightarrow\widetilde{\ell}_{\alpha}^{*}\widetilde{Q}\right)
& -1 & 0 \\
\gamma_{t}^{(2)\alpha} 
\equiv  \gamma\left(N\widetilde{u}\leftrightarrow\widetilde{\ell}_{\alpha}^{*}Q\right)
=\gamma\left(N\widetilde{Q}^{*}\leftrightarrow\widetilde{\ell}_{\alpha}^{*}\overline{u}\right)
& -1 & 0 \\
\gamma_{t}^{(3)\alpha} 
 \equiv  \gamma\left(N\ell_{\alpha}\leftrightarrow Q\overline{u}\right)
& -1 & 0 \\
\gamma_{t}^{(4)\alpha} 
\equiv  \gamma\left(Nu\leftrightarrow\overline{\ell_{\alpha}}Q\right)
=\gamma\left(N\overline{Q}\leftrightarrow\overline{\ell_{\alpha}}\overline{u}\right)
& -1 & 0  \\
\gamma_{t}^{(5)\alpha} 
 \equiv  \gamma\left(\widetilde{N}_{\pm}\ell_{\alpha}\leftrightarrow Q\widetilde{u}^*\right)
=\gamma\left(\widetilde{N}_{\pm}\ell_{\alpha}\leftrightarrow\widetilde{Q}\overline{u}\right)
& 0 & 1  \\
\gamma_{t}^{(6)\alpha} 
\equiv  \gamma\left(\widetilde{N}_{\pm}\widetilde{u}\leftrightarrow\overline{\ell_{\alpha}}Q\right)
=\gamma\left(\widetilde{N}_{\pm}\widetilde{Q}^{*}\leftrightarrow\overline{\ell_{\alpha}}\overline{u}\right)
& 0 & 1 \\
\gamma_{t}^{(7)\alpha} 
 \equiv  \gamma\left(\widetilde{N}_{\pm}\overline{Q}\leftrightarrow\overline{\ell_{\alpha}}\widetilde{u}^*\right)
=\gamma\left(\widetilde{N}_{\pm}u\leftrightarrow\overline{\ell_{\alpha}}\widetilde{Q}\right)
& 0 & 1  \\
\gamma_{t}^{(8)\alpha} 
\equiv  \gamma\left(\widetilde{N}_{\pm}\widetilde{\ell}_{\alpha}^{*}\leftrightarrow\overline{Q}u\right)
& 2 & 1  \\
\gamma_{t}^{(9)\alpha} 
\equiv  \gamma\left(\widetilde{N}_{\pm}Q\leftrightarrow\widetilde{\ell}_{\alpha}u\right)
=\gamma\left(\widetilde{N}_{\pm}\overline{u}\leftrightarrow\widetilde{\ell}_{\alpha}\overline{Q}\right) 
& 2 & 1 \\
\hline
\end{array}
\end{displaymath}
The evolution equation for $Y_{\Delta {R_B}}$ and 
$Y_{\Delta {R_\chi}}$ are: 
\begin{eqnarray}
\dot{Y}_{\Delta {R_B}} 
 &=& \sum_\alpha 
\left(2\widetilde F_\alpha + F_\alpha\right)
-\gamma^{\rm eff}_{\widetilde g}\,{\cal Y}_{\Delta \widetilde g} ,
\label{eq:BE_Delta_RB} 
\\  
\dot{Y}_{\Delta {R_\chi}} 
 & = & \frac{1}{3} \sum_\alpha  
\left(\widetilde G_\alpha - G_\alpha\right) \!\!
-\frac{\gamma^{\rm eff}_{\widetilde g}}{3}\, {\cal Y}_{\Delta \widetilde g}
+\frac{\gamma^{\rm eff}_{\submuh}}{3} \!
\left({\cal Y}_{\Delta\widetilde H_u} \!
+{\cal Y}_{\Delta\widetilde H_d}\right)\!,
\label{eq:BE_Delta_R3} 
\end{eqnarray}
where the SE rates $\gamma^{\rm eff}_{\widetilde g}$
and $\gamma^{\rm eff}_{\submuh}$ have been also included.
$F_{\alpha}$ and $\widetilde F_{\alpha}$ are given by:
\begin{eqnarray}
F_{\alpha} & = & -\frac{\gamma_{N}^\alpha}{4}\left({\cal Y}_{\Delta\ell_\alpha}
+{\cal Y}_{\Delta H_u}\right)
-\left(\gamma_{t}^{\left(3\right)\alpha}\frac{Y_{N}}{Y_{N}^{eq}}+
2\gamma_{t}^{\left(4\right)\alpha}\right)
{\cal Y}_{\Delta\ell_\alpha}
 \nonumber \\ &  & 
-\left(\gamma_{t}^{\left(3\right)\alpha}
+ \left(1+
\frac{Y_{N}}{Y_{N}^{eq}}\right)
\gamma_{t}^{\left(4\right)\alpha}
\right)
{\cal Y}_{\Delta H_u},
\label{eq:BE_F}
\end{eqnarray}
 and 
\begin{eqnarray}
\widetilde F_{\alpha} & = & 
\epsilon^s_\alpha\left(z\right)\frac{\gamma_{\widetilde{N}}}{2}
\left(\frac{Y_{\widetilde{N}_{\rm tot}}}{Y_{\widetilde{N}}^{eq}}-2\right)
-\frac{\gamma_{\widetilde{N}}^{s,\alpha}}{2}
\left({\cal Y}_{\Delta \widetilde \ell_\alpha}+{\cal Y}_{\Delta H_u}\right)
-\frac{\gamma_{N}^\alpha}{8}\left({\cal Y}_{\Delta \widetilde \ell_\alpha}
+{\cal Y}_{\Delta \widetilde H_u}\right)
 \nonumber \\ 
& & 
-\left(\gamma_{t}^{\left(0\right)\alpha}\frac{Y_{N}}{Y_{N}^{eq}}+
\gamma_{t}^{\left(1\right)\alpha}+\gamma_{t}^{\left(2\right)\alpha}+
\frac{1}{2}\gamma_{t}^{\left(8\right)\alpha}
\frac{Y_{\widetilde{N}_{\rm tot}}}{Y_{\widetilde{N}}^{eq}}+
2\gamma_{t}^{\left(9\right)\alpha}\right){\cal Y}_{\Delta \widetilde \ell_\alpha}
\nonumber \\
 &  & -\left(\frac{1}{2}\gamma_{t}^{\left(0\right)\alpha}
+\frac{1}{2}\gamma_{t}^{\left(1\right)\alpha}
\frac{Y_{N}}{Y_{N}^{eq}}+\gamma_{t}^{\left(8\right)\alpha}
+
\left(1+\frac{1}{2}% \gamma_{t}^{\left(9\right)\alpha}
\frac{Y_{\widetilde{N}_{\rm tot}}}{Y_{\widetilde{N}}^{eq}}
\right)
\gamma_{t}^{\left(9\right)\alpha}
\right){\cal Y}_{\Delta H_u}
\nonumber \\
 &  & -\frac{1}{2}\left(\gamma_{t}^{\left(0\right)\alpha}+\gamma_{t}^{\left(1\right)\alpha}+
\gamma_{t}^{\left(2\right)\alpha}\frac{Y_{N}}{Y_{N}^{eq}}\right)
\left(2{\cal Y}_{\Delta \widetilde H_u}-{\cal Y}_{\Delta H_u}\right).
\label{eq:BE_tF}
\end{eqnarray}
For $G_{\alpha}$ and $\widetilde G_{\alpha}$ we have:
\begin{eqnarray}
G_{\alpha} & = & \epsilon^f_\alpha\left(z\right)
\frac{\gamma_{\widetilde{N}}}{2}
\left(\frac{Y_{\widetilde{N}_{\rm tot}}}{Y_{\widetilde{N}}^{eq}}-2\right)
-\frac{\gamma_{\widetilde{N}}^{f,\alpha}}{2}
\left({\cal Y}_{\Delta\ell_\alpha}+{\cal Y}_{\Delta \widetilde H_u}\right)
\nonumber \\
 &  & \hspace{-0.5cm}-\left(2\gamma_{t}^{\left(6\right)\alpha}+
2\gamma_{t}^{\left(7\right)\alpha}+\gamma_{t}^{\left(5\right)\alpha}
\frac{Y_{\widetilde{N}_{\rm tot}}}{Y_{\widetilde{N}}^{eq}}\right)
{\cal Y}_{\Delta\ell_\alpha}
%\nonumber \\
% &  & 
-\left(\gamma_{t}^{\left(5\right)\alpha}+\gamma_{t}^{\left(6\right)\alpha}+
\frac{1}{2}\gamma_{t}^{\left(7\right)\alpha}
\frac{Y_{\widetilde{N}_{\rm tot}}}{Y_{\widetilde{N}}^{eq}}\right)
{\cal Y}_{\Delta H_u}
\nonumber \\
 &  & -\left(\gamma_{t}^{\left(5\right)\alpha}+\gamma_{t}^{\left(7\right)k}+
\frac{1}{2}\gamma_{t}^{\left(6\right)\alpha}
\frac{Y_{\widetilde{_{\rm tot}N}}}{Y_{\widetilde{N}}^{eq}}\right)
\left(2{\cal Y}_{\Delta \widetilde H_u}-{\cal Y}_{\Delta H_u}\right),
\label{eq:BE_G}
\end{eqnarray}
and 
\begin{eqnarray}
\widetilde G_{\alpha} & = & 
\epsilon^s_\alpha\left(z\right)\frac{\gamma_{\widetilde{N}}}{2}
\left(\frac{Y_{\widetilde{N}_{\rm tot}}}{Y_{\widetilde{N}}^{eq}}-2\right)
-\frac{\gamma_{\widetilde{N}}^{s,\alpha}}{2}
\left({\cal Y}_{\Delta \widetilde \ell_\alpha}+{\cal Y}_{\Delta H_u}\right)
\nonumber \\
&  & 
+\left(\gamma_{\widetilde{N}}^{\left(3\right)\alpha} %!%
+\frac{1}{2}\gamma_{22}^{\alpha}\frac{Y_{\widetilde{N}_{\rm tot}}}{Y_{\widetilde{N}}^{eq}}
+2\gamma_{22}^{\alpha}\right)\left({\cal Y}_{\Delta\widetilde{\ell}_\alpha}
+2{\cal Y}_{\Delta\widetilde{H}_u}-{\cal Y}_{\Delta H_u}
\right)
\nonumber \\
&  &  \hspace{-1.2cm}
-\left(\frac{1}{2}\gamma_{t}^{\left(8\right)\alpha}
\frac{Y_{\widetilde{N}_{\rm tot}}}{Y_{\widetilde{N}}^{eq}}+
2\gamma_{t}^{\left(9\right)\alpha}\right){\cal Y}_{\Delta \widetilde \ell_\alpha}
%\nonumber \\
% &  & 
-\left(\gamma_{t}^{\left(8\right)\alpha}
+
\left(1+\frac{1}{2}%\gamma_{t}^{\left(9\right)\alpha}
\frac{Y_{\widetilde{N}_{\rm tot}}}{Y_{\widetilde{N}}^{eq}}
\right)
\gamma_{t}^{\left(9\right)\alpha}
\right){\cal Y}_{\Delta H_u}.\qquad
\label{eq:BE_tG}
\end{eqnarray}

The density asymmetries of the five charges in the BE
\eqref{eq:BE_Delta_alpha}, \eqref{eq:BE_Delta_RB} and
\eqref{eq:BE_Delta_R3} define the basis
$Y_{\Delta_a}=\left\{Y_{\Delta_\alpha},Y_{\Delta R_B},Y_{\Delta
    R_\chi} \right\}$ in terms of which one needs to express the five
fermionic density-asymmetries $Y_{\Delta\psi_a}=
\{Y_{\Delta\ell_\alpha},\,Y_{\Delta\tilde g},\,Y_{\Delta \tilde
  H_u}\}$. The relation is given by a $5\times 5$ matrix defined
according to:
\be
\label{eq:A5}
Y_{\Delta\psi_a} = A_{ab}\,  Y_{\Delta_b}\,, 
\nonumber
\ee
and the numerical values of $A_{ab}$ for different cases are given in
Section~\ref{sec:nse_regime}.

With the inclusion of $\gamma^{\rm eff}_{\widetilde g}$ and
$\gamma^{\rm eff}_{\submuh}$ the BE \eqref{eq:BE_Delta_alpha},
\eqref{eq:BE_Delta_RB} and \eqref{eq:BE_Delta_R3} are also valid in
the SE regime.  To verify this, one can compare the results obtained
with the complete BE given above, assuming large in-equilibrium SE
reactions $\gamma^{\rm eff}_{\widetilde g}$ and $\gamma^{\rm
  eff}_{\submuh}$, with what is obtained from the set of BE specific
for the SE regime (basically \Eqn{eq:BE_L_tot_fla2} with $dY_{L_{\rm
    tot}^\alpha}/dz \to - dY_{\Delta\alpha}/dz$).  Of course, one also
has to use the $A$ matrix \Eqn{eq:A5} and the corresponding $A^\ell$
and $C^{\widetilde H_{u}}$ matrices \Eqn{eq:leptonHiggs2} appropriate
for the specific temperature regime.  For Case I ($h_{e,d}$ Yukawa
equilibrium) $A$ is given in \Eqn{eq:Aedin} while $A^\ell$ and
$C^{\widetilde H_{u}}$ are given in \ref{app:se_new} in
Eqs.~\eqref{eq:ACT1}.  For Case II ($h_{e,d}$ Yukawa non-equilibrium)
$A$ is given in \Eqn{eq:Aedout} and the corresponding SE matrices are
given in Eqs.~\eqref{eq:ACT1} and \eqref{eq:ACT3SE}.

%% file: se.tex
\section{Chemical equilibrium conditions in the SE regime} 
\label{app:se_new}

In Section \ref{sec:general} items (1)--(5) a set of general
constraints on particles/sparticles chemical potentials were given.
At relatively low temperatures additional conditions hold, that 
are listed  here.  To simplify notations, in the following we denote 
chemical potentials with the same symbol that labels the corresponding
fields: $\phi \equiv \mu_\phi$.
\begin{enumerate}
\item[6$_{\rm SE}$-I.]  Equilibration of the particle-sparticle
  chemical potentials $\mu_\phi=\mu_{\widetilde \phi}$ ({\it
    superequilibration} (SE)\cite{Chung:2008gv}) is ensured when
  reactions like $\tilde \ell\tilde \ell \leftrightarrow \ell\ell$ are
  faster than the Universe expansion rate.  These reactions are
  mediated by gaugino exchange but also require a chirality flip on
  the gaugino line, and thus are proportional to the soft mass
  $m_{\tilde g}^2$.

  Fast reactions induced by the superpotential higgsino mixing term $\muh
  \hat H_u \hat H_d$ imply that the sum of the
  up- and down-higgsino chemical potentials vanishes.

  Since $\muh$ is expected to be of the same order than $m_{\tilde g}$
  it is reasonable to assume that both these reactions are in
  equilibrium in the same temperature range. The corresponding rates
  are given approximately by $\Gamma_{\tilde g} \sim m^2_{\tilde g}/T$
  and $\Gamma_{\muh} \sim \muh^2/T$, and are faster than the Universe
  expansion rate up to temperatures
\be
\label{eq:Tgmu2}
 T\lsim 5\cdot 10^7 
\left(\frac{m_{\widetilde g},\,\muh}{500\,{\rm GeV}}\right)^{2/3}\, {\rm GeV}.
\ee
When the temperature is sufficiently low that the limit $m_{\tilde
  g}\to 0$ is not valid, then gauginos must be considered as
Majorana fermions with an associated vanishing chemical potential:
\be
\label{eq:geq0}
\widetilde g  = 0.
\ee
Then, fast reactions $\widetilde \ell \leftrightarrow \ell +
\widetilde g$, $\widetilde Q \leftrightarrow Q + \widetilde g$,
$H_{u,d} \leftrightarrow \widetilde H_{u,d} + \widetilde g$ etc. imply
SE.

Similarly,  when the limit $\muh\to 0$ is not valid, fast $\widetilde H_u 
\leftrightarrow \overline{\widetilde H_d}$ reactions yield:
\be
\label{eq:mueq0}
\widetilde H_u+\widetilde H_d=0.  
\ee
%
%Notice that Eq.~\eqref{eq:geq0} together
%with fast gaugino interactions (Eqs.~\eqref{eq:tQtell}--\eqref{eq:tutdte})
%
%\begin{eqnarray}
%  \label{eq:tQtell2} 
%   \tilde{Q},\tilde \ell &=&    Q,\ell+  \tilde g \nonumber \\
%  \label{eq:HuHd2}
%   H_{u,d} &=&   \tilde H_{u,d}+  \tilde g \nonumber \\
%  \label{eq:tutdte2}
%   \tilde u,\tilde d,\tilde e  &=&   u,d,e-  \tilde g \nonumber,
%\end{eqnarray}
%
%imply superequilibration: equilibration of the particle-sparticle chemical potentials 
%$\mu_\phi=\mu_{\widetilde \phi}$  $\phi$. 

\item[6$_{\rm SE}$-II.]  For temperatures satisfying Eq.~\eqref{eq:Tgmu2} the
  MSSM has the same global anomalies than the SM: the EW
  $SU(2)_L$-$U(1)_{B+L}$ mixed anomaly and the QCD chiral anomaly. 
EW and QCD sphaleron effects can be described 
  by the effective operators $O_{EW}=\Pi_\alpha(QQQ\ell_\alpha)$
  and $O_{QCD}=\Pi_i(QQu^c_{Li}d^c_{Li})$. Above the EWPT 
  reactions induced by these operators are in thermal
  equilibrium and yield the conditions (compare with the 
  corresponding NSE conditions \eqref{eq:tEWmu} 
  and \eqref{eq:tQCDmu}):
\bea
\label{eq:EW}
&&9\,Q+\sum_\alpha \ell_\alpha = 0 \\
\label{eq:QCD}
&&6\,Q-\sum_i\left(u_i+d_i\right)=0\,,
\eea
where the same chemical potential has been assumed for the three quark doublets
(Eq.\eqref{eq:Q}), which is always appropriate below the limit
\eqref{eq:Tgmu2}.

\end{enumerate}

Eqs.~\eqref{eq:leptons} and \eqref{eq:upquarks}--\eqref{eq:Q},
together with the SE conditions \eqref{eq:geq0}-\eqref{eq:mueq0}, the
two anomaly conditions \eqref{eq:EW}-\eqref{eq:QCD} and the
hypercharge neutrality condition \eqref{eq:YtotY}, give $11+2+2+1=16$
constraints for the 18 chemical potentials.  Note however that there
is one redundant constraint, since by summing up
Eqs.~\eqref{eq:upquarks} and \eqref{eq:downquarks} and taking into
account conditions \eqref{eq:Q}, \eqref{eq:geq0}, and \eqref{eq:mueq0}
we obtain precisely the QCD sphaleron condition
Eq.~\eqref{eq:QCD}. Therefore, like in the SM, we
have three independent chemical potentials, which can be conveniently 
taken to be  $Y_{\Delta_\alpha} \equiv Y_{\Delta B}/3 - Y_{\Delta
  L_\alpha}$:
\be 
\label{eq:YDeltaAlpha_ap}
Y_{\Delta_\alpha} = 3\,\left[\frac{1}{3}
    \sum_i\left(2Y_{\Delta Q_i}+Y_{\Delta u_i}+Y_{\Delta d_i}\right)-
  (2Y_{\Delta \ell_\alpha}+Y_{\Delta
    e_\alpha})\right]\,. 
\ee
The density asymmetries of the leptons and higgsino doublets, that
weight the washout terms in the BE, can be expressed in terms of the
densities of the anomaly free charges \Eqn{eq:YDeltaAlpha_ap} by means
of an $A$ matrix\cite{Barbieri:2000} and $C$ vector\cite{Nardi:2006b}
as:
\be
\label{eq:AC}
Y_{\Delta\ell_\alpha}= A^\ell_{\alpha\beta}\,\, Y_{\Delta_\beta}, 
\qquad\qquad 
Y_{\Delta \tilde H_{u,d}}= 
C^{\tilde H_{u,d}}_\alpha\,\,  Y_{\Delta_\alpha}. 
\ee
In the following we give the results for $A^\ell$ and $C^{\tilde
  H_{u,d}}$ that refer to fermion states, and we recall that in the SE
  regime the density asymmetries of the corresponding scalar partners
  are given by $Y_{\Delta s}=2\,Y_{\Delta f}$ with the factor of 2
  from statistics.

%%%%%%%%%%%%%%%%%%%%%%%%%%%%%%%%%%%%%%%%%%%

\subsection{Yukawa reactions in chemical equilibrium}

\begin{enumerate}
\item[(7$_{\rm SE}$-I)] 
{\sl All Yukawa interactions in equilibrium.}\\
% Assuming moderate values of $\tan\beta$, 
At temperatures below the limit in Eq.~\eqref{eq:Te} all Yukawa
interactions are in equilibrium and we have
% , $A$ and $C$ are given by\footnote{To compare with
% the corresponding matrix obtained in the NSE regime 
% see Eq.~\eqref{eq:Aedin}.} 
%
\begin{eqnarray}
A^\ell &=& \frac{1}{9\times 237}\left(
\begin{array}{ccc}
-221 & 16  & 16\\
 16 & -221 &  16 \\
 16 &  16  & -221
\end{array}\right),\nonumber \\ 
C^{\tilde H_u}&=&-C^{\tilde H_d}=\frac{-4}{237}\left(1,\;1,\;1\right).
\label{eq:ACT1}
\end{eqnarray}
In the SE regime 
$Y_{\Delta\ell_{\rm tot}^\alpha}\equiv
Y_{\Delta\ell_\alpha}+Y_{\Delta\tilde\ell_\alpha} = 3
Y_{\Delta\ell_\alpha}$ 
and then the relation between $Y_{\Delta\ell_{\rm tot}^\alpha}$ and
$Y_{\Delta_\alpha}$ is obtained by simply multiplying the $A$ matrix
in Eq.~\eqref{eq:ACT1} by a factor of 3. This gives the same $A$
matrix obtained in the non-supersymmetric case in the same regime (see
e.g. Eq.~(4.13) in Ref.~\refcite{Nardi:2006b}).  The $C$ matrix
(multiplied by the same factor of 3) differs from the $C$ matrix of
the non-supersymmetric result (that is of course given for the scalar
density-asymmetry $Y_{\Delta H}$) by a factor $1/2$. This is simply
because $Y_{\Delta\tilde H_{\rm tot}^u}\equiv Y_{\Delta\tilde
  H_u}+Y_{\Delta H_u} = \frac{3}{2} Y_{\Delta H_u}$. 
These results agree with Ref.~\refcite{Inui:1993wv}, and hold in general for
supersymmetry within the SE regime.

\item[(7$_{\rm SE}$-II)] 
{\sl Electron and up-quark Yukawa reactions out of  equilibrium.} \\
For temperatures above $10^5(1+\tan^2\beta)\,{\rm GeV}$ the
interactions mediated by the electron Yukawa $h_e$ drop out of
equilibrium, and one of the conditions Eq.~\eqref{eq:leptons} is lost.
However, in the effective theory one can then set $h_e\to 0$ and one
global symmetry is gained. This corresponds to chiral symmetry for the
R-handed electron, that here translates into a symmetry
under phase rotations of the $e$ chiral multiplet. 
% that holds in the limit in which of unbroken supersymmetry.  
Conservation of the corresponding charge ensures that $\Delta
n_{e}+\Delta n_{\tilde e} =3 \Delta n_{e}$ is constant, and since
leptogenesis aims to explain dynamically the generation of a lepton
asymmetry we set this constant to zero, implying a vanishing chemical
potential for the R-handed electron $e=0$.  In this way the chemical
equilibrium condition that is lost is replaced by a new condition
corresponding to the conservation of a global charge, and the three
non-anomalous charges \eqref{eq:YDeltaAlpha_ap} are again sufficient
to describe all the density asymmetries.  At temperatures above $T\sim
2\cdot 10^6\,$GeV interactions mediated by the up-quark Yukawa
coupling $h_u$ drop out of equilibrium.  In this case however, by
setting $h_u \to 0$ no new symmetry is obtained, since chiral symmetry
for the R-handed quarks is anomalous, and the corresponding charge is
violated by  QCD sphalerons. However, after
dropping the first condition in Eq.~\eqref{eq:upquarks} for $u_i=u$,
the QCD sphaleron condition Eq.~\eqref{eq:QCD} ceases to be a
redundant constraint, with the result that also in this case no new
chemical potentials are needed to determine all the particle density
asymmetries.  In this case, the $A$ and $C$ are given by
\begin{eqnarray}
A^\ell&=&\frac{1}{3\times 2886}\left(\begin{array}{ccc}
-1221 &\ \,\, 156  &\ \,\, 156 \\
\ \, 111 &-910  &\ \, 52\\
\ \, 111 &\ \, 52  &-910
\end{array}\right),\nonumber \\
 C^{\tilde H_u}&=&-C^{\tilde H_d}=
\frac{-1}{2886}\left(37,\;52,\;52\right).
\label{eq:ACT2}
\end{eqnarray}

\item[(7$_{\rm SE}$-III)] 
{\sl First generation Yukawa reactions out of equilibrium.} \\
At temperatures $T\gsim 4\cdot
10^6(1+\tan^2\beta)\,{\rm GeV,} $ also the $d$-quark Yukawa coupling
can be set to zero (to remain within the SE regime we assume
$\tan\beta \sim 1$).  In this case the equilibrium dynamics is
symmetric under the exchange $u \leftrightarrow d$ (both chemical
potentials enter only the QCD sphaleron condition Eq.~\eqref{eq:QCD} with
equal weights) and so must be any physical solution of the set of
constraints. Thus, the first condition in Eq.~\eqref{eq:downquarks} can be
replaced by the condition $d=u$, and again three independent
quantities suffice to determine all the particle density asymmetries.
The  $A$ and $C$ matrices in this case are:
% \footnote{To compare with
% the corresponding matrix obtained in the non-superequilibration 
% regime see Eq.~\eqref{eq:Aedout}.} :
%
\begin{eqnarray}
A^\ell&=&\frac{1}{3\times 2148}
\left(\begin{array}{ccc}
-906 &\ \,\, 120  &\ \,\, 120\\
\ \, 75 &-688 &\ \, 28\\
\ \, 75 &\ \,28  &-688 
\end{array}\right), \nonumber \\
C^{\tilde H_u}&=&-C^{\tilde H_d}=
\frac{-1}{2148}\left(37,\;52,\;52\right), \qquad
\label{eq:ACT3SE}
\end{eqnarray}
that agrees with what is obtained in the non-supersymmetric case (see
Eq.~(4.12) of Ref.~\refcite{Nardi:2006b}) after the factor of 1/2
relative to the higgsinos discussed above in (7$_{\rm SE}$-I) is taken
into account.
\end{enumerate}

%% file: review.bbl
\begin{thebibliography}{100}
\expandafter\ifx\csname url\endcsname\relax
  \def\url#1{\texttt{#1}}\fi
\expandafter\ifx\csname urlprefix\endcsname\relax\def\urlprefix{URL }\fi
\expandafter\ifx\csname href\endcsname\relax
  \def\href#1#2{#2} \def\path#1{#1}\fi

\bibitem{Steigman:1976ev}
G.~Steigman, {Observational tests of antimatter cosmologies}, Ann. Rev. Astron.
  Astrophys. 14 (1976) 339--372.
\newblock \href {http://dx.doi.org/10.1146/annurev.aa.14.090176.002011}
  {\path{doi:10.1146/annurev.aa.14.090176.002011}}.

\bibitem{Cohen:1997ac}
A.~G. Cohen, A.~De~Rujula, S.~Glashow, {A Matter - antimatter universe?},
  Astrophys.J. 495 (1998) 539--549.
\newblock \href {http://arxiv.org/abs/astro-ph/9707087}
  {\path{arXiv:astro-ph/9707087}}, \href {http://dx.doi.org/10.1086/305328}
  {\path{doi:10.1086/305328}}.

\bibitem{Dolgov:1991fr}
A.~Dolgov, {NonGUT baryogenesis}, Phys.Rept. 222 (1992) 309--386.
\newblock \href {http://dx.doi.org/10.1016/0370-1573(92)90107-B}
  {\path{doi:10.1016/0370-1573(92)90107-B}}.

\bibitem{Iocco:2008va}
F.~Iocco, G.~Mangano, G.~Miele, O.~Pisanti, P.~D. Serpico, {Primordial
  Nucleosynthesis: from precision cosmology to fundamental physics}, Phys.
  Rept. 472 (2009) 1--76.
\newblock \href {http://arxiv.org/abs/0809.0631} {\path{arXiv:0809.0631}},
  \href {http://dx.doi.org/10.1016/j.physrep.2009.02.002}
  {\path{doi:10.1016/j.physrep.2009.02.002}}.

\bibitem{Steigman:2007xt}
G.~Steigman, {Primordial Nucleosynthesis in the Precision Cosmology Era}, Ann.
  Rev. Nucl. Part. Sci. 57 (2007) 463--491.
\newblock \href {http://arxiv.org/abs/0712.1100} {\path{arXiv:0712.1100}},
  \href {http://dx.doi.org/10.1146/annurev.nucl. 56.080805.140437}
  {\path{doi:10.1146/annurev.nucl. 56.080805.140437}}.

\bibitem{Nakamura:2010zzi}
K.~Nakamura, et~al., {Review of particle physics}, J.Phys.G G37 (2010) 075021.
\newblock \href {http://dx.doi.org/10.1088/0954-3899/37/7A/075021}
  {\path{doi:10.1088/0954-3899/37/7A/075021}}.

\bibitem{Steigman:2005uz}
G.~Steigman, {Primordial Nucleosynthesis: Successes And Challenges}, Int. J.
  Mod. Phys. E15 (2006) 1--36.
\newblock \href {http://arxiv.org/abs/astro-ph/0511534}
  {\path{arXiv:astro-ph/0511534}}, \href
  {http://dx.doi.org/10.1142/S0218301306004028}
  {\path{doi:10.1142/S0218301306004028}}.

\bibitem{Cyburt:2004yc}
R.~H. Cyburt, B.~D. Fields, K.~A. Olive, E.~Skillman, {New BBN limits on
  Physics Beyond the Standard Model from He4}, Astropart. Phys. 23 (2005)
  313--323.
\newblock \href {http://arxiv.org/abs/astro-ph/0408033}
  {\path{arXiv:astro-ph/0408033}}, \href
  {http://dx.doi.org/10.1016/j.astropartphys.2005.01.005}
  {\path{doi:10.1016/j.astropartphys.2005.01.005}}.

\bibitem{Olive:1999ij}
K.~A. Olive, G.~Steigman, T.~P. Walker, {Primordial Nucleosynthesis: Theory and
  Observations}, Phys. Rept. 333 (2000) 389--407.
\newblock \href {http://arxiv.org/abs/astro-ph/9905320}
  {\path{arXiv:astro-ph/9905320}}, \href
  {http://dx.doi.org/10.1016/S0370-1573(00)00031-4}
  {\path{doi:10.1016/S0370-1573(00)00031-4}}.

\bibitem{Hu:2001bc}
W.~Hu, S.~Dodelson, {Cosmic Microwave Background Anisotropies}, Ann. Rev.
  Astron. Astrophys. 40 (2002) 171--216.
\newblock \href {http://arxiv.org/abs/astro-ph/0110414}
  {\path{arXiv:astro-ph/0110414}}, \href
  {http://dx.doi.org/10.1146/annurev.astro. 40.060401.093926}
  {\path{doi:10.1146/annurev.astro. 40.060401.093926}}.

\bibitem{Dodelson:2003ft}
S.~Dodelson, {Modern cosmology}Amsterdam, Netherlands: Academic Pr. (2003) 440
  p.

\bibitem{Larson:2010gs}
D.~Larson, J.~Dunkley, G.~Hinshaw, E.~Komatsu, M.~Nolta, et~al., {Seven-Year
  Wilkinson Microwave Anisotropy Probe (WMAP) Observations: Power Spectra and
  WMAP-Derived Parameters}, Astrophys.J.Suppl. 192 (2011) 16.
\newblock \href {http://arxiv.org/abs/1001.4635} {\path{arXiv:1001.4635}},
  \href {http://dx.doi.org/10.1088/0067-0049/192/2/16}
  {\path{doi:10.1088/0067-0049/192/2/16}}.

\bibitem{Sakharov:1967dj}
A.~D. Sakharov, {Violation of CP Invariance, c Asymmetry, and Baryon Asymmetry
  of the Universe}, Pisma Zh. Eksp. Teor. Fiz. 5 (1967) 32--35.

\bibitem{Kuzmin:1985mm}
V.~A. Kuzmin, V.~A. Rubakov, M.~E. Shaposhnikov, {On the Anomalous Electroweak
  Baryon Number Nonconservation in the Early Universe}, Phys. Lett. B155 (1985)
  36.
\newblock \href {http://dx.doi.org/10.1016/0370-2693(85)91028-7}
  {\path{doi:10.1016/0370-2693(85)91028-7}}.

\bibitem{'tHooft:1976up}
G.~'t~Hooft, {Symmetry breaking through Bell-Jackiw anomalies}, Phys. Rev.
  Lett. 37 (1976) 8--11.
\newblock \href {http://dx.doi.org/10.1103/PhysRevLett.37.8}
  {\path{doi:10.1103/PhysRevLett.37.8}}.

\bibitem{Kobayashi:1973fv}
M.~Kobayashi, T.~Maskawa, {CP Violation in the Renormalizable Theory of Weak
  Interaction}, Prog. Theor. Phys. 49 (1973) 652--657.
\newblock \href {http://dx.doi.org/10.1143/PTP.49.652}
  {\path{doi:10.1143/PTP.49.652}}.

\bibitem{Jarlskog:1985ht}
C.~Jarlskog, {Commutator of the Quark Mass Matrices in the Standard Electroweak
  Model and a Measure of Maximal CP Violation}, Phys. Rev. Lett. 55 (1985)
  1039.
\newblock \href {http://dx.doi.org/10.1103/PhysRevLett.55.1039}
  {\path{doi:10.1103/PhysRevLett.55.1039}}.

\bibitem{Gavela:1994ds}
M.~B. Gavela, M.~Lozano, J.~Orloff, O.~Pene, {Standard model CP violation and
  baryon asymmetry. Part 1: Zero temperature}, Nucl. Phys. B430 (1994)
  345--381.
\newblock \href {http://arxiv.org/abs/hep-ph/9406288}
  {\path{arXiv:hep-ph/9406288}}, \href
  {http://dx.doi.org/10.1016/0550-3213(94)00409-9}
  {\path{doi:10.1016/0550-3213(94)00409-9}}.

\bibitem{Gavela:1994dt}
M.~B. Gavela, P.~Hernandez, J.~Orloff, O.~Pene, C.~Quimbay, {Standard model CP
  violation and baryon asymmetry. Part 2: Finite temperature}, Nucl. Phys. B430
  (1994) 382--426.
\newblock \href {http://arxiv.org/abs/hep-ph/9406289}
  {\path{arXiv:hep-ph/9406289}}, \href
  {http://dx.doi.org/10.1016/0550-3213(94)00410-2}
  {\path{doi:10.1016/0550-3213(94)00410-2}}.

\bibitem{Huet:1994jb}
P.~Huet, E.~Sather, {Electroweak baryogenesis and standard model CP violation},
  Phys. Rev. D51 (1995) 379--394.
\newblock \href {http://arxiv.org/abs/hep-ph/9404302}
  {\path{arXiv:hep-ph/9404302}}, \href
  {http://dx.doi.org/10.1103/PhysRevD.51.379}
  {\path{doi:10.1103/PhysRevD.51.379}}.

\bibitem{Rubakov:1996vz}
V.~A. Rubakov, M.~E. Shaposhnikov, {Electroweak baryon number non-conservation
  in the early universe and in high-energy collisions}, Usp. Fiz. Nauk 166
  (1996) 493--537.
\newblock \href {http://arxiv.org/abs/hep-ph/9603208}
  {\path{arXiv:hep-ph/9603208}}, \href
  {http://dx.doi.org/10.1070/PU1996v039n05ABEH000145}
  {\path{doi:10.1070/PU1996v039n05ABEH000145}}.

\bibitem{Trodden:1998ym}
M.~Trodden, {Electroweak baryogenesis}, Rev. Mod. Phys. 71 (1999) 1463--1500.
\newblock \href {http://arxiv.org/abs/hep-ph/9803479}
  {\path{arXiv:hep-ph/9803479}}, \href
  {http://dx.doi.org/10.1103/RevModPhys.71.1463}
  {\path{doi:10.1103/RevModPhys.71.1463}}.

\bibitem{Kajantie:1995kf}
K.~Kajantie, M.~Laine, K.~Rummukainen, M.~E. Shaposhnikov, {The Electroweak
  Phase Transition: A Non-Perturbative Analysis}, Nucl. Phys. B466 (1996)
  189--258.
\newblock \href {http://arxiv.org/abs/hep-lat/9510020}
  {\path{arXiv:hep-lat/9510020}}, \href
  {http://dx.doi.org/10.1016/0550-3213(96)00052-1}
  {\path{doi:10.1016/0550-3213(96)00052-1}}.

\bibitem{Ignatiev:1978uf}
A.~Y. Ignatiev, N.~V. Krasnikov, V.~A. Kuzmin, A.~N. Tavkhelidze, {Universal CP
  Noninvariant Superweak Interaction and Baryon Asymmetry of the Universe},
  Phys. Lett. B76 (1978) 436--438.
\newblock \href {http://dx.doi.org/10.1016/0370-2693(78)90900-0}
  {\path{doi:10.1016/0370-2693(78)90900-0}}.

\bibitem{Yoshimura:1978ex}
M.~Yoshimura, {Unified Gauge Theories and the Baryon Number of the Universe},
  Phys. Rev. Lett. 41 (1978) 281--284.
\newblock \href {http://dx.doi.org/10.1103/PhysRevLett.41.281}
  {\path{doi:10.1103/PhysRevLett.41.281}}.

\bibitem{Toussaint:1978br}
D.~Toussaint, S.~B. Treiman, F.~Wilczek, A.~Zee, {Matter - Antimatter
  Accounting, Thermodynamics, and Black Hole Radiation}, Phys. Rev. D19 (1979)
  1036--1045.
\newblock \href {http://dx.doi.org/10.1103/PhysRevD.19.1036}
  {\path{doi:10.1103/PhysRevD.19.1036}}.

\bibitem{Dimopoulos:1978kv}
S.~Dimopoulos, L.~Susskind, {On the Baryon Number of the Universe}, Phys. Rev.
  D18 (1978) 4500--4509.
\newblock \href {http://dx.doi.org/10.1103/PhysRevD.18.4500}
  {\path{doi:10.1103/PhysRevD.18.4500}}.

\bibitem{Ellis:1978xg}
J.~R. Ellis, M.~K. Gaillard, D.~V. Nanopoulos, {Baryon Number Generation in
  Grand Unified Theories}, Phys. Lett. B80 (1979) 360.
\newblock \href {http://dx.doi.org/10.1016/0370-2693(79)91190-0}
  {\path{doi:10.1016/0370-2693(79)91190-0}}.

\bibitem{Weinberg:1979bt}
S.~Weinberg, {Cosmological Production of Baryons}, Phys. Rev. Lett. 42 (1979)
  850--853.
\newblock \href {http://dx.doi.org/10.1103/PhysRevLett.42.850}
  {\path{doi:10.1103/PhysRevLett.42.850}}.

\bibitem{Yoshimura:1979gy}
M.~Yoshimura, {ORIGIN OF COSMOLOGICAL BARYON ASYMMETRY}, Phys. Lett. B88 (1979)
  294.
\newblock \href {http://dx.doi.org/10.1016/0370-2693(79)90471-4}
  {\path{doi:10.1016/0370-2693(79)90471-4}}.

\bibitem{Barr:1979ye}
S.~M. Barr, G.~Segre, H.~A. Weldon, The magnitude of the cosmological baryon
  asymmetry, Phys. Rev. D20 (1979) 2494.

\bibitem{Nanopoulos:1979gx}
D.~V. Nanopoulos, S.~Weinberg, Mechanisms for cosmological baryon production,
  Phys. Rev. D20 (1979) 2484.

\bibitem{Yildiz:1979gx}
A.~Yildiz, P.~Cox, Net baryon number, \uppercase{CP} violation with unified
  fields, Phys. Rev. D21 (1980) 906.

\bibitem{Riotto:1999yt}
A.~Riotto, M.~Trodden, {Recent progress in baryogenesis}, Ann. Rev. Nucl. Part.
  Sci. 49 (1999) 35--75.
\newblock \href {http://arxiv.org/abs/hep-ph/9901362}
  {\path{arXiv:hep-ph/9901362}}, \href
  {http://dx.doi.org/10.1146/annurev.nucl.49.1.35}
  {\path{doi:10.1146/annurev.nucl.49.1.35}}.

\bibitem{Cline:2006ts}
J.~M. Cline, {Baryogenesis}\href {http://arxiv.org/abs/hep-ph/0609145}
  {\path{arXiv:hep-ph/0609145}}.

\bibitem{Losada:1996ju}
M.~Losada, {High temperature dimensional reduction of the MSSM and other
  multi-scalar models}, Phys. Rev. D56 (1997) 2893--2913.
\newblock \href {http://arxiv.org/abs/hep-ph/9605266}
  {\path{arXiv:hep-ph/9605266}}.

\bibitem{Carena:1996wj}
M.~S. Carena, M.~Quiros, C.~E.~M. Wagner, {Opening the Window for Electroweak
  Baryogenesis}, Phys. Lett. B380 (1996) 81--91.
\newblock \href {http://arxiv.org/abs/hep-ph/9603420}
  {\path{arXiv:hep-ph/9603420}}, \href
  {http://dx.doi.org/10.1016/0370-2693(96)00475-3}
  {\path{doi:10.1016/0370-2693(96)00475-3}}.

\bibitem{Delepine:1996vn}
D.~Delepine, J.~M. Gerard, R.~Gonzalez~Felipe, J.~Weyers, {A light stop and
  electroweak baryogenesis}, Phys. Lett. B386 (1996) 183--188.
\newblock \href {http://arxiv.org/abs/hep-ph/9604440}
  {\path{arXiv:hep-ph/9604440}}.

\bibitem{Affleck:1984fy}
I.~Affleck, M.~Dine, A new mechanism for baryogenesis, Nucl. Phys. B249 (1985)
  361.

\bibitem{Dine:1995kz}
M.~Dine, L.~Randall, S.~D. Thomas, Baryogenesis from flat directions of the
  supersymmetric standard model, Nucl. Phys. B458 (1996) 291--326.
\newblock \href {http://arxiv.org/abs/hep-ph/9507453}
  {\path{arXiv:hep-ph/9507453}}.

\bibitem{Cohen:1987vi}
A.~G. Cohen, D.~B. Kaplan, {Thermodynamic generation of the baryon asymmetry},
  Phys.Lett. B199 (1987) 251.
\newblock \href {http://dx.doi.org/10.1016/0370-2693(87)91369-4}
  {\path{doi:10.1016/0370-2693(87)91369-4}}.

\bibitem{Cohen:1988kt}
A.~G. Cohen, D.~B. Kaplan, {Spontaneous baryogenesis}, Nucl.Phys. B308 (1988)
  913.
\newblock \href {http://dx.doi.org/10.1016/0550-3213(88)90134-4}
  {\path{doi:10.1016/0550-3213(88)90134-4}}.

\bibitem{Fukugita:1986hr}
M.~Fukugita, T.~Yanagida, {Baryogenesis Without Grand Unification}, Phys. Lett.
  B174 (1986) 45.
\newblock \href {http://dx.doi.org/10.1016/0370-2693(86)91126-3}
  {\path{doi:10.1016/0370-2693(86)91126-3}}.

\bibitem{Minkowski:1977sc}
P.~Minkowski, {mu $\to$ e gamma at a Rate of One Out of 1-Billion Muon
  Decays?}, Phys. Lett. B67 (1977) 421.
\newblock \href {http://dx.doi.org/10.1016/0370-2693(77)90435-X}
  {\path{doi:10.1016/0370-2693(77)90435-X}}.

\bibitem{Yanagida:1979as}
T.~Yanagida, {Horizontal gauge symmetry and masses of neutrinos}In Proceedings
  of the Workshop on the Baryon Number of the Universe and Unified Theories,
  Tsukuba, Japan, 13-14 Feb 1979.

\bibitem{Glashow}
S.~Glashow, in quarks and leptons, Carg\`ese Lectures, Plenum, NY (1980) 687.

\bibitem{GellMann:1980vs}
M.~Gell-Mann, P.~Ramond, R.~Slansky, Complex spinors and unified
  theoriesPublished in Supergravity, P. van Nieuwenhuizen and D.Z. Freedman
  (eds.), North Holland Publ. Co., 1979.

\bibitem{Mohapatra:1980yp}
R.~N. Mohapatra, G.~Senjanovic, {Neutrino mass and spontaneous parity
  nonconservation}, Phys. Rev. Lett. 44 (1980) 912.
\newblock \href {http://dx.doi.org/10.1103/PhysRevLett.44.912}
  {\path{doi:10.1103/PhysRevLett.44.912}}.

\bibitem{Khlebnikov:1988sr}
S.~Khlebnikov, M.~Shaposhnikov, {The Statistical Theory of Anomalous Fermion
  Number Nonconservation}, Nucl.Phys. B308 (1988) 885--912.
\newblock \href {http://dx.doi.org/10.1016/0550-3213(88)90133-2}
  {\path{doi:10.1016/0550-3213(88)90133-2}}.

\bibitem{Grossman:2003}
Y.~Grossman, T.~Kashti, Y.~Nir, E.~Roulet, {Leptogenesis from Supersymmetry
  Breaking}, Phys. Rev. Lett. 91 (2003) 251801.
\newblock \href {http://arxiv.org/abs/hep-ph/0307081}
  {\path{arXiv:hep-ph/0307081}}, \href
  {http://dx.doi.org/10.1103/PhysRevLett.91.251801}
  {\path{doi:10.1103/PhysRevLett.91.251801}}.

\bibitem{DAmbrosio:2003}
G.~D'Ambrosio, G.~F. Giudice, M.~Raidal, {Soft leptogenesis}, Phys. Lett. B575
  (2003) 75--84.
\newblock \href {http://arxiv.org/abs/hep-ph/0308031}
  {\path{arXiv:hep-ph/0308031}}, \href
  {http://dx.doi.org/10.1016/j.physletb.2003.09.037}
  {\path{doi:10.1016/j.physletb.2003.09.037}}.

\bibitem{Boubekeur:2002jn}
L.~Boubekeur, {Leptogenesis at low scale}\href
  {http://arxiv.org/abs/hep-ph/0208003} {\path{arXiv:hep-ph/0208003}}.

\bibitem{Luty:1992un}
M.~A. Luty, Baryogenesis via leptogenesis, Phys. Rev. D45 (1992) 455--465.

\bibitem{Gherghetta:1993kn}
T.~Gherghetta, G.~Jungman, {Cosmological consequences of spontaneous lepton
  number violation in SO(10) grand unification}, Phys. Rev. D48 (1993)
  1546--1554.
\newblock \href {http://arxiv.org/abs/hep-ph/9302212}
  {\path{arXiv:hep-ph/9302212}}.

\bibitem{Plumacher:1996kc}
M.~Pl{\"u}macher, {Baryogenesis and lepton number violation}, Z. Phys. C74
  (1997) 549.
\newblock \href {http://arxiv.org/abs/hep-ph/9604229}
  {\path{arXiv:hep-ph/9604229}}, \href
  {http://dx.doi.org/10.1007/s002880050418} {\path{doi:10.1007/s002880050418}}.

\bibitem{Plumacher:1997ru}
M.~Plumacher, Baryon asymmetry, neutrino mixing and supersymmetric
  \uppercase{SO}(10) unification, Nucl. Phys. B530 (1998) 207--246.
\newblock \href {http://arxiv.org/abs/hep-ph/9704231}
  {\path{arXiv:hep-ph/9704231}}, \href
  {http://dx.doi.org/10.1016/S0550-3213(98)00410-6}
  {\path{doi:10.1016/S0550-3213(98)00410-6}}.

\bibitem{Covi:1996wh}
L.~Covi, E.~Roulet, F.~Vissani, {CP violating decays in leptogenesis
  scenarios}, Phys. Lett. B384 (1996) 169--174.
\newblock \href {http://arxiv.org/abs/hep-ph/9605319}
  {\path{arXiv:hep-ph/9605319}}, \href
  {http://dx.doi.org/10.1016/0370-2693(96)00817-9}
  {\path{doi:10.1016/0370-2693(96)00817-9}}.

\bibitem{Buchmuller:2000as}
W.~Buchmuller, M.~Plumacher, Neutrino masses and the baryon asymmetry, Int. J.
  Mod. Phys. A15 (2000) 5047--5086.
\newblock \href {http://arxiv.org/abs/hep-ph/0007176}
  {\path{arXiv:hep-ph/0007176}}.

\bibitem{Covi:1997dr}
L.~Covi, N.~Rius, E.~Roulet, F.~Vissani, {Finite temperature effects on CP
  violating asymmetries}, Phys. Rev. D57 (1998) 93--99.
\newblock \href {http://arxiv.org/abs/hep-ph/9704366}
  {\path{arXiv:hep-ph/9704366}}, \href
  {http://dx.doi.org/10.1103/PhysRevD.57.93}
  {\path{doi:10.1103/PhysRevD.57.93}}.

\bibitem{Giudice:2003jh}
G.~F. Giudice, A.~Notari, M.~Raidal, A.~Riotto, A.~Strumia, {Towards a complete
  theory of thermal leptogenesis in the SM and MSSM}, Nucl. Phys. B685 (2004)
  89--149.
\newblock \href {http://arxiv.org/abs/hep-ph/0310123}
  {\path{arXiv:hep-ph/0310123}}, \href
  {http://dx.doi.org/10.1016/j.nuclphysb.2004.02.019}
  {\path{doi:10.1016/j.nuclphysb.2004.02.019}}.

\bibitem{Barbieri:2000}
R.~Barbieri, P.~Creminelli, A.~Strumia, N.~Tetradis, {Baryogenesis through
  leptogenesis}, Nucl. Phys. B575 (2000) 61--77.
\newblock \href {http://arxiv.org/abs/hep-ph/9911315}
  {\path{arXiv:hep-ph/9911315}}, \href
  {http://dx.doi.org/10.1016/S0550-3213(00)00011-0}
  {\path{doi:10.1016/S0550-3213(00)00011-0}}.

\bibitem{Endoh:2004}
T.~Endoh, T.~Morozumi, Z.~Xiong, {Primordial lepton family asymmetries in
  seesaw model}, Prog. Theor. Phys. 111 (2004) 123--149.
\newblock \href {http://arxiv.org/abs/hep-ph/0308276}
  {\path{arXiv:hep-ph/0308276}}, \href {http://dx.doi.org/10.1143/PTP.111.123}
  {\path{doi:10.1143/PTP.111.123}}.

\bibitem{Abada:2006a}
A.~Abada, S.~Davidson, F.-X. Josse-Michaux, M.~Losada, A.~Riotto, {Flavour
  Issues in Leptogenesis}, JCAP 0604 (2006) 004.
\newblock \href {http://arxiv.org/abs/hep-ph/0601083}
  {\path{arXiv:hep-ph/0601083}}.

\bibitem{Nardi:2006b}
E.~Nardi, Y.~Nir, E.~Roulet, J.~Racker, {The importance of flavor in
  leptogenesis}, JHEP 01 (2006) 164.
\newblock \href {http://arxiv.org/abs/hep-ph/0601084}
  {\path{arXiv:hep-ph/0601084}}.

\bibitem{Abada:2006b}
A.~Abada, S.~Davidson, A.~Ibarra, F.~Josse-Michaux, M.~Losada, A.~Riotto,
  {Flavour matters in leptogenesis}, JHEP 09 (2006) 010.
\newblock \href {http://arxiv.org/abs/hep-ph/0605281}
  {\path{arXiv:hep-ph/0605281}}.

\bibitem{Davidson:2008}
S.~Davidson, E.~Nardi, Y.~Nir, {Leptogenesis}, Phys. Rept. 466 (2008) 105--177.
\newblock \href {http://arxiv.org/abs/0802.2962} {\path{arXiv:0802.2962}},
  \href {http://dx.doi.org/10.1016/j.physrep.2008.06.002}
  {\path{doi:10.1016/j.physrep.2008.06.002}}.

\bibitem{Chen:2007fv}
M.-C. Chen, {TASI 2006 Lectures on Leptogenesis}, hep-ph/0703087\href
  {http://arxiv.org/abs/hep-ph/0703087} {\path{arXiv:hep-ph/0703087}}.

\bibitem{Davidson:2007xu}
S.~Davidson, Flavoured leptogenesis, arXiv:0705.1590 [hep-ph]\href
  {http://arxiv.org/abs/arXiv:0705.1590 [hep-ph]} {\path{arXiv:arXiv:0705.1590
  [hep-ph]}}.

\bibitem{Nardi:2007fs}
E.~Nardi, {Topics in leptogenesis}, AIP Conf. Proc. 917 (2007) 82--89.
\newblock \href {http://arxiv.org/abs/hep-ph/0702033}
  {\path{arXiv:hep-ph/0702033}}.

\bibitem{Nardi:2007cf}
E.~Nardi, {Recent Issues in Leptogenesis}, arXiv:0706.0487 [hep-ph].

\bibitem{Fong:2010qh}
C.~S. Fong, M.~C. Gonzalez-Garcia, E.~Nardi, J.~Racker, {Supersymmetric
  Leptogenesis}, JCAP 1012 (2010) 013.
\newblock \href {http://arxiv.org/abs/1009.0003} {\path{arXiv:1009.0003}},
  \href {http://dx.doi.org/10.1088/1475-7516/2010/12/013}
  {\path{doi:10.1088/1475-7516/2010/12/013}}.

\bibitem{Fong:2010bv}
C.~S. Fong, M.~Gonzalez-Garcia, E.~Nardi, {Early Universe effective theories:
  The Soft Leptogenesis and R-Genesis Cases}, JCAP 1102 (2011) 032.
\newblock \href {http://arxiv.org/abs/1012.1597} {\path{arXiv:1012.1597}},
  \href {http://dx.doi.org/10.1088/1475-7516/2011/02/032}
  {\path{doi:10.1088/1475-7516/2011/02/032}}.

\bibitem{Engelhard:2006yg}
G.~Engelhard, Y.~Grossman, E.~Nardi, Y.~Nir, {The importance of N2
  leptogenesis}, Phys. Rev. Lett. 99 (2007) 081802.
\newblock \href {http://arxiv.org/abs/hep-ph/0612187}
  {\path{arXiv:hep-ph/0612187}}, \href
  {http://dx.doi.org/10.1103/PhysRevLett.99.081802}
  {\path{doi:10.1103/PhysRevLett.99.081802}}.

\bibitem{BahatTreidel:2008}
O.~Bahat-Treidel, Z.~Surujon, {The (ir)relevance of Initial Conditions to Soft
  Leptogenesis}, JHEP 11 (2008) 046.
\newblock \href {http://arxiv.org/abs/0710.3905} {\path{arXiv:0710.3905}},
  \href {http://dx.doi.org/10.1088/1126-6708/2008/11/046}
  {\path{doi:10.1088/1126-6708/2008/11/046}}.

\bibitem{Pilaftsis:1997}
A.~Pilaftsis, {CP violation and baryogenesis due to heavy Majorana neutrinos},
  Phys. Rev. D56 (1997) 5431--5451.
\newblock \href {http://arxiv.org/abs/hep-ph/9707235}
  {\path{arXiv:hep-ph/9707235}}, \href
  {http://dx.doi.org/10.1103/PhysRevD.56.5431}
  {\path{doi:10.1103/PhysRevD.56.5431}}.

\bibitem{Pilaftsis:2004}
A.~Pilaftsis, T.~E.~J. Underwood, {Resonant leptogenesis}, Nucl. Phys. B692
  (2004) 303--345.
\newblock \href {http://arxiv.org/abs/hep-ph/0309342}
  {\path{arXiv:hep-ph/0309342}}, \href
  {http://dx.doi.org/10.1016/j.nuclphysb.2004.05.029}
  {\path{doi:10.1016/j.nuclphysb.2004.05.029}}.

\bibitem{Pilaftsis:2005a}
A.~Pilaftsis, {Resonant tau leptogenesis with observable lepton number
  violation}, Phys. Rev. Lett. 95 (2005) 081602.
\newblock \href {http://arxiv.org/abs/hep-ph/0408103}
  {\path{arXiv:hep-ph/0408103}}, \href {http://dx.doi.org/10.1103/Phys
  RevLett.95.081602} {\path{doi:10.1103/Phys RevLett.95.081602}}.

\bibitem{Pilaftsis:2005b}
A.~Pilaftsis, T.~E.~J. Underwood, {Electroweak-scale resonant leptogenesis},
  Phys. Rev. D72 (2005) 113001.
\newblock \href {http://arxiv.org/abs/hep-ph/0506107}
  {\path{arXiv:hep-ph/0506107}}, \href
  {http://dx.doi.org/10.1103/PhysRevD.72.113001}
  {\path{doi:10.1103/PhysRevD.72.113001}}.

\bibitem{Pilaftsis:2008}
A.~Pilaftsis, {Electroweak Resonant Leptogenesis in the Singlet Majoron Model},
  Phys. Rev. D78 (2008) 013008.
\newblock \href {http://arxiv.org/abs/0805.1677} {\path{arXiv:0805.1677}},
  \href {http://dx.doi.org/10.1103/PhysRevD.78.013008}
  {\path{doi:10.1103/PhysRevD.78.013008}}.

\bibitem{Cutkosky:1960sp}
R.~E. Cutkosky, {Singularities and discontinuities of Feynman amplitudes}, J.
  Math. Phys. 1 (1960) 429--433.

\bibitem{Grossman:2004}
Y.~Grossman, T.~Kashti, Y.~Nir, E.~Roulet, {New ways to soft leptogenesis},
  JHEP 11 (2004) 080.
\newblock \href {http://arxiv.org/abs/hep-ph/0407063}
  {\path{arXiv:hep-ph/0407063}}, \href
  {http://dx.doi.org/10.1088/1126-6708/2004/11/080}
  {\path{doi:10.1088/1126-6708/2004/11/080}}.

\bibitem{Fong:2009}
C.~S. Fong, M.~C. Gonzalez-Garcia, {On Gaugino Contributions to Soft
  Leptogenesis}, JHEP 03 (2009) 073.
\newblock \href {http://arxiv.org/abs/0901.0008} {\path{arXiv:0901.0008}},
  \href {http://dx.doi.org/10.1088/1126-6708/2009/03/073}
  {\path{doi:10.1088/1126-6708/2009/03/073}}.

\bibitem{Buchmuller:2001}
W.~Buchm{\"u}ller, M.~Pl{\"u}macher, {Spectator processes and baryogenesis},
  Phys. Lett. B511 (2001) 74.
\newblock \href {http://arxiv.org/abs/hep-ph/0104189}
  {\path{arXiv:hep-ph/0104189}}, \href
  {http://dx.doi.org/10.1016/S0370-2693(01)00614-1}
  {\path{doi:10.1016/S0370-2693(01)00614-1}}.

\bibitem{Nardi:2006a}
E.~Nardi, Y.~Nir, J.~Racker, E.~Roulet, {On Higgs and sphaleron effects during
  the leptogenesis era}, JHEP 01 (2006) 068.
\newblock \href {http://arxiv.org/abs/hep-ph/0512052}
  {\path{arXiv:hep-ph/0512052}}.

\bibitem{Garayoa:2009}
J.~Garayoa, S.~Pastor, T.~Pinto, N.~Rius, O.~Vives, {On the full Boltzmann
  equations for Leptogenesis}, JCAP 0909 (2009) 035.
\newblock \href {http://arxiv.org/abs/0905.4834} {\path{arXiv:0905.4834}},
  \href {http://dx.doi.org/10.1088/1475-7516/2009/09/035}
  {\path{doi:10.1088/1475-7516/2009/09/035}}.

\bibitem{Nardi:2007}
E.~Nardi, J.~Racker, E.~Roulet, {CP violation in scatterings, three body
  processes and the Boltzmann equations for leptogenesis}, JHEP 09 (2007) 090.
\newblock \href {http://arxiv.org/abs/hep-ph/0707.0378}
  {\path{arXiv:hep-ph/0707.0378}}, \href
  {http://dx.doi.org/10.1088/1126-6708/2007/09/090}
  {\path{doi:10.1088/1126-6708/2007/09/090}}.

\bibitem{Fong:2010bh}
C.~S. Fong, M.~C. Gonzalez-Garcia, J.~Racker, {CP Violation from Scatterings
  with Gauge Bosons in Leptogenesis}, Phys. Lett. B697 (2011) 463--470.
\newblock \href {http://arxiv.org/abs/1010.2209} {\path{arXiv:1010.2209}},
  \href {http://dx.doi.org/10.1016/j.physletb.2011.02.025}
  {\path{doi:10.1016/j.physletb.2011.02.025}}.

\bibitem{Harvey:1990qw}
J.~A. Harvey, M.~S. Turner, {Cosmological baryon and lepton number in the
  presence of electroweak fermion number violation}, Phys.Rev. D42 (1990)
  3344--3349.
\newblock \href {http://dx.doi.org/10.1103/PhysRevD.42.3344}
  {\path{doi:10.1103/PhysRevD.42.3344}}.

\bibitem{Inui:1993wv}
T.~Inui, T.~Ichihara, Y.~Mimura, N.~Sakai, {Cosmological baryon asymmetry in
  supersymmetric Standard Models and heavy particle effects}, Phys. Lett. B325
  (1994) 392--400.
\newblock \href {http://arxiv.org/abs/hep-ph/9310268}
  {\path{arXiv:hep-ph/9310268}}, \href
  {http://dx.doi.org/10.1016/0370-2693(94)90031-0}
  {\path{doi:10.1016/0370-2693(94)90031-0}}.

\bibitem{Chung:2008gv}
D.~J.~H. Chung, B.~Garbrecht, S.~Tulin, {The Effect of the Sparticle Mass
  Spectrum on the Conversion of B-L to B}, JCAP 0903 (2009) 008.
\newblock \href {http://arxiv.org/abs/0807.2283} {\path{arXiv:0807.2283}},
  \href {http://dx.doi.org/10.1088/1475-7516/2009/03/008}
  {\path{doi:10.1088/1475-7516/2009/03/008}}.

\bibitem{Fong:2010a}
C.~S. Fong, J.~Racker, {On fast CP violating interactions in leptogenesis},
  JCAP 1007 (2010) 001.
\newblock \href {http://arxiv.org/abs/1004.2546} {\path{arXiv:1004.2546}},
  \href {http://dx.doi.org/10.1088/1475-7516/2010/07/001}
  {\path{doi:10.1088/1475-7516/2010/07/001}}.

\bibitem{Chun:2004}
E.~J. Chun, {Late leptogenesis from radiative soft terms}, Phys. Rev. D69
  (2004) 117303.
\newblock \href {http://arxiv.org/abs/hep-ph/0404029}
  {\path{arXiv:hep-ph/0404029}}, \href
  {http://dx.doi.org/10.1103/PhysRevD.69.117303}
  {\path{doi:10.1103/PhysRevD.69.117303}}.

\bibitem{Chen:2004}
M.-C. Chen, K.~T. Mahanthappa, {Lepton flavor violating decays, soft
  leptogenesis and SUSY SO(10)}, Phys. Rev. D70 (2004) 113013.
\newblock \href {http://arxiv.org/abs/hep-ph/0409096}
  {\path{arXiv:hep-ph/0409096}}, \href
  {http://dx.doi.org/10.1103/PhysRevD.70.113013}
  {\path{doi:10.1103/PhysRevD.70.113013}}.

\bibitem{Grossman:2005}
Y.~Grossman, R.~Kitano, H.~Murayama, {Natural soft leptogenesis}, JHEP 06
  (2005) 058.
\newblock \href {http://arxiv.org/abs/hep-ph/0504160}
  {\path{arXiv:hep-ph/0504160}}.

\bibitem{Chun:2007}
E.~J. Chun, L.~Velasco-Sevilla, {SO(10) unified models and soft leptogenesis},
  JHEP 08 (2007) 075.
\newblock \href {http://arxiv.org/abs/hep-ph/0702039}
  {\path{arXiv:hep-ph/0702039}}, \href
  {http://dx.doi.org/10.1088/1126-6708/2007/08/075}
  {\path{doi:10.1088/1126-6708/2007/08/075}}.

\bibitem{Buchmuller:2000}
W.~Buchmuller, S.~Fredenhagen, {Quantum mechanics of baryogenesis}, Phys. Lett.
  B483 (2000) 217--224.
\newblock \href {http://arxiv.org/abs/hep-ph/0004145}
  {\path{arXiv:hep-ph/0004145}}, \href
  {http://dx.doi.org/10.1016/S0370-2693(00)00573-6}
  {\path{doi:10.1016/S0370-2693(00)00573-6}}.

\bibitem{DeSimone:2007b}
A.~De~Simone, A.~Riotto, {Quantum Boltzmann Equations and Leptogenesis}, JCAP
  0708 (2007) 002.
\newblock \href {http://arxiv.org/abs/hep-ph/0703175}
  {\path{arXiv:hep-ph/0703175}}, \href
  {http://dx.doi.org/10.1088/1475-7516/2007/08/002}
  {\path{doi:10.1088/1475-7516/2007/08/002}}.

\bibitem{Anisimov:2010aq}
A.~Anisimov, W.~Buchmuller, M.~Drewes, S.~Mendizabal, {Leptogenesis from
  Quantum Interference in a Thermal Bath}, Phys. Rev. Lett. 104 (2010) 121102.
\newblock \href {http://arxiv.org/abs/1001.3856} {\path{arXiv:1001.3856}},
  \href {http://dx.doi.org/10.1103/PhysRevLett.104.121102}
  {\path{doi:10.1103/PhysRevLett.104.121102}}.

\bibitem{Anisimov:2010dk}
A.~Anisimov, W.~Buchmuller, M.~Drewes, S.~Mendizabal, {Quantum Leptogenesis I},
  Annals Phys. 326 (2011) 1998--2038.
\newblock \href {http://arxiv.org/abs/1012.5821} {\path{arXiv:1012.5821}},
  \href {http://dx.doi.org/10.1016/j.aop.2011.02.002}
  {\path{doi:10.1016/j.aop.2011.02.002}}.

\bibitem{Garny:2009rv}
M.~Garny, A.~Hohenegger, A.~Kartavtsev, M.~Lindner, {Systematic approach to
  leptogenesis in nonequilibrium QFT: vertex contribution to the CP-violating
  parameter}, Phys. Rev. D80 (2009) 125027.
\newblock \href {http://arxiv.org/abs/0909.1559} {\path{arXiv:0909.1559}},
  \href {http://dx.doi.org/10.1103/PhysRevD.80.125027}
  {\path{doi:10.1103/PhysRevD.80.125027}}.

\bibitem{Garny:2009qn}
M.~Garny, A.~Hohenegger, A.~Kartavtsev, M.~Lindner, {Systematic approach to
  leptogenesis in nonequilibrium QFT: self-energy contribution to the
  CP-violating parameter}, Phys. Rev. D81 (2010) 085027.
\newblock \href {http://arxiv.org/abs/0911.4122} {\path{arXiv:0911.4122}},
  \href {http://dx.doi.org/10.1103/PhysRevD.81.085027}
  {\path{doi:10.1103/PhysRevD.81.085027}}.

\bibitem{Cirigliano:2009yt}
V.~Cirigliano, C.~Lee, M.~J. Ramsey-Musolf, S.~Tulin, {Flavored Quantum
  Boltzmann Equations}, Phys. Rev. D81 (2010) 103503.
\newblock \href {http://arxiv.org/abs/0912.3523} {\path{arXiv:0912.3523}},
  \href {http://dx.doi.org/10.1103/PhysRevD.81.103503}
  {\path{doi:10.1103/PhysRevD.81.103503}}.

\bibitem{Beneke:2010dz}
M.~Beneke, B.~Garbrecht, C.~Fidler, M.~Herranen, P.~Schwaller, {Flavoured
  Leptogenesis in the CTP Formalism}, Nucl. Phys. B843 (2011) 177--212.
\newblock \href {http://arxiv.org/abs/1007.4783} {\path{arXiv:1007.4783}},
  \href {http://dx.doi.org/10.1016/j.nuclphysb.2010.10.001}
  {\path{doi:10.1016/j.nuclphysb.2010.10.001}}.

\bibitem{DeSimone:2007c}
A.~De~Simone, A.~Riotto, {On Resonant Leptogenesis}, JCAP 0708 (2007) 013.
\newblock \href {http://arxiv.org/abs/0705.2183} {\path{arXiv:0705.2183}},
  \href {http://dx.doi.org/10.1088/1475-7516/2007/08/013}
  {\path{doi:10.1088/1475-7516/2007/08/013}}.

\bibitem{Cirigliano:2008}
V.~Cirigliano, A.~De~Simone, G.~Isidori, I.~Masina, A.~Riotto, {Quantum
  Resonant Leptogenesis and Minimal Lepton Flavour Violation}, JCAP 0801 (2008)
  004.
\newblock \href {http://arxiv.org/abs/0711.0778} {\path{arXiv:0711.0778}},
  \href {http://dx.doi.org/10.1088/1475-7516/2008/01/004}
  {\path{doi:10.1088/1475-7516/2008/01/004}}.

\bibitem{Fong:2008b}
C.~S. Fong, M.~C. Gonzalez-Garcia, {On Quantum Effects in Soft Leptogenesis},
  JCAP 0808 (2008) 008.
\newblock \href {http://arxiv.org/abs/0806.3077} {\path{arXiv:0806.3077}},
  \href {http://dx.doi.org/10.1088/1475-7516/2008/08/008}
  {\path{doi:10.1088/1475-7516/2008/08/008}}.

\bibitem{Fujihara:2005}
T.~Fujihara, et~al., {Cosmological family asymmetry and CP violation}, Phys.
  Rev. D72 (2005) 016006.
\newblock \href {http://arxiv.org/abs/hep-ph/0505076}
  {\path{arXiv:hep-ph/0505076}}, \href
  {http://dx.doi.org/10.1103/PhysRevD.72.016006}
  {\path{doi:10.1103/PhysRevD.72.016006}}.

\bibitem{Vives:2009}
O.~Vives, {Flavoured leptogenesis: A successful thermal leptogenesis with N(1)
  mass below 10**8-GeV}, J. Phys. Conf. Ser. 171 (2009) 012076.
\newblock \href {http://dx.doi.org/10.1088/1742-6596/171/1/012076}
  {\path{doi:10.1088/1742-6596/171/1/012076}}.

\bibitem{Pascoli:2007a}
S.~Pascoli, S.~T. Petcov, A.~Riotto, {Connecting Low Energy Leptonic
  CP-violation to Leptogenesis}, Phys. Rev. D75 (2007) 083511.
\newblock \href {http://arxiv.org/abs/hep-ph/0609125}
  {\path{arXiv:hep-ph/0609125}}, \href
  {http://dx.doi.org/10.1103/PhysRevD.75.083511}
  {\path{doi:10.1103/PhysRevD.75.083511}}.

\bibitem{Pascoli:2007b}
S.~Pascoli, S.~T. Petcov, A.~Riotto, {Leptogenesis and low energy CP violation
  in neutrino physics}, Nucl. Phys. B774 (2007) 1--52.
\newblock \href {http://arxiv.org/abs/hep-ph/0611338}
  {\path{arXiv:hep-ph/0611338}}, \href
  {http://dx.doi.org/10.1016/j.nuclphysb.2007.02.019}
  {\path{doi:10.1016/j.nuclphysb.2007.02.019}}.

\bibitem{Antusch:2006}
S.~Antusch, S.~F. King, A.~Riotto, {Flavour-dependent leptogenesis with
  sequential dominance}, JCAP 0611 (2006) 011.
\newblock \href {http://arxiv.org/abs/hep-ph/0609038}
  {\path{arXiv:hep-ph/0609038}}.

\bibitem{Antusch:2007a}
S.~Antusch, A.~M. Teixeira, {Towards constraints on the SUSY seesaw from
  flavour- dependent leptogenesis}, JCAP 0702 (2007) 024.
\newblock \href {http://arxiv.org/abs/hep-ph/0611232}
  {\path{arXiv:hep-ph/0611232}}.

\bibitem{Branco:2007}
G.~C. Branco, R.~Gonzalez~Felipe, F.~R. Joaquim, {A new bridge between leptonic
  CP violation and leptogenesis}, Phys. Lett. B645 (2007) 432--436.
\newblock \href {http://arxiv.org/abs/hep-ph/0609297}
  {\path{arXiv:hep-ph/0609297}}, \href
  {http://dx.doi.org/10.1016/j.physletb.2006.12.060}
  {\path{doi:10.1016/j.physletb.2006.12.060}}.

\bibitem{Blanchet:2007a}
S.~Blanchet, P.~Di~Bari, {Flavor effects on leptogenesis predictions}, JCAP
  0703 (2007) 018.
\newblock \href {http://arxiv.org/abs/hep-ph/0607330}
  {\path{arXiv:hep-ph/0607330}}.

\bibitem{Blanchet:2007b}
S.~Blanchet, P.~Di~Bari, G.~G. Raffelt, {Quantum Zeno effect and the impact of
  flavor in leptogenesis}, JCAP 0703 (2007) 012.
\newblock \href {http://arxiv.org/abs/hep-ph/0611337}
  {\path{arXiv:hep-ph/0611337}}.

\bibitem{Campbell:1992jd}
B.~A. Campbell, S.~Davidson, J.~R. Ellis, K.~A. Olive, {On the baryon, lepton
  flavor and right-handed electron asymmetries of the universe}, Phys. Lett.
  B297 (1992) 118--124.
\newblock \href {http://arxiv.org/abs/hep-ph/9302221}
  {\path{arXiv:hep-ph/9302221}}, \href
  {http://dx.doi.org/10.1016/0370-2693(92)91079-O}
  {\path{doi:10.1016/0370-2693(92)91079-O}}.

\bibitem{Cline:1993bd}
J.~M. Cline, K.~Kainulainen, K.~A. Olive, {Protecting the primordial baryon
  asymmetry from erasure by sphalerons}, Phys. Rev. D49 (1994) 6394--6409.
\newblock \href {http://arxiv.org/abs/hep-ph/9401208}
  {\path{arXiv:hep-ph/9401208}}, \href
  {http://dx.doi.org/10.1103/PhysRevD.49.6394}
  {\path{doi:10.1103/PhysRevD.49.6394}}.

\bibitem{Fong:2008a}
C.~S. Fong, M.~C. Gonzalez-Garcia, {Flavoured Soft Leptogenesis}, JHEP 06
  (2008) 076.
\newblock \href {http://arxiv.org/abs/hep-ph/0804.4471}
  {\path{arXiv:hep-ph/0804.4471}}, \href
  {http://dx.doi.org/10.1088/1126-6708/2008/06/076}
  {\path{doi:10.1088/1126-6708/2008/06/076}}.

\bibitem{Fong:2010zu}
C.~S. Fong, M.~Gonzalez-Garcia, E.~Nardi, J.~Racker, {Flavoured soft
  leptogenesis and natural values of the B term}, JHEP 1007 (2010) 001.
\newblock \href {http://arxiv.org/abs/1004.5125} {\path{arXiv:1004.5125}},
  \href {http://dx.doi.org/10.1007/JHEP07(2010)001}
  {\path{doi:10.1007/JHEP07(2010)001}}.

\bibitem{Aristizabal:2009a}
D.~Aristizabal~Sierra, M.~Losada, E.~Nardi, {Lepton Flavor Equilibration and
  Leptogenesis}, JCAP 0912 (2009) 015.
\newblock \href {http://arxiv.org/abs/hep-ph/0905.0662}
  {\path{arXiv:hep-ph/0905.0662}}, \href
  {http://dx.doi.org/10.1088/1475-7516/2009/12/015}
  {\path{doi:10.1088/1475-7516/2009/12/015}}.

\bibitem{DeSimone:2007a}
A.~De~Simone, A.~Riotto, {On the impact of flavour oscillations in
  leptogenesis}, JCAP 0702 (2007) 005.
\newblock \href {http://arxiv.org/abs/hep-ph/0611357}
  {\path{arXiv:hep-ph/0611357}}.

\bibitem{Davidson:2008pf}
S.~Davidson, J.~Garayoa, F.~Palorini, N.~Rius, {CP Violation in the SUSY
  Seesaw: Leptogenesis and Low Energy}, JHEP 09 (2008) 053.
\newblock \href {http://arxiv.org/abs/0806.2832} {\path{arXiv:0806.2832}},
  \href {http://dx.doi.org/10.1088/1126-6708/2008/09/053}
  {\path{doi:10.1088/1126-6708/2008/09/053}}.

\bibitem{Ibanez:1992aj}
L.~E. Ibanez, F.~Quevedo, {Supersymmetry protects the primordial baryon
  asymmetry}, Phys. Lett. B283 (1992) 261--269.
\newblock \href {http://arxiv.org/abs/hep-ph/9204205}
  {\path{arXiv:hep-ph/9204205}}, \href
  {http://dx.doi.org/10.1016/0370-2693(92)90017-X}
  {\path{doi:10.1016/0370-2693(92)90017-X}}.

\bibitem{Cline:1993vv}
J.~M. Cline, K.~Kainulainen, K.~A. Olive, {On the erasure and regeneration of
  the primordial baryon asymmetry by sphalerons}, Phys. Rev. Lett. 71 (1993)
  2372--2375.
\newblock \href {http://arxiv.org/abs/hep-ph/9304321}
  {\path{arXiv:hep-ph/9304321}}, \href
  {http://dx.doi.org/10.1103/PhysRevLett.71.2372}
  {\path{doi:10.1103/PhysRevLett.71.2372}}.

\bibitem{Keung:1983nq}
W.-Y. Keung, L.~Littenberg, {TEST OF SUPERSYMMETRY IN e- e- COLLISION}, Phys.
  Rev. D28 (1983) 1067.
\newblock \href {http://dx.doi.org/10.1103/PhysRevD.28.1067}
  {\path{doi:10.1103/PhysRevD.28.1067}}.

\bibitem{Kashti:2004vj}
T.~Kashti, {Phenomenological consequences of soft leptogenesis}, Phys. Rev. D71
  (2005) 013008.
\newblock \href {http://arxiv.org/abs/hep-ph/0410319}
  {\path{arXiv:hep-ph/0410319}}, \href
  {http://dx.doi.org/10.1103/PhysRevD.71.013008}
  {\path{doi:10.1103/PhysRevD.71.013008}}.

\bibitem{D'Ambrosio:2004fz}
G.~D'Ambrosio, T.~Hambye, A.~Hektor, M.~Raidal, A.~Rossi, {Leptogenesis in the
  minimal supersymmetric triplet seesaw model}, Phys. Lett. B604 (2004)
  199--206.
\newblock \href {http://arxiv.org/abs/hep-ph/0407312}
  {\path{arXiv:hep-ph/0407312}}, \href
  {http://dx.doi.org/10.1016/j.physletb.2004.10.056}
  {\path{doi:10.1016/j.physletb.2004.10.056}}.

\bibitem{Chun:2005ms}
E.~J. Chun, S.~Scopel, {Soft leptogenesis in Higgs triplet model}, Phys. Lett.
  B636 (2006) 278--285.
\newblock \href {http://arxiv.org/abs/hep-ph/0510170}
  {\path{arXiv:hep-ph/0510170}}, \href
  {http://dx.doi.org/10.1016/j.physletb.2006.03.061}
  {\path{doi:10.1016/j.physletb.2006.03.061}}.

\bibitem{Chun:2003ej}
E.~J. Chun, K.~Y. Lee, S.~C. Park, {Testing Higgs triplet model and neutrino
  mass patterns}, Phys. Lett. B566 (2003) 142--151.
\newblock \href {http://arxiv.org/abs/hep-ph/0304069}
  {\path{arXiv:hep-ph/0304069}}, \href
  {http://dx.doi.org/10.1016/S0370-2693(03)00770-6}
  {\path{doi:10.1016/S0370-2693(03)00770-6}}.

\bibitem{Akeroyd:2009nu}
A.~G. Akeroyd, M.~Aoki, H.~Sugiyama, {Lepton Flavour Violating Decays tau to
  lll and mu to e gamma in the Higgs Triplet Model}, Phys. Rev. D79 (2009)
  113010.
\newblock \href {http://arxiv.org/abs/0904.3640} {\path{arXiv:0904.3640}},
  \href {http://dx.doi.org/10.1103/PhysRevD.79.113010}
  {\path{doi:10.1103/PhysRevD.79.113010}}.

\bibitem{Akeroyd:2005gt}
A.~G. Akeroyd, M.~Aoki, {Single and pair production of doubly charged Higgs
  bosons at hadron colliders}, Phys. Rev. D72 (2005) 035011.
\newblock \href {http://arxiv.org/abs/hep-ph/0506176}
  {\path{arXiv:hep-ph/0506176}}, \href
  {http://dx.doi.org/10.1103/PhysRevD.72.035011}
  {\path{doi:10.1103/PhysRevD.72.035011}}.

\bibitem{Akeroyd:2011ir}
A.~G. Akeroyd, S.~Moretti, {Production of doubly charged scalars from the decay
  of a heavy SM-like Higgs boson in the Higgs Triplet Model}\href
  {http://arxiv.org/abs/1106.3427} {\path{arXiv:1106.3427}}.

\bibitem{Chun:2005tg}
E.~J. Chun, {TeV leptogenesis in Z-prime models and its collider probe}, Phys.
  Rev. D72 (2005) 095010.
\newblock \href {http://arxiv.org/abs/hep-ph/0508050}
  {\path{arXiv:hep-ph/0508050}}, \href
  {http://dx.doi.org/10.1103/PhysRevD.72.095010}
  {\path{doi:10.1103/PhysRevD.72.095010}}.

\bibitem{Garayoa:2006xs}
J.~Garayoa, M.~C. Gonzalez-Garcia, N.~Rius, {Soft leptogenesis in the inverse
  seesaw model}, JHEP 02 (2007) 021.
\newblock \href {http://arxiv.org/abs/hep-ph/0611311}
  {\path{arXiv:hep-ph/0611311}}.

\bibitem{Medina:2006hi}
A.~D. Medina, C.~E.~M. Wagner, {Soft leptogenesis in warped extra dimensions},
  JHEP 12 (2006) 037.
\newblock \href {http://arxiv.org/abs/hep-ph/0609052}
  {\path{arXiv:hep-ph/0609052}}.

\bibitem{Allahverdi:2004ix}
R.~Allahverdi, M.~Drees, {Leptogenesis from a sneutrino condensate revisited},
  Phys. Rev. D69 (2004) 103522.
\newblock \href {http://arxiv.org/abs/hep-ph/0401054}
  {\path{arXiv:hep-ph/0401054}}, \href
  {http://dx.doi.org/10.1103/PhysRevD.69.103522}
  {\path{doi:10.1103/PhysRevD.69.103522}}.

\bibitem{Boubekeur:2004ez}
L.~Boubekeur, T.~Hambye, G.~Senjanovic, {Low-scale leptogenesis and soft
  supersymmetry breaking}, PHys. Rev. Lett. 93 (2004) 111601.
\newblock \href {http://arxiv.org/abs/hep-ph/0404038}
  {\path{arXiv:hep-ph/0404038}}, \href
  {http://dx.doi.org/10.1103/PhysRevLett.93.111601}
  {\path{doi:10.1103/PhysRevLett.93.111601}}.

\bibitem{Ellis:2005uk}
J.~R. Ellis, S.~K. Kang, {Sneutrino leptogenesis at the electroweak scale}\href
  {http://arxiv.org/abs/hep-ph/0505162} {\path{arXiv:hep-ph/0505162}}.

\bibitem{basboll:2006}
A.~Basboll, S.~Hannestad, {Decay of heavy Majorana neutrinos using the full
  Boltzmann equation including its implications for leptogenesis}, JCAP 0701
  (2007) 003.
\newblock \href {http://arxiv.org/abs/hep-ph/0609025}
  {\path{arXiv:hep-ph/0609025}}.

\bibitem{HahnWoernle:2009}
F.~Hahn-Woernle, M.~Plumacher, Y.~Y.~Y. Wong, {Full Boltzmann equations for
  leptogenesis including scattering}, JCAP 0908 (2009) 028.
\newblock \href {http://arxiv.org/abs/0907.0205} {\path{arXiv:0907.0205}},
  \href {http://dx.doi.org/10.1088/1475-7516/2009/08/028}
  {\path{doi:10.1088/1475-7516/2009/08/028}}.

\bibitem{Kolb:1980}
E.~W. Kolb, S.~Wolfram, {Baryon Number Generation in the Early Universe}, Nucl.
  Phys. B172 (1980) 224.
\newblock \href {http://dx.doi.org/10.1016/0550-3213(80)90167-4}
  {\path{doi:10.1016/0550-3213(80)90167-4}}.

\bibitem{Fry:1980}
J.~N. Fry, K.~A. Olive, M.~S. Turner, {Evolution of Cosmological Baryon
  Asymmetries}, Phys. Rev. D22 (1980) 2953.
\newblock \href {http://dx.doi.org/10.1103/PhysRevD.22.2953}
  {\path{doi:10.1103/PhysRevD.22.2953}}.

\end{thebibliography}
